\pdfoutput=1

\documentclass[bibchap]{cernmono}

\usepackage{booktabs}
\usepackage{array}
\usepackage{placeins}
\usepackage{afterpage}
\usepackage[utf8]{inputenc}
\usepackage{graphicx,amsmath,amsfonts,amscd,amsthm,amssymb,eucal,psfrag,color,subfig,url}
\usepackage{siunitx}
\usepackage{comment}
\DeclareSIUnit{\CHF}{CHF}
\DeclareSIUnit{\kCHF}{kCHF}
\DeclareSIUnit{\MCHF}{MCHF}
\DeclareSIUnit{\kEUR}{kEUR}
\DeclareSIUnit{\MEUR}{MEUR}
\DeclareSIUnit{\PY}{PY}
\usepackage{enumitem}
\usepackage{graphicx} %
\usepackage{rotating} %
\usepackage{tablefootnote}
\usepackage{multicol}
\usepackage{multirow}
\usepackage{makecell}
\usepackage{pdflscape}
\usepackage{longtable}

\def\permille{\ensuremath{{}^\text{o}\mkern-5mu/\mkern-3mu_\text{oo}}}

\usepackage{xcolor}
\newif\ifreviewnotes \reviewnotesfalse%
\newcommand{\chapterauthors}[1]{{\noindent\raggedright\itshape\hyphenpenalty=10000\exhyphenpenalty=10000 #1\par}\bigskip}
\makeatletter

\newcommand{\l@techqline}[2]{\@dottedtocline{1}{0em}{2.6em}{#1}{#2}}
\newcommand{\listoftechq}{\ifreviewnotes
  \chapter*{Open points for technical input}
  \markboth{Open points for technical input}{Open points for technical input}
  \noindent\textit{Review draft only. Questions for technical input are flagged in
  orange at the corresponding place in the text; analyses of existing data that
  are still to be included (IN) are flagged in blue.
  Points settled in the review of version~1.3 have been removed; the remaining ones, including X5 carried over from version~1.3, await input.}\par\bigskip
  \@starttoc{tq}\fi}
\makeatother

\usepackage[hidelinks,bookmarks=true]{hyperref}

\begin{document}

\title{A European high-capacity 100~MeV-per-nucleon heavy-ion facility for electronics irradiation
based at CERN: Conceptual Design Report}

\author{E.C.~Cort\'es Garc\'ia, %
R.~Garc\'ia Al\'ia,
M.A.~Fraser, M.~S{\l}upecki, A.~Waets and M.~Widorski}

\institute{CERN, CH-1211 Geneva 23, Switzerland\\[2ex]
\textit{Version~1.8 --- 25~November~2026}}

\begin{abstract}
This proposal presents a conceptual design for a European high-energy heavy-ion facility dedicated to electronics radiation-effects testing, directly addressing the critical challenges of global beam time shortages and energy limitations in existing facilities. Aligned with the HORIZON-CL4-2026-SPACE-03-85 call under Horizon Europe’s Cluster 4 (Digital, Industry, and Space), the facility will enhance European sovereignty in space technology by providing approximately 1,900 additional hours of high-energy heavy-ion beam time annually, an increase of about 50\% in the total heavy-ion beam time currently available in Europe for single-event-effect testing, and by extending the accessible ion energy range up to 100 MeV/nucleon, significantly improving penetration capabilities for state-of-the-art electronics testing.

The facility will leverage CERN's ion injector chain up to the Low Energy Ion Ring (LEIR) synchrotron, adding a second ion source, a dedicated slow-extraction system, a transfer line and a user experimental area.

The project timeline includes a pre-technical-design (pre-TDR) phase and a technical design phase leading to a Technical Design Report (TDR), followed by infrastructure and hardware production, installation, commissioning, and user operations. Funding is envisaged through a collaborative framework involving CERN and the European Union, with the aim of ensuring sustainable deployment and long-term operational success.
\end{abstract}


\maketitle

\chapter*{Credits and acknowledgements}
\markboth{Credits and acknowledgements}{Credits and acknowledgements}

This edition of the Conceptual Design Report has been compiled and edited by E.~C.~Cort\'es Garc\'ia.

\vspace{1ex}
\noindent R.~Garc\'ia Al\'ia (SY-STI) is at the origin of the HEARTS@LEIR proposal, and M.~A.~Fraser (SY-ABT) supervised the development of its technical side. Both have accompanied the complete study, as its supervisors and as contributors to the whole of the work presented in this report. A.~Waets (SY-STI) accompanied the study, in particular as an editor for Chapters~\ref{sec:BeamSpecs}, \ref{sec:Operations}, \ref{sec:expArea} and \ref{sec:RiskAssessment} and for Section~\ref{sec:SafetyAccess}. 

\vspace{1ex}
\noindent Chapters~\ref{sec:IonSources}, \ref{sec:MachineTests} and~\ref{sec:RadiationProtection} carry their own author lists, given below the chapter titles; the other chapters are by the authors of the report. The chapter authors are:

\begin{itemize}[nosep]
    \item Chapter~\ref{sec:IonSources}, ion sources and beam preparation for injection into LEIR, and the scope and objectives of the Ion Complex Upgrade project --- M.~S{\l}upecki (BE-ABP) and E.~C.~Cort\'es Garc\'ia (SY-ABT).
    \item Chapter~\ref{sec:MachineTests}, machine tests at LEIR, including the slow-extraction test --- E.~C.~Cort\'es Garc\'ia and M.~A.~Fraser (SY-ABT), T.~Argyropoulos and O.~Hans (BE-OP), M.-A.~Jebramcik and M.~S{\l}upecki (BE-ABP), G.~Le~Godec, S.~Joffe and C.~Mutin (SY-EPC), D.~Bodart (TE-MSC), S.~Deleval (EN-CV), E.~Effinger, J.~C.~Esteban Felipe, S.~Morales Vigo and B.~Salvachua Ferrando (SY-BI), R.~Garc\'ia Al\'ia, D.~Prelipcean and A.~Waets (SY-STI), J.-F.~Gruber and M.~Widorski (HSE-RP).
    \item Chapter~\ref{sec:RadiationProtection}, radiation safety and radiation protection --- M.~Widorski (HSE-RP).
\end{itemize}

\vspace{1ex}
\noindent The conceptual design presented here rests on input, data and advice from experts across the Accelerators and Technology sector and other CERN departments and units. The following colleagues contributed to the study or to the review meetings from which this report is derived (not exhaustive nor complete list):

\vspace{1ex}
{\raggedright\small
\begin{description}[leftmargin=2.6cm,style=nextline,font=\normalfont\bfseries,itemsep=0.2ex]
\item[BE-ABP] R.~Alemany Fern\'andez, F.~Asvesta, G.~Bellodi, H.~Deveci, D.~Gamba, M.-A.~Jebramcik, D.~K\"uchler, R.~Scrivens, M.~S{\l}upecki, R.~Tom\'as Garc\'ia
\item[BE-CEM] S.~Danzeca, A.~Masi
\item[BE-EA] J.~Bernhard, N.~Charitonidis, S.~Evrard, M.~Lazzaroni
\item[BE-OP] T.~Argyropoulos, O.~Hans, B.~Mikulec, A.~Huschauer
\item[DG-DI] M.~Dissing
\item[EN-AA] T.~Ladzinski
\item[EN-ACE] D.~del~\'Alamo, J.-P.~Tock
\item[EN-CV] S.~Deleval, G.~Petrika
\item[EN-EL] M.~Parodi, J.~De Voght
\item[EN-THE] C.~Bertone, I.~R\"uhl
\item[EP-ESE] F.~Faccio
\item[HSE-OHS] S.~La~Mendola, S.~Marsh
\item[HSE-RP] J.-F.~Gruber, M.~Widorski
\item[IPT-KT] E.~Chesta, T.~Rimbot
\item[SR-SE] C.~Levointurier-Vajda, P.~F.~L\'opez, S.~Stavrev
\item[SY-ABT] B.~Balhan, C.~Baud, J.~Borburgh, M.~A.~Fraser, F.~Lackner, W.~Bartmann and the BTP section
\item[SY-BI] B.~Salvachua Ferrando, E.~Effinger, J.~C.~Esteban Felipe, G.~Khatri, T.~Lef\`evre, T.~Levens, S.~Morales Vigo, F.~Roncarolo, A.~Rossi
\item[SY-EPC] S.~Joffe, G.~Le~Godec, C.~Mutin
\item[SY-RF] S.~Albright, H.~Damerau, B.~Woolley
\item[SY-STI] A.-P.~Bernardes, M.~Calviani, C.~Duchemin, N.~Emri\v{s}kov\'a, L.~Esposito, R.~Garc\'ia Al\'ia, A.~Lechner, D.~Prelipcean, A.~Waets
\item[TE-CRG] S.~Blanchard
\item[TE-MSC] D.~Bodart
\item[TE-VSC] J.~A.~Ferreira Somoza, A.~Sinturel
\end{description}
}

\vspace{1ex}
\noindent The support of the ATS Management Board (ATSMB) and of the IEFC committee is acknowledged, as is the CERN Knowledge Transfer group for partially funding the studies presented here. The HEARTS project, funded by the European Union under GA No 101082402 through the Space Work Programme of the European Commission, has enabled the development of the HEARTS@CERN facility in the PS; the work and findings of that project have largely motivated the preparation of this report, which is, however, not itself part of the HEARTS project.

\vspace{1ex}
\noindent E.~C.~Cort\'es Garc\'ia used AI tools (Claude Opus 5.5 by Anthropic) to edit the present document. Chapter~\ref{sec:MachineTests} was aided by the agentic skill set developed at CERN\footnote{\url{https://gitlab.cern.ch/cern-agent-skills}}. The aforementioned tools were also used to revise the manuscript. The authors take full responsibility for the content.

\vspace{2ex}
\noindent\textbf{Publication status.} This document is intended for publication as a CERN Yellow Report and will be submitted to the CERN Reports Editorial Board. The present version is a preprint: it does not carry a CERN report number, an ISBN/ISSN or a DOI, which are allocated by the CERN Scientific Information Service only after final approval. The version of record will be published in open access on the CERN Document Server.

\vspace{1ex}
\noindent\textbf{Copyright.} \copyright{} \the\year{} CERN for the benefit of the HEARTS@LEIR study. Published under the Creative Commons Attribution 4.0 International licence (CC-BY-4.0), \url{http://creativecommons.org/licenses/by/4.0/}.

\newpage

\tableofcontents

\chapter*{List of acronyms}
\addcontentsline{toc}{chapter}{List of acronyms}
\markboth{List of acronyms}{List of acronyms}

\begingroup
\small
\setlength{\LTpre}{0pt}
\begin{longtable}{@{}>{\raggedright\arraybackslash}p{2.6cm}>{\raggedright\arraybackslash}p{12.4cm}@{}}
\toprule
\textbf{Acronym} & \textbf{Meaning} \\
\midrule
\endfirsthead
\toprule
\textbf{Acronym} & \textbf{Meaning} \\
\midrule
\endhead
\bottomrule
\endfoot
ASIC    & Application-specific integrated circuit \\
ATS     & Accelerators and Technology Sector (CERN) \\
ATSMB   & ATS Management Board \\
BBQ     & Base-band tune measurement system \\
BCT     & Beam current transformer \\
BLM     & Beam loss monitor \\
BNL     & Brookhaven National Laboratory \\
BPM     & Beam position monitor \\
BSG     & Beam secondary-emission grid \\
BTV     & Beam television (screen monitor) \\
CCC     & CERN Control Centre \\
CDR     & Conceptual Design Report \\
CEA     & Commissariat \`a l'\'energie atomique et aux \'energies alternatives \\
CLV     & Constant linear velocity \\
CNAO    & Centro Nazionale di Adroterapia Oncologica \\
DUT     & Device under test \\
ECR     & Electron cyclotron resonance \\
EIS     & Important safety element (\textit{\'El\'ement important pour la s\'ecurit\'e}) \\
ES      & Electrostatic septum \\
EU      & European Union \\
FTE     & Full-time equivalent \\
FWHM    & Full width at half maximum \\
GA      & Grant agreement \\
GCR     & Galactic cosmic rays \\
GSI     & GSI Helmholtzzentrum f\"ur Schwerionenforschung \\
GTS-LHC & Grenoble Test Source, LHC version (ECR ion source of Linac~3) \\
GUI     & Graphical user interface \\
HEARTS  & High-Energy Accelerators for Radiation Testing and Shielding \\
HEP     & High-energy physics \\
HL-LHC  & High-Luminosity LHC \\
HV      & High voltage \\
HVAC    & Heating, ventilation and air-conditioning \\
ICU     & Ion Complex Upgrade \\
IEFC    & Injectors and Experimental Facilities Committee \\
IPM     & Ionisation profile monitor \\
LEBT    & Low-energy beam transport \\
LEIR    & Low Energy Ion Ring \\
LET     & Linear energy transfer \\
LHC     & Large Hadron Collider \\
LOD     & Level of detail (of a 3D integration model) \\
LS3     & Long Shutdown~3 \\
MAD-X   & Methodical Accelerator Design, version X \\
MD      & Machine development \\
MS      & Magnetic septum; in Chapter~\ref{sec:ResourceRestructuring}, milestone \\
MWPC    & Multi-wire proportional chamber \\
NEG     & Non-evaporable getter \\
NORMA   & CERN database of normal-conducting magnets \\
NSRL    & NASA Space Radiation Laboratory \\
PPM     & Pulse-to-pulse modulation \\
PS      & Proton Synchrotron \\
PY      & Person-year \\
RDT     & Resonance driving term \\
RF      & Radio frequency \\
RF-KO   & Radio-frequency knock-out \\
RFQ     & Radio-frequency quadrupole \\
RMS     & Root mean square \\
RP      & Radiation protection \\
SCADA   & Supervisory control and data acquisition \\
SEE     & Single-event effect \\
SPS     & Super Proton Synchrotron \\
SS      & Straight section \\
TDR     & Technical Design Report \\
TFB     & Transverse feedback \\
TL      & Transfer line \\
TRL     & Technology readiness level \\
TTL     & Transistor--transistor logic \\
VHE     & Very high energy \\
WP      & Work package \\
XSEC    & Secondary-emission chamber \\
\end{longtable}

\vspace{2ex}
\noindent\textbf{CERN departments and groups}

\begin{longtable}{@{}>{\raggedright\arraybackslash}p{2.6cm}>{\raggedright\arraybackslash}p{12.4cm}@{}}
\toprule
\textbf{Group} & \textbf{Name} \\
\midrule
\endfirsthead
\toprule
\textbf{Group} & \textbf{Name} \\
\midrule
\endhead
\bottomrule
\endfoot
BE      & Beams Department \\
BE-ABP  & Accelerator and Beam Physics \\
BE-CEM  & Controls, Electronics and Mechatronics \\
BE-DSO  & Departmental Safety Officer, BE Department \\
BE-EA   & Experimental Areas \\
BE-OP   & Operations \\
EN      & Engineering Department \\
EN-AA   & Access and Alarms \\
EN-ACE  & Accelerator Coordination and Engineering \\
EN-CV   & Cooling and Ventilation \\
EN-EL   & Electrical Engineering \\
EP      & Experimental Physics Department \\
EP-ESE  & Electronic Systems for Experiments \\
HSE     & Occupational Health \& Safety and Environmental Protection Unit \\
HSE-OHS & Occupational Health and Safety \\
HSE-RP  & Radiation Protection \\
IPT-KT  & Knowledge Transfer \\
SY      & Accelerator Systems Department \\
SY-ABT  & Accelerator Beam Transfer \\
SY-BI   & Beam Instrumentation \\
SY-EPC  & Electrical Power Converters \\
SY-RF   & Radio Frequency \\
SY-STI  & Sources, Targets and Interactions \\
TE      & Technology Department \\
TE-CRG  & Cryogenics \\
TE-MSC  & Magnets, Superconductors and Cryostats \\
TE-VSC  & Vacuum, Surfaces and Coatings \\
\end{longtable}
\noindent Where the context is unambiguous, groups are also referred to without the department prefix (e.g.\ ABT, BI, VSC).
\endgroup

\listoftechq

\newpage

\chapter{Executive summary}
\label{sec:ExecSummary}
With the rapidly increasing number of satellites and spacecraft in orbit, combined with the growing complexity of the electronic devices and systems embedded in them, testing against radiation effects, particularly Single Event Effects (SEE), is becoming ever more critical. Such testing relies on access to accelerator infrastructures primarily devoted to medical, radio-biological, or nuclear and high-energy physics research. However, demand for beam time now far exceeds the available supply, and the gap is widening rapidly.

This shortage is not confined to the space sector. The high-energy physics (HEP) community at CERN and its collaborating institutions represent a substantial and largely underleveraged user base for heavy-ion beam testing: custom-designed ASICs deployed in large numbers across HEP experiments must be qualified against SEE at multiple stages of development, from early prototypes through to production readiness. In practice, however, this need has gone largely unmet. The EP-ESE group alone has used only around 40 hours of heavy-ion beam time per year in recent years, a figure driven less by need than by the constraints of current access: travel to external facilities, scheduling limitations, and hourly beam costs.

Existing facilities also fall short on a second front. Conventional cyclotron-based test facilities are limited to energies of 10–20 MeV/nucleon, restricting silicon penetration to depths that are insufficient for testing complex and 3D package structures or at board level, back-side testing of modern devices being one example; a limitation shared by both the space and HEP communities.

We propose to address both challenges, (i) the shortage of beam time and (ii) the energy limitation, through a European high-energy heavy-ion facility dedicated to electronics radiation-effects testing, built by adapting and upgrading CERN's ion injector chain up to the Low Energy Ion Ring (LEIR) heavy-ion synchrotron (Chapters~\ref{sec:IonSources}--\ref{sec:Beamline2HEARTS}). The facility will provide approximately 1,900 additional hours of high-energy heavy-ion beam time per year (Chapter~\ref{sec:Operations}), an increase of about 50\% in the total heavy-ion beam time currently available in Europe for SEE testing, while extending the accessible ion energy up to 100 MeV/nucleon (Chapter~\ref{sec:BeamSpecs}). This energy reach is five times that of conventional cyclotron facilities and will significantly enhance penetration and enable testing of state-of-the-art technologies.

Machine tests carried out with beam in LEIR in 2025 and 2026 have confirmed, within the accuracy reached so far, the main assumptions that the design makes on the existing machine (Chapter~\ref{sec:MachineTests}): a cooled Pb$^{54+}$ beam of more than $10^{9}$ ions was held on a flat top at the maximum energy, on a cycle within the present thermal envelope of the main magnet circuits, and a first slow-extraction test removed this beam in a controlled way with an RF-KO excitation, the losses being localised in the future extraction straight section.
\pagebreak
\section*{Project timeline and budget}

The construction of the facility is estimated at \SI{18.3}{\MCHF} of funded cost (\SI{13.9}{\MCHF} of materials and equipment and \SI{4.4}{\MCHF} of fixed-term personnel) together with \SI{32.5}{\PY} of CERN staff effort. This figure excludes the signal and power cabling of the new beam line and the cooling of the associated power converters, which could not yet be estimated (Section~\ref{sec:ResourceExclusions}). In line with common practice for infrastructure projects at conceptual design stage, these figures carry an uncertainty of the order of $\pm$30--50\,\%. The funding scheme for construction is still to be defined, with contributions from CERN and the European Commission to be explored with high priority. A detailed breakdown of the resource estimate is given in Chapter~\ref{sec:ResourceEstimate}.

The decision sought at this stage concerns only the first phase. 
Advancing from this conceptual design to a Technical Design Report (TDR) requires \SI{2062}{\kCHF} of funded resources, dominated by fixed-term personnel, together with \SI{9.9}{\PY} of CERN staff effort distributed across eleven groups. This phase is now part of the HEARTS++ proposal submitted to the European Commission under the HORIZON-CL4-2026-SPACE-03-85 call of Horizon Europe.
Subject to the grant, the project foresees 54~months until the delivery of the TDR, tentatively until the end of 2031: a pre-TDR phase in the first 30~months, followed by two years devoted to the full technical design of HEARTS@LEIR.
The decision to construct the facility would be taken after completion of the technical design phase.

The CDR study assumed the following tentative schedule:
\begin{itemize}[nosep]
    \item 2027--2029: pre-technical design and preparation;
    \item 2030--2031: technical design;
    \item 2032--2035: procurement, fabrication and testing of infrastructure and hardware;
    \item 2034--2035: installation;
    \item 2035: beam commissioning and validation;
    \item 2036: start of user operations.
\end{itemize}

The project timeline complies with Long Shutdown~3 (LS3) activities and the HL-LHC upgrade it carries, the top priority of the Organization. Within HEARTS++, pre-TDR studies would run until the end of 2029, and the technical design of HEARTS@LEIR follows in 2030--2031; the subsequent phases are tentative and will be confirmed once the funding decision has been made.

\chapter{Introduction}
Access to high-energy heavy-ion beams (>50 MeV/nucleon) is extremely limited worldwide, and particularly so in Europe. Radiation testing of advanced electronic components requires such beams to ensure sufficient penetration of the Device Under Test (DUT) while maintaining a high Linear Energy Transfer (LET), which determines the reliability level of Single Event Effect (SEE) evaluations. Current cyclotron-based facilities, however, operate at significantly lower energies of 10–20 MeV/nucleon, corresponding to silicon penetration depths of only 100--200~\si{\micro\meter}. This forces device de-lidding and potentially unrealistic exposure of the semiconductor die, compromising the representativeness of the test. The limitation is increasingly critical for next-generation, highly integrated commercial electronics used in space systems, where through-package testing of modern 3D and system-on-chip topologies is often the only physically meaningful test configuration.

This shortage is not confined to the space sector. Custom-designed ASICs are deployed in large numbers across high-energy physics (HEP) experiments and must be qualified against SEE at multiple stages of development, from early prototypes through to production readiness. Yet this need has gone largely unmet: CERN's EP-ESE group alone has used only around 40 hours of heavy-ion beam time per year in recent years, a figure driven less by actual need than by the constraints of current access—travel to external facilities, scheduling limitations, and hourly beam costs.

At CERN, existing accelerator infrastructure could enable such high-energy testing if appropriately upgraded. The present report assesses the required accelerator and experimental area modifications, fully integrating radiation-effects testing needs from the outset, with the European Union envisaged as the main stakeholder, to ensure alignment with European space-qualification priorities, and with CERN's EP-ESE-ME section covering electronics for HEP systems. The activity builds on the success of the HEARTS project and of the HEARTS@CERN facility in the PS East Area~\cite{hearts_project} (referred to as HEARTS@PS in the HEARTS++ proposal), currently at a Technology Readiness Level (TRL) of 6--7, which would be raised to TRL~8 with the dedicated beam line and facility proposed here.

Concretely, we propose to establish a uniquely European high-energy heavy-ion testing capability by upgrading CERN's existing accelerator infrastructure up to the Low Energy Ion Ring (LEIR) synchrotron. By exploiting an injector-chain sector that is currently idle during significant parts of the CERN operational year, the project offers an innovative and cost-efficient path to give Europe continuous access to representative radiation environments for space electronics, with selectable LET values spanning low-LET (0.4 to 16~MeV\,cm$^2$\,mg$^{-1}$), high-LET (up to 75~MeV\,cm$^2$\,mg$^{-1}$), and, for specialized tests, very high-LET regimes ($>$75~MeV\,cm$^2$\,mg$^{-1}$).

Achieving this requires advances beyond simply expanding the beam energy range. The study proposes advanced resonant slow-extraction methods, including custom Radio-Frequency Knock-Out (RF-KO) schemes, to enable fine, programmable control of ion flux and energy across multiple species. This is complemented by the integration of a new ion source enabling rapid switching between species, a capability already planned within the Ion Complex Upgrade (ICU) project. Combined, these innovations will establish a facility unparalleled in versatility and realism for SEE testing, one that not only strengthens European technological sovereignty but also ensures seamless alignment with the evolving qualification frameworks of key stakeholders, positioning it as a cornerstone for future radiation-testing needs.

Once operational, the facility will provide irradiation services to institutional and commercial users in the space sector, including spacecraft manufacturers, satellite operators, and electronic component suppliers, alongside access for collaborating and internal CERN users. By enabling qualification under representative high-energy heavy-ion conditions, it will fill a critical gap in Europe's test capabilities and reduce dependence on non-European facilities. Notably, it will offer the strategic advantage of enabling qualification of state-of-the-art 3D and system-on-chip assemblies, unlocking the potential of advanced consumer, industrial, and automotive electronics for space applications.

Beyond space, the same infrastructure will serve internal users and industries relying on radiation-hardened electronics, such as avionics, nuclear energy, high-energy physics, and medical technology. By cultivating this diverse user ecosystem, the facility will drive industrial collaboration and knowledge transfer, accelerating the development of resilient, high-reliability electronic systems. This will bolster Europe's technological sovereignty and competitiveness in strategic sectors, with CERN at the heart of this innovation network.

\section{HEARTS project (2023--2026)}

Currently, the HEARTS (\textit{High-Energy Accelerators for Radiation Testing and Shielding}) project provides access to very high-energy (VHE) heavy-ion beams ($E >100$ MeV/nucleon) that can mimic the effects of Galactic Cosmic Rays (GCR).\
This capability is crucial for radiation testing of advanced microelectronics, shielding and radiobiology research.\
HEARTS is a European Union-funded initiative coordinated by CERN.\
Launched in January 2023 under the Horizon Europe programme, it aims to establish two high-energy heavy-ion irradiation facilities tailored for space applications: one at CERN in Switzerland and another at GSI Helmholtz Centre for Heavy Ion Research in Germany \cite{cern_home}.\

The project brings together several key partners:
\begin{itemize}
    \item CERN: Providing the IRRAD facility for high-energy heavy ion irradiation in the East Area of the Proton Synchrotron in T8.
    \item GSI Helmholtz Centre: Offering complementary capabilities in heavy ion research.
    \item University of Padua: Contributing academic expertise.
    \item Thales Alenia Space, Airbus Defence and Space and CosyLab: Industrial partners experienced in space applications and accelerator systems.
\end{itemize}
The collaboration aims to create a sustainable and autonomous European infrastructure for radiation testing, reducing reliance on non-European facilities.

In November 2024, HEARTS conducted a successful pilot user campaign at CERN's new high-energy heavy-ion irradiation facility.\
The campaign involved 10 companies and institutions, accumulating 168 hours of testing for electronic components and modules intended for space applications.
A follow-up campaign in November 2025 expanded beam time to 200 hours. This access scheme for both industrial and academic users was optimized and consolidated in 2026 before Long Shutdown 3 (LS3) with another 200 hours offered within 1.5 weeks, as opposed to 2.5 weeks in preceding years. Positive feedback from participants confirmed the facility’s readiness to support routine space electronics testing. In addition, 48 hours of industrial beam time was offered at GSI in 2025.

HEARTS plays a vital role in ensuring Europe's autonomous access to space by providing dedicated beam time for the space industry and supporting the development of radiation-resistant technologies.\
The need of high-energy heavy-ion beams for space applications has been identified to be a strategic enabling technology for the development of the aerospace industry in Europe. 
The present conceptual design study is not a deliverable of the HEARTS project, but its preparation has been largely motivated by the work and findings of HEARTS at the PS.

\section{HEARTS@LEIR: A European hub for electronics radiation testing}

Building on the positive feedback and operational experience, HEARTS@CERN seeks to consolidate its position as a leading user facility and expand access for industrial partners and academic institutions.
In this context, the underutilized Low Energy Ion Ring (LEIR) has been identified as a strategic asset for radiation testing. Future activities, beyond the present HEARTS project, may include dedicated user runs with slowly extracted ion beams from LEIR. This approach aligns with findings from the BioLEIR study~\cite{BioLEIR}, which demonstrated the feasibility of delivering diverse ion beam species, originally explored for biomedical applications, to broader scientific and industrial communities.

This report details the beam and machine specifications required for the proposed activity, as outlined comprehensively in Chapter~\ref{sec:BeamSpecs}.\ The operational model for which the facility is designed, seen from the user and within the envelope of the existing accelerators, follows in Chapter~\ref{sec:Operations}.\ 
With these specifications as our foundation, we explore the injection chain systems required to deliver high-performance beams to LEIR, ensuring optimal alignment with operational demands.\
The required modifications are designed and will be implemented to ensure no operational impact on the LHC ion physics programme or on the ion programme of the SPS North Area.
The introduction of a new ion source will further enrich these programmes with new ion species, available within or beyond the scope of HEARTS.

The addition of a second ion source is a pivotal upgrade and will significantly enhance flexibility and throughput, as outlined in Chapter~\ref{sec:IonSources}.\ Additionally it addresses the hardware upgrade considerations for the Low Energy Beam Transport (LEBT), covering the initial acceleration pathway from the source through the RFQ, Linac~3 and transport to injection into LEIR.\

We proceed with a detailed analysis of operational and hardware upgrades to LEIR required to achieve resonant slow extraction of heavy-ion beams. These critical adaptations are thoroughly documented in Chapters~\ref{sec:LEIRaccelerator} and~\ref{sec:SRE}.\
The assumptions that this design makes on the existing machine were tested with beam in LEIR in 2025 and 2026; these machine tests, which include a first slow-extraction test, are presented in Chapter~\ref{sec:MachineTests}.\
The report further outlines the beam transport system to the irradiation area, including the necessary hardware, as described in Chapters~\ref{sec:Beamline2HEARTS} and \ref{sec:expArea}.\
Chapters \ref{sec:InfrastructureAndIntegration} and \ref{sec:RadiationProtection} document respectively the infrastructure and integration and the radiation safety and protection considerations.
Finally, we conclude with a comprehensive summary, recapping the essential hardware upgrades, their estimated costs, and the personnel requirements needed to bring this vision to fruition in Chapter~\ref{sec:ResourceEstimate}. We close with a risk assessment in Chapter~\ref{sec:RiskAssessment}.

\chapter{Beam specification parameters}
\label{sec:BeamSpecs}

The beam requirements are based on information presented in a previous deliverable of the HEARTS project~\cite{HEARTS_D51}.\
A consolidated set of infrastructure and beam specifications has been established and serves as the baseline for the HEARTS@LEIR initiative.\
This section reviews the beam requirements for a controlled radiation testing facility designed to support testing of electronics at Technology Readiness Levels (TRL) 6–7 \cite{ESA-TRLHandbook}.\
These requirements are then translated into machine and beam specifications aligned with the capabilities of the existing LEIR synchrotron.\

\section{Beam requirements at test station}

The beam and machine requirements for testing electronics at TRL 6–7 are outlined in \cite{HEARTS_D51}.\
In this section, we focus on the beam-specific requirements relevant to the planned testing activities, before translating them into corresponding machine and beam specifications. A schematic overview of these requirements and how they relate to the beam specifications described below is shown in Fig. \ref{fig:CDR_schematic}.

\begin{itemize}
    \item \textbf{Requirement A:} the LET values of the available beams provided by LEIR shall range from 0.4 to 75~MeV\,cm$^2$\,mg$^{-1}$.

    \item \textbf{Requirement B}: The particle penetration depth shall exceed \SI{200}{\micro\meter} in silicon across the entire LET range.

    \item \textbf{Requirement C:} the heavy-ion beam LET shall not exceed a +/-10\% spread across the area being irradiated. This should be intended as Full-Width Half Maximum (FWHM) of the LET distribution at the surface of the device under test in the sample position.

    \item \textbf{Requirement D:} the heavy ion accelerator shall be capable of delivering ions with an average variable flux ranging from a few ($\approx$10) ions~cm$^{-2}$~s$^{-1}$ to at least $10^6$ ions~cm$^{-2}$~s$^{-1}$ on the DUT.\
    This should be interpreted as an instantaneous flux, since the beam will be delivered in spills.\
    The maximum instantaneous flux ($10^6$~ions~cm$^{-2}$~s$^{-1}$) is intended for those tests in which users are interested to reach a target of $10^7$ ions~cm$^{-2}$ in the fastest possible way.
    
    \item \textbf{Requirement E:} the radiation field shall be uniform within ±10\% over the area of the DUT(s) in terms of integrated flux.\ The extracted beam profile usually follows a perturbed Gaussian distribution.\ This shall be homogenized to reach the level of uniformity required for testing. 

    \item \textbf{Requirement F:} an irradiation area with a size tunable from 2 cm $\times$ 2 cm to 20 cm $\times$ 20 cm shall be made available, with variable steps, whilst maintaining a +/-10\% integrated flux uniformity.\
    Flux outside the irradiation area should be at least three orders of magnitude lower.\\
    The combination of Requirements D and F is understood to be flexible with respect to the highest fluxes for the broadest irradiation areas. Users seeking maximum instantaneous flux at the DUT will therefore need to balance irradiation area against the total irradiation time required for their tests. 

    \item \textbf{Requirement G:} the temporal structure of the slowly extracted beam shall not exhibit strong fluctuations. The maximum global particle rate over the extraction cycle shall not exceed twice the mean or requested value over a given integration time relevant for the irradiation test.
\end{itemize}

\section{Beam specifications for HEARTS@LEIR}

\begin{figure}
    \centering
    \includegraphics[width=\linewidth]{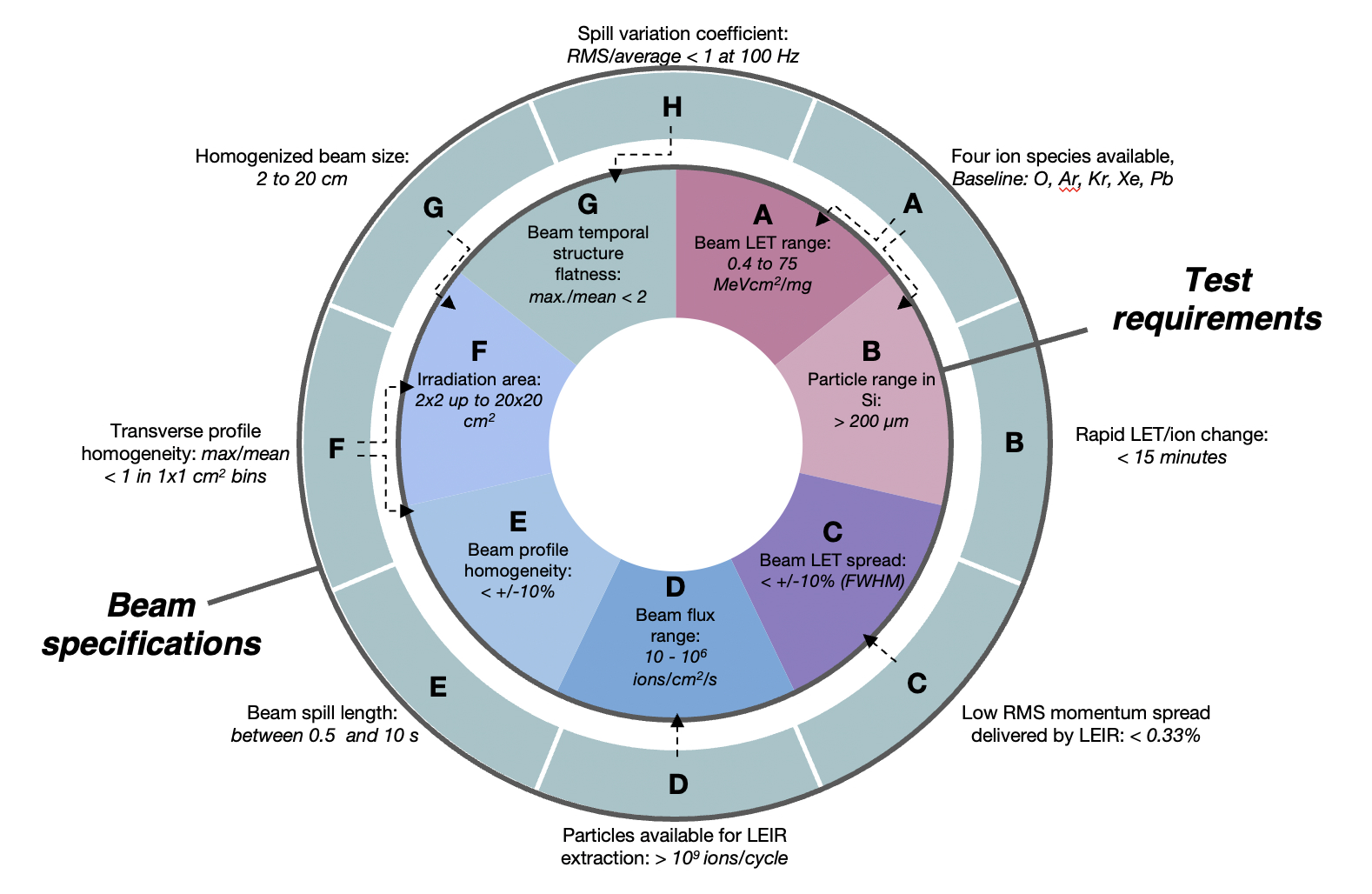}
    \caption{Schematic overview of the requirements, specifications and how they are interconnected, as described in this chapter.}
    \label{fig:CDR_schematic}
\end{figure}

Based on the previously outlined beam requirements (A–G), this section defines the beam characteristics necessary to meet them.\
In addition to the beam and machine specifications, we also outline operational constraints relevant to the competitiveness of the initiative.\
The technical capabilities and operating modes of the LEIR accelerator are described in detail in Chapters~\ref{sec:LEIRaccelerator} and~\ref{sec:Operations}. A schematic overview of these specifications and how they relate to the test requirements described above is shown in Fig. \ref{fig:CDR_schematic}.

\begin{itemize}
    \item \textbf{Specification A}

    At least four ion species shall be made available and accelerated to energies between 20–100 MeV/nucleon to optimally cover the LET range defined in Requirement A.\
    These elements will include a subset of ion species such as B, O, Ne, Mg, Ar, Kr, Xe, and Pb.\
    Tentatively, the baseline is considered to be O, Ar, Kr, Xe and Pb.
    The specific ions to be made available will be determined based on synergies between the HEARTS@LEIR initiative and the Ion Complex Upgrade (ICU) project~\cite{ICUproject}.\\  

    To fulfill Requirement A, a range of ion species must be used, as different elements provide different LET values.\
    The LET ranges accessible with various elements are illustrated in Figure~\ref{fig:LETvsRange}.\ 
    The upper limit of the LET range has been extended to 75~MeV\,cm$^2$\,mg$^{-1}$, exceeding the limit defined in Req. 2.1.a of~\cite{HEARTS_D51}.\\

    Requirement B can also be fulfilled.\
    The diversity and number of ion species offered directly impact the LET range coverage and, consequently, the competitiveness and attractiveness of the facility.\
    Additional improvements, such as lowering the extraction energies of ions like Oxygen (O) and Argon (Ar) to $<$ 20 MeV/nucleon, or extending the palette of available elements, can further enhance LET coverage.\
    Care must be taken to ensure that LET coverage does not compromise penetration depth or beam quality, particularly when adjusting beam energies or introducing new ion species, i.e. requirement B.\\
        
\begin{figure}
    \centering
    \includegraphics[width=0.8\linewidth]{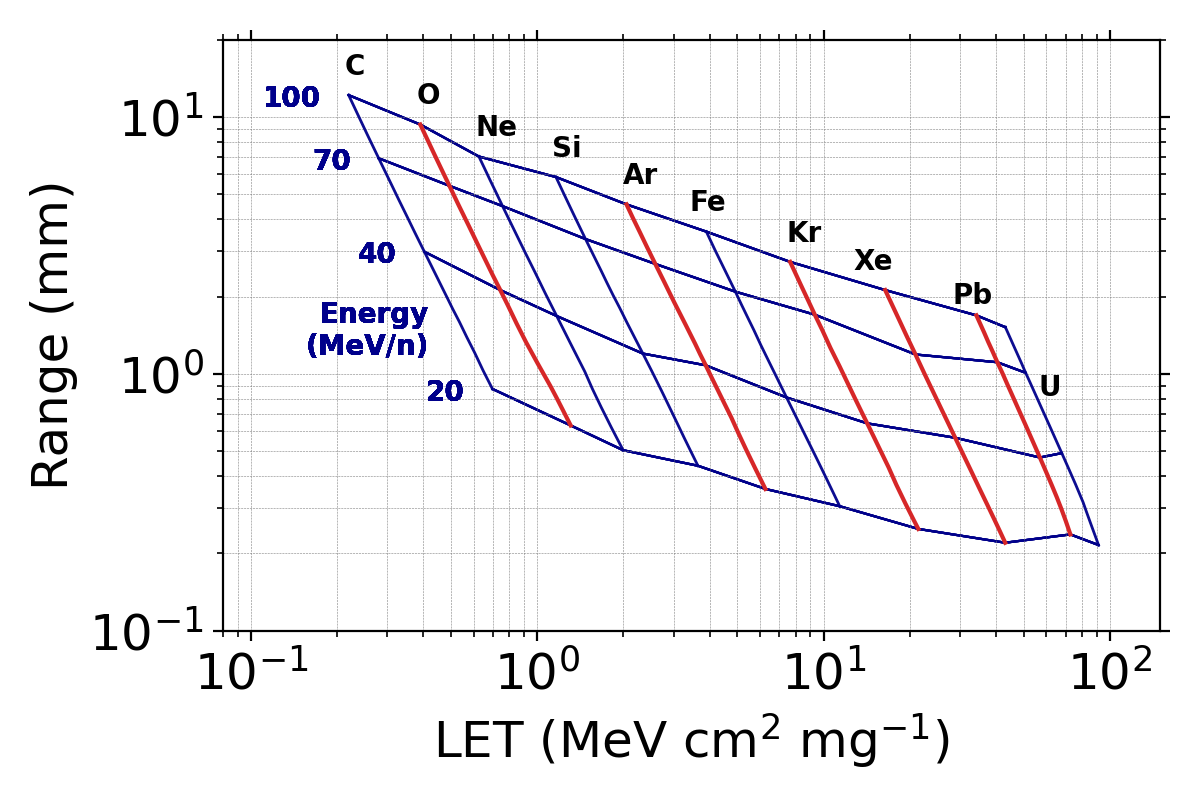}
    \caption{Penetration range as a function of LET in silicon. The coverage provided by five candidate ion species is highlighted in red. The blue lines connect equal kinetic energies (labels in MeV/nucleon) and equal ion species.}
    \label{fig:LETvsRange}
\end{figure}
    \item \textbf{Specification B}
    
    The maximum time allowed for an ion-species change or an LET change shall be less than 15 minutes.\\
    
    A full Single Event Effect (SEE) characterization requires multiple LET values covering the LET range mentioned above.\
    This involves the use of a palette of different ion species.\
    LET values can be adjusted on a cycle-to-cycle basis at LEIR by varying the extraction energy or using degraders in the beamline.
    The 15-minute limit on species change time is primarily determined by ion source limitations.\ 
    Additionally, all elements downstream (Linac~3, LEIR, HEARTS-TL) will need to support pulse-to-pulse modulation (PPM), i.e.\ the ability to change their settings from one machine cycle to the next according to the programmed cycle sequence.\\
    
    \item \textbf{Specification C}

    The RMS relative momentum spread of the beams delivered by LEIR shall not exceed $\pm$0.33\%.\\
    
    The LET spread is determined primarily by the RMS relative momentum spread of the extracted beam.\ 
    In the case of LEIR, this spread is expected to remain well below one $\permille$ (one part per thousand), provided that electron cooling is properly implemented.\
    Therefore, Requirement C is considered neither restrictive nor technically challenging.\\
    
    Nonetheless, LET variations will be monitored via diagnostics and beam instrumentation to ensure that no significant broadening occurs.\
    The main contributor to LET broadening is expected to be the use of dedicated degraders for tailoring LET, rather than intrinsic beam properties.\ Contributions of beam diagnostics are expected as well.\\

    \item \textbf{Specification D}

    The total amount of particles available for extraction in LEIR shall be at least 10$^9$ ions per machine cycle.\\ 

    For operation with different ion species, accumulation and electron cooling will be required for optimal facility operation.\ 
    The available intensity per cycle should be maintained above this threshold in order to achieve high duty cycles, thereby minimising the time required to accumulate 10$^7$ ions/cm$^2$ at the DUT.\
    The specified number of ions per machine cycle is intended to be achieved at the synchrotron flat-top and some flexibility in intensity repeatability is permitted, subject to operational constraints.\\

    \item \textbf{Specification E}
    
    The length of the spills shall range from 0.5~s to 10~s, delivering the average particle fluxes over the whole cycle as defined in Requirement D.\\
    
    When combined with Specification D, these parameters enable the required particle fluxes to be achieved with a reasonable and competitive level of flexibility.\
    The precise spill duration shall be determined on a case-by-case basis, depending on the requirements of the ongoing radiation test campaign, while respecting the electrical and cooling power constraints at LEIR.\\
    
    \item \textbf{Specification F}
    
    The transverse profile of the beam shall be homogenized such that the ratio between max and mean of the measured counts in 1$\times$1 cm$^2$ spatial bins of the dedicated diagnostic shall not be higher than 1.1.\\
    
    The extracted beam typically exhibits a perturbed Gaussian profile in the extraction plane, while maintaining a more Gaussian-like shape in the transverse plane perpendicular to it.\
    Homogenization of these profiles will be implemented using established techniques, as is standard practice at other radiation testing facilities.\ In this document, three techniques are described and compared: blow-up and scrape, octupole tail folding and raster scanning techniques.\ 
    This covers Requirement E.\\
    
    \item \textbf{Specification G}
    
    The homogenized transverse beam size of the beam will range from 2~cm up to 20~cm.\\

    This specification satisfies Requirement F. The technical implementation will be described in detail in Chapter~\ref{sec:Beamline2HEARTS} and will involve the use of quadrupole magnets in the transfer line leading to the irradiation area.\ 
    The final delivery to the experimental area will include a set of collimators and masks to shape the beam.\\ 
    
    \item \textbf{Specification H} 

    The coefficient of variation of the spill, defined as 
    \begin{align}
        c_V = \frac{\sigma}{\mu},
        \label{eq:CoeffVar}
    \end{align}
    where $\sigma$ represents the RMS of the spill and $\mu$ the average and specified particle counts, shall not exceed one when evaluated from particle counts recorded at a sampling rate of 100~Hz.\\

    In recent developments within the accelerator community, the coefficient of variation ($c_V$) has emerged as a key figure of merit for evaluating the stability of slowly extracted beams.\
    This specification is the direct translation of Requirement G.\
    Various techniques described in the literature will be implemented to achieve compliance with this limit.\
    The sampling rate of 100~Hz is a tentative value: the irradiation tests do not require faster time resolution, and the value can be tightened if a spill feedback loop is installed.\
\end{itemize}

\section{Summary}

The beam requirements and corresponding specifications have been reviewed in detail in this section.\
Table~\ref{tab:beamRequirements} presents a summary of the requirements (A–G), and Table~\ref{tab:beamSpecs} outlines the corresponding specifications designed to meet them.\

\begin{table}[h!]
    \begin{center}
    \caption{Summary of beam requirements for a radiation facility based at LEIR for radiation testing of electronics }%
    \begin{tabular}{c c}
       Parameter  & Value\\
       \hline\hline
       LET range (MeV\,cm$^2$\,mg$^{-1}$) &  0.4--75 \\
       Penetration depth in silicon (\si{\micro\meter}) & $\geq$200\\ 
       LET spread & $< |\pm$10\%|\\
       Irradiation area (cm$^2$, H$\times$V) & 2$\times$2 up to 20$\times$20\\ 
       Instantaneous fluxes (ions\,cm$^{-2}$\,s$^{-1}$) & 10--10$^6$\,$^{*}$ \\
       Spatial flux homogeneity & $\leq |\pm$10\%|\\
       Instantaneous temporal\\ flux homogeneity (max/avg)$^{\dagger}$ & $<$ 2\\
    \end{tabular}
    \label{tab:beamRequirements}
    \end{center}
    \footnotesize{*The highest flux range might not be available for wide irradiation fields.\\
    $^{\dagger}$Evaluated at a sampling rate of 100~Hz (tentative value).}  
\end{table}

\begin{table}[h!]
    \begin{center}
    \caption{Summary of beam specifications for a radiation facility based at LEIR that fulfill the requirements for radiation testing of electronics at TRL 6-7.}
    \begin{tabular}{c c c c}
       Parameter  & Unit & Symbol & Value\\
       \hline\hline
       Beam energy  & MeV/nucleon & $E$ & 20--100$^{*}$\\
       Ion species$^1$ & - & - & $^{16}$O$^{8+}$, $^{40}$Ar$^{16+}$, $^{86}$Kr$^{29+}$\\ 
       & & & $^{129}$Xe$^{40+}$, $^{208}$Pb$^{54+}$\\
       Switch time\\
       between ions & min& $T_{\text{ion,switch}}$ & $\leq$15\\ 
       Momentum spread (RMS) & - & $\sigma_p$ & $\leq \pm$0.33\% \\
       Beam intensity & ions & $N_p$ & $\geq$10$^9$\\ 
       Spill length & s & $T_{\text{spill}}$ & 0.5--10\\
       Spill intensity & ions\,s$^{-1}$ & R & 10$^6$--10$^8$\,$^{**}$\\
       Coefficient of variation$^{\dagger}$ & - & $c_V$ & $< 1$\\
       Transverse spatial \\
       homogeneity (max avg$^{-1}$) & - & - & $\leq$1.1
    \end{tabular}
    \label{tab:beamSpecs}
    \end{center}
    \footnotesize{$^*$ $^{208}$Pb$^{54+}$ can only reach 72.1~MeV/nucleon at \SI{4.8}{\tesla\metre}.\\
    $^1$ Ion species and charge state will be defined in cooperation with the ICU project.\\
    $^{**}$ the higher end of the range might be only available for some ion species or irradiation fields.\\
    $^{\dagger}$ Evaluated from particle counts recorded at a sampling rate of 100~Hz (tentative value, Specification~H).}  
\end{table}

\chapter{Desired operational model}
\label{sec:Operations}

A facility of this kind is judged by its users on availability, on the beam conditions, on the ease of changing them within a test, and on the quality of the support they receive. Building on the beam specifications of Chapter~\ref{sec:BeamSpecs}, this chapter sets out the operational model for which the facility is designed: the operational scenario, the envelope within which the existing accelerators are operated, the division of responsibilities between the CERN Control Centre and the facility team, the mode of user operation, and the operational cost that follows from them. These are targets for the technical design described in the following chapters. Where a number is established later in the report --- the beam intensities, the extraction and transport efficiencies, the thermal envelope of the LEIR magnets --- it is quoted here with a reference to the chapter in which it is derived. The operational cost adds to the running costs of the technical infrastructure (electricity, cooling water) given in Chapter~\ref{sec:InfrastructureAndIntegration}. The operational model outlined here is preliminary; it will be developed in full, together with the business model, during the pre-TDR and TDR phases.

\section{Operational scenario}
The facility is designed to deliver approximately 1,900 hours of heavy-ion beam time per year to radiation effects users. The facility will operate on a 24/7 basis during operational periods, with an effective irradiation efficiency of 75\%.
The remaining 25\% accounts for user changeover (estimated at approximately 45 minutes per 8-hour block) and short-duration downtime compensation, consistent with LEIR's high availability of 90–95\%. The HEARTS++ proposal assumes the same potential beam delivery of about 1,900 hours per year and adopts 1,700 hours per year as its design target, keeping a further margin for commissioning, configuration changes and user turnaround. The present report retains the figure of 1,900~hours per year.

Based on these assumptions, the facility is expected to operate for approximately 15 weeks per year, with up to 20 weeks considered feasible depending on demand. Preliminary scheduling considerations presented within the study indicate that this is compatible with the broader ion physics programme at CERN, including LHC ion physics and North Area operations; the sharing of LEIR between HEARTS@LEIR and the other ion users will be studied in detail during the technical design phase.

\section{Compatibility with existing accelerator systems and technical infrastructure}
\label{sec:OpsEnvelope}

A key boundary condition of this conceptual design is that HEARTS@LEIR operation must fit within the existing capacity of the accelerator systems and technical infrastructure, with the obvious exception of the new elements themselves: the extraction system and the beam line to the experimental area. The existing systems are therefore not subject to increased operational stress relative to present conditions, and the project does not depend on their upgrade, but only on the consolidation already foreseen for the ion physics programme. Approval of the CERN ion complex consolidation, expected within the next few years, is consequently a prerequisite for investing in the HEARTS@LEIR upgrade of the related accelerator and experimental-area infrastructure.

Two consequences of this boundary condition are quantified below: the heat dissipated in the LEIR magnets, which sets the achievable duty cycle, and the annual number of magnet pulses, which sets the load on magnets and power converters.

\subsection{Heat dissipation and achievable duty cycle}

The cooling demand of LEIR is dominated by heat dissipation in the magnets, which scales with the RMS of the magnet currents. The operational baseline is therefore to run LEIR at an RMS current compatible with the present envelope. That constraint, rather than the beam physics, defines the admissible combinations of beam energy --- equivalently, magnet flat-top current --- and spill duration.

\begin{figure}[htbp]
    \centering
    \includegraphics[width=\linewidth]{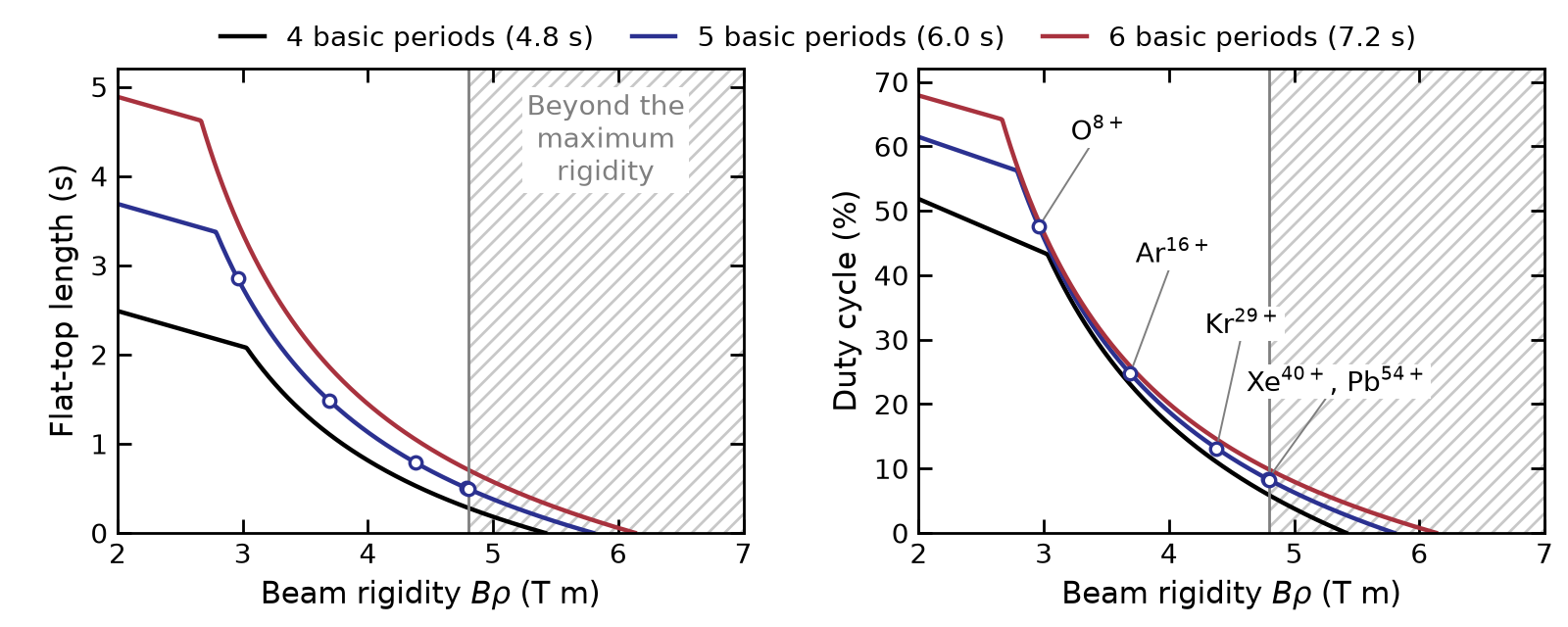}
    \caption{Flat-top length (left) and duty cycle (right) as a function of beam rigidity for a constant RMS magnet current of \SI{1.56}{\kilo\ampere}, the value of the NOMINAL cycle and a conservative estimate of the present LEIR operating envelope. Curves are shown for total cycle lengths of four, five and six basic periods. Circles: the baseline species at their maximum energy for five basic periods (Table~\ref{tab:DutyCycles}). The hatched region lies beyond the maximum operational LEIR rigidity of \SI{4.8}{\tesla\metre}.}
    \label{fig:FlatTopVsRigidity}
\end{figure}

The reference value of the RMS current depends on the cycle over which it is averaged; it was measured during the machine tests (Section~\ref{sec:MTFlattop} and Fig.~\ref{fig:MTDipoleCurrent}). Two cycles of routine operation, both with the same ramp and flat top, bracket it; their length is an integer number of basic periods of \SI{1.2}{\second} (the LEIR cycle is described in Chapter~\ref{sec:LEIRaccelerator}). The EARLY cycle, two basic periods (\SI{2.4}{\second}) long, reaches \SI{1.82}{\kilo\ampere}, in agreement with the value recorded by the converter protection (Table~\ref{tab:MTEPC}). The NOMINAL cycle, three basic periods (\SI{3.6}{\second}) long, has an injection plateau longer by one basic period, and reaches \SI{1.56}{\kilo\ampere}. Both cycles are part of routine operation. The calculations below take the NOMINAL value as the admissible RMS current; since EARLY cycles, with an RMS current \SI{17}{\percent} higher, are routinely played, this choice is conservative: it underestimates rather than overestimates the flat-top length available within the present envelope, which HEARTS@LEIR operation will respect (Section~\ref{sec:MTFlattop}). Taking this operating point of \SI{1.56}{\kilo\ampere}, corresponding to an average dissipated power of \SI{66.3}{\kilo\watt}, and a total cycle length of five basic periods (\SI{6}{\second}), the achievable flat-top lengths and duty cycles at the maximum energy of each baseline species are given in Table~\ref{tab:DutyCycles}. The duty cycle decreases with beam rigidity: it is lowest, approximately 8\,\%, for Xe$^{40+}$ and Pb$^{54+}$ at the maximum LEIR rigidity of about \SI{4.8}{\tesla\meter}, where the flat top is limited to \SI{500}{\milli\second} within the cycle. These values are the maximum achievable for each species at its highest energy; lower energies allow longer flat tops. The general dependence of flat-top length and duty cycle on beam rigidity, at constant RMS current, is shown in Fig.~\ref{fig:FlatTopVsRigidity}~\cite{FlattopExtension2025}.

\begin{table}[htbp]
\centering
\caption{Flat-top length and duty cycle achievable in HEARTS@LEIR at the RMS magnet current of the NOMINAL cycle, \SI{1.56}{\kilo\ampere}, for a total cycle length of five basic periods (\SI{6}{\second}). All species are at $E_{\text{kin}} = 100$~MeV/nucleon except lead, which reaches 72.1~MeV/nucleon at the maximum LEIR rigidity.}
\label{tab:DutyCycles}
\begin{tabular}{lrrr}
\toprule
Ion species & Beam rigidity & Flat-top length & Duty cycle \\
            & (Tm)          & (s)             & (\%) \\
\midrule
$^{16}$O$^{8+}$    & 2.96 & 2.87 & 47.7 \\
$^{40}$Ar$^{16+}$  & 3.69 & 1.49 & 24.8 \\
$^{86}$Kr$^{29+}$  & 4.38 & 0.79 & 13.1 \\
$^{129}$Xe$^{40+}$ & 4.79 & 0.50 &  8.4 \\
$^{208}$Pb$^{54+}$ & 4.80 & 0.50 &  8.3 \\
\bottomrule
\end{tabular}
\end{table}

The spill length actually delivered to a user is the shorter of two limits: the value in Table~\ref{tab:DutyCycles}, set by the magnet RMS current, and the value in Table~\ref{tab:OpsDeliveryTimes}, set by the expected available intensity and the maximum instantaneous flux requested. For the lighter species the RMS-current limit is the binding one; for the heavier species the two are comparable.

By preserving the present heat dissipation of LEIR and adding only the cooling required by the extraction elements and the new beam line, a new cooling plant or a substantial rebuild of the existing one is not necessary, and the HEARTS@LEIR cooling demand can be met with relatively modest modifications to the existing installation, as described in Chapter~\ref{sec:InfrastructureAndIntegration}.

\subsection{Annual cycle budget and economy mode}

The second load on the existing systems is the number of acceleration cycles, which stresses the magnets and their power converters. Analysis of the 2023, 2024 and 2025 cycle records~\cite{CycleComposition2026} gives the figures of Table~\ref{tab:CycleBudget}: on average approximately 3.65~million LEIR pulses per year, of which roughly one million are sent to the PS.

\begin{table}[htbp]
\centering
\caption{LEIR cycles played per year during Run~3, from timing and extraction-septum current records~\cite{CycleComposition2026}.}
\label{tab:CycleBudget}
\begin{tabular}{lrrr}
\toprule
Year & Total cycles & Sent to the PS & Fraction to the PS \\
\midrule
2023 & 3.61\,M & 1.09\,M & 30.2\,\% \\
2024 & 3.66\,M & 1.27\,M & 34.7\,\% \\
2025 & 3.67\,M & 0.79\,M & 20.4\,\% \\
\midrule
\textbf{Average} & \textbf{3.65\,M} & \textbf{1.05\,M} & \textbf{28.4\,\%} \\
\bottomrule
\end{tabular}
\end{table}

Of the cycles not sent to the PS, an upper bound of 360\,k is estimated to be needed for LEIR commissioning, assuming three weeks of five days with two eight-hour shifts and short EARLY cycles. The estimated annual need for HEARTS@LEIR operation, including its own commissioning and contingency, is 1.8~million cycles.

The conceptual design study has shown that introducing an economy mode into LEIR operation, pulsing the elements only when a cycle is actually required, with a reduced pulsing rate at night and at weekends, and an idle mode whenever the machine is unused, keeps the total below the level already reached during Run~3. A zero-like cycle suitable for this purpose, LIN3MEAS, already exists and has been used in operation in 2026. The resulting budget is 360\,k cycles for LEIR commissioning, 1.0~million for the PS (LHC and North Area physics, and HEARTS at the PS) and 1.8~million for HEARTS@LEIR, for a total of 3.16~million cycles per year, which lies below the Run~3 average of 3.65~million. HEARTS@LEIR operation therefore adds no net ageing load to the LEIR magnets and power converters, as long as the idle mode is enabled, as is intended for future operation.

\section{Accelerator operation and support}
Operation of the HEARTS@LEIR accelerator complex, comprising the ion sources, Linac~3, LEIR, and the dedicated transfer line, will be embedded within the CERN Control Centre (CCC) 24/7 operational coverage. Equipment group on-call support for the sources, Linac~3, and LEIR will be maintained at the same level as during standard ion physics periods. In the event of accelerator faults, users will report issues to the CCC, which will either resolve them directly or escalate to the relevant equipment group. The detailed operational model, including the balance between automation and operator intervention, will be developed during the pre-TDR and TDR phases with dedicated involvement from CERN operations team.
\section{Facility staffing}
Three permanent positions at 50–75\% dedication are foreseen for facility operation:

\begin{itemize}
\item A facility scientist (e.g.~within SY-STI), based at the irradiation facility rather than in the CCC, capable of supporting accelerator and beam-related activities;
\item A facility engineer or technical expert (e.g.~within BE-CEM);
\item A facility administrative assistant.
\end{itemize}

This core team will handle user communication, scheduling, and run preparation outside of operational windows, in addition to on-site support during campaigns.

\section{User operation}
Typical user irradiation slots consist of one or two 8-hour shifts. A facility scientist or engineer will be present at the start of each user run, with possible exceptions for experienced users. Following the initial setup, users operate the facility autonomously, selecting from eight predefined accelerator settings comprising combinations of four ion species (e.g. O, Ar, Xe, Pb) at two rigidity settings corresponding to energies of approximately 30 MeV/nucleon and 100 MeV/nucleon. 

Accelerator setting changes, including automated beam validation, are targeted to complete within 15 minutes. Additional local adjustments, including energy and beam size tuning via degraders and masks, DUT positioning and angle of incidence, beam on/off control, and intensity regulation via the slow extraction feedback loop, are performed directly by users at the facility level, on time scales of the order of tens of seconds.
The maximum beam size delivered to the test station is 20 × 20 cm$^2$, preserving a homogeneous central profile. 
Intensity repeatability requirements are not stringent, as only a fraction of the total available beam intensity is utilised and variability is managed through the slow extraction system.
During autonomous operation, users have access to two support escalation paths: accelerator or transfer line issues are directed to the CCC, while facility-level issues are handled by a rotating on-call team composed of the facility scientist, engineer, and other members of the radiation effects community within the Accelerator sector.

\section{Operational costs}
Preliminary operational cost estimates include electricity consumption of approximately \SI{6}{\giga\watt\hour} per year (approx.~\SI{250}{\kCHF}/year), and permanent staff costs of approximately \SI{400}{\kCHF}/year (two level 6–7 positions and one level 3–4 position at 75\% dedication). Additional cost items include CCC operational coverage, equipment group on-call support, and facility consumables, which will be quantified and incorporated into the facility business model. As shown in Section~\ref{sec:OpsEnvelope}, operation in economy mode keeps the annual number of LEIR cycles below the Run~3 level, so that HEARTS@LEIR represents no additional accelerator ageing load.

Table~\ref{tab:OpsDeliveryTimes} translates the operating envelope of Section~\ref{sec:OpsEnvelope} into the quantity a user cares about: how long a campaign takes. It combines the intensities expected at LEIR (Chapter~\ref{sec:IonSources}) with the extraction and transport efficiencies of Chapters~\ref{sec:SRE} and~\ref{sec:Beamline2HEARTS}. The average flux reproduces the values of Table~\ref{tab:fluxes_at_DUT}, the instantaneous flux during the spill is that value divided by the duty cycle.
For oxygen and xenon, the intensity per cycle is set by the radiation-protection limit of Chapter~\ref{sec:RadiationProtection} rather than by the intensity available at LEIR. For the $12 \times 12$~cm$^2$ field the instantaneous flux exceeds the $10^{6}$~ions cm$^{-2}$ s$^{-1}$ of Requirement~D for xenon and lead, and reaches $0.9$ to $1.0 \times 10^{6}$~ions~cm$^{-2}$~s$^{-1}$ for oxygen, argon and krypton. Since the flux scales inversely with the irradiated area, Requirement~D is met for all species for fields up to $10 \times 10$~cm$^2$, in line with the trade-off between field size and flux foreseen in Requirement~F.
Within these limits, the spill length or the extracted intensity can be reduced to match a lower flux requested by a user rather than having to be increased.
A fluence of $10^{7}$~ions\,cm$^{-2}$ is reached in under \SI{90}{\second} for all baseline species and in about \SI{20}{\second} for oxygen. Since beam is delivered once per machine cycle, these times correspond to between four and fifteen cycles and round up to whole cycles in practice. Adopting the pencil-beam scanning option of Section~\ref{sec:TLscanning}, whose transport efficiency is approximately four times higher, would shorten them by the same factor. The consolidated operational cost is reported in Table~\ref{tab:res_opex}.

The operational model thus sets three demands on the accelerator chain, which the following chapters take up in the order in which the beam meets them. The ion sources and Linac~3 must change the ion species within 15~minutes and deliver to LEIR the intensities assumed in Table~\ref{tab:OpsDeliveryTimes} (Chapter~\ref{sec:IonSources}). LEIR must accumulate, cool and accelerate these beams within the RMS envelope of its present operation (Chapter~\ref{sec:LEIRaccelerator}). Finally, the slow extraction and the new beam line must turn the beam on the flat top into spills of 0.5 to \SI{3}{\second} at the device under test, with the efficiencies assumed above (Chapters~\ref{sec:SRE} and~\ref{sec:Beamline2HEARTS}).

\begin{table}[htbp]
    \centering
    \caption{Delivery performance per ion species, derived from the duty cycles of Table~\ref{tab:DutyCycles}. $N^{\mathrm{LEIR-FT}}$ is the accumulated intensity at LEIR flat top from Table~\ref{tab:IonIntDetails}, limited to the radiation-protection limit per cycle of Table~\ref{rp_charge_limits} (oxygen and xenon); the spill length and duty cycle are those imposed by the magnet RMS current. Fluxes are given at the device under test for a $12 \times 12$~cm$^2$ field, assuming an extraction efficiency $\eta_{\text{SRE}} = 0.7$, a beam transport efficiency $\eta_{\text{TL}} = 0.15$ for the baseline blow-up-and-scrape option, and a cycle length of \SI{6}{\second}. The last column is the time needed to accumulate $10^{7}$~ions\,cm$^{-2}$.}
    \label{tab:OpsDeliveryTimes}
    \small
    \setlength{\tabcolsep}{4pt}
    \begin{tabular}{lrrrrrr}
    \toprule
    Ion & $N^{\mathrm{LEIR-FT}}$ & $T_{\text{spill}}$ & Duty & \multicolumn{2}{c}{Flux at DUT} & $T(10^{7}$\,cm$^{-2})$ \\
    \cmidrule(lr){5-6}
    species & ($10^{9}$ ions) & (s) & cycle (\%) & $\bar{\phi}$ & $\phi_{\text{inst}}$ & (s) \\
     & & & & \multicolumn{2}{c}{($10^{5}$\,ions\,cm$^{-2}$\,s$^{-1}$)} & \\
    \midrule
      $^{16}$O$^{8+}$    & 3.88 & 2.87 & 47.7 & 4.7 & 9.9 & 21 \\
      $^{40}$Ar$^{16+}$  &  1.84 & 1.49 & 24.8 &  2.2 &  9.0 & 45 \\
      $^{86}$Kr$^{29+}$  &  0.93 & 0.79 & 13.1 &  1.1 &  8.6 & 89 \\
      $^{129}$Xe$^{40+}$ &  2.30 & 0.50 &  8.4 &  2.8 & 33.5 & 36 \\
      $^{208}$Pb$^{54+}$ &  1.84${}^{a}$ & 0.50 &  8.3 &  2.2 & 26.8 & 45 \\
    \bottomrule
    \end{tabular}\\[3pt]
    \parbox{0.9\linewidth}{\footnotesize ${}^{a}$Eight injections of Table~\ref{tab:IonIntDetails}, the intensity delivered by LEIR for the LHC Pb beams with the LIU upgrades, which is assumed to be available for HEARTS@LEIR as well. The slow-extraction test of 2026 reached $1.29\times10^{9}$ ions on the flat top (Section~\ref{sec:MTSX}), on a cycle that was not optimised for transmission.}
\end{table}

\FloatBarrier

\chapter{Ion sources and beam preparation for injection into LEIR}
\label{sec:IonSources}
\chapterauthors{M.~S{\l}upecki, E.~C.~Cort\'es Garc\'ia}
The operational model of Chapter~\ref{sec:Operations} begins at the ion source: a user changes the ion species within 15~minutes, and the duration of a test campaign scales with the intensity that the source and Linac~3 deliver to LEIR. The source is therefore the first element of the chain that determines whether the beam specifications of Chapter~\ref{sec:BeamSpecs} can be met. Two of them bear directly on the source: Specification~A, which requires at least four ion species spanning the LET range of Requirement~A, and Specification~B, which caps the time needed to change species at 15 minutes. This chapter reviews the present configuration of the source and of Linac~3, examines the options available for multi-species operation, establishes the second ion source as the baseline, and quantifies the beam intensities that the resulting complex is expected to deliver at LEIR.

Five candidate species --- O, Ar, Kr, Xe and Pb --- have been identified and are taken as the operational baseline throughout this report.

\section{Present source and Linac~3 configuration}

Currently, lead ions are produced with the ECR (Electron Cyclotron Resonance) GTS-LHC (Grenoble Test Source), developed at CEA (France) for the Large Hadron Collider beams. The source operates in the afterglow mode at a 10 Hz repetition rate with a 50 ms RF heating pulse. At the end of the RF pulse, once the microwaves are stopped, the ions are released rapidly, producing a short burst with higher peak intensity (afterglow) of higher-charge-states. During this period, the beam intensity varies continuously. A 200~µs slice (extended to 260 µs since 2024) is selected from the afterglow peak and accelerated through the Linac 3 RF cavities at a repetition rate of up to 5 Hz.

\begin{figure}[htb]
    \centering
    \includegraphics[width=\linewidth]{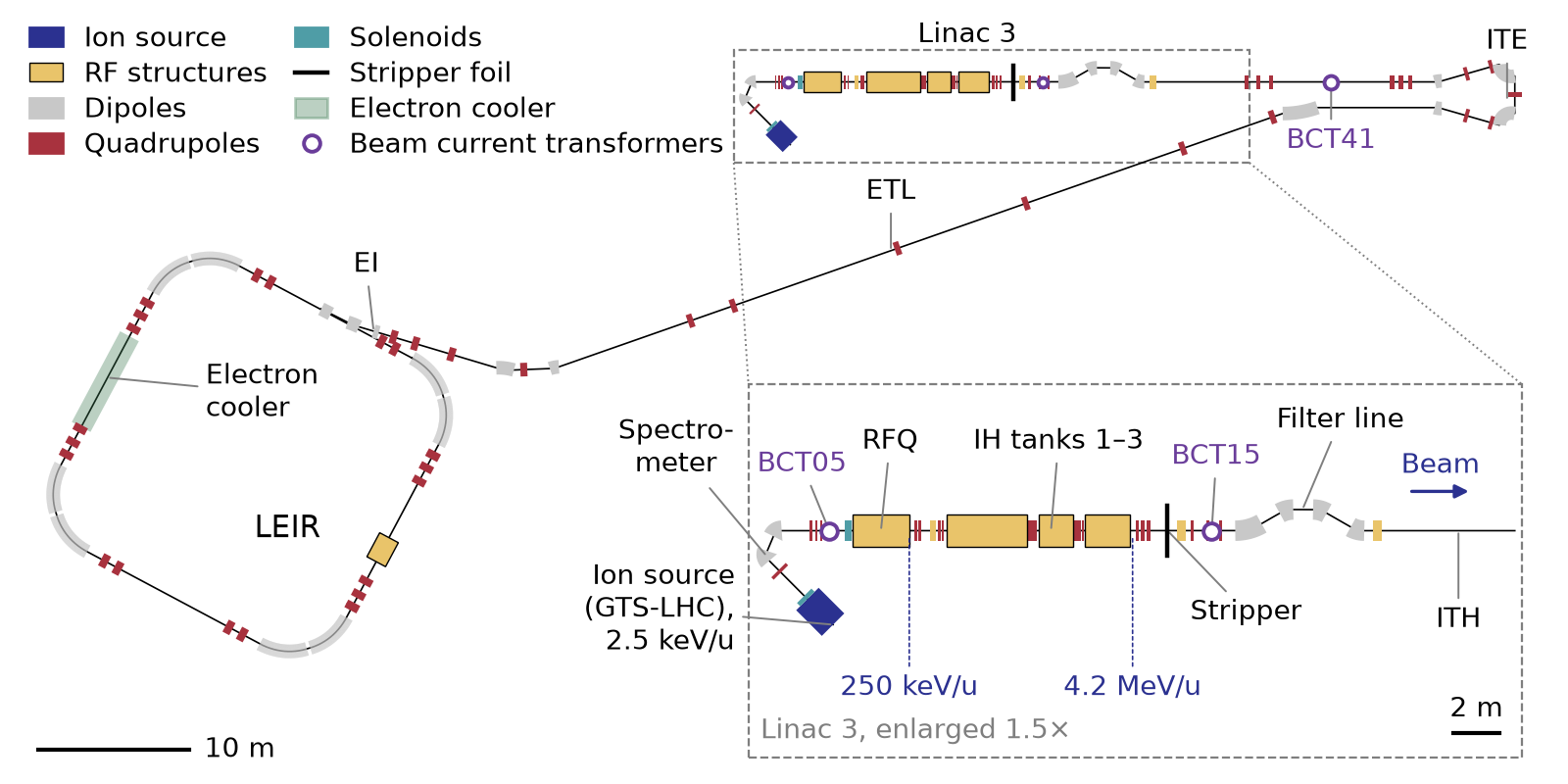}
    \caption{Floor plan from the Linac~3 ion source to the injection into LEIR, to scale. The transfer line (ITH, ITE, ETL, EI) and the LEIR ring are drawn from the surveys of the optics models~\cite{accmodelsTLS, accmodelsLEIR}; Linac~3 is drawn from the element positions in the layout database~\cite{CERNLayoutDB}, upstream of the ITH line, with the $\pm 30^\circ$ dog-leg of the filter line after Ref.~\cite{MontoyaSoto2024}. Inset: Linac~3 enlarged 1.5 times, with the beam energy after the source, the RFQ and the IH tanks. The following RF structures after the stripper-foil are the debunching and fast ramping cavities. Open circles mark the beam current transformers BCT05, BCT15 and BCT41 of Table~\ref{tab:IonIntDetails}.}
    \label{fig:Linac3Layout}
\end{figure}

Figure~\ref{fig:Linac3Layout} shows the path of the ions from the source to the injection into LEIR. The ions are extracted from the source at 2.5~keV/u, and the spectrometer of the low-energy beam transport (LEBT) selects one charge state. The RFQ accelerates the beam to 250~keV/u and three interdigital H-mode (IH) tanks to 4.2~MeV/u. A stripping foil at the exit of the linac raises the charge state (for lead from Pb$^{29+}$ to Pb$^{54+}$), and the filter line selects the charge state sent to LEIR. The transfer line (ITH, the ITE loop, ETL and EI) brings the beam to the injection septum of LEIR; ETL is shared with the ejection towards the PS. A detailed description of the source and of Linac~3 is given in Ref.~\cite{MontoyaSoto2024}.

\section{Multi-species operation}

\subsection{Species change with a single source}
Different ion species can be produced by changing the material supplied to the ion source. Elements or compounds available in gaseous form can be injected via a flow control valve~\cite{ExpwGTSLHC}. Solids are evaporated in one of the two available micro-ovens directly into the plasma chamber of the source. Only solids which have high enough vapour pressure at temperatures reachable by the micro-ovens are feasible.  At present, switching between species for stable long-term operation is only possible on time scales of several days for gases, while switching to or from solids can take several weeks to allow for plasma chamber conditioning and stabilization of the source running parameters.

\subsection{Cocktail beams}
Rapid switching between several species using a single source can be achieved with an operational scheme of cocktail beam~\cite{KALVAS2017205}. It relies on supplying more than one element or compound into the plasma chamber of the source. The beam produced this way consists of a multitude of charge states of all elements present inside the plasma chamber. To select and accelerate only the selected one, a spectrometer setup (dipole and slit) is used in the low energy beam transport line (LEBT) between the source and the first accelerating structure, RFQ. This beam production scheme has three drawbacks.

Firstly, it is important to ensure that the charge-over-mass ratio of the selected charge state of the selected element does not overlap (within resolving power of the Linac 3 spectrometer) with any other charge state of any other element in the cocktail. The resolving power is around 3-4\%. 

Secondly, the maximum beam intensity is fundamentally limited by the space-charge effects, which are most pronounced at low energy, between the source and the spectrometer, where the beam current is also the highest as it comprises of all charge states and elements. By using a cocktail instead of an elemental beam, additional unwanted charges are introduced into the most critical low-energy beam transport region, which results in reduced beam intensity after the spectrometer. The magnitude of this effect in Linac 3 LEBT has not been experimentally quantified due to lack of appropriate beam instrumentation.

Thirdly, the present beam diagnostics in LEBT does not allow for precise and non-beam-destructive evaluation of the source state and beam properties that are required as feedback for any automatic optimization of the source, LEBT and RFQ. Such an automation will be essential to maintain performance and stability after every switch of ions.

\subsection{Baseline: a second ion source}
The limitations described above can be mitigated by measuring the impact of using cocktail beams on beam intensities and implementing Linac 3 upgrades proposed by the Ion Complex Upgrade (ICU)~\cite{Slupecki_2025, Kuechler_2026}. Conversion of the Linac 3 LEBT (or ITL) beamline to pulsed version would enable concurrent measurements of beam properties for all selectable species, either in the dedicated low-energy beam diagnostics line immediately after the source, or after the RFQ for every second source pulse, which is currently not used downstream. These measurements would in turn be used as feedback for the automatic optimization tool, that should maintain performance of all selectable beams regardless if they were requested downstream. It is the only feasible way to guarantee good performance and pulse-to-pulse stability after the switch.

In addition, ICU proposes the installation of a second ion source connected with the existing RFQ and the rest of Linac 3. It is motivated by the difficulty of producing a stable lead beam of sufficient intensity both for the LHC and other users such as HEARTS@LEIR. In other words, it is expected that if lead were mixed with 4 gasses to form a complex cocktail beam, the intensity of resulting lead beam would be negligible, as the lighter gasses are easier to ionize and would likely dominate the space-charge-saturated region of LEBT.

Instead, for HEARTS@LEIR, two ion sources are required: the existing GTS-LHC source for Pb and O mixture (as used presently), and a second, identical source for noble gases (Ar, Kr, Xe, and possibly O as buffer gas). Further details are provided in~\cite{ICUproject} and are aligned with the specifications outlined in this document.

\section{Expected beam intensities}
For the baseline ion species, the pulse intensities have been derived from previous test runs with different ions. A Pb–O mixture provides pulse intensities at LEIR of 0.36$\times 10^{9}$ lead ions, after consideration of injection and stacking efficiency. A high-performance dedicated Oxygen test run yielded 10.2$\times 10^{9}$ ions, with no Lead in the source; this is above the radiation-protection limit per cycle of Table~\ref{rp_charge_limits}, which therefore sets the oxygen intensity delivered to the users (Table~\ref{tab:OpsDeliveryTimes}).\ 
Test runs with $^{40}$Ar$^{11+}$ and $^{129}$Xe$^{39+}$~\cite{Alemany-Fernandez:IPAC2018-TUPAF020} were also performed, and the corresponding single-pulse intensities at LEIR ejection are summarized in Table~\ref{tab:IonIntDetails}.\
A krypton test was performed in Linac~3 in 2023~\cite{Kuchler:2916870}, yielding 83~$\mu$A of $^{86}$Kr$^{22+}$ beam current after the RFQ.\
For HEARTS@LEIR, higher charge states are required to reach $\approx$100 MeV/nucleon. Charge stripping at the end of Linac~3 is therefore foreseen.
Stripping efficiencies for $^{16}$O$^{8+}$, $^{40}$Ar$^{16+}$, $^{86}$Kr$^{29+}$ and $^{129}$Xe$^{40+}$ are calculated with Baron's formula~\cite{BARON1993177}.\

Note that the pulse intensity decreases inversely proportional with the ion charge state. Conversely, achieving accelerations of at least 100 MeV/nucleon with the current LEIR beam rigidities requires a minimum charge-to-mass ratio of $\approx$0.31, necessitating higher charge states at the expense of lower intensities.
The considerations of pulse accumulation, e-cooling and RF capture in LEIR are discussed in Chapter~\ref{sec:LEIRaccelerator}.

\begin{figure}[h!]
    \centering
    \includegraphics[width=0.65\linewidth]{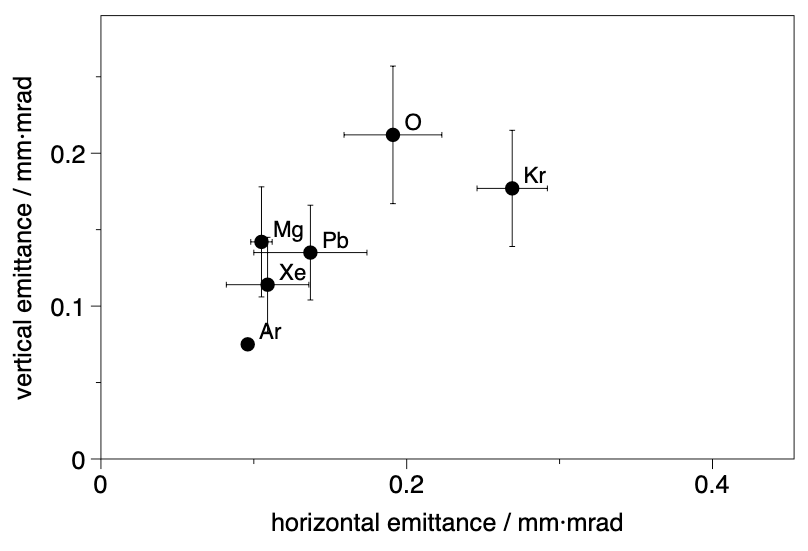}
    \caption{Measured transverse emittances (normalized RMS) of different ion beams in the LEBT. The figure has been taken from \cite{Kuchler:2916870}. For Ar only one measurement was available.}
    \label{fig:IonEmittAfterRFQ}
\end{figure}

As final remark, the transverse emittance after the RFQ has been measured~\cite{Kuchler:2916870}, with results shown in Fig.~\ref{fig:IonEmittAfterRFQ}.\ Note that the values displayed are beams that did not undergo charge stripping.
Therefore the emittances can be taken as guidelines for the lower charge states of Ar, Kr and Xe shown in Table~\ref{tab:IonIntDetails}. 

\begin{table}[htbp]
    \footnotesize
    \renewcommand{\arraystretch}{1.2}
    \setlength{\tabcolsep}{3.5pt}
    \centering
    \caption{Conservative beam intensities through the ion complex for the HEARTS@LEIR baseline ion species. Currents are electrical currents of a single Linac~3 pulse: \textit{BCT05} between the source and the RFQ, \textit{BCT15-NS} after the stripper with the stripper foil retracted (no stripping), and \textit{BCT41} at the Linac~3 output, in the ITH line. $q_{\text{src}}$ is the charge state selected by the LEBT spectrometer from the charge-state distribution extracted from the source, and accelerated in Linac~3. All efficiencies are ratios of numbers of ions: $\varepsilon_{\text{Linac}}$ is the transmission from BCT05 to BCT15-NS; $\varepsilon_{q}$ is the fraction of ions in the selected charge state $q$ after the stripper (Baron's formula unless stated otherwise), or the transmission of the filter line when the beam is not stripped, so that $I_{\text{BCT41}} = I_{\text{BCT15-NS}}\,\varepsilon_{q}\,q/q_{\text{src}}$; $\varepsilon_{\text{LEIR}}$ is the LEIR efficiency from injection to flat top for the given number of injections (\#Inj). $N_{\text{BCT41}} = I_{\text{BCT41}}\,\tau/(q e)$ is the number of ions per Linac~3 pulse, with a pulse length $\tau = \SI{260}{\micro\second}$ unless stated otherwise in the footnotes, and $N_{\text{LEIR}}$ the intensity at the LEIR flat top per injection; the accumulated intensity is \#Inj\,$\times\,N_{\text{LEIR}}$. LEIR cycles are assumed to be \SI{3.6}{\second} long for all species to allow for sufficient cooling time. Bold font indicates measured values.}
    \label{tab:IonIntDetails}
    \begin{tabular}{cc|ccc|ccc|ccc}
      \toprule
        Ion & $q_{\text{src}}$ & BCT05 & $\varepsilon_{\text{Linac}}$ & BCT15-NS & $\varepsilon_{q}$ & BCT41 & $N_{\text{BCT41}}$ & \#Inj & $\varepsilon_{\text{LEIR}}$ & $N_{\text{LEIR}}$ \\
        at LEIR & & (\textmu A) & & (\textmu A) & & (\textmu A) & ($10^{9}$) & & & ($10^{9}$/inj) \\
      \midrule
        $^{16}$O$^{4+}$    & \multirow{2}{*}{4+} & \multirow{2}{*}{\textbf{242}} & \multirow{2}{*}{\textbf{0.38}} & \multirow{2}{*}{\textbf{93}}
                           & \textbf{1} & \textbf{93} & \textbf{37.7} & 1 & \multirow{2}{*}{\textbf{0.40}} & \textbf{15.1}${}^{1a}$ \\
        $^{16}$O$^{8+}$    & & & & & 0.68 & 126 & 25.6 & 1 & & 10.2${}^{1b\dagger}$ \\
      \midrule
        $^{40}$Ar$^{11+}$  & \multirow{2}{*}{11+} & \multirow{2}{*}{\textbf{105}} & \multirow{2}{*}{\textbf{0.55}} & \multirow{2}{*}{\textbf{58}}
                           & \textbf{1} & \textbf{58} & \textbf{6.6}${}^{2a}$ & 1 & \textbf{0.38} & \textbf{2.5} \\
        $^{40}$Ar$^{16+}$  & & & & & 0.32 & 26.7 & 2.71 & 2 & 0.34 & 0.92${}^{2b\dagger}$ \\
      \midrule
        $^{86}$Kr$^{22+}$  & \multirow{2}{*}{22+} & \multirow{2}{*}{\textbf{130}} & \multirow{2}{*}{0.48} & \multirow{2}{*}{62${}^{3a}$}
                           & 1 & 62 & 4.6 & 3 & 0.34 & 1.6 \\
        $^{86}$Kr$^{29+}$  & & & & & 0.20 & 16.1 & 0.90 & 3 & 0.34 & 0.31${}^{3b}$ \\
      \midrule
        $^{129}$Xe$^{39+}$ & \multirow{4}{*}{22+${}^{4}$} & \multirow{4}{*}{\textbf{176}} & \multirow{4}{*}{--} & \multirow{4}{*}{--}
                           & 0.23 & \textbf{34.7} & \textbf{1.09}${}^{4}$ & 1 & \textbf{0.47} & \textbf{0.51}${}^{4a}$ \\
        $^{129}$Xe$^{39+}$ & & & & & 0.23 & \textbf{34.7} & \textbf{1.09}${}^{4}$ & 7 & \textbf{0.14} & \textbf{0.16}${}^{4b}$ \\
        $^{129}$Xe$^{39+}$ & & & & & 0.23 & 34.7 & 1.44 & 5 & 0.34 & 0.49${}^{4c}$ \\
        $^{129}$Xe$^{40+}$ & & & & & 0.22 & 34.1 & 1.38 & 5 & 0.34 & 0.47${}^{4d}$ \\
      \midrule
        $^{208}$Pb$^{54+}$ & \multirow{2}{*}{29+${}^{5}$} & \multirow{2}{*}{\textbf{182}} & \multirow{2}{*}{\textbf{0.57}${}^{*}$} & \multirow{2}{*}{--}
                           & 0.17${}^{\ddagger}$ & \textbf{32.1} & \textbf{1.13} & 1 & \textbf{0.33} & \textbf{0.37}${}^{5a}$ \\
        $^{208}$Pb$^{54+}$ & & & & & 0.17${}^{\ddagger}$ & \textbf{32.1} & \textbf{1.13} & 8 & \textbf{0.20} & \textbf{0.23}${}^{5b}$ \\
      \bottomrule
        \multicolumn{11}{l}{\scriptsize{\makecell[l]{
        ${}^{1a}$Measured during LHC oxygen run in June-July 2025. \\
        ${}^{1b}$Using performance of $^{16}$O$^{4+}$ with the stripping efficiency from Baron's formula. \\
        ${}^{2a}$Measured on 6 Apr 2015. Linac~3 produced \SI{200}{\micro\second} pulses.\\
        ${}^{2b}$Using $^{40}$Ar$^{11+}$ performance and assuming 10\,\% more losses in LEIR due to multiple injections. \\
        ${}^{3a}$Based on test in June 2023: BCT15-NS calculated from the measured\\
        ~~~beam current of \SI{83}{\micro\ampere} in FC3, assuming 75\,\% transport efficiency. \\
        ${}^{3b}$Using performance of $^{86}$Kr$^{22+}$ with the stripping efficiency from Baron's formula. \\
        ${}^{4}$The ion source delivered $^{129}$Xe$^{22+}$. Linac~3 produced \SI{200}{\micro\second} pulses.\\
        ${}^{4a}$Measured on 12 Oct 2017.\\
        ${}^{4b}$Measured on 11 Oct 2017; test beam, LEIR cycle not optimised due to very limited time. \\
        ${}^{4c}$Based on the above, with a Linac~3 pulse length of \SI{260}{\micro\second} and an optimised LEIR cycle.\\
        ${}^{4d}$As 4c, with the stripping correction from Baron's formula.\\
        ${}^{5}$Measured between 17 and 25 Nov 2024. The ion source delivered $^{208}$Pb$^{29+}$.\\
        ${}^{5a}$EARLY cycle not maintained, resulting in degraded injection and cycle efficiency.\\
        ${}^{5b}$Operation at the intensity limit results in low LEIR injection and cycle efficiency.\\
        ${}^{*}$Measured separately on 19 Sep 2024.\\
        ${}^{\ddagger}$Derived from the measured BCT05 and BCT41 currents and $\varepsilon_{\text{Linac}}$.\\
        ${}^{\dagger}$A different charge state could be produced by the source to improve the current gain at stripping.}}}
    \end{tabular}
\end{table}

\pagebreak
 
\section{Key upgrades for HEARTS@LEIR}
\label{sec:SourceUpgrades}

The source-side scope of the project consists of two upgrades, both shared with the Ion Complex Upgrade (ICU) project~\cite{ICUproject} and both prerequisites for Specification~B. %
\begin{figure}[htb]
    \centering
    \includegraphics[width=0.85\linewidth]{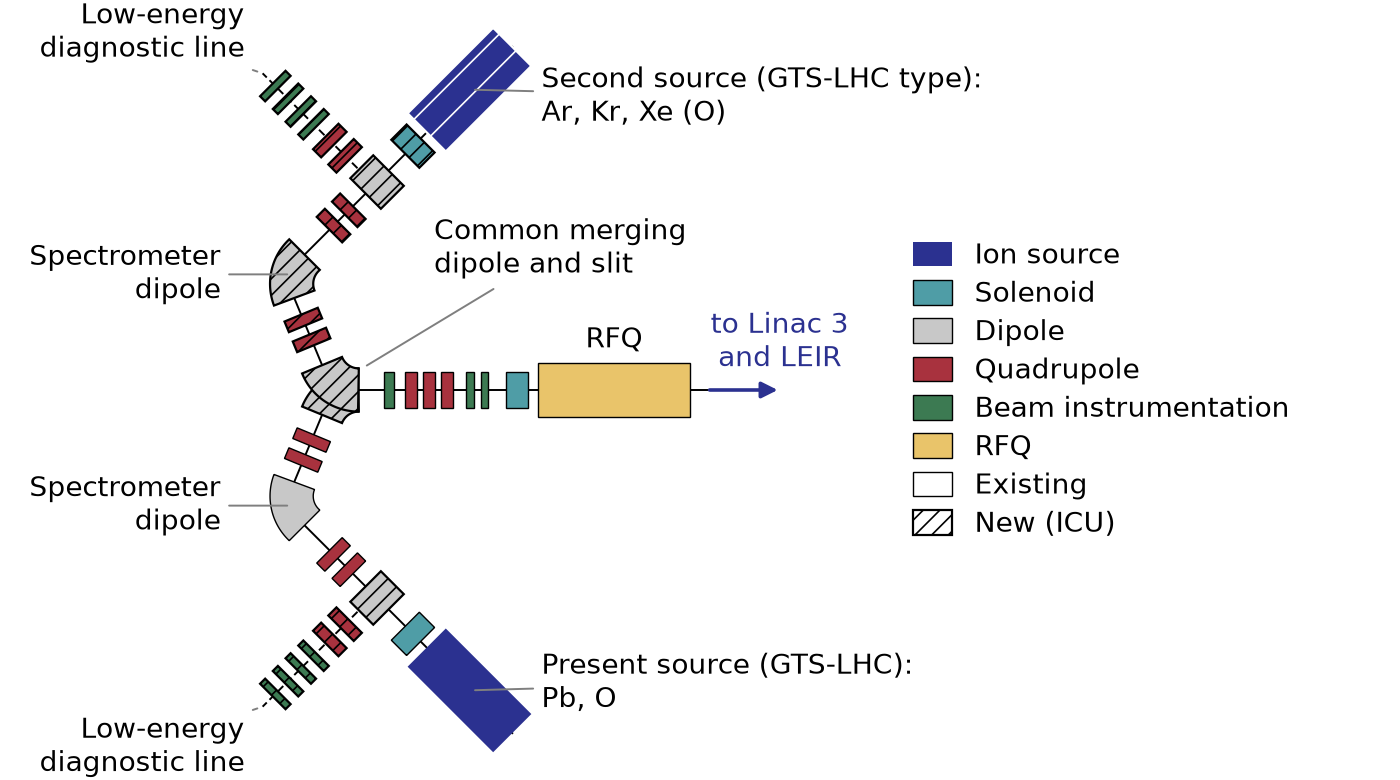}
    \caption{Schematic of the Linac~3 front end with a second ion source, as studied in the ICU project (not to scale, adapted from Ref.~\cite{Kuechler_2026}). A pulsed dipole after each source sends the pulses not used downstream into the low-energy diagnostic line of that source. The two branches merge in a common dipole and slit in front of the RFQ. Hatched: new equipment. The position and connection of the second source are indicative.}
    \label{fig:Linac3TwoSources}
\end{figure}

\subsection{Second ion source and its diagnostic line}
In the present baseline, a second GTS-LHC-type ECR source, based on the existing one with only modest modifications, is installed and connected to the existing RFQ through a new low-energy branch. The existing source continues to serve the Pb--O mixture required by the LHC ion programme, while the second source is dedicated to noble gases (Ar, Kr, Xe, with O available as buffer gas). Splitting the species between two sources removes the space-charge penalty and the plasma-chamber conditioning delays that make a single-source cocktail impractical for the HEARTS palette, and decouples the HEARTS@LEIR operating point from the lead performance required by the LHC. A possible layout of the front end with the two sources is shown in Fig.~\ref{fig:Linac3TwoSources}~\cite{Kuechler_2026}.

{\sloppy The new source comprises a set of five solenoids around the source: two at injection-side (IP2.SOLINJ), one at the center of the plasma chamber (IP2.SOLCEN), and two at extraction-side (IP2.SOLEXT). After extraction, the beamline consists of a solenoid (ITL2.SOL2), two spectrometer dipoles and a horizontal slit for charge-state selection. This new branch merges with the existing LEBT through the second spectrometer dipole and the slit is common with the existing branch. Two low-energy diagnostic branches, one on each source, allow the state of one source and the properties of its beam to be characterised without interrupting delivery from the other. This capability is what makes automatic re-optimisation of a species feasible while another species is being delivered downstream, and is therefore central to meeting the 15-minute switching specification.\par}

\subsection{Pulse-to-pulse modulation compatibility}

To enable fast switching between ion species and stable operation afterwards, the optimal matching of beam coming from a source, LEBT and RFQ need to be maintained, even for the ion species that are not requested downstream. Because the state of the source evolves continuously, the LEBT and RFQ matching needs to follow the resulting beam changes. Considering that the source pulses at 10~Hz, but Linac~3 at 5 Hz, every second pulse is unused downstream. We propose to convert all the equipment between each source and the RFQ to PPM (Specification~B in Chapter~\ref{sec:BeamSpecs}) to devote the presently unused pulses to concurrent measurement and automatic optimisation of the ion beams that are not at the moment requested and delivered. The scope, and the split of responsibilities with the ICU project, will be fixed during the TDR phase.

\section{Resource and cost estimate}
\label{sec:IonSourceCost}

The materials required for the second source, its low-energy branch with the diagnostic line, and the pulse-to-pulse modulated equipment of the low-energy line are listed in Table~\ref{tab:IonSourceCost}. They form work package~WP2 of the consolidated estimate in Chapter~\ref{sec:ResourceEstimate}, which also gives the corresponding personnel, \SI{16.1}{\PY} of Graduate-like and \SI{14.4}{\PY} of staff effort (Table~\ref{tab:res_wp}). The vacuum system and the handling equipment of the source area, described in Chapter~\ref{sec:InfrastructureAndIntegration}, are included here so that the table covers the complete low-energy installation; the cooling sub-station of the sources is booked with the cooling and ventilation of the facility (work package~WP1).

\begin{table}[htbp]
    \centering
    \caption{Preliminary cost estimate for materials and hardware of the second ion source and of the Linac~3 low-energy line. The total corresponds to work package~WP2 of Table~\ref{tab:res_wp}.}
    \label{tab:IonSourceCost}
    \small
    \begin{tabular}{llr}
    \toprule
    Item & Group & Cost (kCHF) \\
    \midrule
    \multicolumn{3}{l}{\textit{Second ion source}}\\
    \quad Microwave generator system & SY-RF & 910 \\
    \quad Source mechanics, electrical and miscellaneous equipment & BE-ABP & 319 \\
    \quad Source solenoids and permanent hexapole & TE-MSC & 200 \\
    \midrule
    \multicolumn{3}{l}{\textit{Low-energy beam transport and diagnostic lines}}\\
    \quad Dipoles: spectrometer, switching and \SI{90}{\degree} bend & TE-MSC & 310 \\
    \quad Solenoids, quadrupoles and correctors & TE-MSC & 385 \\
    \quad Power converters, pulse-to-pulse modulated & SY-EPC & 921 \\
    \quad Beam instrumentation (BPMs, BCT, Faraday cup) & SY-BI & 170 \\
    \quad Slits and stripper modification & SY-STI & --$^{*}$ \\
    \midrule
    \multicolumn{3}{l}{\textit{Services of the source area}}\\
    \quad Vacuum systems of the source and Linac~3 & TE-VSC & 718 \\
    \quad Handling, installation and tooling & EN-THE & 112 \\
    \midrule
    \textbf{Total} & & \textbf{4\,045} \\
    \bottomrule
    \end{tabular}\\[2pt]
    \footnotesize{$^{*}$ Not yet quantified.}
\end{table}

\chapter{The LEIR synchrotron}
\label{sec:LEIRaccelerator}

LEIR is the element of the chain that converts the Linac~3 pulses characterised in Chapter~\ref{sec:IonSources} into the low-emittance, high-intensity and high energy beam required at the entrance of the extraction system. This chapter recalls the machine layout and its nominal optics, and then examines the two processes that set the achievable intensity for each HEARTS@LEIR species: multi-turn accumulation and electron cooling.

\begin{figure}[h!]
    \centering
    \includegraphics[width=0.85\linewidth]{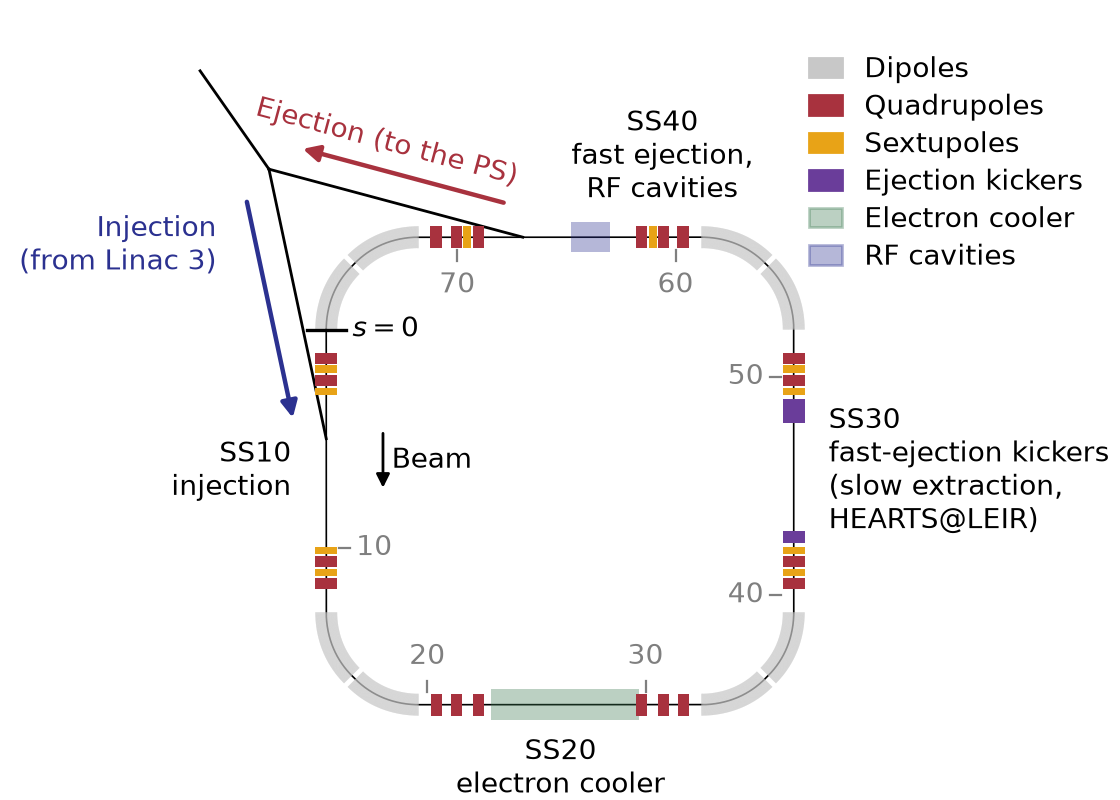}
    \caption{Layout of LEIR from the survey of the 2026 LEIR model, with the beam circulating counter-clockwise. The straight sections are numbered in the direction of the beam from SS10, the injection, to SS40, the fast ejection. SS20 houses the electron cooler, and SS30 the fast-ejection kickers and the space foreseen for the slow-extraction equipment of HEARTS@LEIR (Chapter~\ref{sec:SRE}). The marks inside the ring give the longitudinal coordinate $s$ in metres used in Fig.~\ref{fig:LEIR-NomOptics}. The injection and ejection branches of the common transfer line are drawn schematically~\cite{LEIRDipoleModel2026}.}
    \label{fig:LEIRSchematic}
\end{figure}

\section{Machine layout and nominal operation}

The Low Energy Ion Ring (LEIR) is a key element of the LHC ion injector chain. 
Its main function is to accumulate and compress multiple Linac 3 pulses into short ($\sim$200 ns) bunches.
Currently, eight \SI{260}{\micro\second} pulses are injected via a multi-turn 6D painting scheme and merged using electron cooling. 
The cooler simultaneously reduces transverse and longitudinal emittances and slightly decelerates the incoming beam, allowing efficient merging with previously accumulated bunches and freeing phase space for the next injection~\cite{AccumulationLeadIons}.
Once cooling and stacking are complete, the beam is adiabatically captured in RF buckets and accelerated from 4.2 MeV/nucleon to 72.1~MeV/nucleon at \SI{4.8}{\tesla\metre} (for the $^{208}$Pb$^{54+}$ reference case).

\begin{figure}
    \centering
    \includegraphics[width=\linewidth]{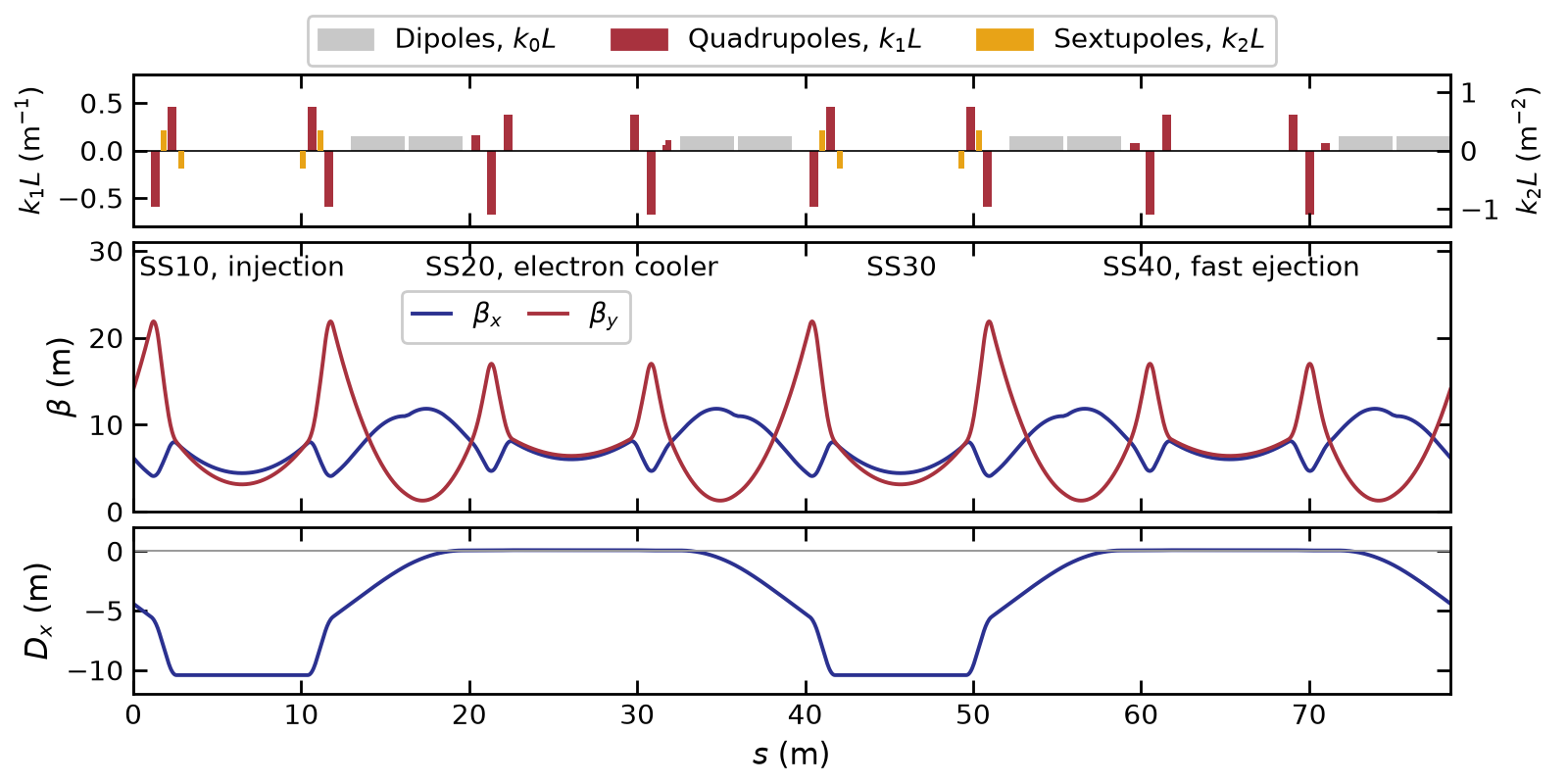}
    \caption{Nominal beam optics of LEIR for accumulation and cooling, computed with the 2026 LEIR model for the quadrupole strengths of the injection plateau, matched to the nominal tunes $Q_x = 1.82$ and $Q_y = 2.72$. Top: positions of the bending magnets (grey) and integrated strengths of the quadrupoles (red, left axis) and of the sextupoles for chromaticity correction (yellow, right axis, nominal settings). Middle: horizontal and vertical $\beta$-functions, with the straight sections indicated. Bottom: horizontal dispersion~\cite{LEIRDipoleModel2026}.}
    \label{fig:LEIR-NomOptics}
\end{figure}

Fast ejection is facilitated by a set of pulsed kickers installed upstream of the magnetic extraction septum. A schematic of the machine is depicted in Fig.~\ref{fig:LEIRSchematic}.
A detailed technical overview of LEIR's operations and hardware is available in \cite{LHCDesignReport}.
The nominal beam optics of the ring is optimized for accumulation and cooling operation.
These are presented in Fig.~\ref{fig:LEIR-NomOptics}.

\begin{figure}
    \centering
    \includegraphics[width=\linewidth]{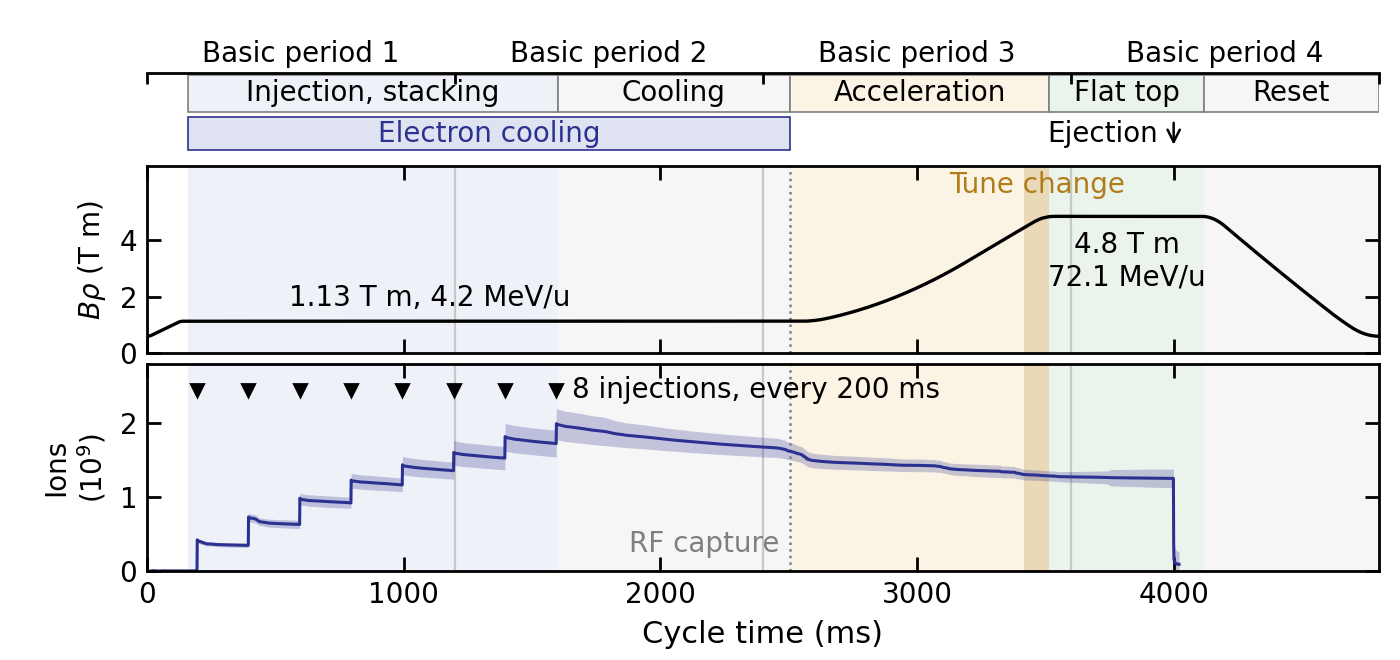}
    \caption{The LEIR cycle of four basic periods (4BP) used for the slow-extraction test in 2026, set-up~(b) (Section~\ref{sec:MT4BP}). Top: phases of the cycle and basic periods of \SI{1.2}{\second}. Middle: programmed magnetic rigidity $B\rho$. Bottom: Pb$^{54+}$ intensity measured by the ring beam-current transformer, mean and RMS spread of \num{1448} cycles of 26 to 28~August~2026; the triangles mark the eight injections. Dotted line: RF capture. Shaded band: change of the working point, in the last \SI{100}{\milli\second} before the flat top, from $(Q_x, Q_y) \approx (1.83, 2.72)$, for accumulation and cooling, to $(1.70, 2.80)$ on the flat top, as measured by the tune monitor~\cite{FourBP2026}.}
    \label{fig:LEIRCycle}
\end{figure}

\paragraph{The LEIR cycle.} Each LEIR cycle lasts an integer number of basic periods of \SI{1.2}{\second}, the time unit of the CERN injector supercycle. Figure~\ref{fig:LEIRCycle} shows the sequence on the cycle of four basic periods (4BP) prepared in 2026 for the slow-extraction test (Section~\ref{sec:MT4BP}), with the programmed magnetic rigidity and the Pb$^{54+}$ intensity measured by the ring beam-current transformer. At the start of the cycle, the main magnets rise to the injection plateau at \SI{1.13}{\tesla\metre} (4.2~MeV/nucleon), which they reach after \SI{0.16}{\second}. Eight Linac~3 pulses are then injected every \SI{200}{\milli\second}, between 195 and \SI{1595}{\milli\second}. Electron cooling merges each pulse with the beam already circulating before the next one arrives, so the intensity grows in steps, to about $2\times10^{9}$ ions after the last injection. Cooling continues until the beam is adiabatically captured by the RF system at \SI{2505}{\milli\second}. The beam is then accelerated in \SI{0.93}{\second} to \SI{4.8}{\tesla\metre} (72.1~MeV/nucleon). In the last \SI{100}{\milli\second} of the ramp, just before the flat top, the working point is moved from the one used for accumulation and cooling, $(Q_x, Q_y) \approx (1.83, 2.72)$, to that of the flat top, $(1.70, 2.80)$, close to the third-integer resonance $3Q_x = 5$ used for slow extraction (Chapter~\ref{sec:SRE}). The flat top lasts about \SI{0.6}{\second}; in the test, the beam, about $1.3\times10^{9}$ ions, was ejected by the fast-extraction kickers at \SI{4000}{\milli\second}, which leaves \SI{0.5}{\second} of flat top for an extraction. In the last \SI{0.7}{\second} of the cycle, the magnets are ramped down to the level at which the next cycle starts. The operational cycles for the LHC, EARLY (two basic periods) and NOMINAL (three basic periods), follow the same sequence with a shorter injection plateau and a flat top of about \SI{0.1}{\second}, during which the beam is ejected towards the PS (Fig.~\ref{fig:MTDipoleCurrent})~\cite{LHCDesignReport, FourBP2026}. The operational LHC cycles do not require a tune change.

\section{Beam stacking and cooling}

As previously noted, beam accumulation with reduced emittance is achievable only at LEIR, owing to its electron cooling capability.
Table~\ref{tab:IonIntDetails} presents the expected intensities for the ion candidates considered for HEARTS@LEIR.

While the existing ion source can meet the requirements for a Pb–O mixture, HEARTS@LEIR operation will depend on multi-pulse accumulation for noble gases (Ar, Kr, Xe).\
A critical advantage of LEIR is its ability to reduce transverse and longitudinal emittance through beam cooling, which can significantly enhance the efficiency of resonant slow extraction.\
However, cooling times for ions lighter than $^{208}$Pb$^{54+}$ are expected to increase, as cooling times scale with $M/Q^2$ (where $Q$ is the charge state and $M$ the ion mass).\ 
The relative cooling times for these ions are detailed in Table~\ref{tab:RelCoolingTimes}.\

Measurements of single $^{208}$Pb$^{54+}$ injections show that \SI{95}{\percent} of the excess emittance is removed within 270 to \SI{300}{\milli\second} after injection, and that the emittance is within \SI{10}{\percent} of its equilibrium after \SI{0.36}{\second} vertically and \SI{0.7}{\second} horizontally (Section~\ref{sec:MTPb}); over the measured range the cooling rate is nearly independent of the amplitude. This cooling-down time, the quantity $3\tau^{\text{e-cool}}$ of the I-LHC design report (an aim of \SI{200}{\milli\second}, against \SI{400}{\milli\second} demonstrated in 1997~\cite{LHCDesignReport}), is well within the injection plateau of the HEARTS@LEIR cycles.\

\begin{table}[h]
    \centering
    \caption{Relative cooling time of proposed ion species for HEARTS@LEIR compared to $^{208}$Pb$^{54+}$. The last column gives the measurement of a pilot run with Ne$^{5+}$ in 2025 (not in the HEARTS baseline), compared with Pb$^{54+}$ over the same range of vertical emittance (see Section~\ref{sec:MTLightIons}).}
    \begin{tabular}{lcc}
        \toprule
        Ion & Scaled relative & Measured relative \\
            & cooling time    & cooling time \\
        \midrule
         $^{129}$Xe$^{40+}$ & 1.13 & --- \\
         $^{86}$Kr$^{29+}$  & 1.43 & --- \\
         $^{40}$Ar$^{16+}$  & 2.19 & --- \\
         $^{16}$O$^{8+}$    & 3.5  & --- \\
        \midrule
         $^{20}$Ne$^{5+}$   & 11.2 & 2.4 to 8 \\
        \bottomrule
    \end{tabular}
    \label{tab:RelCoolingTimes}
\end{table}

Pilot measurements conducted in 2025 with O$^{4+}$ give, from the full width at half maximum of the profiles, normalised transverse emittances between about 290 and \SI{630}{\nano\metre\radian} horizontally and between about 265 and \SI{570}{\nano\metre\radian} vertically on the ramp, preserved during acceleration, for $8.6\times10^{9}$ ions (Section~\ref{sec:MTLightIons}); they can be compared with the emittances after the RFQ in Fig.~\ref{fig:IonEmittAfterRFQ}.\
A dedicated experimental campaign recently investigated the intensity dependence of transverse beam emittances for $^{208}$Pb$^{54+}$: the normalised RMS emittances increase linearly with intensity and reach $\epsilon_{x,n} \approx \SI{245}{\nano\metre\radian}$ and $\epsilon_{y,n} \approx \SI{165}{\nano\metre\radian}$ at $1.5\times10^{9}$ ions on the injection plateau (Section~\ref{sec:MTPb}).\

These results provide a robust foundation for resonant slow extraction beam dynamics studies, as the initial beam conditions are now well-defined. 
These results suggest that cooling of O$^{8+}$ may not be required at all, or, if needed, is expected to be achievable within a cooling-down time of about \SI{0.4}{\second}, scaled with $Q^2/A$ from the Ne$^{5+}$ measurement, or about \SI{1}{\second}, scaled from Pb$^{54+}$; both are within the accumulation plateau of more than \SI{1.5}{\second} (Section~\ref{sec:MTLightIons}).\
The intermediate ions are expected to pose no additional challenge so far.\
However, it remains to be verified whether injection at low rigidity (equivalent to 4.2 MeV/nucleon) can be achieved for all ion species, since all proposed ions are less rigid at injection than $^{208}$Pb$^{54+}$.\

To fully unlock the facility’s potential, a detailed study of beam dynamics, covering accumulation, cooling, and RF-capture, should be prioritized in future phases of the project.\
This will establish precise intensity limits and could reveal advanced operational modes, enabling higher throughput and broader testing capabilities for both industrial partners and academic institutions. Such optimizations would directly enhance the facility’s versatility, efficiency, and value as a leading resource for radiation-effects testing.

\chapter{Beam extraction from LEIR to the experimental area}
\label{sec:SRE}
With the initial beam conditions established, we now turn to the extraction process from LEIR to the irradiation area. To fulfill the requirements outlined in Chapter~\ref{sec:BeamSpecs}, various ion species will be extracted using the resonant slow extraction technique, operated near the third-order resonance and driven by sextupoles~\cite{TOMIZAWA1993}.\
This section provides a detailed implementation overview of the resonant slow extraction scheme, ensuring precise and controlled delivery of diverse ion species.

The discussion begins with the configuration of the extraction optics at LEIR, followed by a conceptual description of the integration of the required components for the beam extraction and delivery systems.\
Beam optics calculations are performed with MADX~\cite{Grote:2003}, if not stated otherwise. Particle tracking simulations are carried out with Xsuite~\cite{Iadarola:2023fuk}.

\section{Resonant slow extraction}

\subsection{Beam optics}
The resonant slow extraction technique, operated near the third-order resonance and driven by dedicated harmonic sextupoles, has a long-standing tradition in synchrotron-based irradiation facilities. It is the method of choice for both heavy-ion cancer therapy centers and many fixed-target experiments.

\begin{figure}
    \centering
    \includegraphics[width=0.6\linewidth]{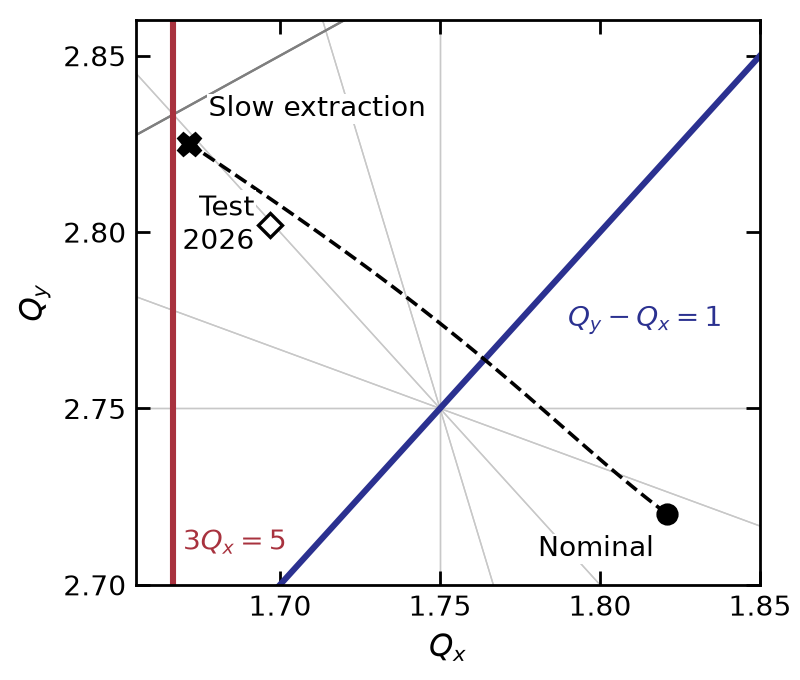}
    \caption{Tune excursion from the nominal working point, $(Q_x, Q_y) = (1.82, 2.72)$, to the slow-extraction working point, $(1.672, 2.825)$, close to the third-order resonance $3Q_x = 5$ (red), for a linear ramp of the five quadrupole families between the strengths of Table~\ref{tab:quadStrengthsSRE}; grey: resonances up to fourth order. On the way, the working point crosses the coupling resonance $Q_y - Q_x = 1$ (blue), which is excited by the solenoids of the electron cooler and corrected with the available skew quadrupoles. Open diamond: working point on the flat top in the slow-extraction test of 2026 (Section~\ref{sec:MTSX}).}
    \label{fig:tuneExcursion}
\end{figure}

\begin{figure}
    \centering
    \includegraphics[width=\linewidth]{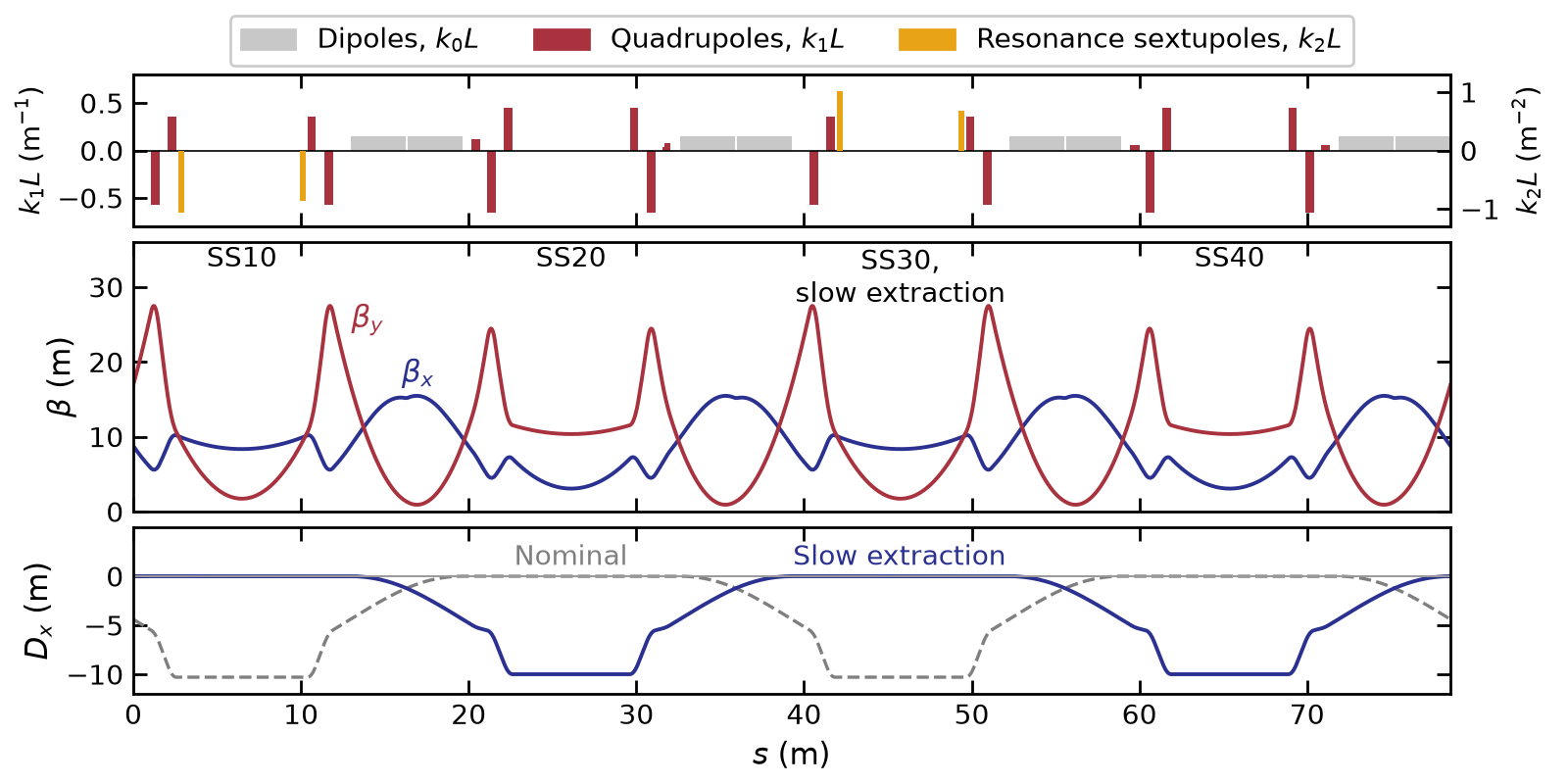}
    \caption{Optics functions of LEIR for slow extraction ($Q_x = 1.672$, $Q_y = 2.825$), design model with the quadrupole strengths of Table~\ref{tab:quadStrengthsSRE}. Top: bending magnets (grey) and integrated strengths of the quadrupoles (red, left axis) and of the resonance sextupoles set for $S = S^*$ (yellow, right axis, Table~\ref{tab:SREsext}). Middle: horizontal and vertical $\beta$-functions. Bottom: horizontal dispersion for slow extraction (solid) and for the nominal optics (dashed). The dispersion wave is shifted by one sector: the dispersion vanishes in SS10 and SS30 and reaches its extremum in SS20 and SS40.}
    \label{fig:LEIR-SREOptics}
\end{figure}

At LEIR, two modifications to the nominal optics are envisaged:
\begin{itemize}
    \item Working point adjustment: The tune is brought close to the third-order resonance, while maintaining at least the same distance to the coupling resonance as in the nominal configuration. The corresponding tune excursion is illustrated in Fig.~\ref{fig:tuneExcursion}.
    
    \item Dispersion-wave shift: The dispersion is shifted by one sector, providing low-dispersion regions at straight sections 10 and 30 (SS10, SS30), as shown in the bottom panel of Fig.~\ref{fig:LEIR-SREOptics}. This configuration is advantageous because the stronger sextupoles used for chromaticity correction act as harmonic sextupoles, enabling control of the resonance driving term. The optics calculations give zero dispersion at the extraction point, the electrostatic septum in SS30 (Fig.~\ref{fig:LEIR-SREOptics}).
\end{itemize}

It is sufficient to adjust the strengths of the focusing and defocusing elements in LEIR to implement the two proposed optics modifications. The required strengths are listed in Table~\ref{tab:quadStrengthsSRE}. The changes are marginal, with the necessary (de)focusing strengths generally lower than nominal, except for the QF2344 quadrupoles.
The resulting horizontal optical functions are displayed in Fig.~\ref{fig:LEIR-SREOptics}.

\begin{table}[]
    \centering
    \caption{ Comparison of quadrupole strengths for preparing LEIR slow-extraction optics.}
    \begin{tabular}{c c c}
    Quadrupole & $k_{\text{start}}$ & $k_{\text{end}}$\\
    family & (m$^{-2}$) & (m$^{-2}$) \\
     \hline\hline
        QD1030 & -1.130 & -1.129\\ 
        QF1030 & 0.904 & 0.704\\
        QD2040 & -1.319 & -1.283\\
        QF2040 & 0.311 & 0.226\\
        QF2344 & 0.715  & 0.884
    \end{tabular}
    \label{tab:quadStrengthsSRE}
\end{table}

The resonance driving term (RDT) can be excited using the sextupoles located in SS10 and SS30. The amplitude $S$ and phase $\phi_S$ of the RDT can be fully controlled via the independently powered focusing sextupoles XFN11, XFN12, XFN31 and XFN32; the defocusing sextupoles XDN11, XDN12, XDN31 and XDN32 in the same straight sections could in principle be used as well. 
The strength needed and the corresponding currents are derived in Section~\ref{sec:SREsext}.

It is foreseen to drive the extraction transversely using the radio-frequency knock-out (RF-KO) technique, by applying a resonant excitation signal through the transverse feedback system. This technique decouples the extracted rate from the magnetic cycle, has proven operationally robust to changes of the extracted momentum, and provides an actuator for a spill feedback loop. The requirements on the LEIR transverse feedback for RF-KO operation are already foreseen within the scope of the consolidation of the transverse feedback system. The excitation scheme, its bandwidth and the feedback architecture will be specified during the technical design phase; the associated risk is carried as~T9 in Chapter~\ref{sec:RiskAssessment}.

\subsection{Strength of the resonance sextupoles}
\label{sec:SREsext}

\begin{figure}[htbp]
    \centering
    \includegraphics[width=\linewidth]{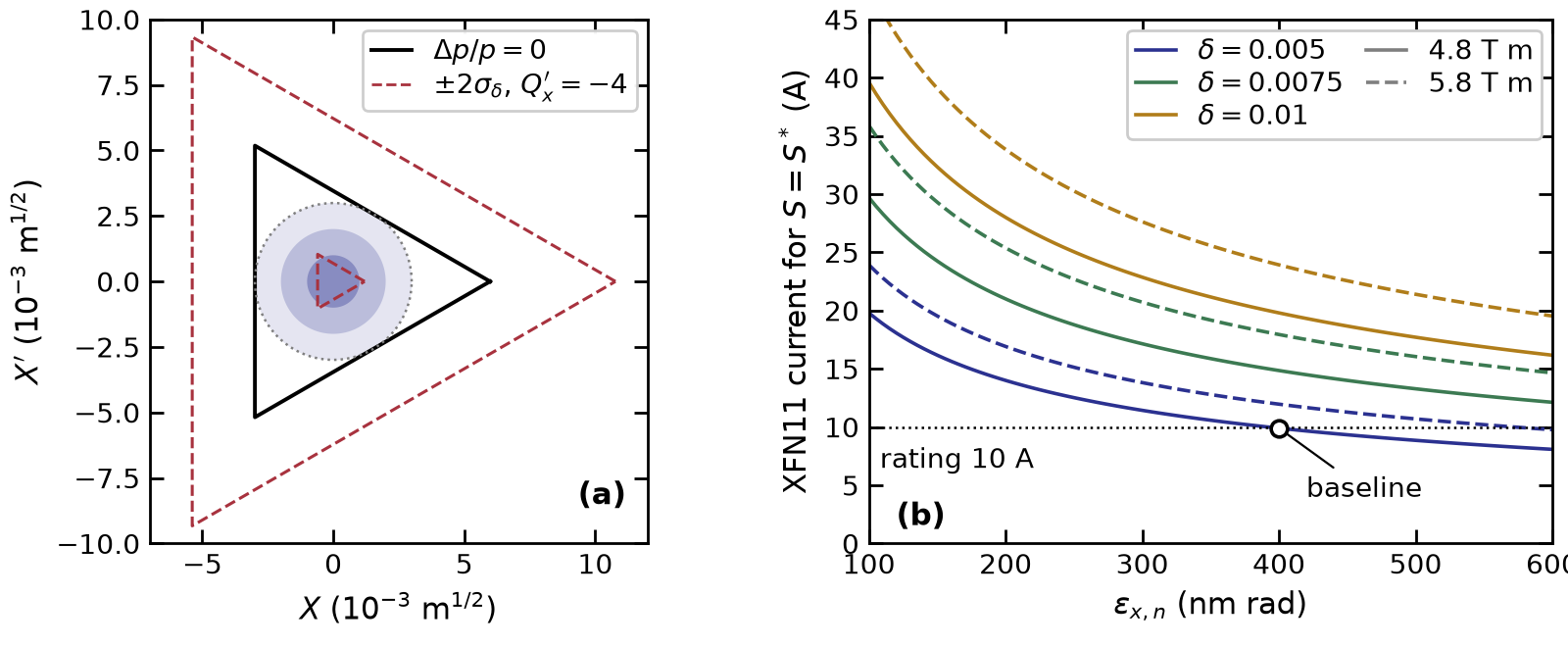}
    \caption{Strength of the resonance sextupoles. (a)~Stable region of Eq.~\eqref{eq:SREKobayashi} for the baseline ($\delta = 0.005$, $S = S^* = \SI{21}{m^{-1/2}}$, black), with its inscribed circle (dotted) and the 1, 2 and 3$\sigma$ circles of the beam ($\epsilon_n = \SI{0.4}{\micro\metre}$, \SI{4.8}{\tesla\metre}); dashed: stable regions of the particles at $\pm2\sigma_\delta$ ($\sigma_\delta = 5\times10^{-4}$) for an uncorrected chromaticity $Q'_x = -4$. (b)~Current of the most loaded sextupole, XFN11, for $S = S^*$ as a function of the normalised horizontal emittance, for three tune distances at 4.8 and \SI{5.8}{\tesla\metre}, with the present rating.}
    \label{fig:SREsext}
\end{figure}

Close to the third-integer resonance $Q_x = 5/3$, the horizontal motion can be effectively described in normalised coordinates $X = x/\sqrt{\beta_x}$, $X' = (\alpha_x x + \beta_x x')/\sqrt{\beta_x}$ by the Kobayashi Hamiltonian
\begin{equation}
H = 3\pi\delta\,(X^2 + X'^2) + \frac{S}{4}\,(3XX'^2 - X^3),
\label{eq:SREKobayashi}
\end{equation}
with
\begin{equation}
\delta = Q_x - \tfrac{5}{3}, \qquad 
S\,e^{3i\phi_S} = \sum_n \tfrac12\,\beta_{x,n}^{3/2}\,(k_2L)_n\,e^{3i\mu_{x,n}},\nonumber
\end{equation}
where $S$ is the strength of the equivalent virtual sextupole, $(k_2L)_n = B''L/B\rho$ the integrated strength of sextupole $n$ and $\mu_{x,n}$ its phase advance in units of $2\pi$. The stable region is an equilateral triangle with the unstable fixed points at $R = 8\pi\delta/S$ from the origin; its inscribed circle has the radius $h = R/2 = 4\pi\delta/S$ (Fig.~\ref{fig:SREsext}a).

Therefore the beam is held inside the triangle if the circle of three rms beam sizes, of radius $3\sqrt{\epsilon}$, which contains \SI{98.9}{\percent} of a Gaussian beam, fits into the inscribed circle:
\begin{equation}
S \le S^* = \frac{4\pi\delta}{3\sqrt{\epsilon}} .
\label{eq:SREStar}
\end{equation}
A weaker $S$ also holds the beam, but in a larger triangle, so that the excitation has to drive the particles further and the separatrix reaches the septum at a larger amplitude. The extraction is therefore operated at $S \approx S^*$. For the baseline, $\delta = 0.005$ and $\epsilon_n = \SI{0.4}{\micro\metre}$, i.e.\ a geometric emittance $\epsilon = \SI{1.0}{\micro\metre}$ at \SI{4.8}{\tesla\metre} ($\beta\gamma = 0.40$), Eq.~\eqref{eq:SREStar} gives $S^* = \SI{21.0}{m^{-1/2}}$.

\paragraph{Momentum spread and chromaticity.} Equation~\eqref{eq:SREStar} holds for each momentum separately: through the chromaticity $Q'_x$, a particle with the momentum offset $\Delta p/p$ sees the tune distance $\delta + Q'_x\,\Delta p/p$, and the size of its stable triangle changes in proportion (Fig.~\ref{fig:SREsext}a). For the condition to hold across the momentum distribution, $\delta$ in Eq.~\eqref{eq:SREStar} is replaced by the smallest tune distance within it, $\delta - |Q'_x|\,n\,\sigma_\delta$, with $n \approx 2$. For a momentum spread $\sigma_\delta$ of about $5\times10^{-4}$ on the flat top and an uncorrected chromaticity of a few units, as in the present machine (Chapter~\ref{sec:MachineTests}), $|Q'_x|\,2\sigma_\delta$ is comparable with the baseline distance. The horizontal chromaticity therefore has to be corrected, to $|Q'_x| \lesssim 0.5$ for a reduction of $S^*$ below \SI{10}{\percent}.

XFLS41 and XFLS42 are skew sextupoles and do not act on the linear chromaticity, and the pole-face windings of the main dipoles are not foreseen for this purpose since they also change the optics. The correction therefore has to come from the harmonic sextupoles, through a residual dispersion at their location: the design optics has none in SS10 and SS30, but the optics established in 2026 has about \SI{-1.5}{\metre} there (\SI{-1.8}{\metre} measured in SS30, Section~\ref{sec:MTOptics}). With the present rating of \SI{10}{\ampere}, which the setting $S = S^*$ alone already uses up (Table~\ref{tab:SREsext}), the harmonic sextupoles cannot provide the chromaticity correction in addition. Full control of both the chromaticity and the resonance driving term requires sextupoles with a higher rating; a rating of \SI{20}{\ampere} is sufficient in the model.

\paragraph{Settings and operating range.} The four harmonic sextupoles XFN11, XFN12, XFN31 and XFN32 are individually powered, so that the amplitude and the phase of $S$ can be set independently. With the calibration $B''L/I = \SI{0.52}{\tesla\per\metre\per\ampere}$, the current is $I = |k_2L|\,B\rho/0.52$. Table~\ref{tab:SREsext} gives the settings for $S = S^*$ with the orientation of the triangle used in the extraction simulations. The most loaded sextupole, XFN11, needs \SI{9.9}{\ampere} at \SI{4.8}{\tesla\metre}, within its rating of \SI{10}{\ampere}, and \SI{11.9}{\ampere} at \SI{5.8}{\tesla\metre}.

The defocusing sextupoles XDN11, XDN12, XDN31 and XDN32 remain available but add little: with the dispersion vanishing in SS10 and SS30 they have no lever on the chromaticity, and, being within about \SI{20}{\degree} in resonance phase of the neighbouring harmonic sextupoles of Table~\ref{tab:SREsext}, they can share the load of these but provide almost no independent control of the phase of~$S$, while their larger vertical $\beta$-function makes them drive vertical resonances more strongly.
In the baseline settings their strengths are therefore set to zero. They are kept as a reserve should a larger resonance driving term be needed than foreseen in Table~\ref{tab:SREsext}, which already brings XFN11 close to its \SI{10}{\ampere} rating: if the working point cannot be set as close as $\delta = 0.005$ to the resonance, or if the horizontal chromaticity cannot be fully corrected, so that $\delta$ has to be increased by about $|Q'_x|\,2\sigma_\delta$ to keep the whole momentum distribution away from the resonance. Since $S^*$ is proportional to $\delta$, the required currents grow in proportion. With the four defocusing sextupoles powered as well, the attainable $S$ increases by a few tens of percent, depending on the orientation of the triangle; this covers a moderately larger tune distance, but not an uncorrected chromaticity of the present machine ($Q'_x \approx -4.7$, Chapter~\ref{sec:MachineTests}), which would nearly double $S^*$.

\begin{table}[htbp]
\centering
\caption{Optics at the harmonic sextupoles in the slow-extraction optics ($Q_x = 1.672$, $L = \SI{0.335}{\metre}$), settings for $S = S^* = \SI{21}{m^{-1/2}}$ ($\delta = 0.005$, $\epsilon_n = \SI{0.4}{\micro\metre}$, chromaticity corrected) and corresponding currents at the baseline \SI{4.8}{\tesla\metre} and future upgrade scenario of \SI{5.8}{\tesla\metre}; rating \SI{10}{\ampere}.}
\label{tab:SREsext}
\begin{tabular}{cccccc}
\toprule
Sextupole & $\beta_x$ (m) & $\mu_x$ ($2\pi$) & $k_2L$ (m$^{-2}$) &  $I$ at \SI{4.8}{\tesla\metre} (A) & $I$ at \SI{5.8}{\tesla\metre} (A) \\
\midrule
XFN11 & 9.90 & 0.792 & $-1.070$ & 9.9 & 11.9 \\
XFN12 & 9.60 & 0.926 & $-0.860$ & 7.9 & 9.6 \\
XFN31 & 9.90 & 1.628 & $+1.016$ & 9.4 & 11.3 \\
XFN32 & 9.60 & 0.091 & $+0.674$ & 6.2 & 7.5 \\
\bottomrule
\end{tabular}
\end{table}

Since $I \propto S^* B\rho \propto \delta\,B\rho/\sqrt{\epsilon_n}$, the current needed for $S = S^*$ grows with the tune distance and the rigidity and with decreasing emittance (Fig.~\ref{fig:SREsext}b). The present rating covers the baseline at \SI{4.8}{\tesla\metre}. A beam with a smaller emittance than the baseline, a larger tune distance or a higher rigidity needs more current; with the present rating such a beam is still held with $S$ below $S^*$, at the cost of a larger triangle and a stronger excitation. Optics errors change the contribution of each sextupole, which scales as $\beta^{3/2}$: a $\beta$-beating of \SI{25}{\percent} rms changes $S$ by about \SI{30}{\percent} for given currents, so that the settings have to be established with beam. The measured emittance, optics and sextupole effects that determine the operating point of the present machine are reported in Chapter~\ref{sec:MachineTests}. Since the chromaticity correction also has to be provided by these sextupoles (see above), a higher current rating is required; the thermal margin of the present sextupoles is discussed in Section~\ref{sec:MTFlattop}.

\subsection{Extraction efficiency}

\begin{figure}
    \centering
    \includegraphics[width=0.75\linewidth]{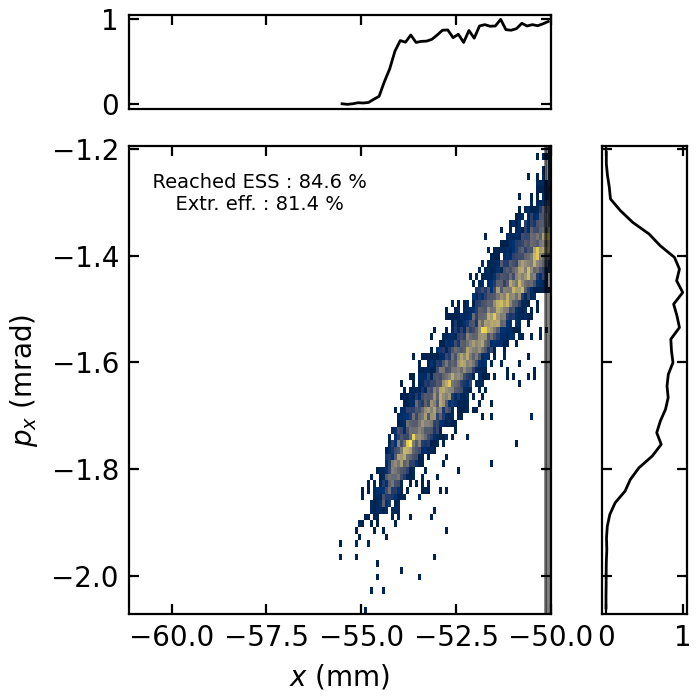} 
    \caption{Exemplary extracted beam distribution at electro-static septum blade. Blade thickness is \SI{200}{\micro\meter}. Upper and right panels illustrate the projection of the particle distribution. }
    \label{fig:lineOfChargeExample}
\end{figure}

\begin{figure}
    \centering
    \includegraphics[width=0.6\linewidth]{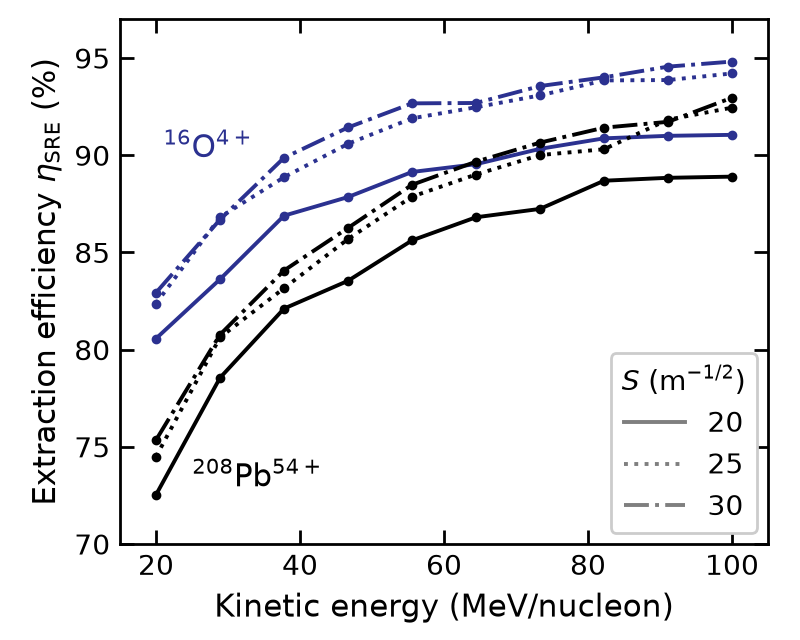} 
    \caption{Simulated extraction efficiency as a function of the kinetic energy for $^{16}$O$^{4+}$ (blue) and $^{208}$Pb$^{54+}$ (black), for three amplitudes $S$ of the resonance driving term (line styles), with normalised RMS emittances of 300~nm\,rad (O) and 400~nm\,rad (Pb).}
    \label{fig:extractionEfficiency}
\end{figure}

The extraction efficiency was evaluated through dedicated simulation campaigns using an updated optics model incorporating three extraction devices: a set of bumper magnets, a thin electrostatic septum, and a downstream magnetic septum placeholder. The machine is configured with extraction optics, and the bumper system induces a localized beam displacement toward the electrostatic septum, which imposes the most restrictive normalized aperture constraint in LEIR. The aperture model is taken from the LEIR optics repository~\cite{accmodelsLEIR}; a first beam-based aperture measurement (Section~\ref{sec:MTAperture}) confirmed its narrow elements, the injection septum and the entrance of SS20, but showed restrictions missing elsewhere, most likely in the bending magnets, and bounds the half-aperture at the loss point in SS30 between about 47 and \SI{75}{\milli\metre}. Locating the missing restrictions and measuring the SS30 aperture with the extraction bump are foreseen before the septum design is frozen (Section~\ref{sec:MTProgramme}).
A representative particle distribution at the electrostatic septum blade is shown in Fig.~\ref{fig:lineOfChargeExample}. An energy and resonance driving term (RDT) scan was performed to characterize the dependence illustrated in Fig.~\ref{fig:extractionEfficiency}. The normalized transverse RMS emittances are 400~nm\,rad for Pb ions and 300~nm\,rad for O ions.
Two effects govern the extraction efficiency in these simulations: (1) beam loss due to interaction with the septum blade, and (2) emittance truncation driven by the increase in RDT amplitude. High-action particles are lost as they cross the stability boundary defined by the RDT amplitude.
At high energies, the adiabatic damping reduces the geometric emittance enough to avoid truncation through RDT amplitude. Then, the extraction efficiency is dominated by optics settings.

\section{Conceptual integration of the extraction system}

The design and feasibility of the extraction system are grounded in the beam specifications outlined in Chapter~\ref{sec:BeamSpecs}.\ 
The primary objective is to identify a viable configuration of extraction devices that effectively directs extracted particles into a transfer beamline leading to the irradiation area.\
Given the constraint that main optical elements must remain unaltered, we evaluated the repositioning of the fast ejection kickers KFH3234 and its spare, KFH31. This adjustment aims to relax the technical requirements for the extraction devices.

A feasible configuration was identified, requiring the relocation of both KFH3234 and KFH31, along with a polarity change for KFH3234. However, to streamline implementation, we adopted a simpler, previously validated approach~\cite{BioLEIR}: modifying the KFH3234 tank to accommodate the vacuum pipe of the extraction channel.
This solution was verified and selected due to its resource efficiency and minimal impact on nominal operations, ensuring both practicality and reliability.

\begin{figure}[h]
    \centering
    \includegraphics[width=\linewidth]{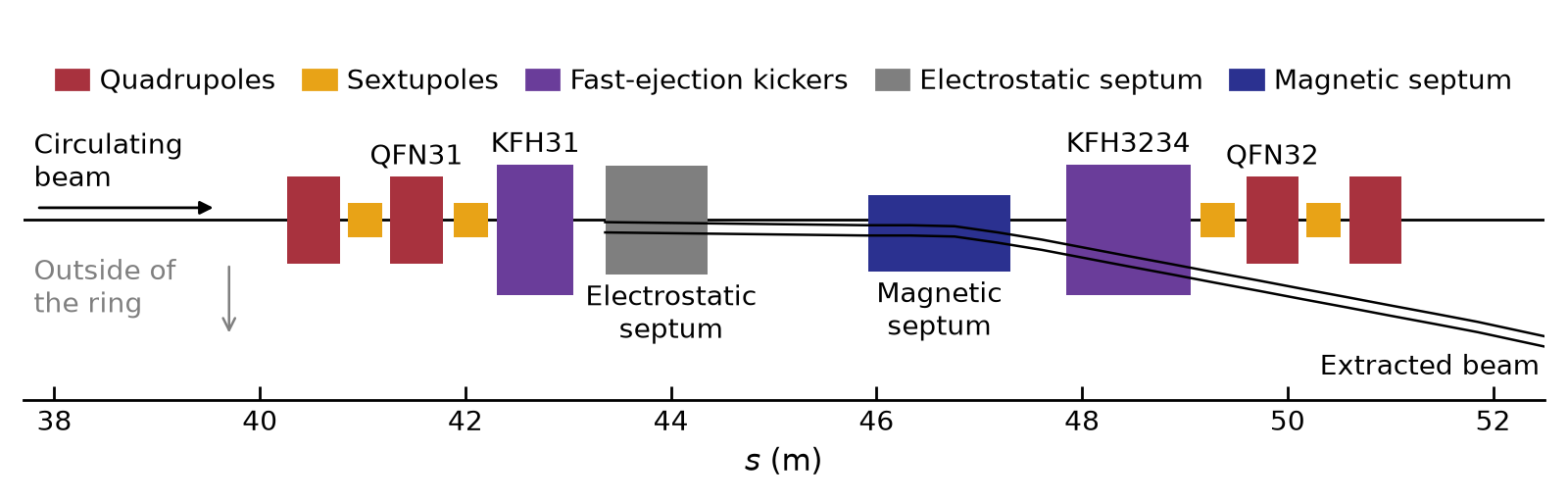}
    \caption{Straight section 30 (SS30) with the slow-extraction devices, to scale along the beam: quadrupoles (red) and sextupoles (yellow) of the lattice, fast-ejection kickers KFH3234 and its spare KFH31 (purple), electrostatic septum (grey) and magnetic septum (blue) with the physical lengths of Table~\ref{tab:SeptaSpecs}. Both directions are to scale. The extracted beam passes through the tank of KFH3234, which must be modified accordingly.}
    \label{fig:StraightSection30-LEIR}
\end{figure}

Figure~\ref{fig:StraightSection30-LEIR} illustrates the geometry of Straight Section 30 and the extraction devices.\ 
The positions of the fast ejection kickers are marked, and a trajectory line through the KFH3234 tank highlights the proposed modification for accommodating the extraction channel.\ Another possible approach is to produce a new tank for the kicker KFH3234.\ 
Based on these considerations, tentative specifications for the extraction devices are summarized in Table~\ref{tab:SeptaSpecs}. The design includes reserved space for:
\begin{itemize}
    \item Two Beam Position Monitors (BPMs) for precise beam alignment with the extraction devices and,
    \item Two orbit bumpers to create an extraction bump, ensuring that the extraction septa define the global horizontal aperture restriction.
\end{itemize}
The final specifications and detailed implementation will be confirmed during the technical design phase.\ 
A relocation downstream of spare kicker KFH31 is being considered to free space and relax the specifications for the magnetic septum; a test in 2026 showed that the fast extraction to the PS works without KFH31 (Section~\ref{sec:MTKFH}).
To aid commissioning and optimization of the resonant slow extraction system, three additional beam instrumentation systems are foreseen as baseline: two Beam Secondary Emission Grids (BSG) and one Beam Television (BTV).

\begin{table}[h]
    \centering
    \caption{Preliminary specifications of electrostatic (ES) and magnetic septa (MS). The beam rigidity at the conceptual level allows for an upgrade scenario where Pb$^{54+}$ reaches energies of 100 MeV/nucleon. This is not baseline and will be determined in future phases of the project.}
    \begin{tabular}{l c c c }
    Parameter & Unit & ES & MS \\
           \hline\hline
    Beam rigidity & Tm & \multicolumn{2}{c}{5.8}\\
    Bend angle & mrad & 5.4  & 180.3 \\
    B-field    &   T  & -  &  0.829   \\
    Integrated B-field &  Tm  & -  & 1.044 \\
    E-field    &   MV/m  & 5.85  & - \\
    Field length & mm  & 700  &  1259  \\
    Physical length & mm  & 1000 &  1385 \\
    Septum thickness &   mm  & 0.2  & 11 \\
    Horizontal full aperture & mm  & 34 &  30 \\
    Vertical full aperture & mm  & - &  20   \\
    Inductance  &  mH  & -  &  10.3 \\
    Resistance &  m$\Omega$   & -  & 0.41\\
    RMS power & kW & - & 12.6\\
    Water-cooling & l/min & - & 18.2\\
    \hline
    \end{tabular}
    \label{tab:SeptaSpecs}
\end{table} 

\section{Resource and cost estimate}
\label{sec:SREcost}

The hardware required for slow extraction comprises the two septa and their ancillary systems, the modification of the KFH3234 tank, two orbit bumpers with their power converters, and the beam instrumentation needed for commissioning and routine operation. The corresponding materials cost is given in Table~\ref{tab:ExtDevicesCost} and the personnel demand in Table~\ref{tab:ExtDevicesPersonel}. Both feed work package~WP3 of the consolidated estimate in Chapter~\ref{sec:ResourceEstimate}, where they are combined with the magnet and instrumentation effort attributed to TE-MSC and SY-BI.

\begin{table}[]
    \centering
    \caption{Preliminary cost estimate for materials and hardware for slow extraction at LEIR. The total corresponds to work package~WP3 of Table~\ref{tab:res_wp}.}
    \small
    \begin{tabular}{llr}
    \toprule
    Item & Group & Cost (kCHF) \\
    \midrule
    \multicolumn{3}{l}{\textit{Electrostatic septum ER.SEH31}}\\
    \quad Septum, spare parts and vacuum equipment & SY-ABT & 500 \\
    \quad Power supply, spares, resistors and HV cabling & SY-ABT & 250 \\
    \quad Controls & SY-ABT & 100 \\
    \midrule
    \multicolumn{3}{l}{\textit{Magnetic septum ER.SMH31}}\\
    \quad Septum and spare coils & SY-ABT & 400 \\
    \quad Vacuum chamber & SY-ABT & 100 \\
    \quad Power converter & SY-EPC & 500$^{*}$ \\
    \quad Controls and signal cabling & SY-ABT & 50 \\
    \midrule
    \multicolumn{3}{l}{\textit{Extraction bump and instrumentation}}\\
    \quad Two orbit bumpers & TE-MSC & 110 \\
    \quad Two power converters & SY-EPC & 122 \\
    \quad Two shoebox BPMs & SY-BI & 160 \\
    \quad Two BSGs and one BTV & SY-BI & 200 \\
    \quad Beam loss monitors (loan for tests) & SY-BI & 3 \\
    \midrule
    \multicolumn{3}{l}{\textit{Harmonic sextupoles}}\\
    \quad Upgrade of XFN11, XFN12, XFN31, XFN32 to \SI{20}{\ampere} & TE-MSC, SY-EPC & --$^{\dagger}$ \\
    \midrule
    \textbf{Total} & & \textbf{2\,495} \\
    \bottomrule
    \end{tabular}\\[2pt]
    \footnotesize{$^{*}$ A 14\,kA / 8\,V converter is required. Recovery of a unit from LHC spare stock is being investigated and would remove this item from the estimate.\\ $^{\dagger}$ Not yet quantified: the magnetic and thermal performance of the present sextupoles at \SI{20}{\ampere} is under study; the item covers either their modification or replacement and the corresponding power converters.}
    \label{tab:ExtDevicesCost}  
\end{table}

\begin{table}[]
    \centering
    \caption{Preliminary personnel estimate for the design, construction, installation and commissioning of the extraction devices, corresponding to the SY-ABT contribution. The complementary effort from SY-BI and TE-MSC is reported in Table~\ref{tab:res_personnel}.}
    \begin{tabular}{lrr}
    \toprule
    Activity & Staff (PY) & Graduate-like (PY) \\
    \midrule
    Beam dynamics          & 0.8 & 2.0 \\
    Septa                  & 3.0 & 3.0 \\
    Controls               & 0.3 & 1.0 \\
    Test and installation  & 0.3 & 1.0 \\
    Commissioning          & 0.5 & 2.0 \\
    \midrule
    \textbf{Total}         & \textbf{4.9} & \textbf{9.0} \\
    \bottomrule
    \end{tabular}
    \label{tab:ExtDevicesPersonel}
\end{table}

\chapter{Machine tests at LEIR}
\label{sec:MachineTests}
\chapterauthors{E.~C.~Cort\'es Garc\'ia, T.~Argyropoulos, D.~Bodart, S.~Deleval, E.~Effinger, J.~C.~Esteban Felipe, M.~A.~Fraser, R.~Garc\'ia Al\'ia, J.-F.~Gruber, O.~Hans, M.-A.~Jebramcik, S.~Joffe, G.~Le~Godec, S.~Morales Vigo, C.~Mutin, D.~Prelipcean, B.~Salvachua Ferrando, M.~S{\l}upecki, A.~Waets, M.~Widorski}

The conceptual design presented in the preceding chapters relies on a number of assumptions about the existing LEIR synchrotron. The main magnet circuits and the cooling plant must sustain flat tops several times longer than in present operation; the beam delivered to the flat top must have the intensity, emittance and momentum spread assumed in the extraction simulations of Chapter~\ref{sec:SRE}; the aperture and optics models used in these simulations must represent the real machine; and the beam must be driven into the third-integer resonance in a controlled way. All of these assumptions can be verified with beam in the present machine, before any hardware for HEARTS@LEIR is designed or procured.

A programme of machine tests addressing these points was carried out between May~2025 and the end of the run in August~2026. The tests used machine-development (MD) time and dedicated or cloned operational cycles, and required no modification of the machine hardware. 
The campaigns are summarised in Table~\ref{tab:MTOverview}. They are presented in the order in which the extraction process depends on them: the thermal envelope of the main circuits (Section~\ref{sec:MTFlattop}), the beam conditions on the flat top (Section~\ref{sec:MTEmittance}), the aperture (Section~\ref{sec:MTAperture}) and the optics (Section~\ref{sec:MTOptics}) of the slow-extraction configuration, the response of the beam to a transverse excitation (Section~\ref{sec:MTRFKO}) and, finally, a test of the slow-extraction process on a dedicated long-flat-top cycle (Section~\ref{sec:MTSX}). All measurements are analysed with the data recorded until the end of the 2026 run; where these data do not allow a firm conclusion, the limitation is stated, and the corresponding measurements are part of the programme of Section~\ref{sec:MTProgramme}.

The level of detail of this chapter is that needed to establish the technical feasibility. A reader interested in the conceptual design at a high level can read the overview of Table~\ref{tab:MTOverview} and the summary of the results in Section~\ref{sec:MTSummary}, and continue with Chapter~\ref{sec:Beamline2HEARTS}; Sections~\ref{sec:MTFlattop} to~\ref{sec:MTKFH} give the measurements and analyses on which the results rest.

\begin{table}[htbp]
\centering
\caption{Machine tests carried out at LEIR for HEARTS@LEIR between May~2025 and August~2026, with the main result of each; details in the sections given and in Section~\ref{sec:MTSummary}.}
\label{tab:MTOverview}
\footnotesize
\begin{tabular}{@{}>{\raggedright\arraybackslash}p{3.6cm}>{\raggedright\arraybackslash}p{2.0cm}>{\raggedright\arraybackslash}p{2.2cm}>{\raggedright\arraybackslash}p{5.6cm}c@{}}
\toprule
Test & Date & Cycle and beam & Main result & Section \\
\midrule
Flat-top extension of the main circuits, without beam & 22~Sep~2025 & NOMINAL with \SI{1.5}{\second} flat top & \SI{1.5}{\second} flat top within the converter protection; magnet temperatures as calculated & \ref{sec:MTFlattop} \\
Transverse emittance during accumulation and ramp & 22~May~2025 & EARLY, O$^{4+}$ & $8.6\times10^{9}$ ions after capture; emittance preserved on the ramp & \ref{sec:MTEmittance} \\
Electron cooling and emittance of a light ion & 8~Jul~2025 & 3BP, Ne$^{5+}$ & Cooling faster than the $Q^2/A$ scaling: O$^{8+}$ cooled within the accumulation plateau & \ref{sec:MTEmittance} \\
Electron cooling and emittance as a function of intensity, along the cycle & 14~Nov~2025, 7~Aug~2026 & MD cycle, 4BP, Pb$^{54+}$ & Single injection cooled in 0.4 (vertical) and \SI{0.7}{\second} (horizontal); flat-top emittances 180 and \SI{355}{\nano\metre\radian}, below the \SI{400}{\nano\metre\radian} assumed & \ref{sec:MTEmittance} \\
Beam-based horizontal aperture & 31~Oct, 3--4~Nov~2025 & dedicated cycle, Pb$^{54+}$ & Narrow elements of the aperture model confirmed; effective model for the rest of the ring; SS30 between 47 and \SI{75}{\milli\metre} & \ref{sec:MTAperture} \\
Slow-extraction working point & 8~Jul and 26~Sep~2025 & 3BP (Ne$^{5+}$), MDNOM (Pb$^{54+}$) & Reached, with the RF-capture efficiency of nominal operation & \ref{sec:MTOptics} \\
Optics of the slow-extraction configuration (orbit response, dispersion) & Sep~2025 -- Aug~2026 & MDNOM, 4BP, Pb$^{54+}$ & $\beta$-beating 14 to \SI{17}{\percent} rms; dispersion in SS30 \SI{-1.8}{\metre} (\SI{2.5}{\percent}) & \ref{sec:MTOptics} \\
Transverse blow-up with RF-KO excitation & 5 and 7~Aug~2026 & 4BP, Pb$^{54+}$ & About \SI{80}{\percent} of a \SI{0.5}{\second} spill at constant flux at the rated $2\times$\SI{100}{\watt} & \ref{sec:MTRFKO} \\
Slow-extraction test, including loss monitoring and radiation survey & Jun -- Aug~2026 & 4BP, Pb$^{54+}$ & Beam extracted from the flat top in a controlled and reproducible way, losses localised in SS30; $1.29\times10^{9}$ ions on the flat top & \ref{sec:MTSX} \\
Fast extraction with two kicker modules & 26~Aug~2026 & MDEARLY, Pb$^{54+}$ & Extraction to the PS without KFH31 & \ref{sec:MTKFH} \\
\bottomrule
\end{tabular}
\end{table}

\FloatBarrier
\section{Summary of the results}
\label{sec:MTSummary}

The machine tests of 2025 and 2026, described in the following sections, confirm, on the present machine and within the accuracy reached, the main assumptions of the conceptual design, and they give the technical design a solid experimental basis:
\begin{itemize}
    \item \emph{Magnets.} The main circuits sustain a \SI{1.5}{\second} flat top with margin to the converter protection, and the measured magnet temperatures confirm the calculation. The harmonic sextupoles provide the baseline strength within the range in which their temperature is stable; a rating of \SI{20}{\ampere} reaches the limit of their cooling. The 4BP cycle, with the flat-top length needed for the heaviest species at the maximum energy, was played \num{27855} times within the RMS envelope of present operation, without any interlock.
    \item \emph{Beam.} A cooled Pb$^{54+}$ beam of $1.3\times10^{9}$ ions, above the $10^{9}$ required, is delivered to a \SI{0.5}{\second} flat top at the maximum energy, with a momentum spread a factor of seven below the specification and a horizontal emittance preserved from the injection plateau. The electron cooling brings a single Pb$^{54+}$ injection to within \SI{10}{\percent} of its equilibrium emittance in about \SI{0.4}{\second} vertically and \SI{0.7}{\second} horizontally, in agreement with the LEIR literature. Compared with lead at equal emittance and corrected for the electron current, Ne$^{5+}$ cools faster than the $Q^2/A$ scaling predicts; the scaling is therefore conservative, and O$^{8+}$ is expected to be cooled well within the accumulation plateau. O$^{4+}$ is available at intensities far above the requirement.
    \item \emph{Machine model.} The slow-extraction working point is reached with the RF-capture efficiency of nominal operation. The beam-based aperture measurement confirms the narrow elements of the aperture model, provides an effective aperture model for the rest of the ring, and bounds the aperture in SS30. The $\beta$-functions measured with two independent methods agree with the 2026 model within 14 to \SI{17}{\percent} rms. The dispersion wave has the shape of the model, and the dispersion in the extraction straight section is measured with an uncertainty of about \SI{2.5}{\percent} in the configuration used for the extraction.
    \item \emph{Extraction process.} A transverse RF-KO excitation removes the beam within the flat top in a controlled and reproducible way, and an orbit bump localises the losses in SS30, as predicted by the tracking; the radiation survey, the beam-loss monitors and the Timepix3 monitor confirm this location, and the Timepix3 count rate follows the loss rate of the BCT throughout the spill. At its rated power the present excitation covers about \SI{80}{\percent} of the shortest specified spill, \SI{0.5}{\second}, at constant flux, and most of a \SI{1}{\second} spill; longer spills lie within its linear regime. The fast extraction to the PS works without the kicker module KFH31, as the extraction layout requires.
\end{itemize}
Beyond these confirmations, the tests revealed effects that are of interest for the design and, in part, for the physics of high-brightness ion beams. The losses observed with the resonance sextupoles alone hint at a dynamic aperture smaller than the single-particle model predicts, for beams accumulated close to the space-charge limit of LEIR. The chromaticity on the flat top differs from the model by about 3 units horizontally and by 1 to 2 units, with the opposite sign, vertically, which points to sextupole components of the main dipoles that the model does not yet contain, and the vertical emittance grows by a factor of about two while the beam coasts without cooling before RF capture, a phase that the set-up used for the slow-extraction test avoids. If the reduced dynamic aperture is confirmed, the extraction will be operated at a larger distance from the resonance than assumed in Chapter~\ref{sec:SRE}, or with a debunched beam on the flat top, both established options of resonant slow extraction. The thermal behaviour of longer flat tops at lower rigidities, the vertical aperture and the aperture of SS30 beyond the bounds given here, the origin of the dispersion deviations outside SS30 and of the reduced dynamic aperture, a model-based correction of the momentum drift, the calibration of the excitation and its coverage of the tune distribution, the rms emittance of light ions and the resonant transfer across the separatrix with sextupoles at the design strength are the subject of the measurement programme of Section~\ref{sec:MTProgramme}.

\section{Thermal envelope of the main magnet circuits}
\label{sec:MTFlattop}

The flat-top length available for a spill is limited by the RMS current of the main magnet circuits, which determines the heat dissipated in the magnet coils and, through it, the load on the cooling plant and on the thermal protection of the power converters. The impact of longer flat tops was first evaluated by calculation~\cite{Bodart2025}. The main dipoles are the most loaded magnets: for a flat top extended by \SI{5}{\second}, their RMS current rises from \SI{1.56}{\kilo\ampere} to \SI{2.71}{\kilo\ampere} and their temperature rises by about \SI{8}{\celsius} for a cooling flow of \SI{91}{\litre\per\minute} (Table~\ref{tab:MTBendCalc}). The main quadrupoles and sextupoles, which operate well below their ratings, rise by about \SI{1}{\celsius} or less. Summed over the 32 main magnets, the dissipated power increases by a factor of 2.8 with respect to present operation. The calculation was updated in 2026~\cite{SLAWGMagnets2026}.

\begin{table}[htbp]
\centering
\caption{Calculated RMS current, dissipated power and temperature rise of the four LEIR main dipoles as a function of the flat-top extension, for the NOMINAL cycle lengthened by the extension of its flat top and a cooling flow of \SI{91}{\litre\per\minute}~\cite{Bodart2025, SLAWGMagnets2026}. The total length of the cycle is 3.6s plus the flat-top extension.}
\label{tab:MTBendCalc}
\begin{tabular}{cccc}
\toprule
Flat-top extension (s) & $I_\mathrm{RMS}$ (kA) & Dissipated power (kW) & $\Delta T$ (\si{\celsius}) \\
\midrule
--- & 1.56 &  66.3 & 3 \\
0.5 & 1.86 &  94.3 & 4 \\
1.0 & 2.07 & 116.2 & 4 \\
1.5 & 2.22 & 133.9 & 5 \\
5.0 & 2.71 & 200.0 & 8 \\
\bottomrule
\end{tabular}
\end{table}

These estimates were verified on 22~September~2025 in a test without beam, in which the main circuits were pulsed with a \SI{1.5}{\second} flat top in a \SI{3.6}{\second} cycle (total length) while the diagnostics of the power converters, the magnets and the cooling plant were monitored. The magnet covers were removed during a planned access to allow an infrared inspection of the coils. The RMS currents recorded by the converter protection are listed in Table~\ref{tab:MTEPC}~\cite{EPC2025Flattop}. All circuits remained below their RMS fault threshold; the main dipoles, at \SI{2.65}{\kilo\ampere}, reached \SI{78}{\percent} of it. At the cooling plant, the water temperature rose by about \SI{1}{\celsius} and the supply-return temperature difference by about \SI{0.5}{\celsius}, which corresponds to an additional thermal load of about \SI{100}{\kilo\watt}~\cite{FlattopMinutes2025}.

\begin{table}[htbp]
\centering
\caption{RMS currents of the LEIR main circuits recorded by the converter RMS protection (thermal time constant \SI{10}{\second}) for the EARLY cycle and for the extended flat top of 22~September~2025~\cite{EPC2025Flattop}. The last column gives the maximum RMS current of the test as a fraction of the fault threshold.}
\label{tab:MTEPC}
\small
\begin{tabular}{lrrrr}
\toprule
Circuit & RMS fault & \multicolumn{2}{c}{$I_\mathrm{RMS,cycle}$ / $I_\mathrm{RMS,max}$ (A)} & Fraction of \\
        & level (A) & EARLY (\SI{2.4}{\second}) & Test (\SI{1.5}{\second} FT, \SI{3.6}{\second}) & fault level \\
\midrule
RPAHG.250.ER.RBHN     & 3410 & 1820 / 1870 & 2500 / 2650   & 0.78 \\
RPAGS.250.ER.RQFN2040 &  330 &   41 / 43   & 60.6 / 62.25  & 0.19 \\
RPAGR.250.ER.RQDN1030 &  396 &  148 / 152  &  173 / 217    & 0.55 \\
RPAGR.250.ER.RQFN1030 &  396 &  120 / 124  &  167 / 172.5  & 0.44 \\
RPAGR.250.ER.RQDN2040 &  396 &  120 / 124  &  249 / 258    & 0.65 \\
RPAGR.250.ER.RQFN2344 &  396 &  100 / 103  &  144 / 149    & 0.38 \\
\bottomrule
\end{tabular}
\end{table}

\paragraph{Magnet temperatures.} During the test, the temperatures of the four main dipoles, of three main quadrupoles and of four sextupoles were measured on the magnets with an infrared camera, at the inlet and at the outlet side of each magnet, on the NOMINAL cycle and in the test, with the main quadrupoles at up to \SI{249}{\ampere} and the sextupoles at \SI{10}{\ampere} (Table~\ref{tab:MTTemp})~\cite{SLAWGMagnets2026}. In the test, the flat top was extended to \SI{1.5}{\second} within the \SI{3.6}{\second} length of the NOMINAL cycle, and the RMS current of the main dipoles reached \SI{2.5}{\kilo\ampere} (Table~\ref{tab:MTEPC}). This value lies between those calculated for flat-top extensions of 1.5 and \SI{5}{\second} in Table~\ref{tab:MTBendCalc}, 2.22 and \SI{2.71}{\kilo\ampere}, for which the calculated temperature rise is 5 and \SI{8}{\celsius}; interpolated in the dissipated power, the calculation gives about \SI{6.5}{\celsius} for the test. The measured temperature difference across the main dipoles rises from \SI{2.1}{\celsius} on average on the NOMINAL cycle to \SI{8.3}{\celsius} in the test, with values between 7.0 and \SI{9.9}{\celsius} for the four magnets. The highest temperatures are recorded at the bus-bar connections, which are expected to experience the most critical temperature rise. 
The temperature of the coils is close to the calculated value. The main quadrupoles rise by \SI{0.9}{\celsius} on average and by \SI{1.7}{\celsius} at most, in agreement with the calculation. The sextupoles, whose current does not depend on the flat-top length, rise by \SI{1.3}{\celsius} on average in both cases. The calculation of the main dipoles is therefore confirmed by the measurement, and the main magnets are compatible with a slow-extraction cycle with a \SI{1.5}{\second} flat top~\cite{SLAWGMagnets2026}. A measurement campaign with a stabilisation period of at least \SI{12}{\hour} will confirm the accuracy and the repeatability of these values~\cite{SLAWGMagnets2026}.

\begin{table}[htbp]
\centering
\caption{Temperatures (\si{\celsius}) measured with an infrared camera at the inlet and outlet side of the main dipoles, of three main quadrupoles and of four sextupoles, and their difference $\Delta T$, on the NOMINAL cycle and in the test of 22~September~2025 (\SI{1.5}{\second} flat top within the \SI{3.6}{\second} cycle; RMS current of the main dipoles \SI{2.5}{\kilo\ampere}, quadrupoles at up to \SI{249}{\ampere}, sextupoles at \SI{10}{\ampere})~\cite{SLAWGMagnets2026}. The highest dipole temperatures are those of the bus-bar connections.}
\label{tab:MTTemp}
\small
\begin{tabular}{ccccccc}
\toprule
Magnet & \multicolumn{3}{c}{NOMINAL cycle} & \multicolumn{3}{c}{Test, \SI{2.5}{\kilo\ampere} RMS} \\
       & Inlet & Outlet & $\Delta T$ & Inlet & Outlet & $\Delta T$ \\
\midrule
BHN10 & 27.3 & 28.9 & 1.6 & 28.0 & 36.4 & 8.4 \\
BHN20 & 27.4 & 29.0 & 1.6 & 27.5 & 35.3 & 7.8 \\
BHN30 & 27.3 & 29.1 & 1.8 & 28.0 & 35.0 & 7.0 \\
BHN40 & 27.8 & 31.0 & 3.2 & 28.5 & 38.4 & 9.9 \\
Dipoles, average & & & 2.1 & & & 8.3 \\
\midrule
QDN41 & 25.6 & 26.3 & 0.7 & 28.8 & 30.5 & 1.7 \\
QFN41 & 25.8 & 26.2 & 0.4 & 26.5 & 27.5 & 1.0 \\
QFN43 & 26.0 & 26.3 & 0.3 & 26.7 & 27.1 & 0.4 \\
Quadrupoles, average & & & 0.5 & & & 0.9 \\
\midrule
XFLS41 & 25.6 & 27.0 & 1.4 & 26.3 & 28.5 & 2.2 \\
XDN21 & 26.6 & 26.8 & 0.2 & 28.1 & 29.4 & 1.3 \\
XFN31 & 26.5 & 29.7 & 3.2 & 27.8 & 29.4 & 1.6 \\
XDN22 & 26.6 & 26.8 & 0.2 & 28.2 & 28.3 & 0.1 \\
Sextupoles, average & & & 1.3 & & & 1.3 \\
\bottomrule
\end{tabular}
\end{table}

The following conclusions were reached~\cite{FlattopMinutes2025}: the converters of the main dipoles, quadrupoles and sextupoles can sustain flat tops of up to \SI{5}{\second} for MD purposes, for instance with one such cycle per supercycle. 
The limiting concern is the thermo-mechanical ageing of the power semiconductors under extended cycling, which applies to all LEIR converters, and continuous operation with \SI{5}{\second} flat tops is not recommended.
The cooling plant is specified for an inlet temperature of \SI{27}{\celsius} but operated at \SI{25}{\celsius}; raising the operating value by \SI{2}{\celsius} would provide additional margin. 
It was therefore agreed that HEARTS@LEIR operation will respect the RMS envelope of present operation, so as not to add stress on ageing equipment: no further extension of the flat top at the maximum rigidity is required, and longer flat tops at lower rigidities must remain compatible with this envelope.

\begin{figure}[htbp]
    \centering
    \includegraphics[width=0.6\linewidth]{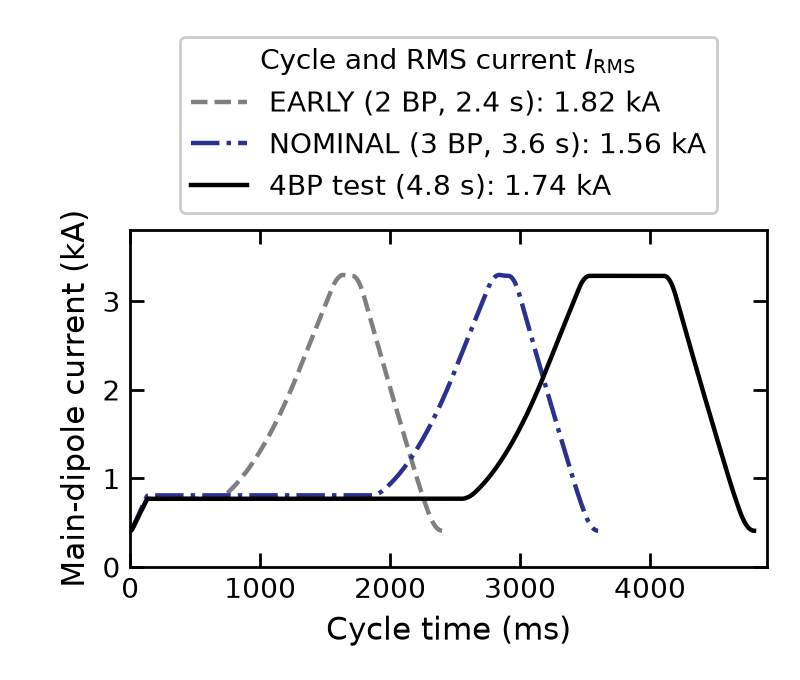}
    \caption{Current programme of the LEIR main dipoles for the EARLY and NOMINAL operational cycles and for the 4BP cycle used for the slow-extraction test. The legend gives the RMS current over each cycle.}
    \label{fig:MTDipoleCurrent}
\end{figure}

Figure~\ref{fig:MTDipoleCurrent} compares the current programme of the main dipoles for the two operational cycles, EARLY (two basic periods) and NOMINAL (three basic periods), with that of the 4BP cycle, extended by one basic period to four and used for the slow-extraction test (Section~\ref{sec:MTSX}). The two operational cycles have the same ramp and flat top and differ in the length of the injection plateau; their RMS currents are \SI{1.82}{\kilo\ampere} and \SI{1.56}{\kilo\ampere}, respectively, the former in agreement with the value recorded by the converter protection (Table~\ref{tab:MTEPC}). The 4BP cycle has a flat top of about \SI{0.6}{\second} at the maximum rigidity, of which \SI{0.5}{\second} are usable before the fast-extraction kicker fires. Its RMS current of \SI{1.74}{\kilo\ampere} lies between those of the two operational cycles, that is, within the present operating envelope. The cycle was commissioned in June~2026 and played regularly from mid-July to the end of the run: \num{27855} times in total, \num{24192} times with beam, and about \num{10000} times in August alone. This represents the first sustained operation of LEIR with the flat-top length foreseen for the heaviest species at the maximum energy. 

No RMS-protection or thermal interlock of the main circuits was triggered while the 4BP cycle was played in July and August~2026, as confirmed by the power-converter group. During the slow-extraction test of 21~July~2026, the cycle ran for three hours with continuous beam loss without any interlock or vacuum spike (LEIR logbook).

\begin{figure}[htbp]
    \centering
    \includegraphics[width=0.7\linewidth]{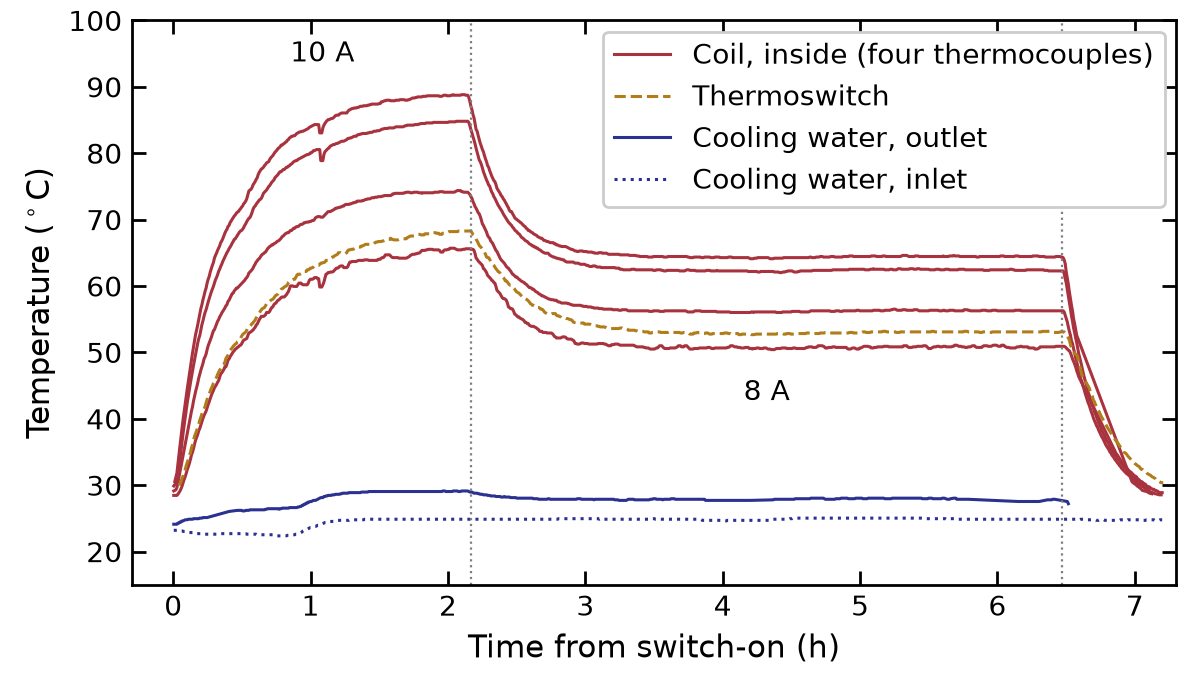}
    \caption{Bench test of a harmonic sextupole of the XFN/XDN type with a constant current of \SI{10}{\ampere} and then \SI{8}{\ampere}: temperatures measured with thermocouples inside the coil, at the thermoswitch and in the cooling water at the inlet and at the outlet~\cite{SLAWGMagnets2026}. A constant current of \SI{10}{\ampere} corresponds to the RMS current of a sextupole set to about \SI{20}{\ampere} on the flat top.}
    \label{fig:MTXFNBench}
\end{figure}

\paragraph{Harmonic sextupoles.} The resonance of the slow extraction is driven by the four individually powered harmonic sextupoles XFN11, XFN12, XFN31 and XFN32, rated at \SI{10}{\ampere}. For the strength needed at \SI{4.8}{\tesla\metre} with the baseline tune distance and emittance (Section~\ref{sec:SREsext}), the most loaded sextupole requires \SI{9.9}{\ampere}, \SI{99}{\percent} of its rating, and \SI{11.9}{\ampere} at \SI{5.8}{\tesla\metre}; a rating of \SI{20}{\ampere} would double the available range. On the cycle considered in the calculation, a current of \SI{10}{\ampere} on the flat top corresponds to an RMS current of \SI{5.4}{\ampere}, and \SI{20}{\ampere} to \SI{9.8}{\ampere}~\cite{SLAWGMagnets2026}. The thermal behaviour at these RMS values was tested on a bench with a sextupole of the same type (XFN/XDN, \SI{0.52}{\tesla\per\metre\per\ampere}), powered with a constant current of \SI{10}{\ampere} for about two hours and then of \SI{8}{\ampere} for about four hours (Fig.~\ref{fig:MTXFNBench}). At \SI{8}{\ampere} the coil temperatures stabilise between 51 and \SI{64}{\celsius}, that is, up to \SI{39}{\celsius} above the cooling water at the inlet (\SI{25}{\celsius}). At \SI{10}{\ampere} the hottest point of the coil reaches \SI{89}{\celsius} after two hours and still rises by about \SI{2}{\celsius} per hour. The temperature rise of the coil grows slightly faster than the square of the current, while that of the cooling water grows more slowly (4 and \SI{3}{\celsius}), which shows that the cooling efficiency saturates at this current~\cite{SLAWGMagnets2026}. The currents needed for the baseline at 4.8 and \SI{5.8}{\tesla\metre} (Table~\ref{tab:SREsext}) correspond to RMS currents of about 5 to \SI{6}{\ampere}, below the \SI{8}{\ampere} at which the coil temperature is stable. A rating of \SI{20}{\ampere} reaches the limit of the present cooling; before it is adopted, a two-dimensional magnetic model will verify that the sextupole field remains linear at \SI{20}{\ampere}, and a measurement with a stabilisation period of at least \SI{12}{\hour} will confirm the temperatures~\cite{SLAWGMagnets2026}.

\section{Beam conditions on the flat top}
\label{sec:MTEmittance}

The transverse emittance, the intensity and the momentum spread of the circulating beam at the start of the flat top are the initial conditions of the slow-extraction simulations of Chapter~\ref{sec:SRE} and of the design of the transfer line. They were measured with the ionisation profile monitors (IPM), the beam current transformer (BCT) and the longitudinal Schottky monitor for the species available in the 2025 and 2026 runs.

\paragraph{Profile data and method.} The IPMs record the horizontal and vertical profiles on 64 channels with a pitch of \SI{1}{\milli\metre}. A profile is delivered every half integration time: every \SI{10}{\milli\second} for the \SI{20}{\milli\second} integration used with Pb$^{54+}$, and every \SI{20}{\milli\second} for the \SI{40}{\milli\second} used with the light ions. In both planes the channels above $+\SI{15}{\milli\metre}$ carry no signal, so that the usable range extends from $-25$ to $+\SI{15}{\milli\metre}$ around the nominal centre. The analyses presented here use the raw profiles of the shots with beam: the background recorded before injection is subtracted, and the beam size is the width of a Gaussian fit with a free baseline, or, where the profile is not Gaussian, its rms width. The emittance is $\epsilon_{n} = \beta\gamma\,\sigma^2/\beta_\mathrm{IPM}$. The $\beta$-functions at the monitors are taken from the 2026 model of LEIR (Section~\ref{sec:MTOptics}), rematched to the tunes measured in each part of the cycle: \SI{11.6}{\metre} horizontally and \SI{7.0}{\metre} vertically on the injection plateau ($q_x = 0.823$, $q_y = 0.734$), and \SI{11.5}{\metre} and \SI{8.3}{\metre} on the flat top in the slow-extraction optics ($q_x = 0.689$, $q_y = 0.811$), where the dispersion at the horizontal monitor is $D_x = \SI{-2.9}{\metre}$. On the flat top the rematched model agrees with the $\beta$-functions measured from the orbit response within 15 to \SI{17}{\percent} rms, and with the \SI{11.7}{\metre} measured at the BPM next to the horizontal monitor. Depending on which quadrupole families absorb the difference between the model and the measured tunes, the $\beta$-functions at the monitors vary by about \SI{10}{\percent} on the flat top and by about \SI{15}{\percent} on the injection plateau, where no $\beta$ measurement is available; this is the optics uncertainty of the emittances below. At injection energy the dispersive contribution at the horizontal monitor ($D_x \approx \SI{-0.4}{\metre}$) is negligible.

The signal of the monitors per circulating ion varies by up to a factor of three between acquisitions, at an almost constant voltage of the microchannel plates, and decreases by 10 to \SI{30}{\percent} as the beam is cooled. If this saturation depends on the local charge density, it broadens the profiles, so the emittances below are upper bounds with respect to the response of the monitors. Along the 4BP cycle of 7~August~2026, the vertical beam size does not change with the voltage of the microchannel plates between 1100 and \SI{1200}{\volt}, which indicates that this effect is small.

\subsection{Lead ions: cooling and intensity dependence of the emittance}
\label{sec:MTPb}

\begin{figure}[htbp]
    \centering
    \includegraphics[width=\linewidth]{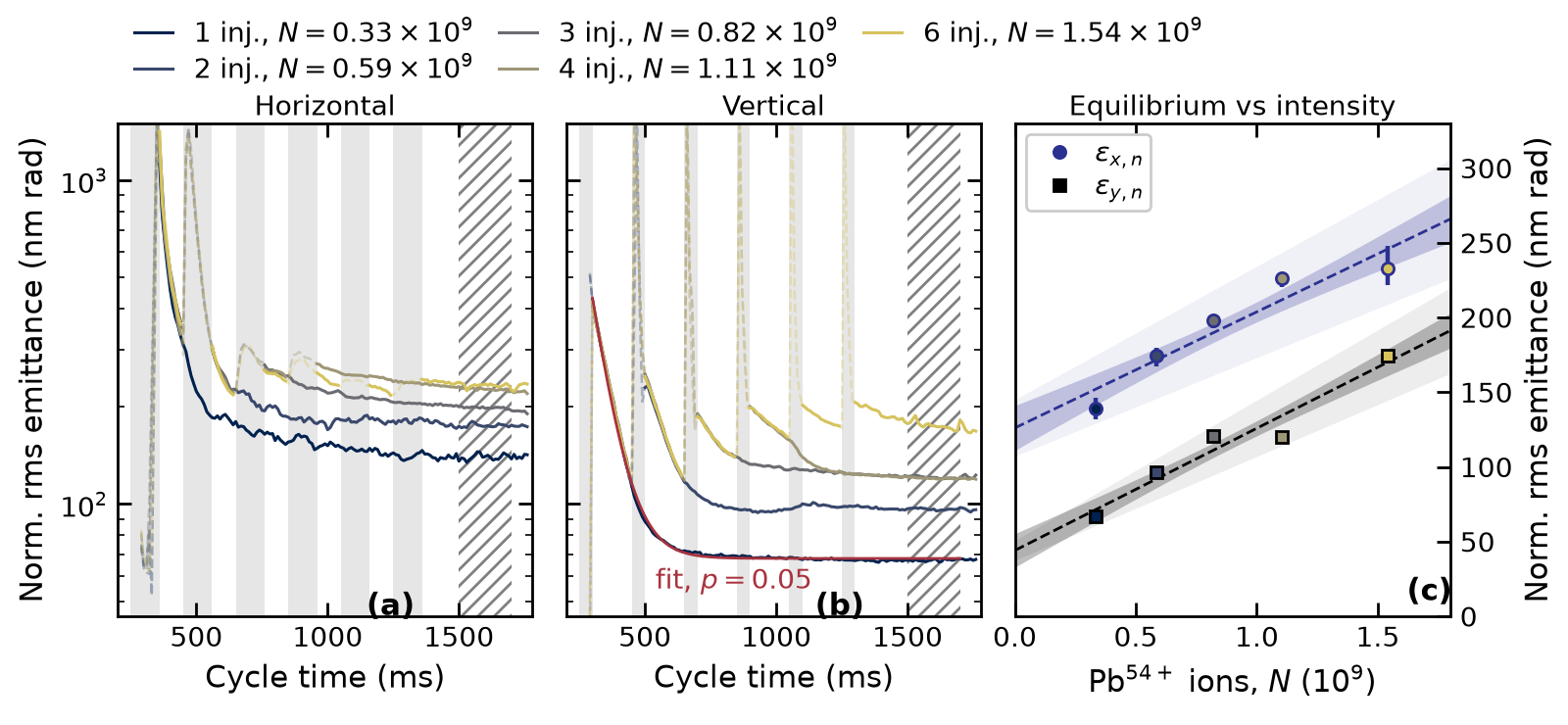}
    \caption{Pb$^{54+}$ emittance on the injection plateau, 14~November~2025, for five settings with 1 to 6 injections (every \SI{200}{\milli\second} from \SI{250}{\milli\second}; $N$: number of ions on the plateau, BCT). (a)~Horizontal and (b)~vertical normalised rms emittance, median over the 17 to 23 shots of each setting. Grey bands: first \SI{110}{\milli\second} after each injection in the horizontal plane, where the injected beam is hollow, and first \SI{50}{\milli\second} in the vertical plane, where the profiles contain the injection flash; in these intervals the width of the Gaussian fit is not meaningful (dashed). Red: fit of the amplitude-dependent cooling model to the single injection ($p = 0.05$, see text). Hatched: interval of the equilibrium values of~(c). (c)~Equilibrium emittance, 1500--\SI{1700}{\milli\second} (interquartile range over the shots; colours as in (a) and (b)), as a function of the number of ions, with linear fits: dark band, \SI{68}{\percent} confidence band of the fit; light band, optics uncertainty of \SI{15}{\percent}.}
    \label{fig:MTPbEmitt}
\end{figure}

The cooling and the equilibrium emittance of Pb$^{54+}$ were measured on 14~November~2025 on an MD cycle whose injection plateau extends to about \SI{1800}{\milli\second}. The number of Linac~3 pulses accumulated in LEIR, injected every \SI{200}{\milli\second} from \SI{250}{\milli\second}, was increased in five steps from one to six, with the electron-cooler settings adjusted to obtain an efficient cooling drag together with an acceptable accumulation performance; 17 to 23 shots were recorded for each setting. After the last injection the beam is cooled for more than \SI{0.5}{\second}, so that each setting gives both the cooling of the beam and its equilibrium emittance at the corresponding intensity, between 0.33 and $1.54\times10^{9}$ ions (BCT, Fig.~\ref{fig:MTPbEmitt}). The beam size is the width of a Gaussian fit with a free baseline. It is not defined during the first \SI{50}{\milli\second} after each injection in the vertical plane, where the profiles contain the injection flash, and during the first \SI{110}{\milli\second} in the horizontal plane, where the injected beam is hollow, with two maxima 10 to \SI{13}{\milli\metre} on either side of the centre of the cooled beam (bands in Fig.~\ref{fig:MTPbEmitt}a and~b). Afterwards the profiles are single-peaked. At equilibrium the horizontal profiles show, besides the Gaussian core, a flat pedestal of 1 to \SI{2}{\percent} of the peak signal on the side of the full detector range, which grows during the plateau and is attributed to the baseline of the monitor; it is absorbed by the baseline of the fit.

\paragraph{Electron cooling.} With a single injection of $0.33\times10^{9}$ ions, the beam reaches its equilibrium in both planes (Fig.~\ref{fig:MTPbEmitt}a and~b):
\begin{itemize}
    \item \emph{Vertical plane.} The emittance decreases from about \SI{440}{\nano\metre\radian}, \SI{50}{\milli\second} after the injection, to its equilibrium of \SI{67}{\nano\metre\radian}. It is within \SI{10}{\percent} of the equilibrium \SI{360}{\milli\second} after the injection and within \SI{5}{\percent} after \SI{460}{\milli\second}.
    \item \emph{Horizontal plane.} The emittance is about \SI{1150}{\nano\metre\radian} when the profile becomes single-peaked, \SI{110}{\milli\second} after the injection. It falls to about \SI{185}{\nano\metre\radian} within the following \SI{200}{\milli\second}, and then more slowly, as the core of the profile continues to narrow, to its equilibrium of \SI{139}{\nano\metre\radian}. It is within \SI{10}{\percent} of the equilibrium \SI{700}{\milli\second} after the injection and within \SI{5}{\percent} after \SI{940}{\milli\second}.
\end{itemize}
Measured from the first usable profile, \SI{95}{\percent} of the excess emittance is removed \SI{270}{\milli\second} (vertical) and \SI{300}{\milli\second} (horizontal) after the injection. These values are comparable with the cooling-down time $3\tau^\text{e-cool}$ of \SI{200}{\milli\second} aimed at in the I-LHC design, which improved on the \SI{400}{\milli\second} demonstrated in the LEAR test of 1997~\cite{LHCDesignReport}. They agree with the LEIR measurements of Ref.~\cite{BiancacciZampetakis2024}, which report overall cooling times of about \SI{200}{\milli\second} vertically and \SI{600}{\milli\second} horizontally for a beam of similar intensity.

The design of the LEIR electron cooler relies on the dependence of the cooling time on the angle $\Theta$ between the ion and electron trajectories, $\tau \propto \Theta^{3}$~\cite{LHCDesignReport}, which applies to ions whose angle exceeds the angular spread of the electron beam: a large injected emittance would then be cooled slowly at first and faster as the beam shrinks. The single injection shows no such slow initial phase. Fitted with $\mathrm{d}\epsilon/\mathrm{d}t = -(\epsilon-\epsilon_\mathrm{eq})/\tau_0 \cdot (\epsilon_\mathrm{eq}/\epsilon)^{p}$, where $p = 1.5$ corresponds to $\tau \propto \Theta^{3}$, the vertical emittance gives $p = 0.05 \pm 0.02$ and $\tau_0 = 70 \pm 4$~ms (Fig.~\ref{fig:MTPbEmitt}b), and the horizontal one, from the second moment of the profile, which is defined also while the injected beam is hollow, $p = 0.13 \pm 0.02$ and $\tau_0 = 83 \pm 15$~ms. The cooling rate is therefore nearly independent of the amplitude between the first profile and the equilibrium, for ion angles in the cooler from about \SI{1.6}{\milli\radian} down to \SI{0.3}{\milli\radian}, which indicates that the ions are within the angular spread of the electrons from the first profile on. The regime $\tau \propto \Theta^{3}$ may be reached with larger injected emittances or other settings of the cooler, which were not explored here. Only the final approach of the horizontal core to its equilibrium is slower, as expected where intensity-dependent heating balances the cooling (see below).

Each further injection is merged into the cooled beam within a similar time: in the setting with two injections, the emittance is within \SI{10}{\percent} of its equilibrium \SI{260}{\milli\second} (vertical) and \SI{380}{\milli\second} (horizontal) after the second injection. A cycle with two injections recorded on 31~October~2025 with the same monitor gives 250 and \SI{380}{\milli\second}, and equilibrium emittances of 105 and \SI{184}{\nano\metre\radian}, close to the 96 and \SI{175}{\nano\metre\radian} of 14~November, which shows that the measurement is reproducible from day to day. For HEARTS@LEIR, where a single injection or a few injections are cooled on an injection plateau of more than \SI{1.5}{\second}, the measured cooling of Pb$^{54+}$ leaves a margin of at least a factor of two, also in the slower horizontal plane.

\paragraph{Emittance and intensity.} The equilibrium emittances were evaluated from 1500 to \SI{1700}{\milli\second}, after cooling and before RF capture, and the intensity was taken from the BCT over the same interval (Fig.~\ref{fig:MTPbEmitt}c). Both emittances increase linearly with the number of ions $N$, expressed in units of $10^{9}$:
\begin{equation}
    \epsilon_{x,n} = \left[(126 \pm 15) + (78 \pm 15)\,N \right]~\text{nm\,rad}, \qquad
    \epsilon_{y,n} = \left[(44 \pm 11) + (82 \pm 12)\,N \right]~\text{nm\,rad},
\end{equation}
where the errors are those of the fit, without the optics uncertainty. At $1.5\times10^{9}$ ions, the emittances are $\epsilon_{x,n} \approx \SI{245}{\nano\metre\radian}$ and $\epsilon_{y,n} \approx \SI{165}{\nano\metre\radian}$. The increase of the equilibrium emittance with intensity indicates intensity-dependent heating that counteracts the electron cooling. Intra-beam scattering is expected to dominate, while the contributions of space charge and impedance have not yet been quantified for the HEARTS@LEIR cycles.

\begin{figure}[htbp]
    \centering
    \includegraphics[width=\linewidth]{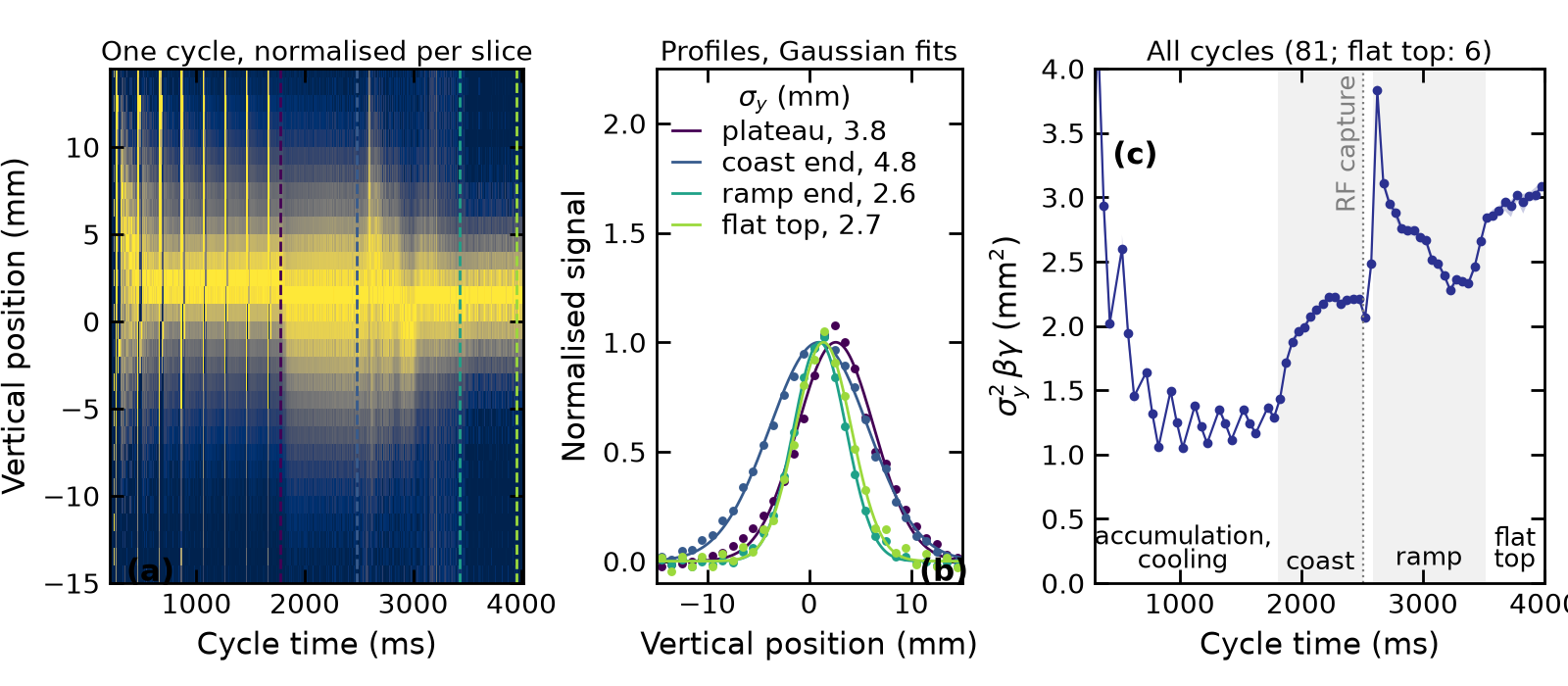}
    \caption{Vertical IPM data along the 4BP cycle, 7~August~2026 (set-up~(a)). (a)~Signal of one cycle without excitation, background subtracted, each profile normalised to its maximum to show the width at all intensities; the injections every \SI{200}{\milli\second} up to about \SI{1.6}{\second} appear as vertical lines; dashed: times of the profiles in~(b). (b)~Profiles of the same cycle averaged over \SI{50}{\milli\second} at four stages, normalised, with Gaussian fits (width $\sigma_y$ in the legend); the peak signals are 15\,700, 5200, 3300 and 2600 counts. (c)~$\sigma_y^2\beta\gamma$, constant for a preserved emittance at fixed optics: median and interquartile range over the 81 cycles (six without excitation after \SI{3450}{\milli\second}), \SI{50}{\milli\second} bins; bins with an injection or a saturated profile are excluded. The cooling stops at \SI{1805}{\milli\second}; shaded: coast without cooling with the tunes near the difference resonance, and ramp; dotted: RF capture. The peak at the start of the ramp coincides with the crossing of the difference resonance.}
    \label{fig:MTIPM4BP}
\end{figure}

\paragraph{Emittance on the flat top.} If the normalised emittance were preserved during acceleration, the values above would correspond, at the maximum energy of Pb$^{54+}$ ($\beta\gamma = 0.40$), to geometric rms emittances of about \SI{600}{\nano\metre\radian} horizontally and \SI{420}{\nano\metre\radian} vertically at $1.5\times10^{9}$ ions. On 7~August~2026 the IPMs recorded the whole 4BP cycle in set-up~(a) (Section~\ref{sec:MT4BP}), from the injection plateau to the flat top, so that the evolution of the emittance can be followed within the same cycles (Fig.~\ref{fig:MTIPM4BP}). Of the 81 cycles recorded, six were not excited on the flat top (Section~\ref{sec:MTRFKO}); the BCT gives $1.1\times10^{9}$ ions at the end of the accumulation, $0.85\times10^{9}$ before RF capture and $0.49\times10^{9}$ on the flat top. The horizontal profiles of the injection plateau saturate the monitor, so only the vertical plane can be followed through the cycle. The ratio of $\sigma_y^2\beta\gamma$ between two times of the same cycle does not depend on the gain of the monitor and, at fixed optics, equals the ratio of the emittances; it is used below as the median of the ratios over the cycles.
\begin{itemize}
    \item \emph{Cooled plateau.} Between the last injection, at about \SI{1.6}{\second}, and the end of the cooling, at \SI{1805}{\milli\second}, the vertical emittance is about \SI{185}{\nano\metre\radian}. It is larger than the equilibrium of 14~November at this intensity, about \SI{135}{\nano\metre\radian}, since the last injection is cooled for only about \SI{0.2}{\second}.
    \item \emph{Coast without cooling.} During the \SI{700}{\milli\second} in which the beam coasts without cooling, with the tunes close to the difference resonance ($q_x \approx 0.77$, $q_y \approx 0.75$), $\sigma_y^2\beta\gamma$ grows by a factor of 1.71 (interquartile range 1.69 to 1.74), at the same energy and with nearly the same optics. The vertical emittance grows by a factor of 1.4 to 1.9, depending on the quadrupole families used in the rematch, to about \SI{310}{\nano\metre\radian}, while \SI{24}{\percent} of the beam is lost.
    \item \emph{Capture and ramp.} From the end of the coast to the start of the flat top, across RF capture, the crossing of the difference resonance and the ramp, $\sigma_y^2\beta\gamma$ increases by 11 to \SI{29}{\percent} and varies by up to \SI{15}{\percent} along the ramp. With the change of $\beta_y$ at the monitor between the two optics, from 7.1 to \SI{8.3}{\metre}, the vertical emittance is preserved within about \SI{10}{\percent}.
    \item \emph{Flat top.} Without excitation, $\sigma_y^2\beta\gamma$ grows by \SI{6}{\percent} in \SI{0.5}{\second}. Averaged from 3550 to \SI{4000}{\milli\second}, the six cycles without excitation give $\sigma_y^2 = \SI{7.4}{\milli\metre\squared}$ and $\epsilon_{y,n} \approx \SI{355}{\nano\metre\radian}$.
\end{itemize}
Between the cooled plateau and the end of the flat top, the vertical emittance thus grows by a factor of about 1.9 (1.3 to 2.3 over the range of $\beta$-functions of the rematched optics), almost entirely while the beam coasts without cooling near the coupling resonance. In this phase the heating by intra-beam scattering is no longer compensated, and the coupling can transfer to the vertical plane part of the horizontal emittance, which is the larger one on the injection plateau; the data do not separate the two. In the horizontal plane, $\sigma_x^2 = \SI{7.2}{\milli\metre\squared}$ on the flat top, of which \SI{2.0}{\milli\metre\squared} is the dispersive contribution of the measured momentum spread ($D_x = \SI{-2.9}{\metre}$, $\sigma_\delta = 0.49\times10^{-3}$), gives $\epsilon_{x,n} \approx \SI{180}{\nano\metre\radian}$, against about \SI{215}{\nano\metre\radian} expected on the injection plateau for the $1.1\times10^{9}$ ions accumulated (Fig.~\ref{fig:MTPbEmitt}c): the horizontal emittance is preserved within the optics uncertainty.

Both flat-top emittances, about 180 and \SI{355}{\nano\metre\radian}, are below the \SI{400}{\nano\metre\radian} assumed for Pb$^{54+}$ in the extraction studies of Chapter~\ref{sec:SRE}, the vertical one only slightly. Set-up~(b), which delivered $1.3\times10^{9}$ ions to the flat top at the end of August~2026 (Section~\ref{sec:MT4BP}), keeps the cooling on until RF capture and moves the tune change to the end of the ramp, so that it avoids the coast without cooling during which the vertical emittance grows in set-up~(a). The microchannel plates of the IPMs were, however, switched off for almost all cycles of set-up~(b), so its emittances were not measured. With the intensity dependence of Fig.~\ref{fig:MTPbEmitt}c and about $1.6\times10^{9}$ ions accumulated, the horizontal emittance is expected at about \SI{250}{\nano\metre\radian}, and the vertical one between about \SI{175}{\nano\metre\radian} without growth and \SI{330}{\nano\metre\radian} with the growth of set-up~(a). The vertical value is not critical for the extraction, which acts in the horizontal plane, and broad beams are favourable for the irradiation fields. It is likely that the vertical growth is caused by linear coupling: as the tunes approach the difference resonance during the coast and cross it at the start of the ramp, emittance can be shared or exchanged between the two planes, the horizontal emittance being the larger one on the injection plateau~\cite{Franchi2007}. The crossing of the coupling resonance has been used in this way at the SIS18 synchrotron of GSI, to reduce the horizontal emittance before a slow extraction and improve the spill quality~\cite{YangIPAC2022}. This remains a conjecture: a quantitative account requires the skew-quadrupole and solenoid components of the ring, which set the strength of the coupling, and the time spent by the tunes at each distance from the resonance.

\subsection{Light ions: Ne\texorpdfstring{$^{5+}$}{5+} and O\texorpdfstring{$^{4+}$}{4+}}
\label{sec:MTLightIons}

The electron-cooling rate scales with $Q^2/A$, so that the lighter species of the HEARTS@LEIR palette are expected to cool more slowly than lead (Table~\ref{tab:RelCoolingTimes}). Light ions are regularly accumulated and accelerated in LEIR: Table~\ref{tab:IonIntDetails} includes the intensities measured with O$^{4+}$ during the LHC oxygen run of 2025 and with Ar$^{11+}$ in 2015. Within the HEARTS@LEIR studies, the cooling time and the transverse emittances were measured with a Ne$^{5+}$ beam on 8~July~2025, and the emittance along the cycle with O$^{4+}$ on 22~May~2025.

The cooling of a single Ne$^{5+}$ injection of $7\times10^{9}$ ions was recorded in 17 shots after the injection at \SI{240}{\milli\second}~\cite{MD20250708}. The vertical profiles are single-peaked and contained in the usable range, and the analysis of the Pb$^{54+}$ data applies unchanged; the horizontal profile is not Gaussian during the cooling and its centre moves by about \SI{10}{\milli\metre} towards the electron beam, so it is not used. The vertical emittance decreases exponentially, with an emittance cooling time between 440 and \SI{470}{\milli\second}, from \SI{700}{\nano\metre\radian} at \SI{280}{\milli\second} to \SI{160}{\nano\metre\radian} at the optics change at about \SI{1.4}{\second}; \SI{95}{\percent} of the excess emittance would be removed after about \SI{1.3}{\second}. At \SI{1.3}{\second} the normalised emittances are about \SI{140}{\nano\metre\radian} horizontally and \SI{170}{\nano\metre\radian} vertically, with the $\beta$-functions of the model rematched to the tunes measured on the plateau ($q_x = 0.822$, $q_y = 0.707$). After RF capture, $3.9\times10^{9}$ ions were accelerated to \SI{4.8}{\tesla\metre}. At the start of the flat top, in the last profile recorded (\SI{2800}{\milli\second}), the vertical emittance is about \SI{240}{\nano\metre\radian}, between 210 and \SI{270}{\nano\metre\radian} for the five optics settings of the MD, with $\beta_y$ from the tunes measured in each; the horizontal profile carries no signal there.

Since the cooling rate depends on the amplitude, Ne$^{5+}$ and Pb$^{54+}$ are compared over the same range of vertical emittance, with the same monitor and the same energy, and corrected for the electron current, which was \SI{18.4}{\milli\ampere} for Ne$^{5+}$ and \SI{23.1}{\milli\ampere} and \SI{42.2}{\milli\ampere} for Pb$^{54+}$ on 10 and 14~November~2025 (electron-cooler acquisitions). Reducing the emittance by a factor of two, within the range from about 700 to \SI{170}{\nano\metre\radian}, takes 370 to \SI{640}{\milli\second} for Ne$^{5+}$, against 100 to \SI{150}{\milli\second} (10~November) and 75 to \SI{90}{\milli\second} (14~November) for Pb$^{54+}$. The ratio, 2.4 to 5.4 and 7 to 8 respectively, is two and a half to five times smaller than the $Q^2/A$ scaling corrected for the electron current, 13 to 14 and 20 to 26. The result does not depend on the estimator of the width (Gaussian or rms) and is conservative with respect to the time base of the light-ion profiles, which is the longest that the number of profiles per cycle allows. Its origin is not identified: the measurement is a single MD, with a cooler set-up not optimised for this species, and the heating by intra-beam scattering, which scales as $NQ^4/A^2$ and was six to eight times stronger for the Pb$^{54+}$ beams, is not included in the scaling. For the design the conclusion is nevertheless clear: there is no indication that light ions cool more slowly than the scaling from lead predicts, which is therefore conservative. Scaled with $Q^2/A$ from the Ne$^{5+}$ measurement, the cooling-down time of O$^{8+}$ at the same electron current is about \SI{0.4}{\second}; scaled from the Pb$^{54+}$ measurements, in which \SI{95}{\percent} of the excess emittance is removed within 270 to \SI{300}{\milli\second}, it is about \SI{1}{\second}. Both estimates are within the accumulation plateau of more than \SI{1.5}{\second} (Chapter~\ref{sec:LEIRaccelerator}).

\begin{figure}[htbp]
    \centering
    \includegraphics[width=\linewidth]{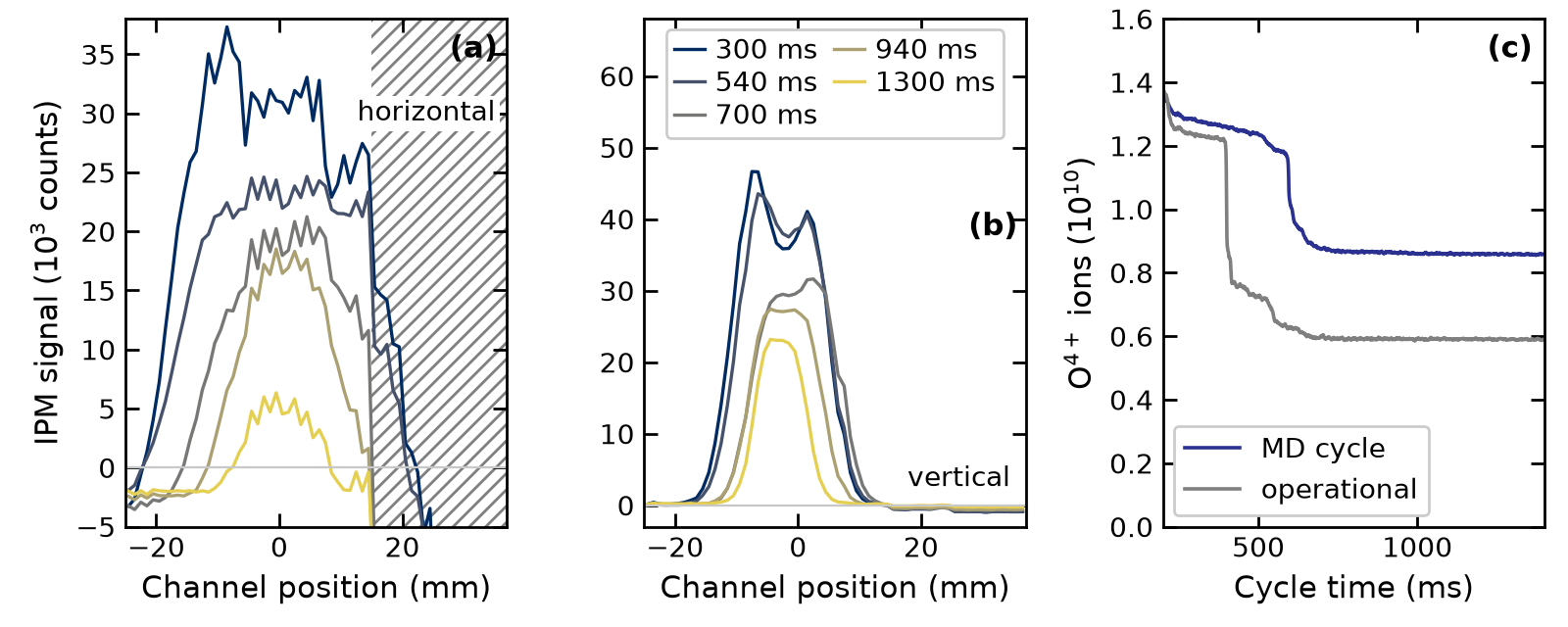}
    \caption{O$^{4+}$ on the MD version of the EARLY cycle, 22~May~2025 (RF capture at \SI{450}{\milli\second}). (a)~Horizontal and (b)~vertical IPM profiles, background subtracted, median over 149 shots, at five cycle times; hatched: channels without signal in the horizontal plane. (c)~BCT intensity in O$^{4+}$ ions, median over the cycles, for the MD cycle and for the operational O$^{4+}$ cycle.}
    \label{fig:MTO4}
\end{figure}

The emittance of O$^{4+}$ was measured along an MD version of the EARLY cycle, on which the RF capture had been advanced to \SI{450}{\milli\second}; 149 shots were recorded. About $1.3\times10^{10}$ O$^{4+}$ ions ($5\times10^{10}$ charges) were accumulated, in the MD cycle as in the operational cycle (BCT, Fig.~\ref{fig:MTO4}c). The profiles are not Gaussian (Fig.~\ref{fig:MTO4}):
\begin{itemize}
    \item the horizontal profiles are flat-topped and, up to about \SI{650}{\milli\second}, extend into the channels without signal;
    \item the vertical profiles show two maxima and a flat top. This is the signature of a saturated microchannel plate, but it may also be the shape of the beam.
\end{itemize}
A Gaussian fit is therefore not appropriate, but the full width at half maximum (FWHM) is well defined from \SI{760}{\milli\second} horizontally and throughout the cycle vertically. During the ramp it decreases as $1/\sqrt{\beta\gamma}$, with a spread of \SI{3}{\percent} horizontally and \SI{1}{\percent} vertically from 760 to \SI{1300}{\milli\second}, while $\beta\gamma$ increases by a factor of 2.5. This is the adiabatic damping expected for a preserved normalised emittance, which suggests that the profile shape is not an artefact of the monitor. Between a flat-topped and a Gaussian distribution of the same FWHM, the normalised rms emittances on the ramp lie between about 290 and \SI{630}{\nano\metre\radian} horizontally and between about 265 and \SI{570}{\nano\metre\radian} vertically, for $0.86\times10^{10}$ ions and the $\beta$-functions of the nominal optics, a range that includes the \SI{300}{\nano\metre\radian} assumed for oxygen in the extraction studies of Chapter~\ref{sec:SRE}. If saturation broadens the profiles, these values are upper bounds. A measurement of the rms emittance would require IPM settings (microchannel-plate gain) adapted to light ions.

The BCT shows that the beam loss after RF capture is large. In the MD cycle, $0.86\times10^{10}$ ions remain after capture, a loss of \SI{32}{\percent}. In the operational cycle, $0.59\times10^{10}$ ions remain, a loss of about \SI{50}{\percent}. In both cycles the loss occurs within about \SI{300}{\milli\second} after capture (Fig.~\ref{fig:MTO4}c). Even after this loss, the number of O$^{4+}$ ions on the ramp exceeds the $10^{9}$ ions required on the flat top (Table~\ref{tab:beamSpecs}) by almost an order of magnitude.

\subsection{Momentum spread and flat-top stability}
\label{sec:MTMomentum}

The longitudinal Schottky monitor is logged only during the first \SI{2.8}{\second} of the 4BP cycle, that is, until shortly after RF capture~\cite{FourBP2026}. A preliminary analysis of the spectra gives, for Pb$^{54+}$ at injection energy just before capture, an rms momentum spread of about $1.2\times10^{-3}$ for set-up~(a) of the 4BP cycle, in which the beam coasts without cooling for \SI{0.7}{\second} before capture, and of about $0.85\times10^{-3}$ for set-up~(b), in which the cooling is kept on until capture (Section~\ref{sec:MT4BP}). The larger momentum spread of set-up~(a) is reflected in its capture loss, which is three times larger than that of set-up~(b). The momentum spread on the flat top was measured by longitudinal tomography on 26~August~2026, on the 4BP cycle of the slow-extraction test (set-up~(b)): $0.49\times10^{-3}$ rms (LEIR logbook). At the time of the measurement the bunch performed a coherent dipole oscillation with an amplitude of at least its rms length, which the longitudinal Schottky spectrum recorded on the flat top the following day also shows as a strong central line; the value therefore characterises an oscillating bunch rather than a matched distribution. These Schottky spectra were not logged, and their screenshots do not resolve the momentum spread. The tomography value is consistent with the Schottky measurements at injection energy once the adiabatic damping of the momentum spread during acceleration is taken into account. It is a factor of about seven below the rms momentum spread of $3.3\times10^{-3}$ allowed by the beam specifications (Table~\ref{tab:beamSpecs}).

During the commissioning of the 4BP cycle in June~2026, the revolution frequency was observed to drift along the flat top by about \SI{3.5}{\kilo\hertz}, which corresponds to a momentum change of about $3.5\times10^{-3}$ between 3600 and \SI{4050}{\milli\second}. With the horizontal chromaticity measured on the flat top, $Q'_x = -4.7$ (Section~\ref{sec:MTOptics}), such a drift shifts the horizontal tune by about 0.016, two thirds of the distance between the flat-top working point and the resonance, and would by itself move part of the beam onto the resonance. The drift is attributed to the decay of eddy currents in the main bending magnets after the end of the ramp. Without correction, a strong orbit drift at the start of the flat top was observed (LEIR logbook, 24~July~2026). In later operation it was compensated by a counteracting ramp of the RF frequency: on 26~August~2026 the revolution frequency rises by \SI{1.4}{\kilo\hertz} between 3700 and \SI{3990}{\milli\second} while the orbit at the high-dispersion monitors stays constant within \SI{0.6}{\milli\metre} (Section~\ref{sec:MTOptics}), and the cycles without excitation and without resonance sextupoles lose only \SI{1.9}{\percent} of their beam on the flat top (Section~\ref{sec:MTSX}). The momentum on the flat top can therefore be held constant with the existing RF system; a correction derived from a model of the field decay is part of the programme of Section~\ref{sec:MTProgramme}.

\subsection{Beam parameters on the flat top}
\label{sec:MTBeamFT}

Table~\ref{tab:MTBeamFT} collects the beam parameters measured on the flat top, or as close to it as the data allow, together with the number of ions at which each was measured. The emittances of Pb$^{54+}$ are below the \SI{400}{\nano\metre\radian} assumed in the extraction studies of Chapter~\ref{sec:SRE}, the vertical one only slightly, and the range of O$^{4+}$ includes the \SI{300}{\nano\metre\radian} assumed for oxygen. The momentum spread of Pb$^{54+}$ is a factor of about seven below the specification; for the light ions it was not measured on the flat top.

\begin{table}[htbp]
\centering
\caption{Beam parameters measured on the flat top or closest to it: normalised rms emittances and rms momentum spread, with the cycle, the stage of the cycle ($\beta\gamma$) and the number of ions at the time of the measurement (Sections~\ref{sec:MTPb} to~\ref{sec:MTMomentum}). The emittances use the $\beta$-functions of the model rematched to the measured tunes (optics uncertainty \SI{10}{\percent} on the flat top of Pb$^{54+}$, about \SI{15}{\percent} otherwise). Assumed in the extraction studies of Chapter~\ref{sec:SRE}: \SI{400}{\nano\metre\radian} for Pb$^{54+}$ and \SI{300}{\nano\metre\radian} for oxygen; specification of the momentum spread: $3.3\times10^{-3}$ rms (Table~\ref{tab:beamSpecs}).}
\label{tab:MTBeamFT}
\small
\begin{tabular}{@{}ccccccc@{}}
\toprule
Ion & Cycle, date & Stage ($\beta\gamma$) & Ions & $\epsilon_{x,n}$ & $\epsilon_{y,n}$ & $\sigma_\delta$ \\
 & & & ($10^{9}$) & (nm\,rad) & (nm\,rad) & ($10^{-3}$) \\
\midrule
Pb$^{54+}$ & 4BP (a), 7~Aug~2026 & flat top (0.40) & 0.49 & 180 & 355 & -- \\
Pb$^{54+}$ & 4BP (b), 26~Aug~2026 & flat top (0.40) & 1.3 & --$^{a}$ & --$^{a}$ & 0.49$^{b}$ \\
Ne$^{5+}$ & 3BP, 8~Jul~2025 & flat top (0.39) & 3.9 & --$^{c}$ & 240 (210--270) & -- \\
O$^{4+}$ & EARLY, 22~May~2025 & end of ramp (0.31--0.39) & 8.6 & 290--630$^{d}$ & 265--570$^{d}$ & -- \\
\bottomrule
\end{tabular}\\[2pt]
\footnotesize{$^{a}$ Not measured (microchannel plates off); expected about 250 (horizontal) and 175 to \SI{330}{\nano\metre\radian} (vertical), Section~\ref{sec:MTPb}. $^{b}$ Longitudinal tomography, bunch with a coherent dipole oscillation. $^{c}$ No horizontal signal on the flat top. $^{d}$ From the FWHM, between a flat-topped and a Gaussian distribution; nominal optics assumed.}
\end{table}

\section{Beam-based aperture measurement}
\label{sec:MTAperture}

With the beam conditions on the flat top established, the next question is where the beam is lost. The aperture model determines the loss pattern in the ring and whether the extraction septa define the global horizontal aperture restriction, as the extraction method of Chapter~\ref{sec:SRE} requires. The LEIR aperture model~\cite{accmodelsLEIR} had not been validated with beam, and the injection and extraction septa were not included in it. A first beam-based campaign was therefore carried out with Pb$^{54+}$ on 31~October and 3--4~November~2025~\cite{SXMockupPrep2026}.

\begin{figure}[htbp]
    \centering
    \includegraphics[width=\linewidth]{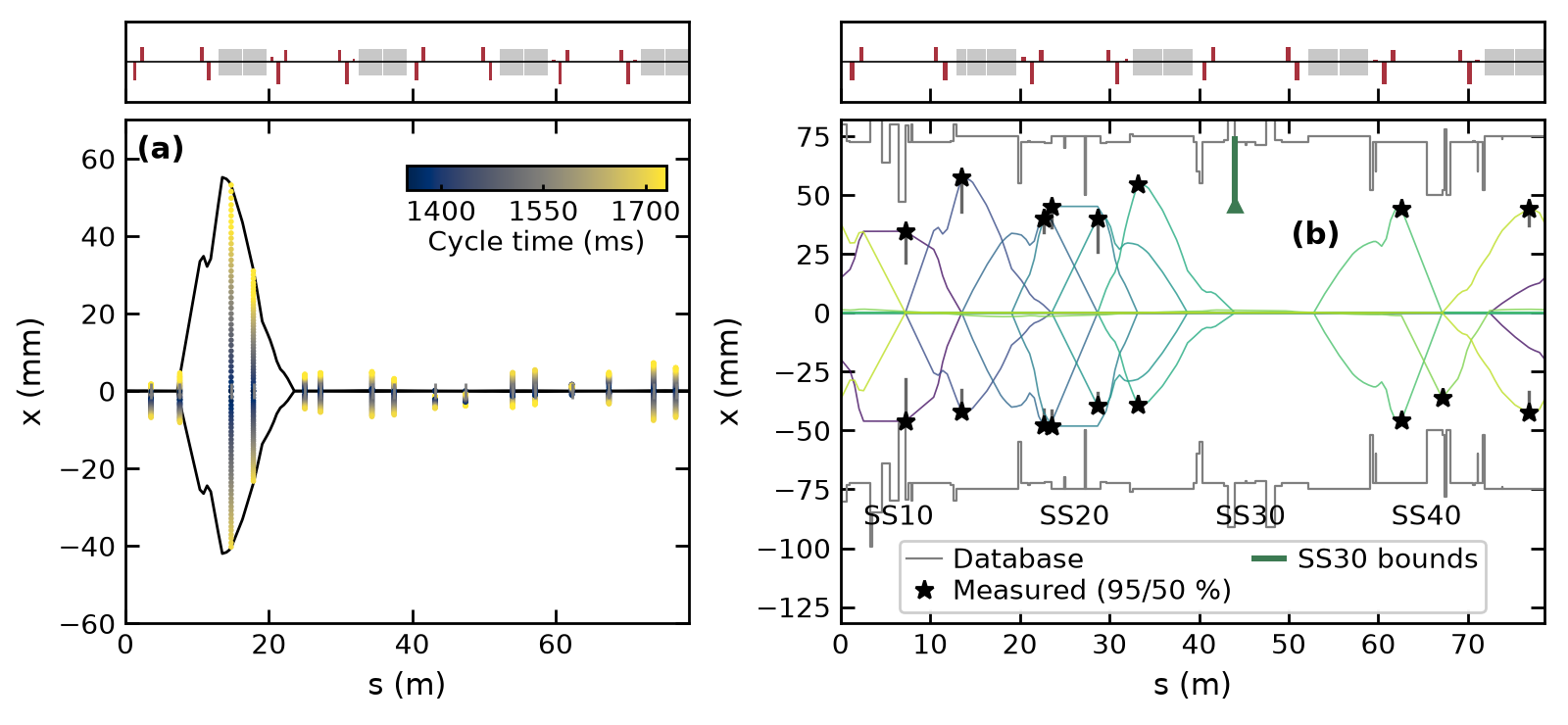}
    \caption{Beam-based horizontal aperture measurement, November~2025. The top strips show the dipoles (grey) and quadrupoles (red). (a)~Example of a closed four-corrector bump ramped over \SI{500}{\milli\second} until beam loss: BPM readings with the closed orbit subtracted, coloured by cycle time, and model bump shape scaled to the readings at the time of loss, for one shot of each polarity. (b)~Mechanical aperture of the LEIR aperture database~\cite{accmodelsLEIR} (grey; the larger chambers of the monitor tanks, 85 to \SI{100}{\milli\metre}, are off scale) and model bump of each scan scaled to the excursion at which \SI{95}{\percent} of the beam is lost (thin lines, one per bump and polarity, coloured by the position of the bump maximum); stars: excursion at the bump maxima, mean over the shots; bars: down to the excursion at which \SI{50}{\percent} is lost; green: lower and upper bounds on the aperture at the loss point in SS30 from the slow-extraction test (Section~\ref{sec:MTSX}).}
    \label{fig:MTAperture}
\end{figure}

\paragraph{Method.} For each section of the ring, a closed and localised orbit bump was computed with MAD-X for a group of four consecutive orbit correctors out of the 14 usable ones. The measurements used a dedicated cycle, on which the RF capture was set to \SI{1305}{\milli\second} and the electron cooler and the orbit bumps of the cycle were switched off at that time. On the flat bottom, the corrector strengths were then scaled up over \SI{500}{\milli\second} until the beam was lost (Fig.~\ref{fig:MTAperture}a). The beam intensity was recorded with the BCT and the orbit with the BPMs, both at a sampling rate of \SI{1}{\kilo\hertz}. For each shot, the time of loss is defined as the time at which the BCT signal has fallen to \SI{5}{\percent} of its value before the ramp. The model bump shape is scaled to the BPM readings at that time by least squares, and the maximum excursion of the scaled bump gives the limit. Ten bumps were scanned in both directions, with about eleven shots each. Two are not used: one at the end of SS40, whose outward scan never removed more than about \SI{80}{\percent} of the beam, and a second bump with its maximum at the injection septum, whose scans did not reach the \SI{95}{\percent} loss level.

\paragraph{Results.} The measurement succeeded in three of the four straight sections: the injection region, the electron cooler and the fast-extraction region. The limits are listed in Table~\ref{tab:MTAperture}. At the \SI{95}{\percent} loss level they range from 35 to \SI{57}{\milli\metre} on the outside and from $-39$ to \SI{-48}{\milli\metre} on the inside of the ring. The tightest restriction, about $+$\SI{35}{\milli\metre}, is located at the injection septum. Together with the restrictions at the vacuum valve VVS22 and at the fast-extraction septum, this is the first experimental verification of the septum apertures in LEIR, which the aperture model either lacked or had not verified. The asymmetry between the outer and inner apertures in the bending magnets is consistent with the recently revised aperture model. For the tracking simulations that prepared the slow-extraction test (Section~\ref{sec:MTSX}), these limits were combined with conservative values where no measurement is available; the simulations of Chapter~\ref{sec:SRE} are based on the previous model.

\begin{table}[htbp]
\centering
\caption{Horizontal limits from the beam-based aperture measurement of November~2025, at the maximum of each bump: excursion at which \SI{95}{\percent} and \SI{50}{\percent} of the beam are lost, mean over the shots (the standard deviation over the shots is 0.1 to \SI{0.8}{\milli\metre}); $D_x$: model dispersion at the bump maximum.}
\label{tab:MTAperture}
\small
\begin{tabular}{cccccc}
\toprule
$s$ (m) & Location & $D_x$ (m) & \multicolumn{2}{c}{Outer limit (mm)} & Inner limit (mm) \\
 & & & \SI{95}{\percent} & \SI{50}{\percent} & \SI{95}{\percent} / \SI{50}{\percent} \\
\midrule
7.2  & SS10, injection septum & $-10.3$ & 34.6 & 21.2 & $-46.0$ / $-28.0$ \\
13.5 & bending magnet, sector 1 & $-3.8$ & 57.5 & 42.8 & $-42.0$ / $-32.7$ \\
22.6 & QFN23, sector 2 & 0.0 & 40.2 & 34.2 & $-47.7$ / $-41.1$ \\
23.6 & SS20, entrance & 0.0 & 45.2 & 36.5 & $-48.1$ / $-41.6$ \\
28.7 & SS20, electron cooler & 0.0 & 40.3 & 25.9 & $-39.4$ / $-34.1$ \\
33.1 & bending magnet, sector 2 & 0.0 & 54.9 & 53.3 & $-39.1$ / $-36.8$ \\
62.6 & SS40, upstream end & 0.0 & 44.2 & 42.5 & $-45.8$ / $-44.9$ \\
76.8 & sector 4, end & $-2.6$ & 44.1 & 37.1 & $-42.5$ / $-33.6$ \\
\bottomrule
\end{tabular}
\end{table}

\begin{figure}[htbp]
    \centering
    \includegraphics[width=\linewidth]{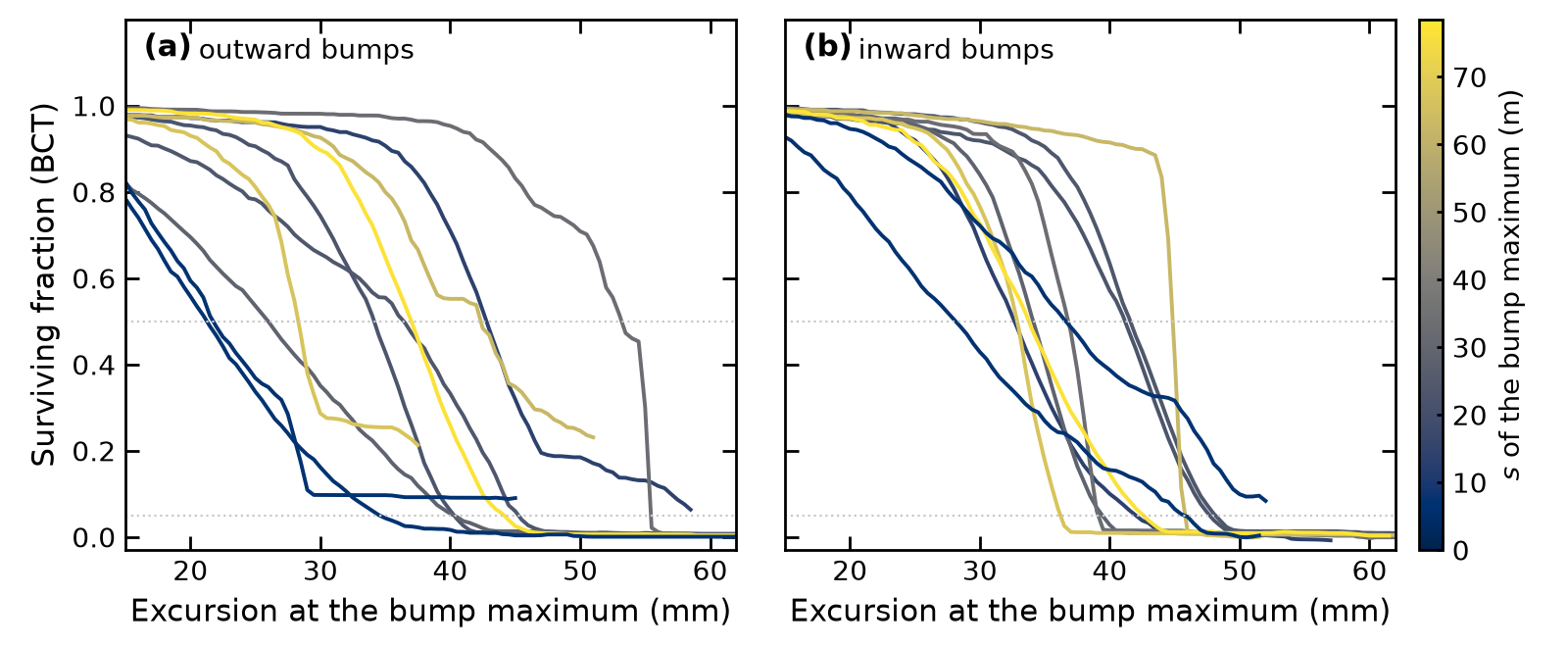}
    \caption{Loss curves of the beam-based aperture measurement, November~2025: surviving fraction of the beam (BCT, normalised to its value before the bump ramp) against the excursion of the scaled model bump at its maximum, median over the shots of each bump, coloured by the position of the bump maximum. (a)~Outward bumps; (b)~inward bumps. Dotted: \SI{50}{\percent} and \SI{5}{\percent} surviving fraction.}
    \label{fig:MTApertureLoss}
\end{figure}

\paragraph{Uncertainties.} Figure~\ref{fig:MTApertureLoss} shows, for each bump, the intensity measured by the BCT as a function of the bump excursion. The limits of the individual shots of a bump agree within 0.1 to \SI{0.8}{\milli\metre}, and the first analysis of the same data, made with averaged BPM readings, is reproduced within \SI{0.4}{\milli\metre}. The beam is not lost at a sharp edge: between \SI{50}{\percent} and \SI{95}{\percent} loss the excursion increases by 1.6 to \SI{18}{\milli\metre} (Table~\ref{tab:MTAperture}). For a beam scraped by an edge, the lost fraction is $\exp(-d^2/2\sigma^2)$, where $d$ is the distance between the beam centre and the edge and $\sigma$ the rms beam size. At \SI{95}{\percent} loss the beam centre is therefore within $0.3\sigma$ of the edge, so the \SI{95}{\percent} level is the better estimate of the limit, and from \SI{50}{\percent} to \SI{95}{\percent} loss the excursion should increase by $0.86\sigma$. With the emittance measured on the aperture cycle, \SI{184}{\nano\metre\radian} (Section~\ref{sec:MTPb}), the rms beam size at the bump maxima is 3.7 to \SI{4.4}{\milli\metre}, and about \SI{11}{\milli\metre} at the injection septum, where the dispersion ($D_x = \SI{-10.4}{\metre}$) and the momentum spread of about $1\times10^{-3}$ after RF capture dominate. The measured increase corresponds to 0.4 to 1.5 rms beam sizes at five of the eight locations, but to 2.3 to 3.9 at the entrance of SS20, at the electron cooler and in the bending magnet of sector~1. There, part of the beam is lost before the bump maximum reaches the limit: the restriction does not lie at the bump maximum, or the beam touches more than one restriction. Since the limit is quoted at the maximum of the model bump, the physical aperture at a restriction located elsewhere within the bump is smaller than the quoted value.

\paragraph{Comparison with the aperture model.} The measured limits lie systematically inside the mechanical aperture of the database (Fig.~\ref{fig:MTAperture}b). For each bump we determine the excursion at the bump maximum at which the scaled bump first touches the database aperture, outwards and inwards. The ratio of the measured to this excursion ranges from 0.52 to 0.98, and two groups appear:
\begin{itemize}
    \item Where the database contains a narrow element within the bump, the agreement is close. At the injection septum (database \SI{47}{\milli\metre}) the measured full aperture is \SI{14}{\percent} smaller, and at the entrance of SS20 (database \SI{50}{\milli\metre} at the monitor UEH22) \SI{7}{\percent} smaller. The two polarities of a bump then correspond to a scale factor $k$ of the BPM readings of 1.16 and 1.07 and to an offset $c$ of the orbit at the bump maximum of 6.6 and \SI{1.5}{\milli\metre}. These are compatible with a calibration error of the BPMs, with a misalignment of the septum, and with the closed orbit at the bump maximum, which is not measured.
    \item Where the database contains only the vacuum chamber of the bending magnets and of the straight sections (half-aperture 72 to \SI{75}{\milli\metre}), the measured full apertures are 34 to \SI{42}{\percent} smaller. Reconciling them would require $k = 1.5$ to 1.7, which is not plausible for a calibration error; the offsets, up to \SI{13}{\milli\metre} in the bending magnets, reproduce the asymmetry between the outer and inner limits there.
\end{itemize}
The optics model does not explain the difference. The bumps were computed, and extrapolated from the BPMs to the bump maximum, with the previous model of LEIR. Repeating the extrapolation with the 2026 model rematched to the measured tunes (Section~\ref{sec:MTEmittance}), whose $\beta$-functions differ from the previous ones by up to \SI{30}{\percent}, changes the limits by at most \SI{1.5}{\percent}, since each bump is fitted to the readings of the 16 BPMs. The measurement therefore confirms the narrow elements of the database within the uncertainties of the BPM calibration and of the alignment, and shows that the database lacks restrictions of about 40 to \SI{55}{\milli\metre} elsewhere, most likely in the bending magnets. For the extraction, which relies on the septa defining the global aperture, the measured limits are used as an effective aperture model. Locating the missing restrictions requires a beam-based calibration of the BPMs, the alignment of the septa and of the monitors, bumps with their maximum at the suspected elements, and a review of the aperture database against the mechanical drawings (Section~\ref{sec:MTProgramme}).

\paragraph{SS30.} In the free straight section SS30, where the extraction septa are to be installed, the bumps could not be localised. The orbit correctors inside SS30 have a small effect on the orbit at the SS30 BPMs, as the near-zero entries of the measured orbit-response matrix confirm, so the closed-orbit leakage of the bump becomes significant. The slow-extraction test of Section~\ref{sec:MTSX} nevertheless bounds the aperture at its loss point, the horizontal corrector DHV31 at the start of SS30 ($\beta_x \approx \SI{8.6}{\metre}$ in the slow-extraction optics; dispersion of about \SI{-1.8}{\metre} measured in SS30, Section~\ref{sec:MTOptics}):
\begin{itemize}
    \item With the SS30 bump of 41 to \SI{44}{\milli\metre} at the first SS30 monitor, the beam without excitation loses less than \SI{2}{\percent} on the flat top. For a beam with Gaussian betatron amplitudes this places the aperture at least 2.8 rms beam sizes beyond the orbit. With $\sigma_x \approx \SI{2.2}{\milli\metre}$, from $\epsilon_{x,n} \approx \SI{180}{\nano\metre\radian}$ (Section~\ref{sec:MTPb}) and the dispersive contribution of the momentum spread, the half-aperture at DHV31 is at least about \SI{47}{\milli\metre}.
    \item With the excitation, the losses are concentrated at DHV31, so the distance to the aperture there, normalised to $\sqrt{\beta_x}$, is smaller than at all the locations measured above. The smallest normalised limit among those is 11.4~mm\,m$^{-1/2}$, which gives an upper bound of about \SI{75}{\milli\metre}.
\end{itemize}
The aperture of the database at DHV31, \SI{72.5}{\milli\metre}, lies within these bounds.

\section{Optics of the slow-extraction configuration}
\label{sec:MTOptics}

The aperture measurements were made with the nominal optics. The slow-extraction optics described in Chapter~\ref{sec:SRE} brings the horizontal working point close to the third-integer resonance and displaces the dispersion wave by one sector, so that the dispersion is small in SS10 and SS30 and maximum in SS20 and SS40. Three types of measurement test whether this configuration can be established in the machine and whether the model describes it.

\begin{figure}[htbp]
    \centering
    \includegraphics[width=\linewidth]{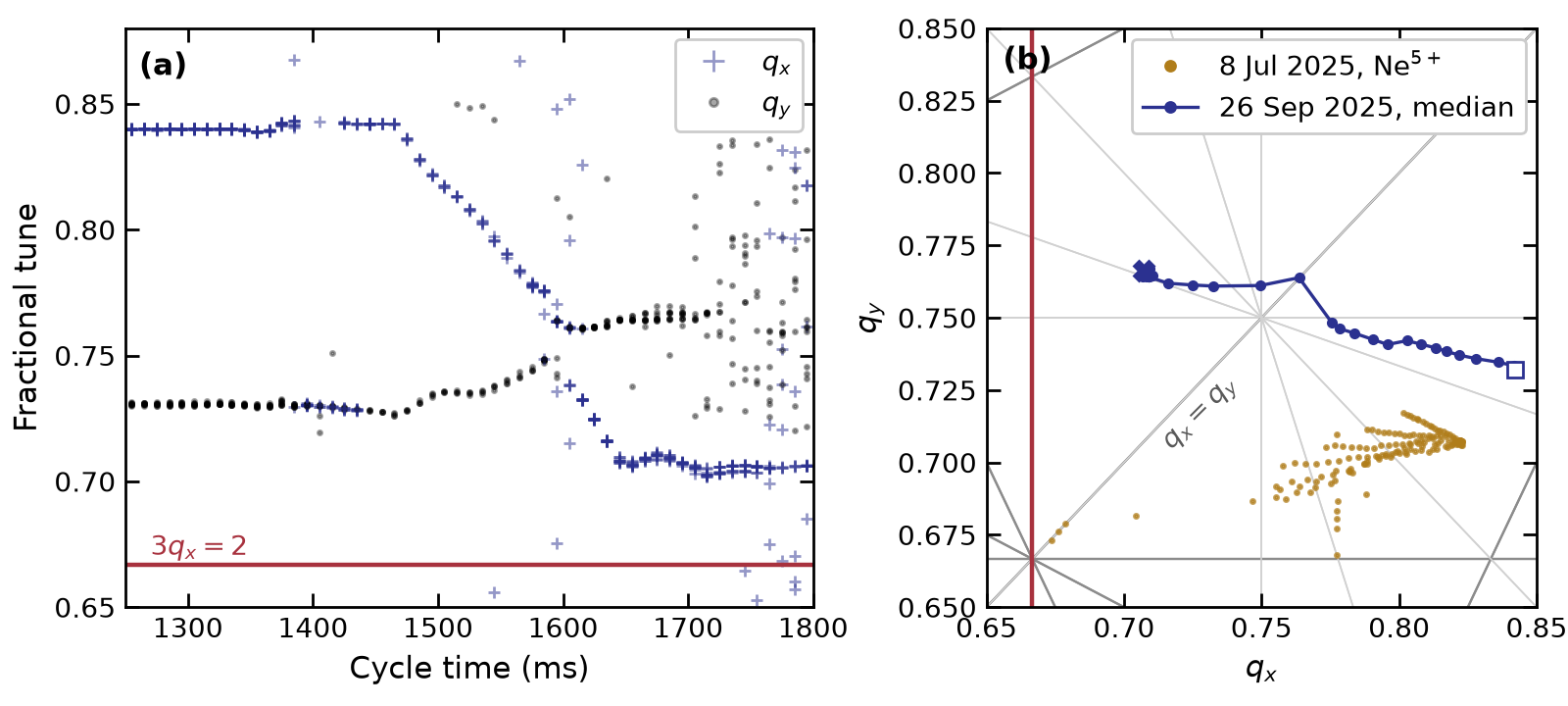}
    \caption{Move to the slow-extraction working point. (a)~Fractional tunes measured with the BBQ during the optics change of 26~September~2025 (ten cycles). (b)~Tune diagram with resonance lines up to fourth order: median path of 26~September~2025 (open square: start; cross: end) and tunes measured on 8~July~2025 with a Ne$^{5+}$ beam for different scaling factors of the optics change. The red line marks the third-integer resonance.}
    \label{fig:MTTunes}
\end{figure}

\paragraph{Working point.} A first attempt to establish the slow-extraction working point was made on 8~July~2025 with a Ne$^{5+}$ beam~\cite{MD20250708}. The strengths of all five quadrupole families were scaled in steps from the nominal to the slow-extraction values, with a \SI{188}{\milli\second} ramp at flat bottom after \SI{1.4}{\second} of injection and cooling. No beam loss was observed, and the trim procedure was validated. The tunes, however, locked onto the difference resonance $q_x = q_y$, which could not be crossed (Fig.~\ref{fig:MTTunes}b). On 26~September~2025 the path in the tune diagram was modified so as to cross the difference resonance as perpendicularly as possible: the vertical tune was raised to about 0.76 while the horizontal tune was lowered from 0.84 to about 0.70 within about \SI{180}{\milli\second} (Fig.~\ref{fig:MTTunes}a). The working point was reached with the RF-capture efficiency of nominal operation and with the dispersion wave displaced by one sector, as confirmed by the measured orbit response to an energy change. The same strategy was applied to both set-ups of the 4BP cycle (Section~\ref{sec:MTSX}).

\begin{figure}[htbp]
    \centering
    \includegraphics[width=\linewidth]{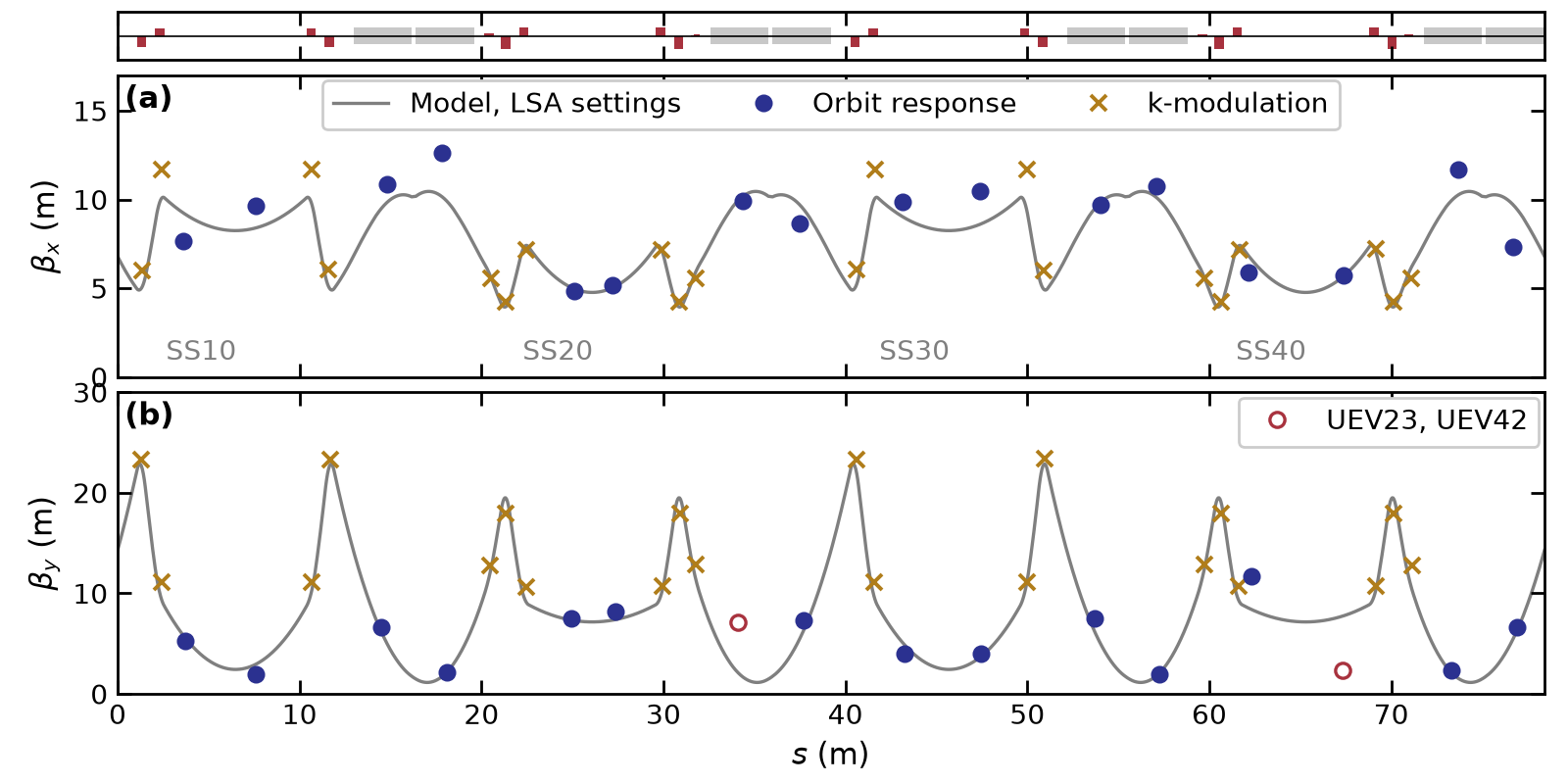}
    \caption{Horizontal (a) and vertical (b) $\beta$-functions of the slow-extraction optics on the flat top of the 4BP cycle (\SI{3700}{\milli\second}), from the orbit response matrix and from k-modulation of the quadrupole families (shown at each quadrupole of the family), compared with the model of August~2026 with the quadrupole strengths set in LSA~\cite{SLAWGOptics2026}. The top strip shows the dipoles (grey) and quadrupoles (red). The readings of UEV23 and UEV42 (open circles) are excluded from the statistics. The measured values are digitised from the figures of Ref.~\cite{SLAWGOptics2026}.}
    \label{fig:MTBeta}
\end{figure}

\paragraph{Optics functions.} The $\beta$-functions of the slow-extraction optics were measured on the flat top of the 4BP cycle, at \SI{3700}{\milli\second}, with two independent methods: from the orbit response matrix, at the 16 monitors of each plane, and by k-modulation of the five quadrupole families, which gives the average $\beta$-function over the quadrupoles of each family~\cite{SLAWGOptics2026}. The measurements are compared in Fig.~\ref{fig:MTBeta} with the model of August~2026, which contains the multipole components of the main dipoles derived from their field maps, evaluated at the injection field~\cite{LEIRDipoleModel2026}, with the quadrupole strengths set in LSA. With these components the model gives smaller horizontal $\beta$-functions in the dipoles, in better agreement with the measurements: the horizontal $\beta$-beating of the orbit-response data, \SI{24}{\percent} rms with respect to the previous model of the slow-extraction optics, is \SI{14}{\percent} rms with respect to the new model, with a maximum of \SI{25}{\percent}. In the vertical plane, UEV23 and UEV42 give readings inconsistent with their neighbours ($+230$ and \SI{-70}{\percent}); UEV42 was also anomalous in September~2025, which points to faulty monitors. Excluding them, the vertical $\beta$-beating is \SI{17}{\percent} rms, with a maximum of \SI{38}{\percent}. The k-modulation confirms the model within \SI{11}{\percent} for all families in the vertical plane and within \SI{4}{\percent} for three of the five families in the horizontal plane. The exceptions are the families QDN1030 and QFN1030 in SS10 and SS30, where the measured horizontal $\beta$-functions exceed the model by 11 and \SI{25}{\percent}, in agreement with the orbit-response values at the SS30 monitors, UEH31 and UEH32 ($+9$ and \SI{+22}{\percent}). The measurement at \SI{3700}{\milli\second} is affected by the transients of the main dipoles on the flat top, which make the tunes and the beam energy drift (Section~\ref{sec:MTMomentum}), and the analysis is preliminary~\cite{SLAWGOptics2026}. For the extraction, a $\beta$-beating of this size changes the resonance driving term of given sextupole currents by about \SI{20}{\percent} (\SI{30}{\percent} for \SI{25}{\percent} rms, Section~\ref{sec:SREsext}), so the sextupole settings will be established with beam.

\begin{figure}[htbp]
    \centering
    \includegraphics[width=\linewidth]{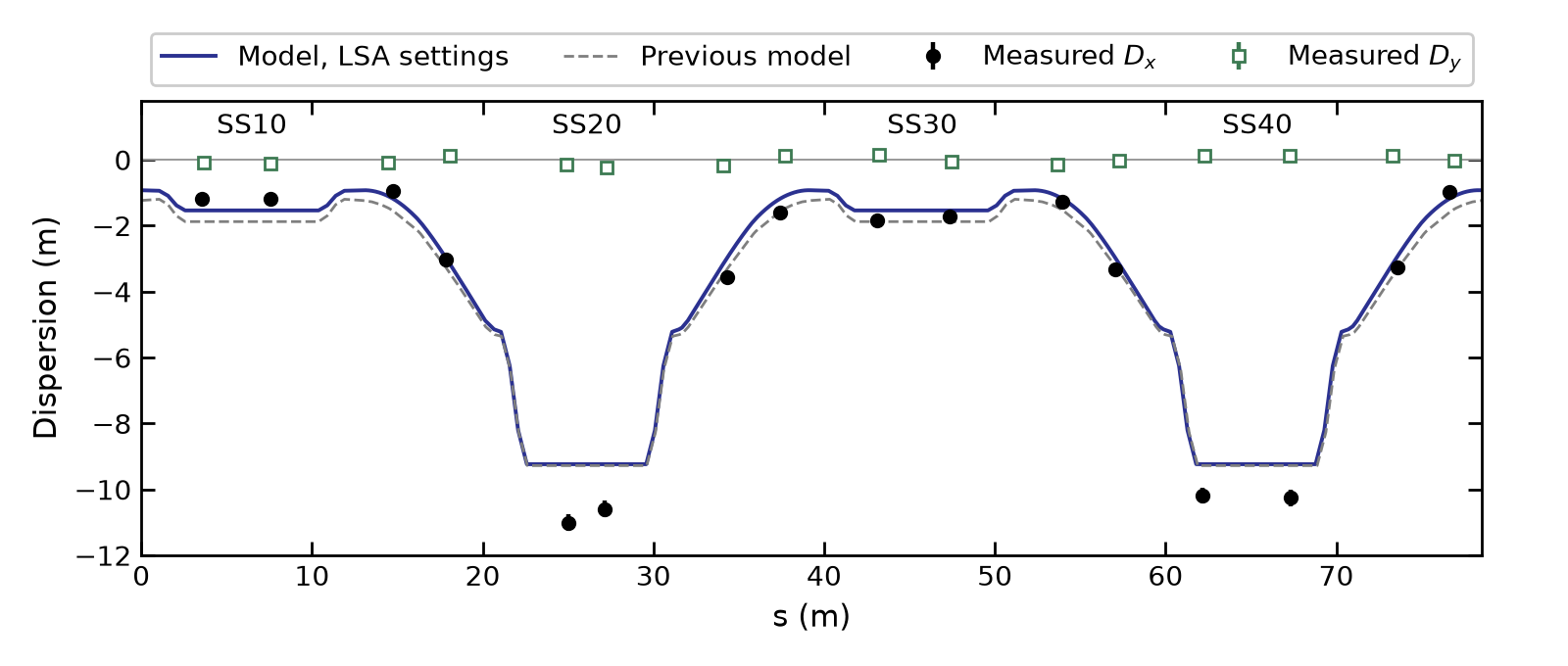}
        \caption{Horizontal and vertical dispersion measured on the flat top of the 4BP cycle on 26~August~2026, with the SS30 bump applied and the resonance sextupoles powered, compared with the model of the slow-extraction optics of August~2026 with the quadrupole strengths set in LSA (model tunes 1.744 and 2.792, not rematched to the measured 1.691 and 2.80) and with the previous model. Error bars: standard error of the slope from the residuals of the fit to the four RF settings (one cycle each).}
    \label{fig:MTDispersion}
\end{figure}

\paragraph{Dispersion.} The dispersion was measured on the flat top of the 4BP cycle on 26~August~2026 from the orbit shift produced by changes of the RF frequency of $\Delta f/f = +7.5\times10^{-4}$, $-7.5\times10^{-4}$ and $-1.5\times10^{-3}$ with respect to a reference setting, i.e.\ momentum offsets of about $\pm 1\times10^{-3}$ and $-2\times10^{-3}$ (Fig.~\ref{fig:MTDispersion}). For each BPM, the orbit averaged from 3700 to \SI{3990}{\milli\second} was fitted with a straight line against $\Delta f/f$, and the slope was converted into dispersion with the slip factor of the model, $\eta = \alpha_c - 1/\gamma^2 = -0.738$. The measurement has the following limitations:
\begin{itemize}
    \item \emph{Statistics.} Each RF setting is a single cycle, and the reference cycle was recorded about \SI{30}{\minute} before the other three. The fit has a free offset, so a slow orbit drift between the cycles enters only through the cycle-to-cycle reproducibility. The uncertainty quoted below is the standard error of the slope derived from the residuals of the four-point fit: about \SI{2.5}{\percent}, or 0.03 to \SI{0.26}{\metre}.
    \item \emph{Linearity and beam loss.} The cycle with the largest offset lost \SI{7}{\percent} of its beam during the averaging window, against at most \SI{1.5}{\percent} for the others. Omitting it changes the dispersion by about \SI{1}{\percent}.
    \item \emph{Slip factor.} An error of \SI{10}{\percent} on the model value of $\alpha_c$ would change all values by \SI{1.7}{\percent}, alike.
    \item \emph{Machine conditions.} The four cycles were recorded in the configuration of the slow-extraction test, with the SS30 bump applied and the resonance sextupoles powered (Section~\ref{sec:MTSX}); the exciter was disabled, and the intensity on the flat top was between 1.37 and $1.46\times10^{9}$ ions. To compare with the model in the same configuration, the bump was added to the MAD-X model by fitting the orbit correctors to the measured orbit change (residual \SI{1.8}{\milli\metre} rms over the 16 monitors), with the sextupoles at their settings and the tunes rematched to the measured values. The bump and the quadrupolar feed-down of the sextupoles in it change the model dispersion at the monitors by less than \SI{0.17}{\metre}, that is, by less than \SI{2}{\percent} where the dispersion is largest. During the averaging window the revolution frequency rises by \SI{1.4}{\kilo\hertz} in all four cycles, while the orbit at the high-dispersion monitors stays constant within \SI{0.6}{\milli\metre} rms: this is the RF compensation of the field drift described in Section~\ref{sec:MTMomentum}, under which the beam momentum remains constant.
    \item \emph{BPM calibration.} The calibration of the individual monitors has not been checked; a scale error of a monitor enters its dispersion value directly.
\end{itemize}
The measurement is compared with the model of the slow-extraction optics of August~2026 with the quadrupole strengths set in LSA for the flat top. With these strengths the model tunes are $(1.744, 2.792)$, against $(1.691, 2.80)$ measured with the BBQ; the strengths are used as set, without rematching the tunes, because the dispersion of a rematched model depends on which quadrupole families are adjusted. Over the 16 horizontal monitors the measured dispersion deviates from the model by \SI{18}{\percent} rms, which is larger than the uncertainties listed above, but the measured dispersion wave has the shape of the model, with the extremes in the expected straight sections. In SS30, where the extraction septa are foreseen, the measured values, $-1.84 \pm 0.04$ and $-1.73 \pm 0.04$~m, are 12 to \SI{20}{\percent} larger in magnitude than the model value of \SI{-1.54}{\metre}. In SS20 and SS40, where the dispersion is largest, the measured values exceed the model by 10 to \SI{19}{\percent} ($-10.2$ to \SI{-11.0}{\metre} against \SI{-9.23}{\metre}), and in SS10 and at the adjacent monitors UEH13 and UEH44 they are 18 to \SI{23}{\percent} smaller. The vertical dispersion is below \SI{0.25}{\metre} everywhere. The deviations are not caused by the bump or by the sextupoles, which change the model by much less, and the difference between SS10 and SS30, which the model treats alike, points to the calibration of individual monitors or to field and alignment errors of the ring. An error of the slip factor would scale all values alike and cannot, by itself, explain the pattern. For the extraction, the dispersion in SS30 is now known from the measurement, with an uncertainty of about \SI{2.5}{\percent}, in the configuration in which the extraction will be operated; a calibration of the monitors with the largest deviations (UEH11 to UEH13, UEH44 and UEH21) would separate the remaining causes.

\paragraph{Chromaticity on the flat top.} The same four cycles give the horizontal and vertical chromaticity on the flat top, from the tunes measured by the BBQ between 3600 and \SI{3990}{\milli\second}: $Q'_x = -4.7 \pm 0.3$ and $Q'_y = -2.9 \pm 0.2$, with $Q' = \mathrm{d}Q/(\Delta p/p)$ and the standard error of a linear fit to the four settings (one cycle each). Both tunes follow the linear dependence within \num{7e-4}, the cycle-to-cycle scatter of the BBQ, over the range of momentum offsets from $-2\times10^{-3}$ to $+1\times10^{-3}$. Like the dispersion, they were measured with the SS30 bump applied and the resonance sextupoles powered. The 2026 model of the slow-extraction optics without chromatic sextupoles, which were not powered in 2026~\cite{LEIRNLO2026}, gives $Q'_x$ between $-1.4$ and $-1.7$ and $Q'_y$ between $-4.0$ and $-4.6$, depending on whether and how the tunes are rematched. The resonance sextupoles do not change these values: at the dispersion of about \SI{-1.5}{\metre} at their location, the contributions of the SS10 and SS30 pairs cancel. The measured horizontal chromaticity is therefore about 3 units more negative, and the vertical one 1 to 2 units less negative, than in the model. Shifts of opposite sign in the two planes are the signature of sextupolar fields in dispersive regions that the model does not contain, for example in the main dipoles: their field maps show a sextupole component that depends on the excitation level~\cite{LEIRDipoleModel2026}, which the 2026 model does not include, and their transients after the ramp do not appear to decay on the flat top~\cite{LEIRNLO2026}.

\section{Transverse excitation of the circulating beam}
\label{sec:MTRFKO}

As described in Chapter~\ref{sec:SRE}, the slow extraction of HEARTS@LEIR is foreseen to be driven by RF knockout (RF-KO). A transverse excitation near the betatron frequency increases the amplitudes of the particles inside the stable region of the third-integer resonance until they cross the separatrix, and the extracted flux is controlled through the amplitude of the excitation. In LEIR the excitation can be applied with the transverse feedback (TFB) system, which would also be the actuator of a spill feedback loop. This section reports a first measurement of the response of the beam to such an excitation (Sections~\ref{sec:MTRFKOSetup} to~\ref{sec:MTRFKOKick}) and compares the measured excitation strength with the strength needed for the shortest spills of the specification (Section~\ref{sec:MTRFKOReq}).

The measurement relies on the horizontal ionisation profile monitor, whose profiles of the excited beam are of modest quality (Section~\ref{sec:MTRFKOMethod}). The results are therefore mainly qualitative: they establish how the beam responds to the excitation and where the present system saturates, while the diffusion coefficients derived below are indicative. Their uncertainty is about \SI{25}{\percent}, from the choice of the fit window (\SI{22}{\percent}), the $\beta$-function at the monitor (\SI{10}{\percent}) and the statistics of the slopes (\SI{2}{\percent}), added in quadrature. In addition, they are lower bounds as far as the measured profiles are incomplete, whereas a saturation of the monitor in the dense core of the beam would act in the opposite direction (Section~\ref{sec:MTRFKOMethod}).

\subsection{Excitation system and measurement}
\label{sec:MTRFKOSetup}

The horizontal kicker of the TFB, KDHV21, is a pair of strip-line electrodes, \SI{218}{\milli\metre} long and \SI{66}{\milli\metre} wide, separated by a gap of \SI{110}{\milli\metre}~\cite{LEIRTFB2008}. Each electrode is driven by a \SI{100}{\watt} amplifier, with the wave travelling against the beam so that the electric and magnetic forces add; the dump loads of the system are rated for the same power. The excitation signal is generated by the BBQ system and fed into the TFB. It is a linear frequency sweep (chirp) that covers a band of fractional tunes of width $\Delta q = 0.01$ in \SI{1}{\milli\second} and is repeated every millisecond. Its amplitude is set by the dimensionless gain $A$ of the BBQ output, $0 \leq A \leq 1$. On the cycle used for the measurement the feedback function of the TFB was gated from 205 to \SI{2800}{\milli\second}, so the damper was off during the excitation and did not counteract it. The four resonance sextupoles were at zero current during this scan: the beam was not close to a separatrix, and particles driven to large amplitudes are lost on the machine aperture. The same excitation with the sextupoles powered is discussed in Section~\ref{sec:MTSX}.

The measurement was made on the flat top of the 4BP cycle at \SI{4.8}{\tesla\metre} with about $0.5\times10^{9}$ Pb$^{54+}$ ions. The excitation was applied from \SI{3500}{\milli\second} to the end of the cycle at \SI{4049}{\milli\second}. On 7~August~2026, with the chirp covering $q_x$ from 0.685 to 0.695, the gain was scanned in seven steps from 0.05 to 1.0, with 6 to 14 cycles per step and six cycles without excitation ($A = 0$ and 0.01) as reference. Three instruments were used:
\begin{itemize}
    \item The horizontal IPM delivered a profile integrated over \SI{20}{\milli\second} every \SI{10}{\milli\second}; five consecutive profiles are averaged here into bins of \SI{50}{\milli\second}.
    \item The BCT recorded the intensity.
    \item During the excitation the BBQ records the spectrum of the coherent response of the beam in windows of 1024 turns, every millisecond, and returns the frequency of its maximum.
\end{itemize}

An earlier scan on 5~August~2026, on a copy of the 4BP cycle with the excitation starting at \SI{3600}{\milli\second}, was analysed with the same procedure (Section~\ref{sec:MTRFKOMethod}). With the chirp below the horizontal tune ($q_x$ from 0.667 to 0.685 and narrower bands within it), the second moment did not grow beyond that of the unexcited beam for any gain between 0.01 and 1.0: the excess slopes are within $\pm\SI{20}{\milli\metre\squared\per\second}$, the scatter between cycles, and the BCT loss between 3600 and \SI{4000}{\milli\second} is below \SI{0.5}{\percent}. When the band was moved to $q_x$ from 0.710 to 0.726 at $A = 1.0$ at the end of the scan, the slope rose to $252 \pm 3$~\si{\milli\metre\squared\per\second} (three cycles), about \SI{70}{\percent} of the saturated value of 7~August~2026, and \SI{27}{\percent} of the beam was lost within \SI{400}{\milli\second}; with $q_x$ from 0.720 to 0.730 it was \SI{138}{\milli\metre\squared\per\second} and \SI{12}{\percent}. The excitation therefore acts only when the chirp covers the tune distribution of the beam. This makes the position of the band, and not only the gain, a parameter of the spill control.

\subsection{Profile data and analysis method}
\label{sec:MTRFKOMethod}

\begin{figure}[htbp]
    \centering
    \includegraphics[width=\linewidth]{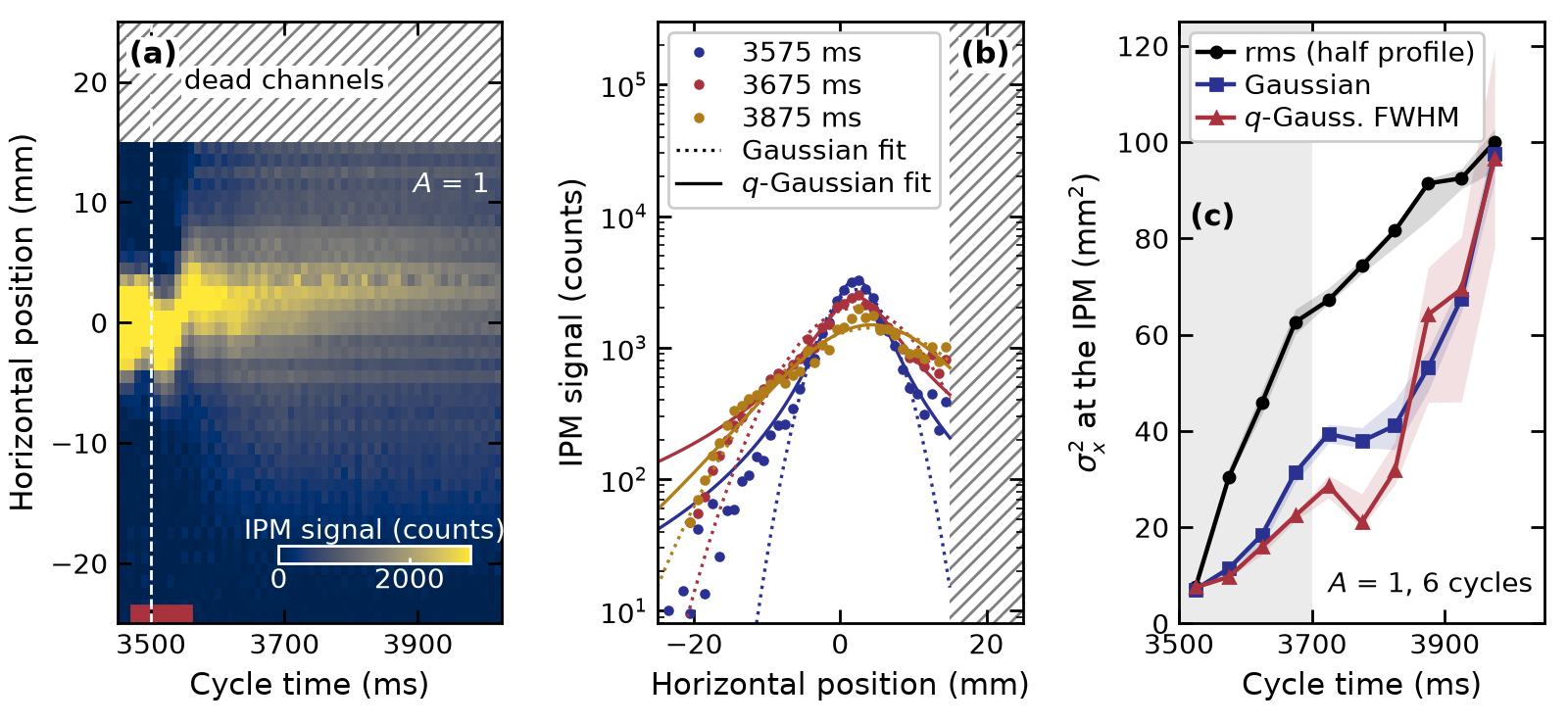}
    \caption{IPM data for the exciter gain $A = 1.0$, 7~August~2026. (a)~Horizontal profile of one cycle, delivered every \SI{10}{\milli\second}, after subtraction of the baseline measured on the outermost channels, on a linear colour scale saturated at the 99.8th percentile of the signal after \SI{3570}{\milli\second}. The excitation starts at \SI{3500}{\milli\second} (dashed line). Hatched: channels without signal on the flat top. Red bar: transient at the end of the ramp, common to all cycles. (b)~Profiles of the same cycle averaged over \SI{50}{\milli\second}, on a logarithmic scale (points at or below zero not shown), with the fits of a Gaussian (dotted) and of a $q$-Gaussian (solid); the logarithmic scale shows the halo that dominates the second moment of the profile. (c)~Beam size of the six cycles at this gain, median and interquartile range: second moment of the half of the profile on the side of the full detector range (reference), squared width of the Gaussian fit, and squared full width at half maximum of the $q$-Gaussian fit divided by $2.355^2$. Shaded: fit window of Table~\ref{tab:MTRFKO}.}
    \label{fig:MTRFKOIPMhigh}
\end{figure}

\begin{figure}[htbp]
    \centering
    \includegraphics[width=\linewidth]{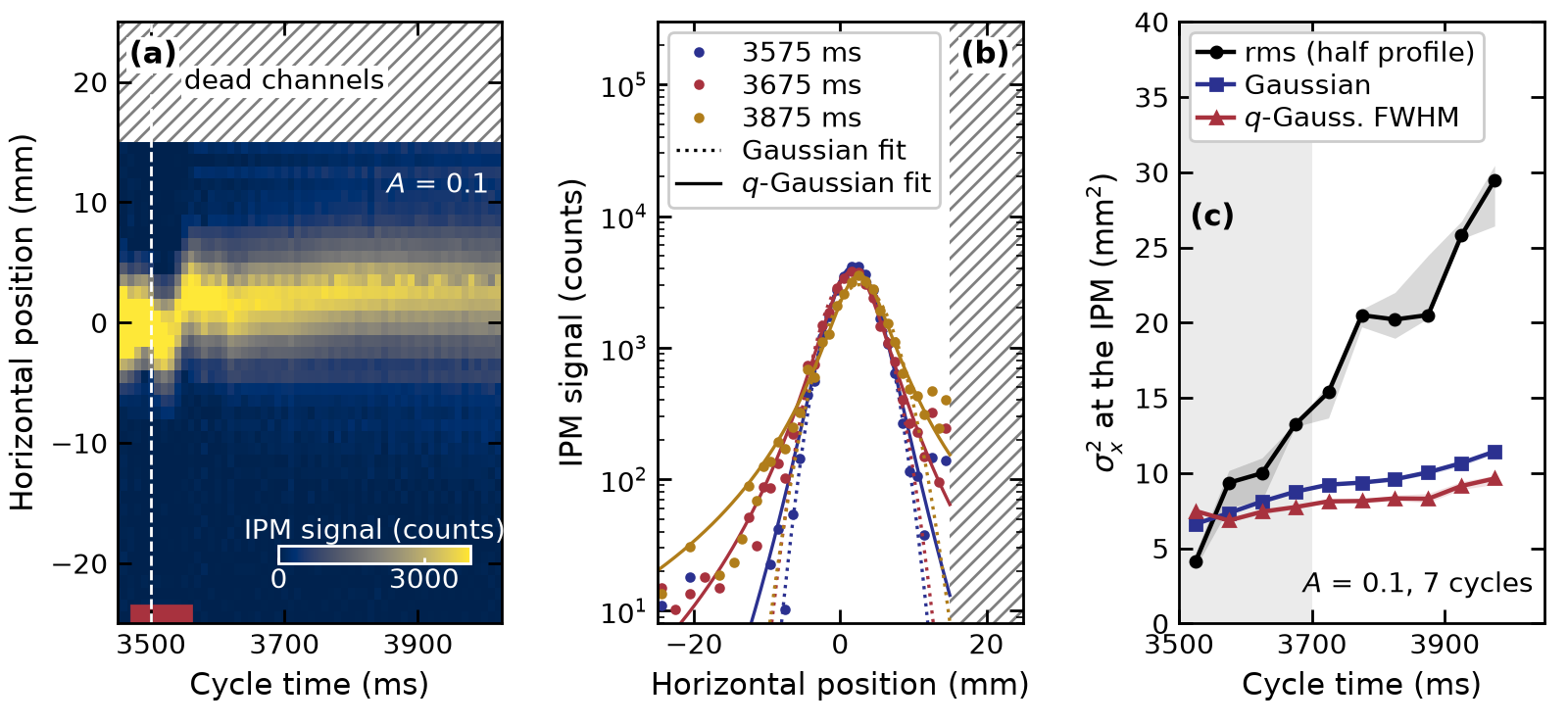}
    \caption{IPM data for the exciter gain $A = 0.1$ (linear regime), presented as in Fig.~\ref{fig:MTRFKOIPMhigh}; (c)~seven cycles.}
    \label{fig:MTRFKOIPMlow}
\end{figure}

Figures~\ref{fig:MTRFKOIPMhigh} and~\ref{fig:MTRFKOIPMlow} illustrate the data for a saturated gain ($A = 1.0$) and for a gain in the linear regime ($A = 0.1$). Panel~(a) shows the raw signal of one cycle around the start of the excitation, panel~(b) the profiles of the same cycle at three times with two fits, and panel~(c) three measures of the beam size for all cycles of the gain.

\paragraph{Quality of the profiles.} Four properties of the monitor limit the analysis:
\begin{itemize}
    \item \emph{Usable range.} On the flat top, the channels above $+\SI{15}{\milli\metre}$ carry no signal. With the beam centred at about $+\SI{2.5}{\milli\metre}$, the profile is measured up to \SI{27}{\milli\metre} from the centre on one side and up to only \SI{12.5}{\milli\metre} on the other.
    \item \emph{Transient at the end of the ramp.} From 3470 to \SI{3560}{\milli\second}, every cycle, with and without excitation, shows a baseline shift of several hundred counts, a displacement of the horizontal profile by up to \SI{4}{\milli\metre} and a drop of the integrated horizontal signal by up to a factor of three (panel~a). The profiles of the first \SI{50}{\milli\second} of the excitation are affected by it. The transient coincides with the end of the ramp, and it is also present on 5~August~2026, when the excitation started only at \SI{3600}{\milli\second}. The displacement corresponds, at $D_x = \SI{-2.9}{\metre}$, to a momentum change of about $1.4\times10^{-3}$, consistent with the momentum drift at the start of the flat top caused by the decay of eddy currents in the main magnets (Section~\ref{sec:MTMomentum}); it is therefore a property of the beam. The vertical profile, at zero dispersion, keeps its position and its integrated signal. The drop of the horizontal signal is therefore not a change of the beam intensity but an effect on the horizontal monitor, which is located close to a bending magnet, plausibly caused by the change of the magnetic field at the end of the ramp.
    \item \emph{Systematic errors.} Even the narrow profile of the unexcited beam is described by a Gaussian only to $\chi^2/\mathrm{ndf} = 4$ to 10 relative to the statistical errors, which reflects channel-to-channel variations of the response.
    \item \emph{Response.} For the wide profiles of the large gains the integrated signal is not proportional to the beam intensity (Section~\ref{sec:MTRFKOLate}).
\end{itemize}

\paragraph{Profile shape.} The excitation does not broaden the profile uniformly. Already at $A = 0.1$ the core keeps its initial width while tails develop (Fig.~\ref{fig:MTRFKOIPMlow}b). At $A = 1.0$ the core is depleted within a few hundred milliseconds, and \SI{175}{\milli\second} after the start of the excitation 10 to \SI{16}{\percent} of the peak signal is still found 15 to \SI{20}{\milli\metre} from the centre (Fig.~\ref{fig:MTRFKOIPMhigh}b). Four descriptions were compared on all profiles:
\begin{itemize}
    \item A Gaussian describes the unexcited beam, but not the excited one: for $A \geq 0.3$ its $\chi^2/\mathrm{ndf}$ exceeds that of the unexcited beam typically by a factor of five from \SI{100}{\milli\second} on.
    \item A $q$-Gaussian, $[1+(q-1)u^2]^{-1/(q-1)}$ with $u = (x-x_0)/w$, which becomes a Gaussian for $q \to 1$ and develops power-law tails for $q > 1$, describes the profiles with a two to three times lower $\chi^2$. The fitted $q$ is about 1 without excitation, increases to 1.4 for $A = 0.05$ and to 1.8 for $A = 0.1$ over the excitation window, and lies between 1.9 and 2.4 for $A \geq 0.3$ from 50 to \SI{350}{\milli\second}; later, when the core is depleted, it decreases again. With the baseline taken from the outermost channels, these values are lower than those of an earlier analysis of the same data ($q = 5$--6), where the baseline was partly estimated on channels that still carry beam.
    \item Different widths on the two sides of the peak reduce $\chi^2$ by only 10 to \SI{25}{\percent}: the asymmetry of the profiles is a secondary effect.
    \item The best description is a core with the width of the unexcited beam plus a wide component. For $A = 1.0$ the wide component holds \SI{54}{\percent} of the signal after \SI{75}{\milli\second}, \SI{79}{\percent} after \SI{175}{\milli\second} and \SI{90}{\percent} after \SI{475}{\milli\second}; for $A = 0.1$ it grows from 16 to \SI{54}{\percent} over the same time.
\end{itemize}

\paragraph{Diffusion coefficient.} The strength of the excitation is characterised by the diffusion coefficient $D$. In normalised phase space $(X, P)$, a sequence of kicks with uncorrelated phases increases the mean squared amplitude as $2Dt = \langle X^2+P^2\rangle$~\cite{ArrutiaDiffusive}, so that the geometric rms emittance grows as $\mathrm{d}\varepsilon_x/\mathrm{d}t = D$. This relation holds for the second moment of any distribution, Gaussian or not, and the rms beam size at the IPM, $\sigma_x^2 = \beta_x\varepsilon_x + D_x^2\sigma_\delta^2$, grows as
\begin{equation}
    \frac{\mathrm{d}\sigma_x^2}{\mathrm{d}t} = \beta_x D .
    \label{eq:MTRFKOslope}
\end{equation}
The momentum spread is not changed by a transverse excitation, so the dispersive term ($D_x = \SI{-2.9}{\metre}$ at the IPM) does not contribute to the slope. The conversion to $D$ uses $\beta_x = \SI{11.5}{\metre}$ at the monitor, from the model rematched to the measured tunes (Section~\ref{sec:MTEmittance}), with an uncertainty of \SI{10}{\percent}.

For profiles with a narrow core and wide tails, the width of a Gaussian fit follows the core and is not the rms size: in Figs.~\ref{fig:MTRFKOIPMhigh}c and~\ref{fig:MTRFKOIPMlow}c it grows two to five times more slowly than the second moment of the profile, and so does the width derived from the $q$-Gaussian fit. The logarithmic scale of Figs.~\ref{fig:MTRFKOIPMhigh}b and~\ref{fig:MTRFKOIPMlow}b shows the reason: the halo holds a small fraction of the signal but, through the weight $x^2$, a large part of the second moment. At \SI{3875}{\milli\second}, \SI{5}{\percent} of the signal lies more than \SI{10}{\milli\metre} from the centre for $A = 0.1$ and carries \SI{45}{\percent} of the second moment; for $A = 1.0$, \SI{28}{\percent} of the signal carries \SI{78}{\percent}. The unexcited beam has no measurable halo, and the baseline of the monitor is the same for all gains, so the halo is a property of the excited beam. A $q$-Gaussian with $q \geq 5/3$ has no finite variance, so the width derived from its FWHM characterises the core by construction. If all particles diffused with the same coefficient, the profile would remain Gaussian and the three widths would agree; their divergence shows that the excitation acts unevenly on the beam (Section~\ref{sec:MTRFKOLate}). The second moment is therefore computed directly from the data, without a model of the profile, on the side of the profile that extends to \SI{27}{\milli\metre} from the centre, assuming a symmetric distribution. The baseline is taken from the five outermost channels on that side; an error of the baseline shifts the second moment by an almost constant amount and leaves its slope unchanged. The result is a lower bound of the rms size of the beam, for three reasons:
\begin{itemize}
    \item the particles lost at the aperture, which had the largest amplitudes, are missing;
    \item beyond about \SI{8}{\milli\metre} from the centre the measured side carries less signal than the other;
    \item the signal beyond \SI{27}{\milli\metre} is not measured.
\end{itemize}
A saturation of the microchannel-plate gain in the dense core, one possible cause of the change in the response of the monitor discussed in Section~\ref{sec:MTRFKOLate}, would act in the opposite direction, since it reduces the weight of the core. For $A \leq 0.1$ the response stays within \SI{10}{\percent}; for the saturated gains it changes mainly after the fit window.

For each \SI{50}{\milli\second} bin, the second moment of the unexcited cycles is subtracted from that of every excited cycle, which also removes most of the effect of the common transient. The excess is fitted with a straight line over the first four bins, 25 to \SI{175}{\milli\second} after the start of the excitation, and the median and the standard deviation of the slopes are taken over the cycles of each gain. A straight line fitted from 50 to \SI{250}{\milli\second} gives slopes 20 to \SI{25}{\percent} smaller for the large gains, and one fitted over the first \SI{100}{\milli\second} about \SI{20}{\percent} larger (Table~\ref{tab:MTRFKO}); this spread is taken as a systematic uncertainty. The fit is limited to the first \SI{200}{\milli\second} because afterwards a large part of the beam is lost and the response of the monitor to the wide profiles is no longer proportional to the intensity (Section~\ref{sec:MTRFKOLate}).

\subsection{Response to the exciter gain}
\label{sec:MTRFKOGain}

\begin{figure}[htbp]
    \centering
    \includegraphics[width=\linewidth]{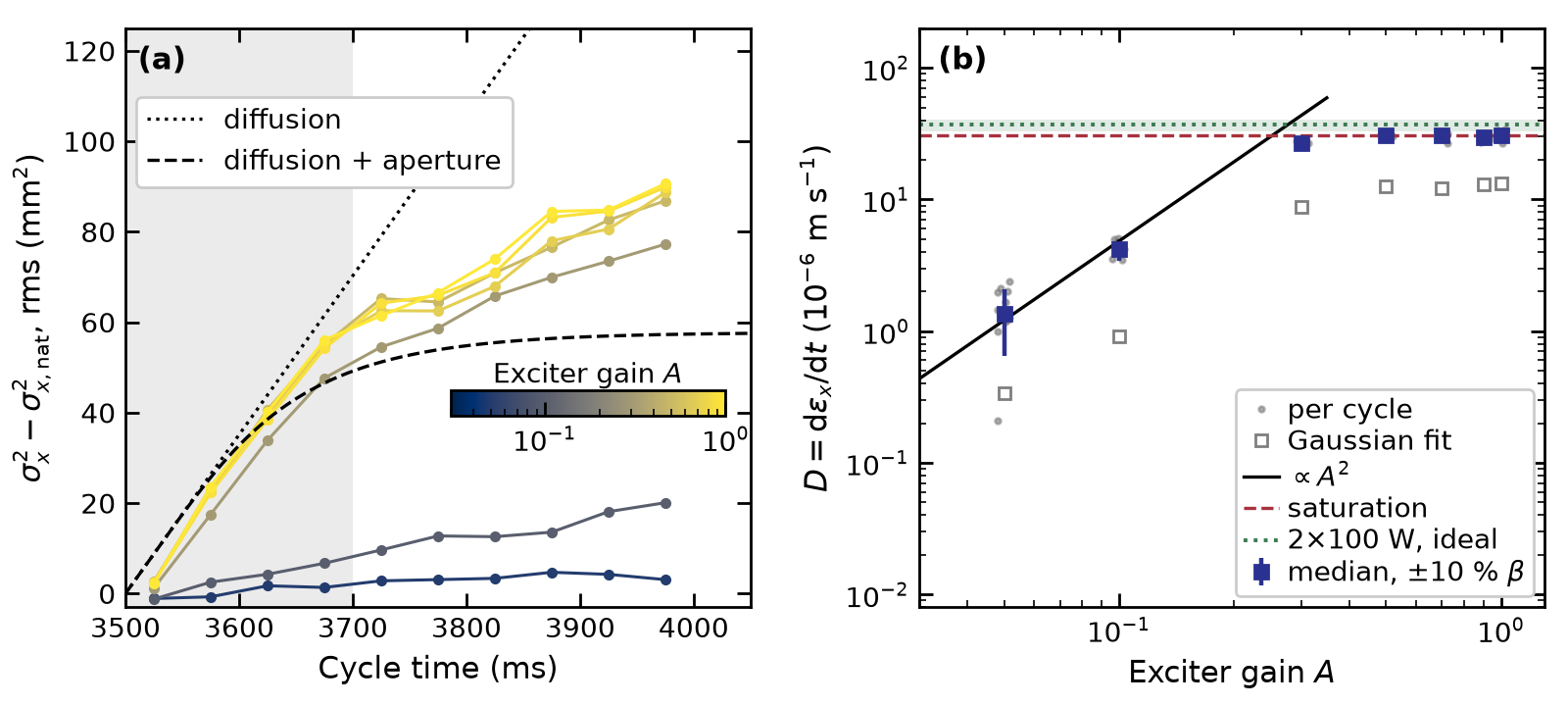}
    \caption{Horizontal blow-up with RF-KO excitation on the flat top of the 4BP cycle, 7~August~2026. (a)~Excess of the second moment of the profile over the unexcited beam at the IPM, median per gain (colour scale). Shaded: fit window. Black: diffusion with the coefficient of the saturated gains, without aperture (dotted) and with an absorbing aperture fitted to the BCT (dashed; Section~\ref{sec:MTRFKOLate}). (b)~Diffusion coefficient $D = \mathrm{d}\varepsilon_x/\mathrm{d}t$ from the first \SI{200}{\milli\second}, per cycle (grey dots) and median per gain (squares), the latter with the statistical spread and the \SI{10}{\percent} uncertainty of the $\beta$-function added in quadrature; open squares: from the width of a Gaussian fit, for comparison. Solid: $D \propto A^2$ fitted to $A \leq 0.1$; dashed: mean of $A \geq 0.5$; dotted: $2\times\SI{100}{\watt}$ with an ideal kicker (Section~\ref{sec:MTRFKOKick}), with a band of $\pm\SI{10}{\percent}$.}
    \label{fig:MTRFKO}
\end{figure}

Figure~\ref{fig:MTRFKO}a shows the median excess of the second moment for each gain as a function of cycle time; the black curves are the model of Section~\ref{sec:MTRFKOLate}. Figure~\ref{fig:MTRFKO}b shows the diffusion coefficient derived from the first \SI{200}{\milli\second}, for each cycle and as the median of each gain, together with the value obtained from the width of a Gaussian fit as in earlier analyses. It is compared with:
\begin{itemize}
    \item a quadratic dependence on the gain, fitted to $A \leq 0.1$ (solid line);
    \item the mean of the saturated gains $A \geq 0.5$ (dashed);
    \item the value expected for $2\times\SI{100}{\watt}$ with an ideal kicker, derived in Section~\ref{sec:MTRFKOKick} (dotted).
\end{itemize}
The values are listed in Table~\ref{tab:MTRFKO}.

\begin{table}[htbp]
\centering
\caption{Horizontal blow-up on the flat top of the 4BP cycle on 7~August~2026 as a function of the exciter gain $A$. $\mathrm{d}\sigma_x^2/\mathrm{d}t$: slope of the excess of the second moment of the profile over the unexcited beam, over the first \SI{200}{\milli\second} of the excitation (median and standard deviation over the cycles) and from 50 to \SI{250}{\milli\second} (median), and slope of the squared width of a Gaussian fit over the first \SI{200}{\milli\second} (median, for comparison). $D=\mathrm{d}\varepsilon_x/\mathrm{d}t$: diffusion coefficient from the second moment over the first \SI{200}{\milli\second}, a lower bound, with an additional uncertainty of \SI{10}{\percent} from the $\beta$-function at the monitor. Loss: median fraction of the beam lost 200 and \SI{450}{\milli\second} after the start of the excitation (BCT).}
\label{tab:MTRFKO}
\small
\begin{tabular}{rrrrrrrr}
\toprule
 & & \multicolumn{3}{c}{$\mathrm{d}\sigma_x^2/\mathrm{d}t$ (mm$^2$\,s$^{-1}$)} & & \multicolumn{2}{c}{Loss (\%)} \\
\cmidrule(lr){3-5}\cmidrule(lr){7-8}
 & & \multicolumn{2}{c}{Second moment} & Gaussian & & & \\
$A$ & Cycles & 0--\SI{200}{\milli\second} & 50--\SI{250}{\milli\second} & 0--\SI{200}{\milli\second} & $D$ ($10^{-6}$\,m\,s$^{-1}$) & \SI{200}{\milli\second} & \SI{450}{\milli\second} \\
\midrule
0, 0.01 & 6 & -- & -- & -- & -- & 1 & 3 \\
0.05 & 14 & $16 \pm 8$ & 17 & 3.9 & 1.4 & 1.8 & 5.4 \\
0.10 &  7 & $49 \pm 8$ & 46 & 10.6 & 4.3 & 2.4 & 7.0 \\
0.30 &  6 & $311 \pm 12$ & 243 & 100 & 27 & 14 & 35 \\
0.50 &  7 & $355 \pm 13$ & 276 & 146 & 31 & 18 & 41 \\
0.70 &  6 & $355 \pm 26$ & 263 & 140 & 31 & 19 & 43 \\
0.90 &  6 & $341 \pm 20$ & 284 & 151 & 30 & 18 & 43 \\
1.00 &  6 & $354 \pm 31$ & 261 & 154 & 31 & 20 & 44 \\
\bottomrule
\end{tabular}
\end{table}

Three regimes can be distinguished.
\begin{itemize}
    \item \emph{Without excitation}, the second moment grows by about \SI{6}{\milli\metre\squared\per\second}, and the $1/e$ lifetime of the beam on the flat top is about \SI{15}{\second}.
    \item \emph{For $A \leq 0.1$}, the response increases steeply with the gain. A chirp with a peak deflection $\theta$ at a kicker with $\beta$-function $\beta_k$ gives~\cite{ArrutiaDiffusive}
\begin{equation}
    D = \frac{\beta_k\,\theta^2}{16\,\Delta q\,T_0},
    \label{eq:MTRFKOD}
\end{equation}
where $T_0$ is the revolution period. Since $\theta$ is proportional to the voltage on the electrodes and the gain sets this voltage, $D$ is expected to scale as $A^2$, and the power delivered to the kicker as well. The slopes of the two lowest gains differ by a factor of $3.1 \pm 0.5$, against 4 expected, i.e.\ $\mathrm{d}\sigma_x^2/\mathrm{d}t = KA^2$ with $K \approx \SI{5600}{\milli\metre\squared\per\second}$ (4900 to 6300). With only two gains in this regime, and a growth carried by a halo of a few percent of the beam (Section~\ref{sec:MTRFKOMethod}), this scaling characterises the halo rather than the whole beam and is not established precisely.
    \item \emph{For $A \geq 0.3$}, the growth saturates at $(351 \pm 7)$~mm$^2$\,s$^{-1}$ in the first \SI{200}{\milli\second}, or about \SI{270}{\milli\metre\squared\per\second} from 50 to \SI{250}{\milli\second}. The value at $A = 0.3$ is already \SI{88}{\percent} of the saturated one. The amplifiers are the most likely cause. The same \SI{100}{\watt} amplifiers are used at the AD, where the BBQ output has to be attenuated by \SI{6}{\deci\bel} to keep them out of saturation. As far as is known, the LEIR signal path has no such attenuation, so the amplifiers reach their full output at $A \approx 0.5$, with compression setting in below this value. The saturated value is therefore the diffusion coefficient of the present system at \SI{100}{\watt} per electrode:
\begin{equation}
    D_{100} \geq 3.1\times10^{-5}~\mathrm{m\,s^{-1}} \quad (2.4\text{ to }3.7\times10^{-5}~\mathrm{m\,s^{-1}}\text{ for the different fit windows}),
\end{equation}
with an additional \SI{10}{\percent} from the $\beta$-function at the monitor. The width of a Gaussian fit gives a value 2.4 times lower, \SI{1.3e-5}{\metre\per\second}, which characterises the core of the beam rather than its rms size. Two observations confirm that the limit lies in the excitation chain and not in the beam or in the monitor: the intensity of the cycles with $A = 0.5$ to 1.0 decays identically (Fig.~\ref{fig:MTRFKOLoss}a), and their profiles coincide from the first \SI{50}{\milli\second}, before the beam size has changed.
\end{itemize}

\subsection{Beam loss and evolution of the beam beyond 200~ms}
\label{sec:MTRFKOLate}

\begin{figure}[htbp]
    \centering
    \includegraphics[width=\linewidth]{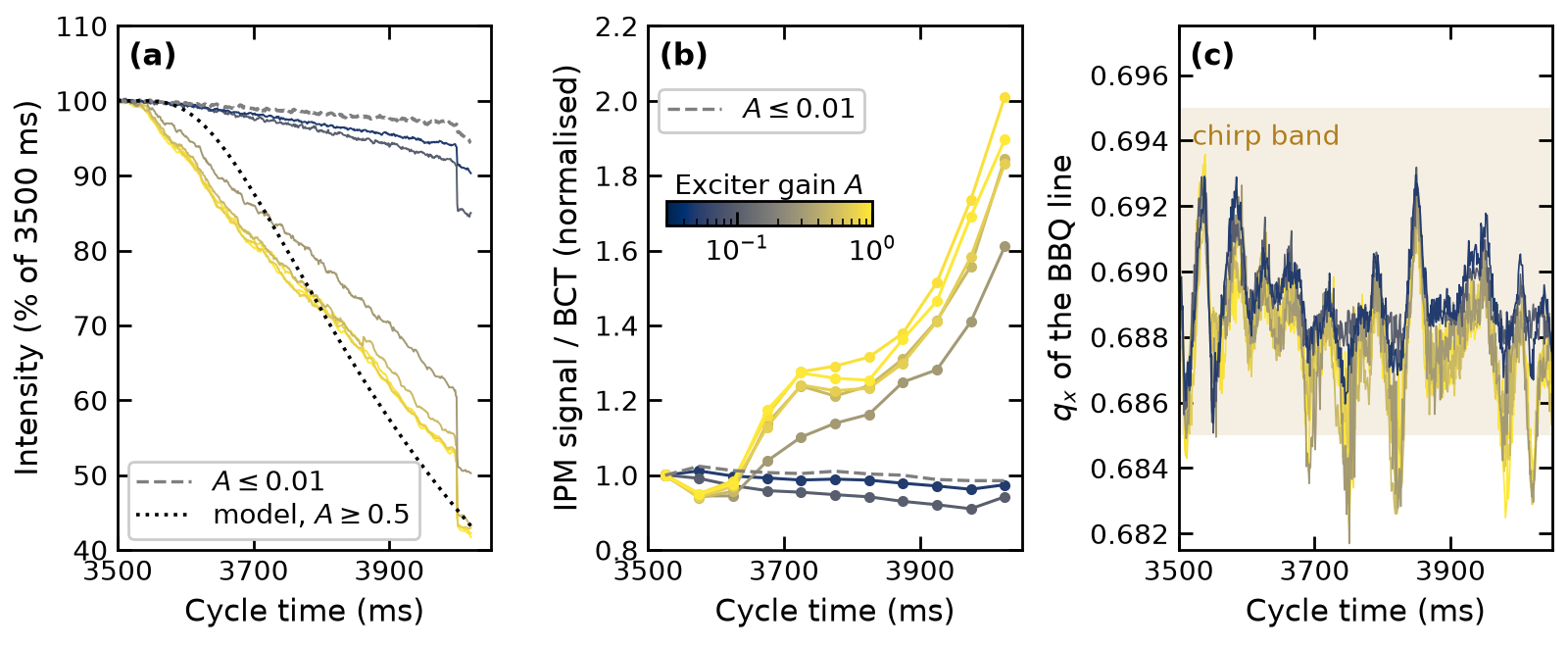}
    \caption{Beam intensity and diagnostics during the gain scan of Fig.~\ref{fig:MTRFKO}, median per gain, coloured by the exciter gain as in Fig.~\ref{fig:MTRFKO}a; dashed: cycles without excitation. (a)~Intensity measured by the BCT at \SI{1}{\kilo\hertz}, normalised at the start of the excitation, with the diffusion model with aperture for $A \geq 0.5$ (dotted). The drop common to all gains after \SI{3980}{\milli\second} belongs to the end of the flat top. (b)~Integrated IPM signal divided by the BCT intensity, normalised to the first bin. (c)~Fractional tune of the beam response measured by the BBQ during the excitation, with the band covered by the chirp; the BBQ delivers a tune every \SI{1}{\milli\second} (FFT over 1024 or 512 turns).}
    \label{fig:MTRFKOLoss}
\end{figure}

Figure~\ref{fig:MTRFKOLoss} shows, for the same cycles, three quantities needed to interpret the evolution of the beam after the fit window:
\begin{itemize}
    \item (a) the intensity measured by the BCT, normalised to the start of the excitation;
    \item (b) the integrated IPM signal divided by the BCT intensity, normalised to the first bin, i.e.\ the response of the IPM per circulating ion;
    \item (c) the frequency of the beam response measured by the BBQ, every millisecond, as the median over the cycles of each gain.
\end{itemize}

\paragraph{Beam loss.} For $A \geq 0.5$ the beam is lost from about \SI{50}{\milli\second} after the start of the excitation: \SI{18}{\percent} within \SI{200}{\milli\second} and \SI{43}{\percent} within \SI{450}{\milli\second}. Without excitation the losses are \SI{1}{\percent} and \SI{3}{\percent} (Fig.~\ref{fig:MTRFKOLoss}a, Table~\ref{tab:MTRFKO}). The cycles with $A = 0.3$ lose slightly less, and those with $A \leq 0.1$ lose 2 to \SI{7}{\percent} within \SI{450}{\milli\second}.

\paragraph{Diffusion with an aperture.} To relate the loss to the beam size, the diffusion equation in action $J$,
\begin{equation}
    \frac{\partial f}{\partial t} = D\,\frac{\partial}{\partial J}\!\left(J\,\frac{\partial f}{\partial J}\right),
\end{equation}
was solved numerically with an absorbing boundary at an action $J_\mathrm{ap}$, beyond which particles are lost. The initial state is the Gaussian beam with the measured size of the unexcited cycles, and the diffusion coefficient is $D_{100}$. The only free parameter, $J_\mathrm{ap}$, was fitted to the BCT signal of the saturated cycles up to \SI{450}{\milli\second}. The fit gives a limiting amplitude of \SI{20.5}{\milli\metre} at the IPM, 7.8 times the rms size of the unexcited beam, and reproduces the measured intensity within \SI{8}{\percent} of the initial value (Fig.~\ref{fig:MTRFKOLoss}a, dotted). The measured loss begins earlier and is more linear in time than in the model. The BCT signal alone does not determine $D$: the Gaussian-fit value with an aperture at \SI{13.7}{\milli\metre} describes it equally well, but is excluded by the signal observed beyond \SI{15}{\milli\metre} in the profiles.

In Fig.~\ref{fig:MTRFKO}a, the dotted line is the growth of the second moment without aperture, $\beta_x D_{100}\,t$, and the dashed curve the growth of the surviving beam with the aperture. The data follow both up to about \SI{200}{\milli\second}. Afterwards the model levels off at about \SI{57}{\milli\metre\squared}, because the particles reaching large amplitudes are lost as fast as the diffusion supplies them, whereas the measured second moment keeps growing to about \SI{90}{\milli\metre\squared} by \SI{475}{\milli\second}. The aperture model is therefore an effective description of the loss rather than a quantitative model of the beam: its limiting amplitude corresponds to at most \SI{21}{\milli\metre} at the largest horizontal $\beta$-function of the optics, well inside the apertures measured in Section~\ref{sec:MTAperture}.

\paragraph{Response of the monitor.} For $A \leq 0.1$ the integrated IPM signal follows the BCT intensity within \SI{10}{\percent}. For $A \geq 0.3$ it rises relative to the BCT intensity by a factor of 1.3 to 2 from about \SI{150}{\milli\second} (Fig.~\ref{fig:MTRFKOLoss}b). The response of the monitor is thus not proportional to the number of ions for wide, low-density profiles; a saturation of the microchannel-plate gain in the dense core of the narrow initial beam is one possible cause. The growth of the second moment after about \SI{250}{\milli\second}, and its disagreement with the aperture model, are therefore not interpreted quantitatively.

\paragraph{Profile shape.} A diffusion with the same coefficient for all particles keeps the projected profile of an initially Gaussian beam Gaussian, and an absorbing aperture makes its tails lighter than Gaussian ($q < 1$ when the model profiles are fitted like the data). The core with tails observed at all gains ($q \approx 2$) shows that the excitation acts unevenly: a part of the beam is driven to large amplitudes, while the rest stays near its initial size and is transferred to the tails progressively. A plausible origin is that the chirp does not cover the whole tune distribution, and the chromatic tune spread contributes to this. With the horizontal chromaticity measured on the flat top, $Q'_x = -4.7$ (Section~\ref{sec:MTOptics}), and the momentum spread of $0.49\times10^{-3}$ measured by tomography during the slow-extraction test (Section~\ref{sec:MTMomentum}), the rms tune spread is $2.3\times10^{-3}$, and $\pm3\sigma$ of the tune distribution span $\pm6.9\times10^{-3}$, more than the $\pm5\times10^{-3}$ of the band. Moreover, the band, from 0.685 to 0.695, is not centred on the unexcited beam at $q_x \approx 0.688$: it reaches only $1.3\sigma$ below the centre of the distribution, so that, for the same momentum spread on 7~August, about \SI{10}{\percent} of the particles, those with the largest momenta, never come into resonance with the excitation. The momentum spread of 0.85 to $1.2\times10^{-3}$ measured with the Schottky monitor before capture, taken as an upper bound, would leave a larger fraction outside the band. No Schottky signal was recorded during the excitation. Other candidates are a residual drift of the momentum along the flat top, which moves the tune of the whole beam across the band, and the strong dependence of the response on the position of the band (Section~\ref{sec:MTRFKOSetup}), which shows that the band and the tune distribution are close to each other. The diffusive model of Section~\ref{sec:MTRFKOReq} assumes that all particles are excited equally; the coverage of the tune distribution by the excitation is therefore a parameter of the spill control in its own right, to be measured together with the tune spread.

\paragraph{Tune.} Without excitation the BBQ sees no line, since it measures the coherent response to the chirp (Fig.~\ref{fig:MTRFKOLoss}c).
\begin{itemize}
    \item For $A \leq 0.1$ the line lies at $q_x = 0.689$, near the centre of the band.
    \item For $A \geq 0.3$ it lies 1 to $2\times10^{-3}$ lower.
    \item In the first \SI{100}{\milli\second} of the flat top the line oscillates with a period of about \SI{50}{\milli\second} and a peak-to-peak amplitude of $6.5\times10^{-3}$ at all gains. With $Q'_x = -4.7$ this corresponds to a momentum oscillation of about $1.4\times10^{-3}$ peak to peak, as the displacement of the horizontal profile at the end of the ramp (Section~\ref{sec:MTRFKOMethod}).
    \item Later on, the line makes short excursions of 3 to $5\times10^{-3}$, lasting a few milliseconds, at the same cycle times in every cycle, for example at 3850 and \SI{4020}{\milli\second}. Their origin, a programmed function of the machine or the tune measurement itself, has not been identified.
    \item Between 3600 and \SI{4040}{\milli\second} the line drifts by $-0.6$ to $-0.8\times10^{-3}$ for $A \leq 0.1$, which corresponds to a momentum change of about $1.5\times10^{-4}$: the ramp of the RF frequency compensates the drift of $3.5\times10^{-3}$ observed without correction (Section~\ref{sec:MTMomentum}) to within about \SI{5}{\percent}. For $A \geq 0.3$ the line drifts by $-1.4$ to $-2.0\times10^{-3}$ over the same time.
    \item Lines at the mains frequency of \SI{50}{\hertz} and its harmonics do not appear in the tune: their amplitude is below $1\times10^{-4}$, the noise level of the measurement. They are visible in the BCT signal up to \SI{450}{\hertz}, with a relative amplitude of $1.8\times10^{-3}$ at \SI{50}{\hertz}, but equally in the cycles without excitation, in which the beam is not lost. They therefore belong to the intensity measurement, and a modulation of the loss rate by the ripple of the power converters cannot be resolved with this monitor.
\end{itemize}
The line stays at least $1.5\times10^{-3}$ inside the band. Since the BBQ sees the beam only where the chirp drives it, it cannot show particles outside the band; a measurement of the tune distribution, or a scan with a band wider than the expected spread, is needed to settle this point.

\subsection{Kick strength of the transverse feedback}
\label{sec:MTRFKOKick}

A strip-line pair of length $L$ and gap $d$, driven in push-pull with a peak voltage $V$ per electrode and a wave travelling against the beam, deflects a particle with velocity $\beta c$ by
\begin{equation}
    \theta = \kappa\, g\, \frac{2V}{d}\, \frac{L\,(1+\beta)}{\beta c\, B\rho} ,
\end{equation}
where the factor $1+\beta$ accounts for the magnetic force and $g$ for the field coverage of the electrodes. For \SI{100}{\watt} into \SI{50}{\ohm}, $V = \SI{100}{\volt}$. The electrodes of KDHV21 are narrower than the gap ($w/d = 0.6$), so the field on axis is lower than $2V/d$. For two parallel plates of finite width, $g = \tanh(\pi w/2d) = 0.74$; this estimate is to be replaced by an electromagnetic simulation of the actual geometry. With $\beta = 0.372$ and $B\rho = \SI{4.8}{\tesla\metre}$, the ideal kicker ($\kappa = 1$) gives $\theta = \SI{0.75}{\micro\radian}$ and, with Eq.~\eqref{eq:MTRFKOD} and $\beta_k = \SI{7.4}{\metre}$ in the rematched model, $D = \SI{3.7e-5}{\metre\per\second}$.

The measured $D_{100}$ corresponds to a kick efficiency $\kappa = \sqrt{D_{100}/D_\mathrm{ideal}} \geq 0.91$ (0.80 to 1.00 over the fit windows), with a further \SI{7}{\percent} from the uncertainty of the $\beta$-functions at the IPM and at the kicker. Within the uncertainties of the measurement and of the field-coverage estimate, the kick of the present system at full power is thus consistent with the expectation, and the saturation of the response is explained by the rating of the amplifiers. The linear regime does not allow a finer statement: extrapolated as $A^2$ to $A = 0.5$, it would exceed the ideal kicker by a factor of about three, which indicates either that the amplifiers compress already from $A \approx 0.25$, or that the response at low gain is not purely diffusive; since the low-gain response is carried by the halo, the latter is the more likely. A direct calibration requires the voltage on the electrodes as a function of the gain, which has not been measured; the analogue monitor of the kicker drive (EA.TFB-KICK21) is available in OASIS but is not logged.

\subsection{Diffusion coefficient required for HEARTS@LEIR}
\label{sec:MTRFKOReq}

The excitation needed for the specification follows from the diffusive model of RF-KO extraction~\cite{ArrutiaDiffusive}. The particles inside the stable region of the resonance perform a random walk in normalised phase space under the excitation. The stable triangle is replaced by a circle of the same area, with radius $R = \sqrt{48\sqrt{3}\,\pi}\,\delta Q/S$, where $\delta Q$ is the distance of the tune to the resonance and $S$ the normalised strength of the resonance driving term. Particles reaching $R$ are extracted, which makes $R$ an absorbing boundary. The density is expanded in the eigenmodes of the diffusion equation in the circle, $J_0(\zeta_k r/R)$, which decay as $\exp(-\zeta_k^2 \int D\,\mathrm{d}t/R^2)$ ($\zeta_k$ being the zeros of the Bessel function $J_0$). The extracted flux is the flux through the boundary.

The amplitude modulation of the excitation is obtained by requiring the flux to stay at the constant value $N_0/t^*$, where $N_0$ is the stored beam and $t^*$ the spill duration, and solving this condition for $D(t)$ step by step. The kicker voltage is then proportional to $\sqrt{D(t)}$.
\begin{itemize}
    \item At the start, the tails of the initial Gaussian beam lie close to the boundary, and a small $D$ suffices.
    \item As the beam is depleted, $D$ has to increase.
    \item Once the lowest mode dominates, the requirement diverges as
\begin{equation}
    D(f) \rightarrow \frac{R^2}{\zeta_1^2\, t^*\,(1-f)} ,
\end{equation}
where $f$ is the fraction of the beam already extracted.
\end{itemize}
When the available excitation is limited to a maximum $D_\mathrm{max}$, the spill consists of a short rise, a plateau at constant flux while $D(t) < D_\mathrm{max}$, and a decaying tail. The model assumes that all particles are excited equally, which the profiles of Section~\ref{sec:MTRFKOLate} do not support; the comparison below is therefore a guide to the order of magnitude.

\begin{figure}[htbp]
    \centering
    \includegraphics[width=\linewidth]{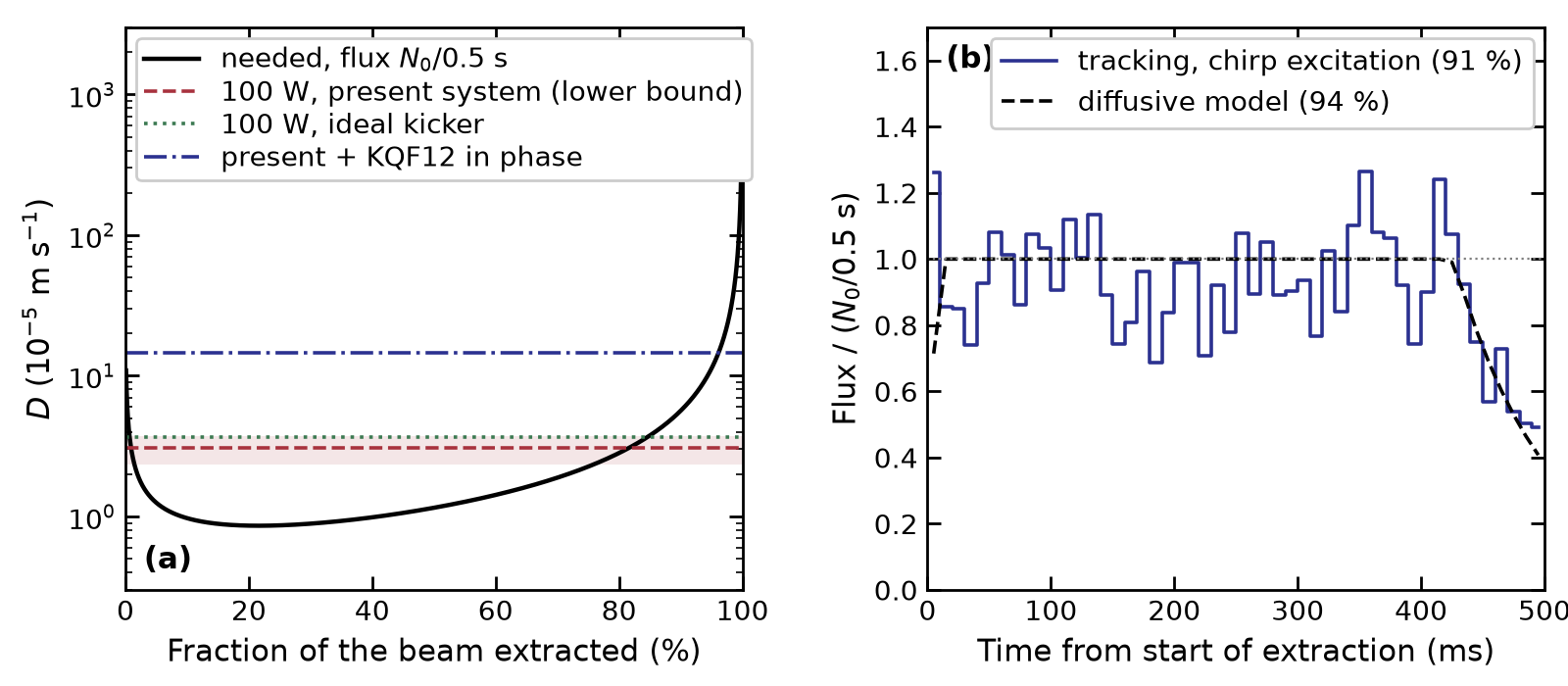}
    \caption{Excitation needed for HEARTS@LEIR at \SI{4.8}{\tesla\metre} ($S = \SI{20}{\metre}^{-1/2}$, $\delta Q = \num{5e-3}$, $\varepsilon_n = \SI{0.4}{\micro\metre}$). (a)~Diffusion coefficient needed to extract the whole beam at constant flux in \SI{0.5}{\second}, diffusive model, as a function of the fraction already extracted. Horizontal lines: available excitation at \SI{100}{\watt} per electrode for the present system (lower bound; band: fit windows), for an ideal kicker, and for the present system with the second strip-line kicker driven in phase. (b)~Spill in \SI{10}{\milli\second} bins for an excitation limited to the ideal kicker, from tracking with a chirp excitation (random start phase of each sweep) and from the diffusive model; the percentages are the fractions extracted within \SI{0.5}{\second}.}
    \label{fig:MTRFKOReq}
\end{figure}

For the shortest spills of the specification (\SI{0.5}{\second}, Table~\ref{tab:beamSpecs}) at \SI{4.8}{\tesla\metre}, with $S = \SI{20}{\metre}^{-1/2}$, $\delta Q = \num{5e-3}$ and a normalised emittance of \SI{0.4}{\micro\metre}, the radius of the stable region is $R = \SI{4.0e-3}{\metre}^{1/2}$ in normalised coordinates, four times the rms size of the beam. The required diffusion coefficient is shown in Fig.~\ref{fig:MTRFKOReq}a as a function of the fraction of the beam already extracted. It reaches \SI{2.8e-5}{\metre\per\second} at \SI{80}{\percent}, \SI{5.7e-5}{\metre\per\second} at \SI{90}{\percent} and \SI{1.1e-4}{\metre\per\second} at \SI{95}{\percent}, i.e.\ 0.9, 1.8 and 3.7 times the lower bound of $D_{100}$. The fraction of the beam delivered at constant flux and the fraction extracted within \SI{0.5}{\second} are listed in Table~\ref{tab:MTRFKOSpill} for the available excitations.

\begin{table}[htbp]
\centering
\caption{Spill of the whole beam requested in \SI{0.5}{\second} at \SI{4.8}{\tesla\metre} when the excitation is limited to $D_\mathrm{max}$ (diffusive model; in brackets: tracking with a chirp excitation).}
\label{tab:MTRFKOSpill}
\begin{tabular}{lrrr}
\toprule
Available excitation & $D_\mathrm{max}$ ($10^{-5}$\,m\,s$^{-1}$) & Constant flux (\%) & Extracted (\%) \\
\midrule
Present system, \SI{100}{\watt} (lower bound) & 3.1 & 81 & 93 (90) \\
Ideal kicker ($\kappa = 1$), \SI{100}{\watt} & 3.7 & 84 & 94 (91) \\
Present system + second kicker & 15 & 96 & 99 (97) \\
\bottomrule
\end{tabular}
\end{table}

The second strip-line kicker of the ring is \SI{300}{\milli\metre} long. Driving it with the amplifiers of the vertical plane, in phase with KDHV21, increases $D$ by a factor of 4.8 for the same kick efficiency, without exceeding the rating of the amplifiers.

Figure~\ref{fig:MTRFKOReq}b shows that a chirp excitation reproduces the model. The figure compares the spill of the model with a tracking simulation of the extraction: a H\'enon map with $10^6$ particles and a chromatic tune spread of \num{9.5e-4}, which corresponds to the natural chromaticity of the design model, $|Q'_x| = 1.9$ (the measured $Q'_x = -4.7$ gives \num{2.3e-3}, Section~\ref{sec:MTRFKOLate}, and Chapter~\ref{sec:SRE} requires $|Q'_x| \lesssim 0.5$ after correction), excited by the chirp described above with a random start phase of each sweep and an amplitude following the modulation of the model. The excitation is limited to the ideal kicker, $D_\mathrm{max} = \SI{3.7e-5}{\metre\per\second}$. The tracking reproduces the rise, the plateau and the tail of the model, and 91~\% of the beam is extracted within \SI{0.5}{\second}, against 94~\% in the model. The flux fluctuates from sweep to sweep, with a coefficient of variation of about 0.3 in \SI{1}{\milli\second} bins. With identical sweeps repeated every millisecond, the excitation is a line spectrum: most particles do not resonate with it, and less than a fifth of the beam is extracted at the same amplitude. The signal generation should therefore randomise the phase from sweep to sweep.

The requirement scales as $(\delta Q/S)^2/t^*$. For a \SI{1}{\second} spill it halves, and the present system delivers about \SI{90}{\percent} of the beam at constant flux; a \SI{5}{\second} spill needs ten times less than a \SI{0.5}{\second} spill, which lies within the linear regime of the present system. For the shortest spills at the highest rigidity, the present system at its rated power covers the bulk of the spill (about \SI{80}{\percent} at constant flux) but not its end; a margin of a factor of two to four is needed for the last 10 to \SI{5}{\percent} of the beam and for a spill feedback loop. Within the rating of the amplifiers, this margin can come from the second kicker, from the field coverage of the electrodes, and from a better coverage of the tune distribution by the excitation band. Given the qualitative nature of the present data, these options are inputs to the consolidation of the transverse feedback system rather than a quantitative specification.

\section{Slow-extraction test on the 4BP cycle}
\label{sec:MTSX}

The measurements described so far characterise the ingredients of the extraction separately; the test presented in this section combines them on a single cycle. Since LEIR has no slow-extraction septum, resonant slow extraction cannot be demonstrated end to end in the present machine. The test therefore validates all steps upstream of the septum: the preparation of a cooled, high-intensity beam on a long flat top, the move to the slow-extraction working point, and the controlled removal of particles from the circulating beam by a transverse excitation. Particles that leave the stable region are lost on the machine aperture instead of being extracted, and the loss rate, measured with the BCT and with beam-loss and radiation detectors, takes the place of the spill.

\subsection{Preparation and instrumentation}
\label{sec:MTSXPrep}

\paragraph{Loss localisation with an orbit bump.} The test was prepared with tracking simulations in Xsuite, using the aperture model updated with the measurements of Section~\ref{sec:MTAperture} and the slow-extraction optics~\cite{SXMockupPrep2026}. Without a local bump, the simulated losses are shared between two locations in the bending magnets, 13 and \SI{32}{\metre} downstream of the start of SS30, with \SI{77}{\percent} and \SI{23}{\percent} of the lost particles. With a closed bump of \SI{40}{\milli\metre} in SS30, \SI{96}{\percent} of the losses are concentrated within the first metre of SS30, at the horizontal corrector DHV31, and the remaining \SI{4}{\percent} at the second location. Concentrating the losses at a single, known point makes them accessible to loss monitors and to radiation measurements, and turns the test into a check of the aperture model in the extraction straight section, where the beam-based measurement of Section~\ref{sec:MTAperture} could not be made.

\begin{figure}[htbp]
    \centering
    \includegraphics[width=0.72\linewidth]{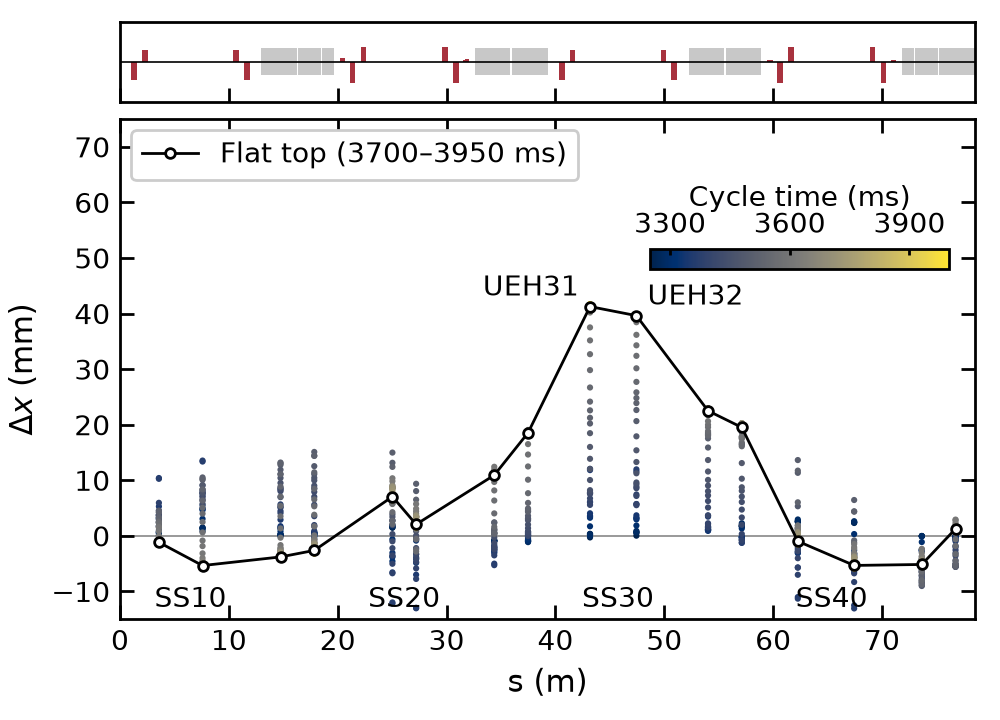}
    \caption{SS30 orbit bump of the slow-extraction test, 26~August~2026 (set-up~(b)). The top strip shows the dipoles (grey) and quadrupoles (red). Horizontal BPM readings with the orbit before the bump (2700--\SI{3200}{\milli\second}) subtracted, coloured by cycle time while the bump is ramped up and held, and average over the flat top (open circles). Data of the reference RF frequency of the dispersion measurement.}
    \label{fig:MTSS30Bump}
\end{figure}

\paragraph{Construction of the bump.} A bump computed with the MAD-X model could not be localised in SS30 (Section~\ref{sec:MTAperture}), because the orbit correctors inside SS30 have little effect at the SS30 monitors. The bump was therefore constructed from a measured orbit response. In July~2026 each usable horizontal orbit corrector was kicked in turn by \SI{0.5}{\milli\radian} on the flat top of an MD cycle, from 3500 to \SI{4050}{\milli\second}, and the orbit was recorded for three cycles per configuration; the corrector strengths that close a bump at SS30 were then computed from the measured response matrix. The bump is ramped up at the end of the acceleration, between about 3300 and \SI{3600}{\milli\second}. On 26~August~2026 it displaced the orbit on the flat top by 41 and \SI{40}{\milli\metre} (readings of 44 and \SI{41}{\milli\metre}) at UEH31 and UEH32, the two horizontal monitors of SS30 (Fig.~\ref{fig:MTSS30Bump}); in July~2026, about \SI{36}{\milli\metre} had been recorded at the same monitors. The bump is not fully closed: at several monitors of sectors~2 and~3 the orbit is displaced by 10 to \SI{25}{\milli\metre}, while elsewhere in the ring it changes by less than about \SI{5}{\milli\metre}. The residual displacement in sectors~2 and~3 is not large enough to move the bottleneck away from SS30, as the loss measurements below confirm, but it has to be taken into account in the comparison with the dispersion model (Section~\ref{sec:MTOptics}).

\paragraph{Loss monitors.} As LEIR had no beam-loss monitors, five ionisation-chamber BLMs of the type used in the LHC were installed in spring~2026: two next to the fast-extraction kickers in SS30, for the slow-extraction losses, and one each at the fast extraction, the injection and the electron cooler~\cite{SLAWGBLM2026}. They were read out with a portable acquisition card connected by optical fibre to a computer at the rack, which recorded twelve running sums with integration times from \SI{40}{\micro\second} to about \SI{84}{\second}; the data were saved on demand and did not go to the logging database. To increase the chance of detecting the losses, one BLM in SS30 was moved on 23~June~2026, and the BLM of the electron cooler was moved on 31~July~2026 to the loss location found by the radiation survey (see below), in line with the two others. A FLUKA estimate predicted a response of about $10^{-14}$~Gy per lost Pb$^{54+}$ ion for a monitor placed about \SI{50}{\centi\metre} from the impact point, that is, of the order of \SI{10}{\micro\gray\per\second} for a loss rate of $10^{9}$ ions per second, provided that the impact location is known. In addition, a Timepix3 radiation monitor~\cite{Timepix1, Timepix2} was installed in the middle of SS30. Its detection threshold is four orders of magnitude below that of the ionisation-chamber BLMs, so that it can record losses that remain below the BLM threshold, and its time resolution of \SI[group-digits=false]{1.5625}{\nano\second} allows the time structure of the losses to be measured.

\subsection{The 4BP cycle}
\label{sec:MT4BP}

\begin{figure}[htbp]
    \centering
    \includegraphics[width=0.6\linewidth]{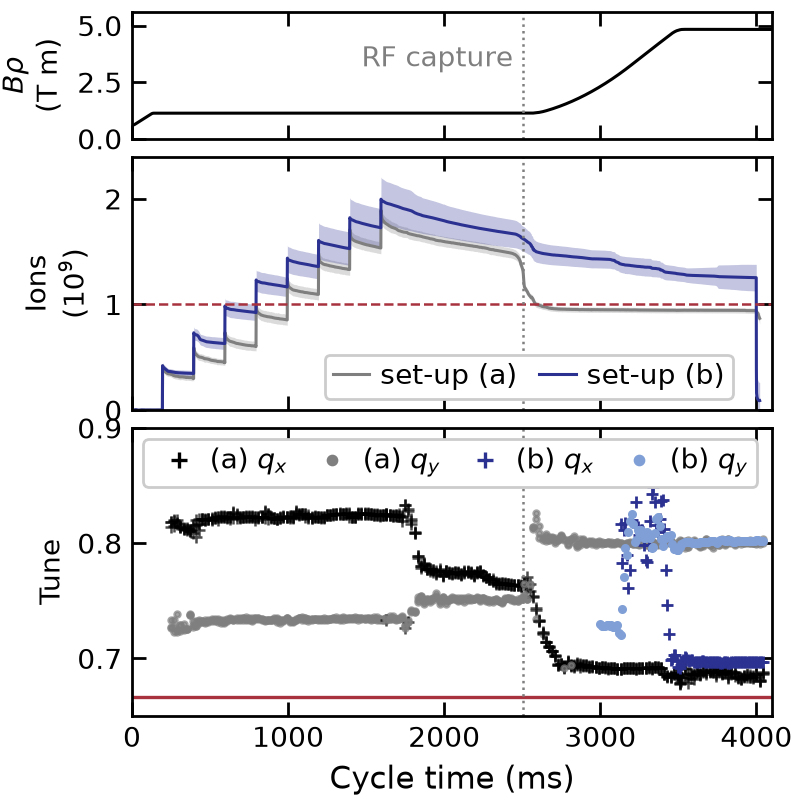}
    \caption{The 4BP cycle used for the slow-extraction test. From top to bottom: beam rigidity; BCT intensity (mean and RMS spread) for set-up~(a), 16~June~2026, and set-up~(b), 26--28~August~2026, with the HEARTS@LEIR requirement (dashed); fractional tunes measured on 16~June~2026 for set-up~(a) ($q_x$ black, $q_y$ grey) and median of the measurements of 26~August~2026 for set-up~(b) ($q_x$ dark blue, $q_y$ light blue), with the third-integer resonance (red). The dotted line marks the RF capture.}
    \label{fig:MT4BPCycle}
\end{figure}

A dedicated cycle, extended by one basic period to four (4BP), with a flat top of about \SI{0.5}{\second} at the maximum energy of Pb$^{54+}$ (72.1~MeV/nucleon, \SI{4.8}{\tesla\metre}), was commissioned in June~2026~\cite{FourBP2026}. Two set-ups were used (Fig.~\ref{fig:MT4BPCycle}):
\begin{enumerate}[label=(\alph*)]
    \item the nominal cooling time. The electron cooling stops at \SI{1805}{\milli\second}, at the end of the accumulation, together with a first step of the tune change that brings the tunes close to the difference resonance ($q_x \approx 0.77$, $q_y \approx 0.75$). The beam then coasts without cooling for \SI{700}{\milli\second} until RF capture at \SI{2505}{\milli\second}. The difference resonance is crossed after the capture, at the start of the ramp, when the tunes are moved to the slow-extraction working point ($q_x \approx 0.69$, $q_y \approx 0.80$). On 16~June~2026, $1.35\times10^{9}$ ions were captured and $0.94\times10^{9}$ reached the flat top;
    \item the electron cooling extended until RF capture, and the whole tune change moved to the end of the ramp. The tunes measured on 26~August~2026 show the vertical tune raised at about \SI{3150}{\milli\second} and the horizontal tune lowered at about \SI{3420}{\milli\second}, shortly before the flat top, where they settle at $q_x \approx 0.697$ and $q_y \approx 0.80$. On 26--28~August~2026, $1.63\times10^{9}$ ions were captured and $1.29\times10^{9}$ reached the flat top.
\end{enumerate}
For the slow-extraction test, the four harmonic sextupoles in SS10 and SS30 (XFN11, XFN12, XFN31 and XFN32), which are individually powered, were ramped from zero to $+3.1$, $+1.9$, $-1.9$ and \SI{-3.1}{\ampere} between 3500 and \SI{3600}{\milli\second}, the SS10 and SS30 pairs with opposite signs, to drive the third-integer resonance. The tune shift caused by the quadrupolar feed-down of the sextupoles was corrected ad hoc to keep the tunes constant on the flat top.

\paragraph{Strength of the resonance sextupoles.} With the calibration of the XFN sextupoles, $B''L/I = \SI{0.52}{\tesla\per\metre\per\ampere}$, and $L = \SI{0.335}{\metre}$, the currents correspond at \SI{4.8}{\tesla\metre} to $k_2 = \pm\SI{1.00}{m^{-3}}$ for XFN11 and XFN32 and $\pm\SI{0.60}{m^{-3}}$ for XFN12 and XFN31, i.e.\ to integrated strengths $k_2L = \pm0.34$ and $\pm\SI{0.20}{m^{-2}}$. In the slow-extraction optics they excite a resonance driving term, expressed as the strength of the equivalent virtual sextupole, of $S \approx \SI{4.5}{m^{-1/2}}$, independent of the sign convention of the currents; the chromatic sextupoles, which were not powered in 2026~\cite{LEIRNLO2026}, are not included. This is about a fifth of the design value $S^* = \SI{21}{m^{-1/2}}$ (Section~\ref{sec:SREsext}). With an rms $\beta$-beating of \SI{25}{\percent}, larger than the 14 to \SI{17}{\percent} measured (Section~\ref{sec:MTOptics}), the value of $S$ obtained from the model lies between 4.3 and \SI{7.6}{m^{-1/2}} (\SI{68}{\percent} range). In the SS30 bump the sextupoles XFN31 and XFN32 also act as quadrupoles: at \SI{40}{\milli\metre} this feed-down would shift the horizontal tune by about $-0.017$. With the sextupoles on and the exciter disabled, the tune monitor (BBQ) measured $q_x = 0.691$ on the flat top (interquartile range from 0.690 to 0.692, 119 cycles), against 0.688 just before the sextupoles were switched on, so that the feed-down was compensated. The tune lies $\delta = 0.024$ from the third-integer resonance, five times the design distance. In the single-particle model the stable region of the resonance, with an inscribed radius $4\pi\delta/S \approx \SI{0.07}{m^{1/2}}$ in normalised coordinates, is therefore much larger than the aperture, and the sextupoles do not create a separatrix within the aperture of the machine.

\begin{figure}[htbp]
    \centering
    \includegraphics[width=0.75\linewidth]{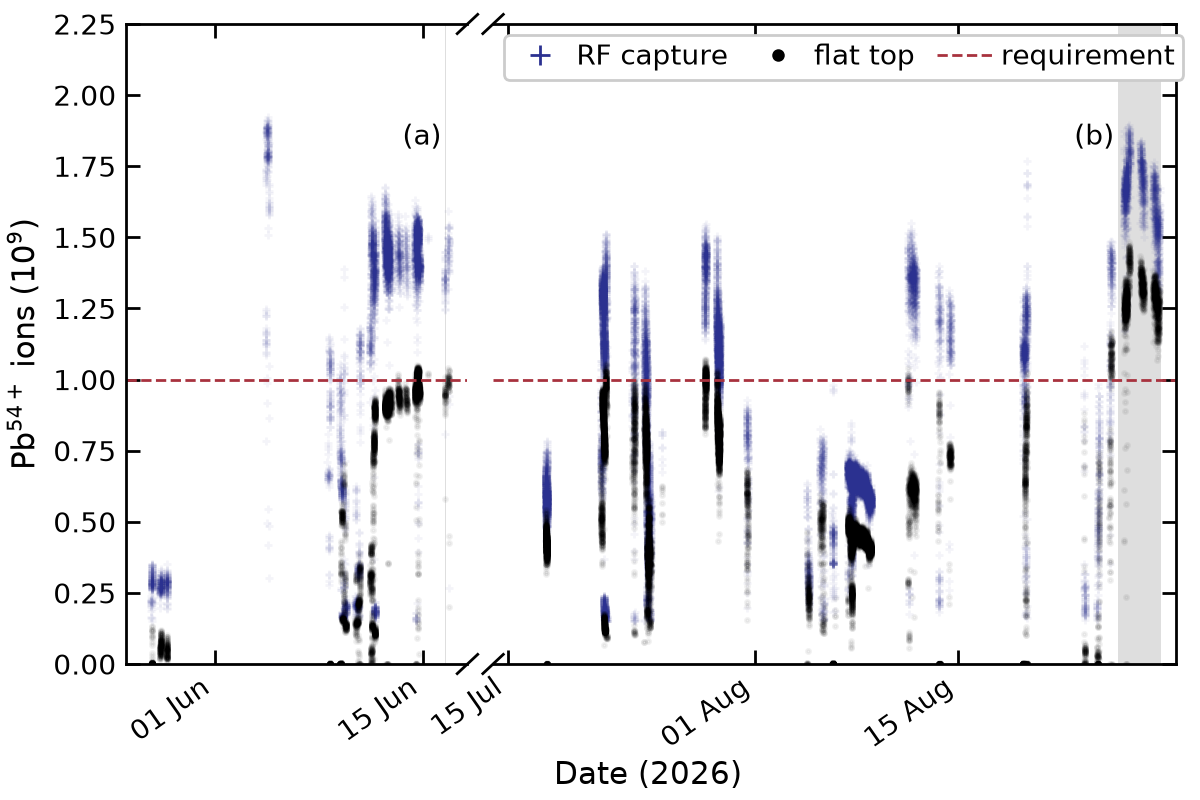}
    \caption{Pb$^{54+}$ intensity at RF capture (\SI{2500}{\milli\second}) and at the start of the flat top (\SI{3500}{\milli\second}) for the \num{24192} 4BP cycles with beam in 2026 (BCT), against the HEARTS@LEIR requirement. Cycles whose BCT acquisition ended before the flat top (\num{288}) have no flat-top point. The shaded periods mark the data of set-ups~(a) and~(b) shown in Fig.~\ref{fig:MT4BPCycle}.}
    \label{fig:MT4BPRun}
\end{figure}

\paragraph{Intensity over the run.} The 4BP cycle was played \num{27855} times between May and August~2026, \num{24192} times with beam ($>0.15\times10^{9}$ ions at RF capture). Figure~\ref{fig:MT4BPRun} shows the intensity at RF capture and at the start of the flat top for all cycles with beam. In set-up~(b), \SI{99}{\percent} of the cycles of 26--28~August~2026 delivered more than $10^{9}$ ions to the flat top, with a median of $1.28\times10^{9}$. Set-up~(a) is therefore close to, and set-up~(b) exceeds with margin, the intensity of $10^{9}$ ions per cycle on the flat top required for HEARTS@LEIR (Table~\ref{tab:beamSpecs}).

\begin{figure}[htbp]
    \centering
    \includegraphics[width=\linewidth]{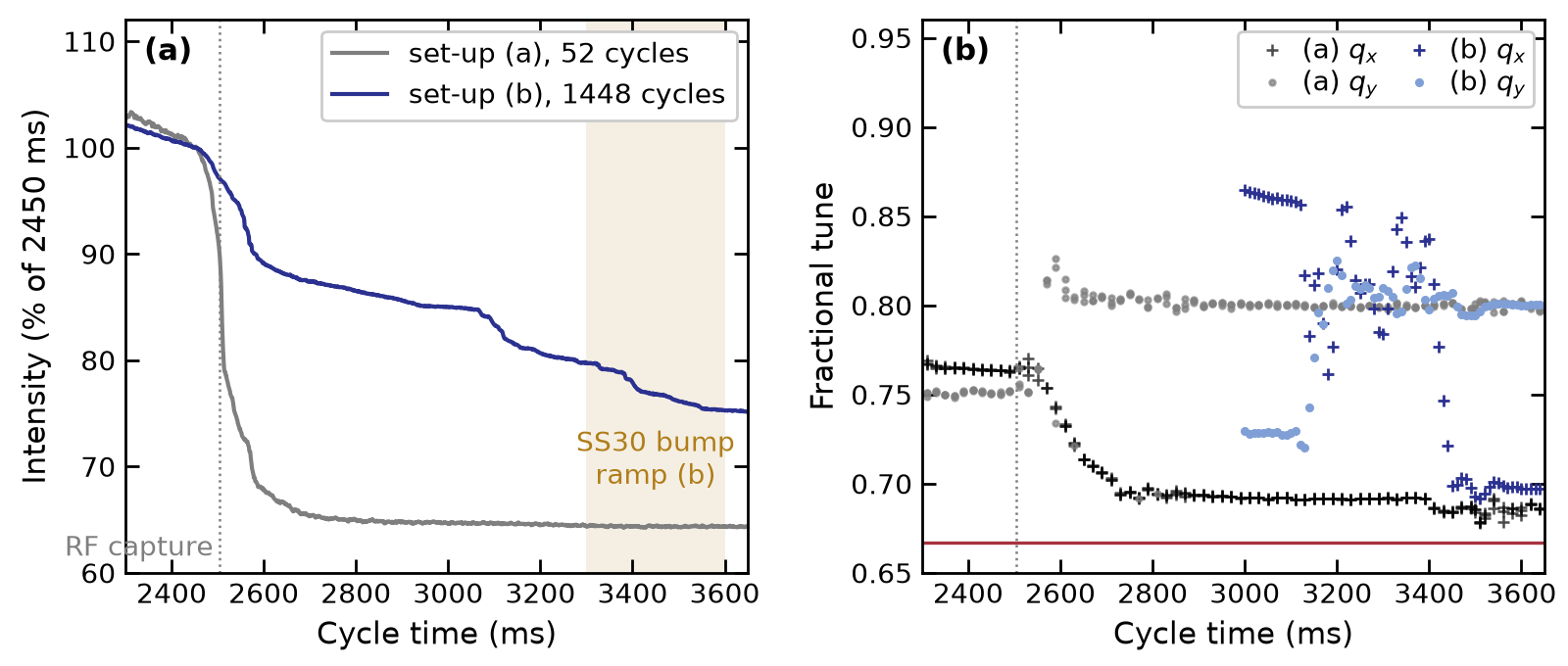}
    \caption{Loss between RF capture and the flat top. (a)~Median BCT intensity normalised at \SI{2450}{\milli\second}, just before RF capture, for set-up~(a), 16~June~2026 (52 cycles), and set-up~(b), 26--28~August~2026, cycles without excitation (1448 cycles); shaded: ramp of the SS30 bump in set-up~(b). (b)~Fractional tunes over the same window: set-up~(a), tune-meter acquisitions of 16~June~2026; set-up~(b), median of the BBQ acquisitions of 26~August~2026, available from \SI{3000}{\milli\second}. Red: third-integer resonance. Dotted: RF capture.}
    \label{fig:MT4BPTransmission}
\end{figure}

\paragraph{Loss between RF capture and the flat top.} Between \SI{2450}{\milli\second}, just before RF capture, and the start of the flat top, set-up~(a) loses \SI{36}{\percent} of its beam and set-up~(b) \SI{24}{\percent}. Set-up~(a) also loses beam earlier, during the coast without cooling: \SI{24}{\percent} on 7~August~2026 (Section~\ref{sec:MTPb}). Figure~\ref{fig:MT4BPTransmission} shows where in the cycle this loss occurs, together with the tune path of each set-up.
\begin{itemize}
    \item In set-up~(a), 32 of the 36 percentage points are lost within \SI{150}{\milli\second} of RF capture, between 2450 and \SI{2600}{\milli\second}, while the tunes are still at $q_x = 0.764$, $q_y = 0.752$, before the crossing of the difference resonance. Only \SI{2.5}{\percent} are lost while it is crossed, between 2600 and \SI{2700}{\milli\second}. The loss is therefore a capture loss: after \SI{700}{\milli\second} without cooling, the momentum spread of about $1.2\times10^{-3}$ measured with the Schottky monitor before capture (Section~\ref{sec:MTMomentum}) exceeds the acceptance of the RF bucket for part of the beam.
    \item In set-up~(b), with cooling until capture and a momentum spread of about $0.85\times10^{-3}$, the capture loss falls to \SI{11}{\percent}. The remaining \SI{13}{\percent} are lost along the ramp: \SI{5.8}{\percent} between 2600 and \SI{3100}{\milli\second}, \SI{3.6}{\percent} between 3100 and \SI{3300}{\milli\second}, when the vertical tune is raised and the tunes stay close to the difference resonance ($q_x \approx 0.816$, $q_y \approx 0.809$), and \SI{3.6}{\percent} between 3300 and \SI{3500}{\milli\second}, when the horizontal tune is lowered and the SS30 bump is ramped up.
\end{itemize}
The capture efficiency is thus the main lever on the delivered intensity, and the cooling until capture of set-up~(b) the main reason for its higher flat-top intensity; the losses at the tune change of set-up~(b) are a second, smaller lever.

\subsection{Results}
\label{sec:MTSXResults}

\begin{figure}[htbp]
    \centering
    \includegraphics[width=\linewidth]{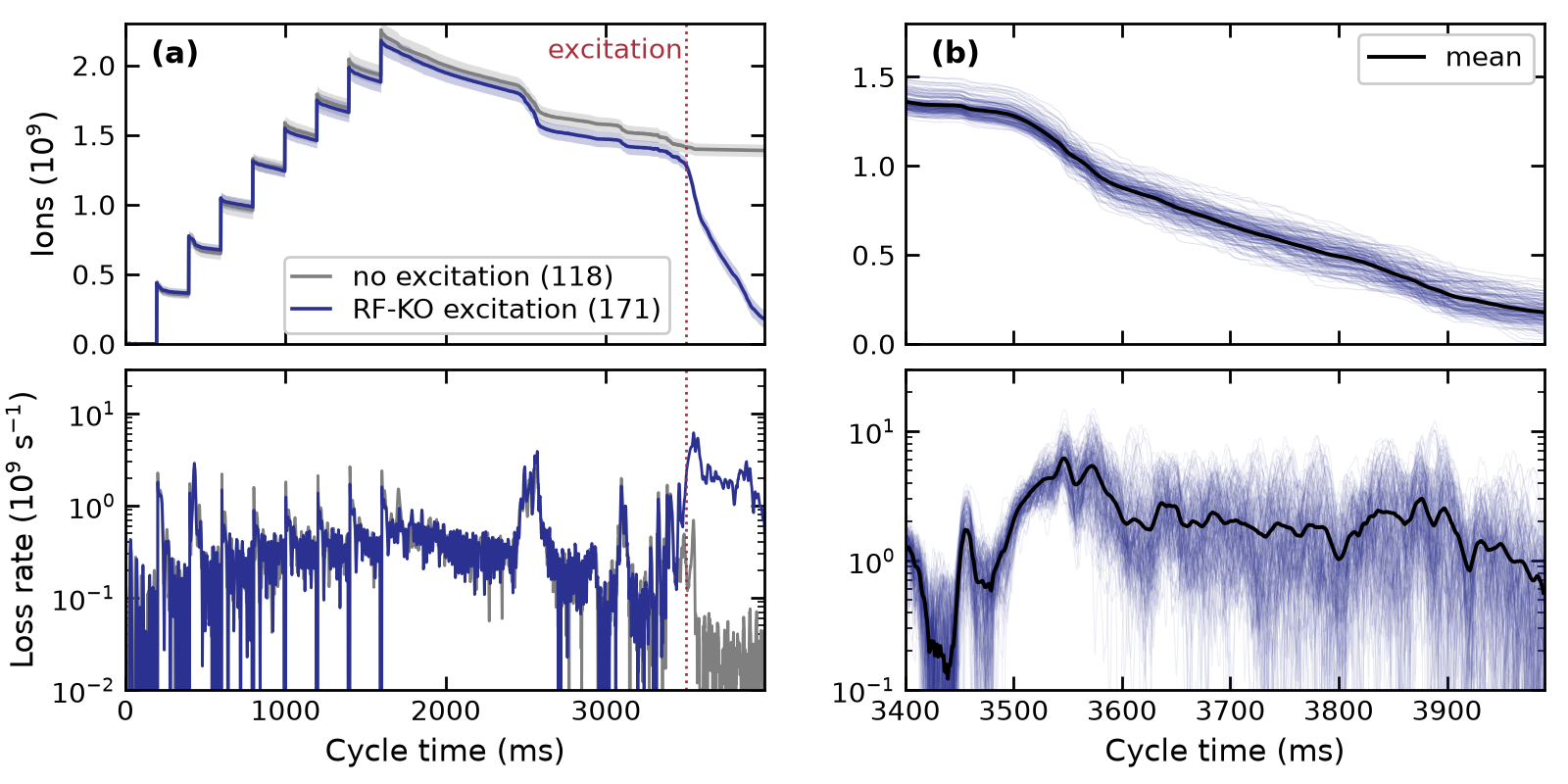}
    \caption{Slow-extraction test on the 4BP cycle, 26~August~2026: cycles with more than $0.8\times10^{9}$ ions at \SI{3500}{\milli\second} and the resonance sextupoles on, with RF-KO excitation ($A = 1.0$, chirp $q_x = 0.686$--0.693, 171 cycles) and with the exciter disabled (118 cycles). (a)~BCT intensity (mean and RMS spread) and loss rate $-\mathrm{d}I/\mathrm{d}t$, smoothed over \SI{10}{\milli\second}. (b)~Intensity and loss rate during the excitation for the individual excited cycles, with their mean.}
    \label{fig:MTSpill}
\end{figure}

On the flat top, two actions are applied from \SI{3500}{\milli\second}: the resonance sextupoles are ramped up within \SI{100}{\milli\second}, and the RF-KO excitation of Section~\ref{sec:MTRFKO} is switched on, here with a constant gain $A = 1.0$ and a chirp covering $q_x$ from 0.686 to 0.693. To separate their effects, the cycles of 26~August~2026 with more than $0.8\times10^{9}$ ions on the flat top were classified using the per-cycle acquisitions of the sextupole currents and of the exciter state:
\begin{itemize}
    \item \emph{Sextupoles and excitation} (171 cycles): \SI{16}{\percent} of the beam is lost within \SI{50}{\milli\second}, \SI{31}{\percent} within \SI{100}{\milli\second} and \SI{86}{\percent} by \SI{3990}{\milli\second}, shortly before the end of the flat top. The loss rate peaks at $6.2\times10^{9}$ ions per second at \SI{3546}{\milli\second}, during the ramp of the sextupoles, and then settles at about $2\times10^{9}$ ions per second (Fig.~\ref{fig:MTSpill}).
    \item \emph{Sextupoles without excitation} (118 cycles with the same settings, the exciter disabled): \SI{1.6}{\percent} of the beam is lost by \SI{3990}{\milli\second}, most of it during the ramp of the sextupoles; the loss rate on the plateau is $0.02\times10^{9}$ ions per second (Fig.~\ref{fig:MTSpill}).
    \item \emph{Neither} (424 cycles): \SI{1.9}{\percent}, the natural lifetime of the beam on the flat top.
    \item \emph{Excitation without sextupoles}, from the gain scan of 7~August~2026 (Section~\ref{sec:MTRFKO}; $A \geq 0.5$, chirp $q_x$ from 0.685 to 0.695, all four sextupoles at zero current): \SI{2}{\percent} within \SI{50}{\milli\second}, \SI{19}{\percent} within \SI{200}{\milli\second} and \SI{46}{\percent} by \SI{3990}{\milli\second}, at $0.5\times10^{9}$ ions.
\end{itemize}
On 26~August~2026 the excitation with this chirp band was applied only together with the sextupoles, so the comparison without sextupoles relies on the scan of 7~August~2026. That scan differs in the intensity, the chirp band and the settings of the SS30 bump, which were changed on 26~August~2026 before the sextupoles were switched on, so the comparison is qualitative. With the exciter disabled, the tune measured with the sextupoles on lies at the upper edge of the chirp band (0.686 to 0.693). The peak of the loss rate occurs while the sextupoles are ramped. With the gain held constant, the flux decreases slowly after the peak as the beam is depleted, as expected in the diffusive model with a constant diffusion coefficient (Section~\ref{sec:MTRFKOReq}). The excitation without sextupoles also removes beam, but more slowly and onto the machine aperture, as discussed in Section~\ref{sec:MTRFKOLate}. In July~2026, at a lower intensity of about $0.8\times10^{9}$ ions, \SI{78}{\percent} of the excited cycles lost more than \SI{70}{\percent} of their beam within the flat top.

\begin{figure}[htbp]
    \centering
    \includegraphics[width=0.75\linewidth]{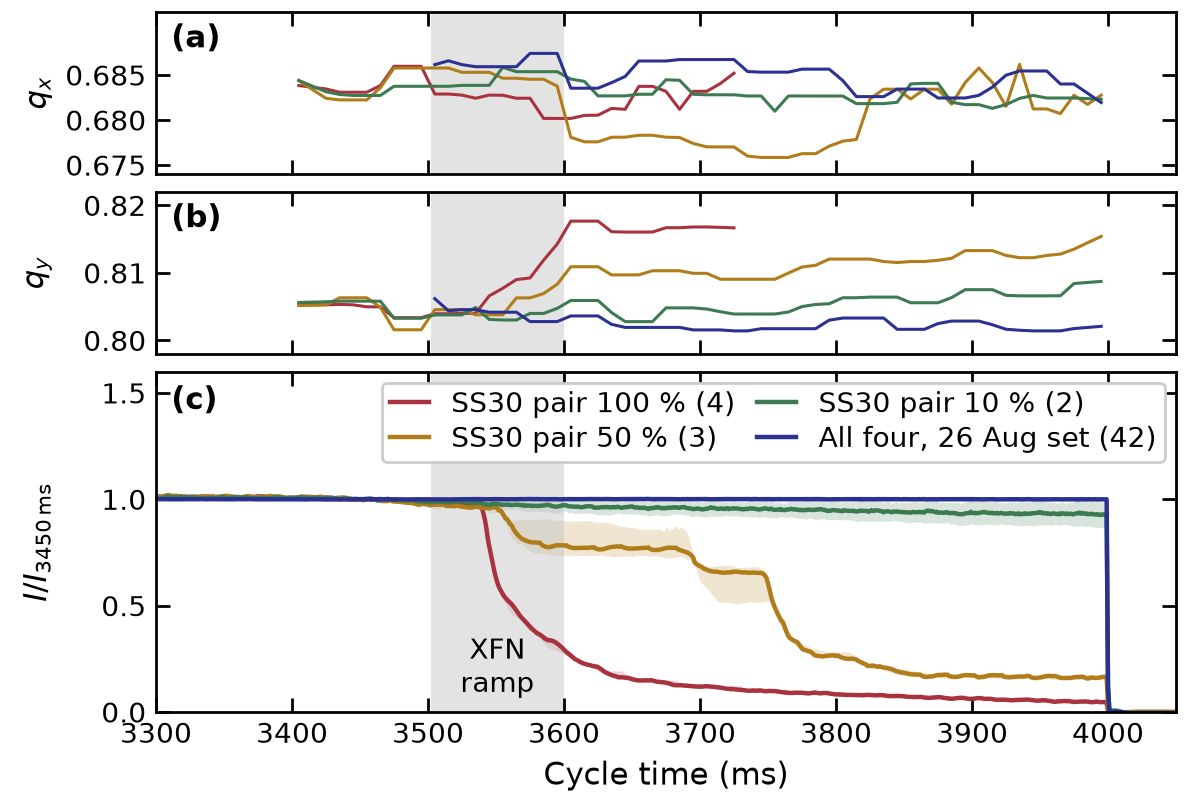}
    \caption{Effect of the resonance sextupoles without effective excitation. SS30 pair at 100, 50 and \SI{10}{\percent}: XFN31 and XFN32 at these fractions of the values of 26~August, XFN11 and XFN12 at \SI{10}{\percent}, on 24~July~2026 ($0.4\times10^{9}$ ions; exciter of the tune monitor at amplitude 0.01, chirp from 0.675 to 0.690). All four, 26~August set: the four sextupoles at the values of 26~August, on 28~July~2026 ($1.0\times10^{9}$ ions, exciter disabled). (a)~Horizontal and (b)~vertical tune measured by the BBQ, median over the cycles in \SI{10}{\milli\second} bins and running median over \SI{50}{\milli\second}, shown while more than \SI{10}{\percent} of the beam remains. (c)~BCT intensity on the flat top, normalised to \SI{3450}{\milli\second} (median and range over the cycles, number of cycles in brackets). Grey band: sextupole ramp.}
    \label{fig:MTSXSext}
\end{figure}

\paragraph{Indication of a dynamic-aperture limitation.} The beam was also lost on the flat top with the resonance sextupoles powered and without effective excitation (Fig.~\ref{fig:MTSXSext}). On 24~July~2026, with \num{0.4e9} ions and $q_x = 0.684$, $q_y = 0.805$ before the sextupole ramp, the exciter of the tune monitor ran at its minimum amplitude, 0.01, with a chirp from 0.675 to 0.690, which does not blow up the beam (Section~\ref{sec:MTRFKO}). Setting XFN31 and XFN32 to \num{-1.86} and \SI{-3.1}{A}, with XFN11 and XFN12 at \SI{10}{\percent} of the values of 26~August, removed \SI{95}{\percent} of the beam, most of it during the sextupole ramp (four cycles). Halving the two currents removed \SI{83}{\percent} (three cycles), and returning them to \SI{10}{\percent} of their value a few per cent (median \SI{3}{\percent}, 11 cycles). With the RF-KO excitation, lowering the sextupole currents reduced the losses (28~July and 7~August~2026). These few cycles hint at a reduction of the dynamic aperture by the sextupoles. The size of the effect is not explained by the single-particle model: with the resonance driving term computed from the optics model, $S \approx \SI{3.5}{m^{-1/2}}$ for the settings of 24~July, the stable region at this tune distance is about five times larger than the physical clearance, and the tracking of a Gaussian beam with the measured tunes over 8192 turns gives no loss. The emittances used in the tracking, from the previous optics model, are 15 to \SI{30}{\percent} smaller than those of Section~\ref{sec:MTPb}; this does not change the result, since the beam extends to less than a tenth of the stable region.

The beam conditions do not explain the losses either. On 28~July~2026 the full set of 26~August, with the exciter disabled and at a similar tune distance ($q_x = 0.685$), removed a median of \SI{0.1}{\percent} (55 cycles, at most about \SI{6}{\percent}) of a beam with 2.5 times the intensity, $1.0\times10^{9}$ ions, and with larger emittances at the end of the ramp, $\epsilon_{x,n} = 215$ and \SI{430}{\nano\metre\radian} against 160 and \SI{305}{\nano\metre\radian} on 24~July (IPM, from 3380 to \SI{3460}{\milli\second}). Its incoherent space-charge tune shift was therefore about twice as large, for an equal bunch length; the bunch length was not logged for these cycles. The only logged difference is the evolution of the tunes during the sextupole ramp (Fig.~\ref{fig:MTSXSext}a and~b). Before the ramp, the tunes of all cycles of 24~July coincide. The feed-down of XFN31 and XFN32 in the SS30 bump was not compensated on that day: during the ramp, $q_y$ rose from 0.805 to 0.817 with the pair at full strength and to 0.811 at half strength, while $q_x$ fell by about 0.003 and 0.006, and the beam was lost while the tunes moved. On 28~July the tunes stayed at $q_x = 0.685$ and $q_y = 0.802$ to 0.804 over the whole flat top. With $q_x + q_y \approx 1.50$ at full strength, the working point of 24~July reached the fourth-order resonance $2Q_x + 2Q_y = 9$, which sextupoles drive in second order. The other cycles near this resonance are too few to confirm the correlation, and it does not explain the loss with the SS30 pair at half strength, at $q_x + q_y \approx 1.489$.

The present data therefore do not identify the missing ingredient. Possible explanations are a resonance driving term not included in the model, for example from the sextupole component of the main dipoles combined with beta-beating, resonances of higher order excited by the sextupoles near the working point, and the incoherent space-charge tune shift, which the BBQ does not measure. The chromaticity on the flat top differs from the model by about 3 units horizontally and by 1 to 2 units vertically, with opposite signs (Section~\ref{sec:MTOptics}), which indicates that the sextupolar content of the model is incomplete. If one of these explanations is confirmed, the distance to the resonance assumed in Chapter~\ref{sec:SRE} ($\delta = 0.005$) is too small to be realistic, or the beam has to be extracted debunched. Both are established options of resonant slow extraction; a larger distance requires proportionally stronger sextupoles, which is the motivation of the \SI{20}{\ampere} rating discussed in Section~\ref{sec:MTFlattop}.

\begin{figure}[htbp]
    \centering
    \includegraphics[width=0.6\linewidth]{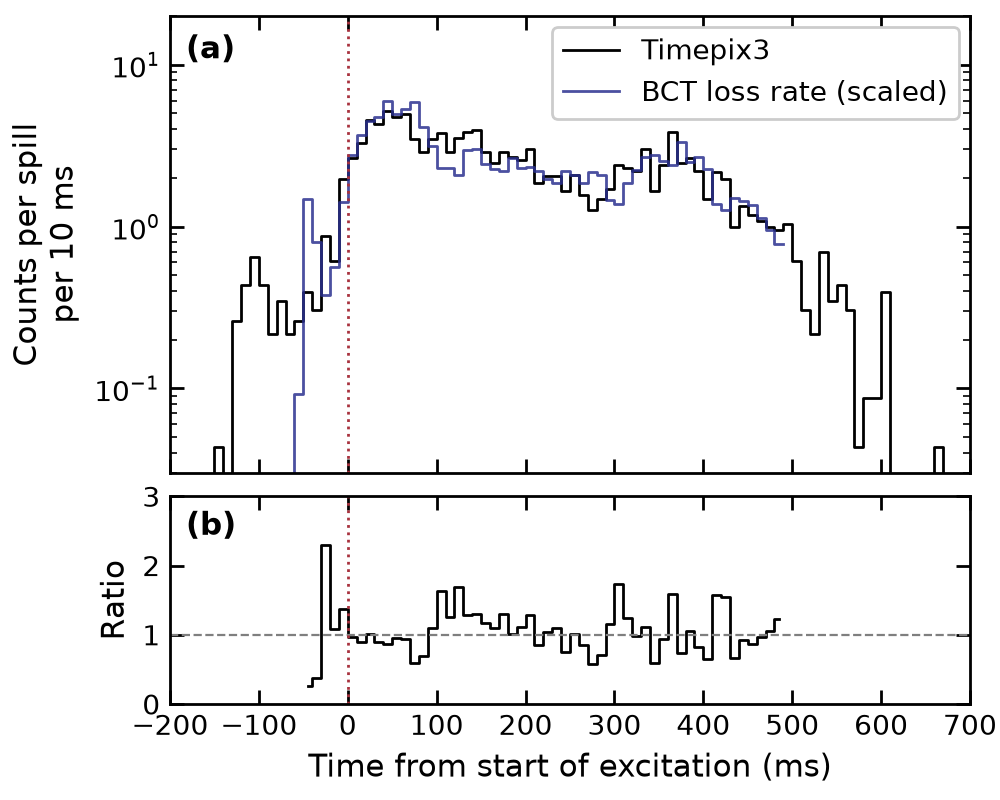}
    \caption{Time structure of the losses in the slow-extraction test of August~2026. (a)~Average count rate per spill of the Timepix3 monitor in SS30 for the excited cycles of 26--28~August~2026~\cite{SLAWGTimepix2026}, and loss rate from the BCT for the excited cycles of 26~August~2026 after subtraction of the cycles without excitation, scaled to the same number of counts between $-50$ and \SI{490}{\milli\second}; \SI{10}{\milli\second} bins. The BCT curve ends at \SI{3990}{\milli\second}, before the fast-extraction kicker removes the remaining beam. (b)~Ratio of the two signals.}
    \label{fig:MTTimepix}
\end{figure}

\paragraph{Time structure of the losses.} The Timepix3 monitor in SS30 records the secondary particles produced by the ions lost on the vacuum chamber, and its count rate is taken to be proportional to the loss rate~\cite{SLAWGTimepix2026}. The spills of the slow-extraction test were identified in its data with the timestamps of the cycles: the clock of the monitor follows the timing of the machine with a constant offset and a drift of \SI{-16.55}{\micro\second} per second, and spills that deviate by more than three standard deviations ($\sigma = \SI{14}{\milli\second}$) from this relation are discarded. A single spill gives only 100 to 150 counts, too few for its time profile. Averaged over the spills, the count rate has the same time structure as the loss rate derived from the BCT (Fig.~\ref{fig:MTTimepix}): in \SI{10}{\milli\second} bins, the ratio of the two signals is 1.05 on average between 50 and \SI{450}{\milli\second} after the start of the excitation, with a scatter of \SI{30}{\percent}. Both show the peak during the ramp of the sextupoles, a second maximum about \SI{360}{\milli\second} after the start of the excitation and structures on a time scale of 50 to \SI{100}{\milli\second}, whose origin has not been identified yet. The agreement shows that the losses measured by the BCT occur at the aperture bottleneck in SS30 throughout the spill. The tracks recorded by the monitor show a mixed radiation field; with the present statistics, no conclusion can be drawn on the direction of incidence or on the linear energy transfer of the particles. With more statistics, both would become available and would allow the loss location to be reconstructed~\cite{SLAWGTimepix2026}.

\paragraph{Spill quality from the BCT.} The BCT gives a first, bandwidth-limited figure of merit of the spill. Over the plateau, from 3600 to \SI{3900}{\milli\second}, the loss rate of the cycles with sextupoles and excitation, computed in \SI{10}{\milli\second} bins, has a mean of $2.0\times10^{9}$ ions per second and a coefficient of variation $c_V = 0.69$ (0.62 to 0.77 over the cycles). The same estimator applied to cycles without loss gives a noise of $0.2\times10^{9}$ ions per second per bin, so that the noise-corrected coefficient of variation is 0.68, which corresponds to a duty factor $1/(1+c_V^2) = 0.68$. This value characterises the fluctuations on the \SI{10}{\milli\second} scale. It is about three times larger than the sweep-to-sweep fluctuations of the tracking of Section~\ref{sec:MTRFKOReq} ($c_V \approx 0.2$ in \SI{10}{\milli\second} bins), which points to the structures on a time scale of 50 to \SI{100}{\milli\second} mentioned above. Faster fluctuations cannot be resolved with the present loss detectors: the Timepix3 monitor records too few counts per spill, and the signal of the BLMs is close to their noise level (see below). The spill quality of Specification~H will therefore be measured on the extracted beam.

\paragraph{Radiation survey.} A radiation survey was carried out on 29~July~2026, after the 4BP cycle was played every \SI{7.2}{\second} for 4.7~hours. During this period, 1761 cycles carried beam, with $(0.81 \pm 0.07)\times10^{9}$ ions at the flat top, so that about $1.4\times10^{12}$ ions were driven onto the aperture. During the survey, the residual dose rate was measured with a portable monitor to localise the beam losses. The highest value was found about \SI{30}{\centi\metre} downstream of UEHV31, the first BPM in SS30, close to the impact point predicted by the tracking: about \num{20000} counts per second with a portable BGO scintillator, corresponding to approximately \SI{30}{\micro\sievert\per\hour} on contact, centred vertically, on the right side of the vacuum chamber. A BLM was moved to this location on 31~July~2026.

\paragraph{Beam-loss monitors.} In June~2026, no signal was observed on the two BLMs next to the fast-extraction kickers~\cite{SLAWGBLM2026}. The spikes seen on the other monitors were also present without beam and on a channel without detector, which identifies them as interference of a machine device with the acquisition electronics. After the BLM of the electron cooler was moved to the loss location, the signals of all monitors became too noisy to be used. At the end of August~2026, after this BLM had been repositioned so that it no longer touched the surrounding equipment, the electronics had been reset and the intensity of the spill had been increased (set-up~(b)), it recorded signals coincident with the spills. In the running sum with an integration time of \SI{82}{\milli\second}, closest to the acquisition interval of \SI{0.1}{\second}, the dose rate rises from a baseline of about \SI{0.3}{\micro\gray\per\second} to about \SI{10}{\micro\gray\per\second} for 0.6 to \SI{0.7}{\second}, the duration of the spill. This is of the order of the FLUKA estimate for the measured loss rate of about $2\times10^{9}$ ions per second. The analysis is preliminary: the interference is still visible, the signal is barely above the noise, and the conversion of the dose into a number of lost ions remains to be calibrated~\cite{SLAWGBLM2026}. The ionisation chambers, whose lower limit of about \SI{0.2}{\micro\gray\per\second} lies less than two orders of magnitude below the expected signal, are therefore not adequate to measure the loss pattern. For HEARTS@LEIR, a permanent installation with its own acquisition and detectors more sensitive to these losses will be studied~\cite{SLAWGBLM2026}; the Timepix3 monitor, for example, has a lower limit of about \SI{6e-12}{\gray\per\second}~\cite{SLAWGTimepix2026}.

\paragraph{Scope of the test.} The test was designed to validate the steps upstream of the septum, not to reproduce the operating conditions of HEARTS@LEIR. The spill lasts about \SI{0.5}{\second}, at the lower end of the specified range of 0.5 to \SI{10}{\second}, and its rate exceeds the specified range of $10^{6}$ to $10^{8}$ ions per second by one to three orders of magnitude. The excitation was applied with a constant gain, so the flux decreases along the flat top instead of being held constant by an amplitude modulation (Section~\ref{sec:MTRFKOReq}). Without a septum, neither the extraction efficiency nor the properties of the extracted beam can be measured, and a figure of merit for Specification~H requires a detector with an adequate dynamic range and bandwidth. What the test establishes is that a cooled beam of more than $10^{9}$ Pb$^{54+}$ ions can be prepared on a \SI{0.5}{\second} flat top at the slow-extraction working point, and that the RF-KO excitation removes it in a controlled and reproducible way within the flat top, with the losses localised by an orbit bump in the future extraction straight section. With the sextupole strengths used, a fifth of the design value at five times the design distance from the resonance, the single-particle model predicts no separatrix within the aperture, so the test does not probe the resonant transfer itself; the losses observed with the sextupoles alone indicate, however, that the dynamic aperture is smaller than this model predicts.

\section{Fast extraction with two out of three kicker modules}
\label{sec:MTKFH}

The extraction scheme of Chapter~\ref{sec:SRE} requires a modification of the tank of the fast-extraction kicker KFH3234 and possibly the relocation of the module KFH31. The fast extraction of LEIR to the PS uses three kicker modules, KFH31, KFH32 and KFH34. On 26~August~2026, on the MDEARLY cycle with Pb$^{54+}$, the timing of KFH31 was delayed so that the beam no longer saw its pulse. With KFH32 and KFH34 alone, the extraction worked after their voltage was reduced from 55 to \SI{50}{\kilo\volt}; with the original voltage the extraction efficiency read about \SI{105}{\percent}, and the reduction restored the nominal value. The trajectory in the transfer line was similar to that of operation. KFH31 therefore acts as a steering element for the line to the PS and is not needed for the deflection itself, in agreement with the optics calculation of the fast extraction.

\chapter{Beam line to the experimental area}
\label{sec:Beamline2HEARTS}

The slow extraction of Chapter~\ref{sec:SRE}, whose feasibility in the present machine is examined in Chapter~\ref{sec:MachineTests}, delivers the beam at the exit of the magnetic septum; its phase-space distribution there is the input to the transfer line. The task of the line is to transport the beam to the experimental area and to shape it into the irradiation field required by the users. This chapter presents the beam optics and the homogenisation schemes that were studied, compares them, and derives the magnet and hardware requirements of the retained baseline and their cost.

The layout of the line is constrained by the geometry of the experimental hall, within which the entire beam-line footprint must fit. The design drivers are Specifications~F and~G of Chapter~\ref{sec:BeamSpecs}:
\begin{itemize}
    \item a transverse flux uniformity within $\pm 10$\,\% (ratio of maximum to mean),
    \item irradiation field sizes from $2 \times 2$ to $20 \times 20$~cm$^2$.
\end{itemize}

Three homogenisation schemes, the three techniques announced in Chapter~\ref{sec:BeamSpecs}, have been studied. The first relies on blowing up the beam and scraping it to homogenise the transverse distribution at the device under test (DUT). The second follows the standard established at the NASA Space Radiation Laboratory (NSRL) at BNL~\cite{NSRL-Octupoles} and folds the tails of the distribution with octupoles. The third sweeps a small beam spot across the field in a spiral pattern (Section~\ref{sec:TLscanning}); it needs no additional focusing elements, but introduces a time structure in the delivered dose.

\begin{figure}
    \centering
    \includegraphics[width=\linewidth]{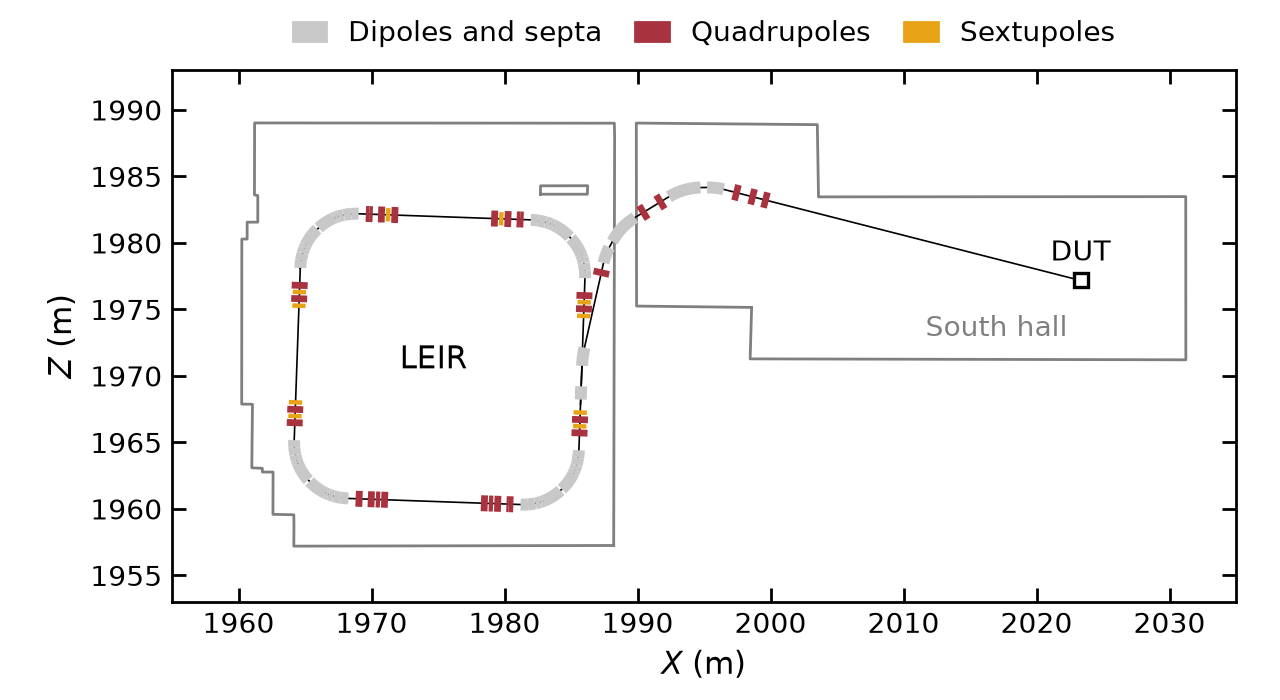}
    \includegraphics[width=\linewidth]{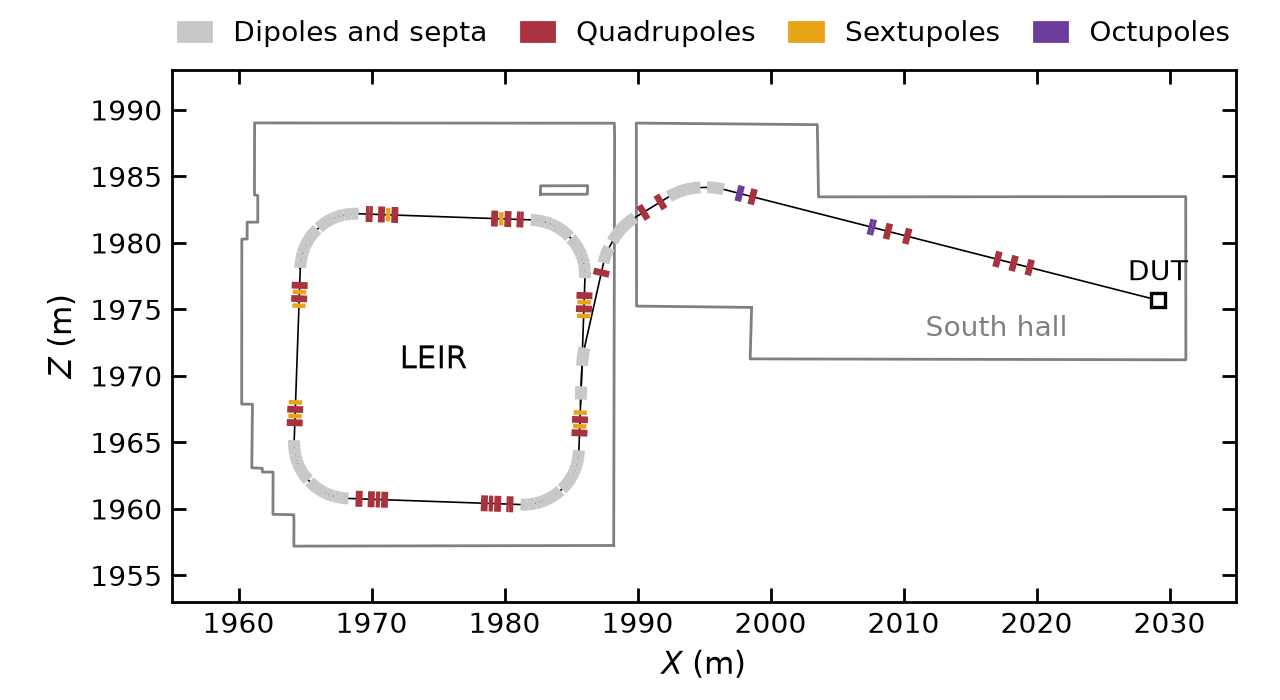}
    \caption{Schematic layout of the two beam transfer line options to the HEARTS@LEIR irradiation bench. The top panel shows the baseline minimalistic design; the bottom panel shows the octupole-based design. Magnetic elements are color-coded: grey — dipoles (bending magnets), red — quadrupoles (focusing/defocusing), yellow — sextupoles (chromatic correction), purple — octupoles (transverse distribution homogenization). Grey lines: outlines of the LEIR hall and of the South hall; square: device under test (DUT).}
    \label{fig:TL2HEARTS-Options}
\end{figure}

The first two schemes require different lattices, shown in Fig.~\ref{fig:TL2HEARTS-Options}; the third runs on the lattice of the first. Blow-up and scraping is adopted as the baseline. The three schemes are compared in Section~\ref{sec:TLcomparison}, and the cost of the baseline is given in Section~\ref{sec:TLcost}.

\section{Beam transport and optics}

The beam transport and optics have been conceptually designed to incorporate an achromatic cell, effectively canceling dispersion and thus momentum deviation contributions.\
A highly flexible beam optics system has been developed, featuring triplet focusing systems.\

The initial conditions for the beamline design are obtained from slow extraction simulations and emittance measurements. For the horizontal plane, the extracted beam distribution at the electrostatic septum is used. The vertical emittance is based on measurements performed with the Ionisation Profile Monitor (IPM). The parameters are listed in Table~\ref{tab:TLinitCond}.

\begin{table}[]
    \centering
    \caption{Initial optical functions considered for the transfer line design. For Pb$^{54+}$ the vertical emittance measured on the flat top is larger than the value assumed here: about \SI{890}{\nano\metre\radian} geometric RMS at $0.5\times10^{9}$ ions with set-up~(a) of the 4BP cycle (Section~\ref{sec:MTPb}); the vertical acceptance of the line is to be verified with it.}
    \begin{tabular}{l c c c}
    Quantity & Unit & Horizontal & Vertical\\
    \hline\hline
    Emittance & nm rad & 100 & 250 \\ 
    (RMS geo.)\\
    $\beta_{x,y}$ & m & 100 & 5\\
    $\alpha_{x,y}$ & - & -5 & 0\\
    \end{tabular}
    \label{tab:TLinitCond}
\end{table}

\subsection{Baseline minimalistic version}

\begin{figure}
    \centering
    \includegraphics[width=\linewidth]{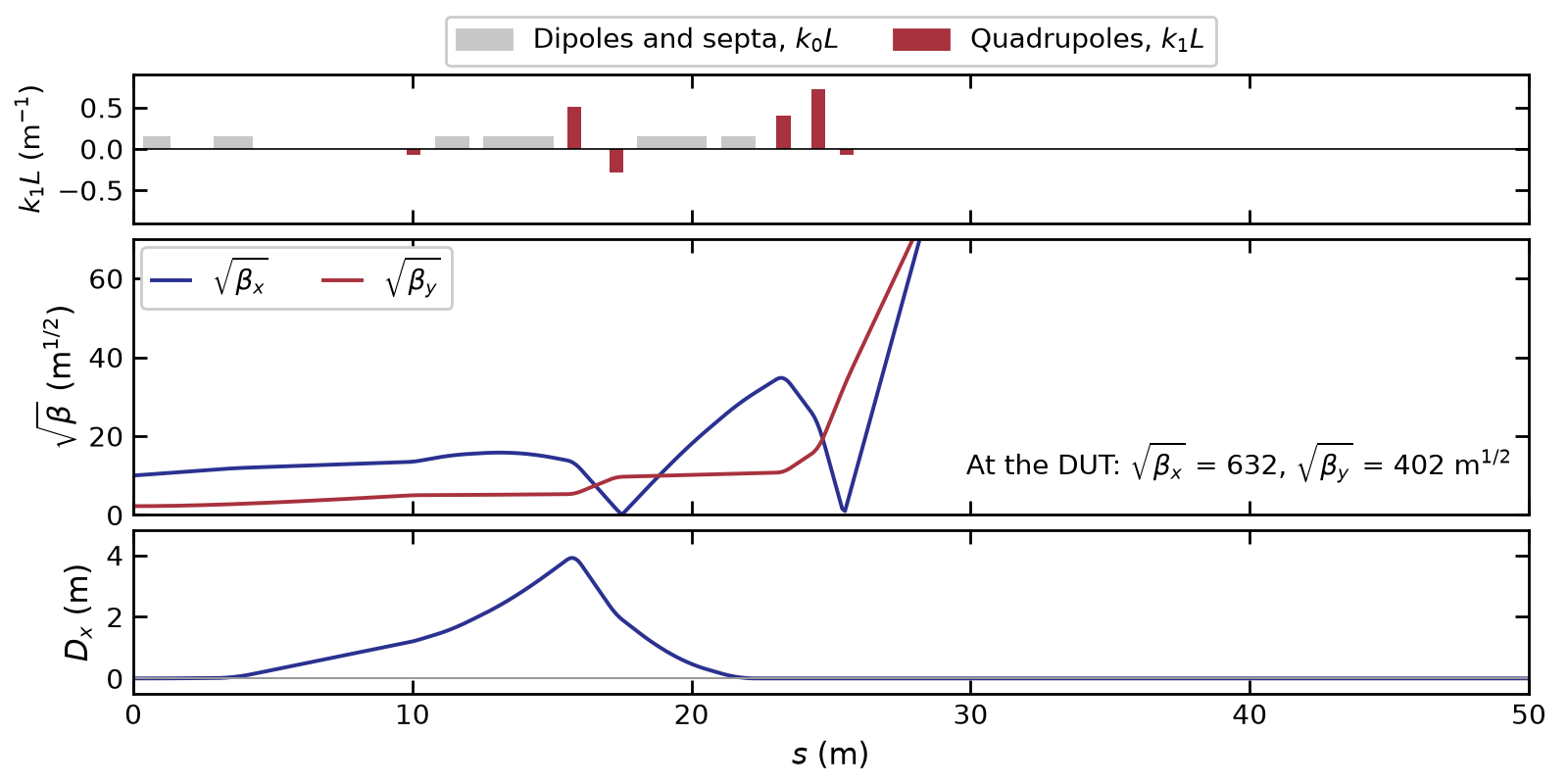}
    \caption{Beam optics of the minimalistic baseline version for irradiation of electronics, with the final focus for the $20 \times 20$~cm$^2$ field. Top: dipoles and septa (grey) and integrated quadrupole strengths (red), as in Fig.~\ref{fig:TL2HEARTS-Options}. Middle: square root of the $\beta$-functions, proportional to the beam size, shown up to \SI{70}{m^{1/2}}; downstream of the final focus it grows linearly in the drift to the DUT (values at the DUT in the panel). Bottom: horizontal dispersion.}
    \label{fig:OpticsTL-OptA}
\end{figure}

The baseline design incorporates a bending magnet assembly providing a total deflection angle of $\approx90^{\circ}$, equivalent in function to a sector magnet in LEIR. The proposed transfer line requires dedicated (de)focusing elements to ensure efficient beam transmission. The current total length of the transfer line is 50 m, subject to revision as additional engineering constraints and boundary conditions are established.
The optical lattice is designed to support fully achromatic beam transport: the dispersion function is suppressed to zero at the exit of the bending arc section. This achromaticity condition is particularly advantageous in mitigating the impact of momentum spread inherent to the slowly extracted beam. Downstream of the arc, a quadrupole triplet is implemented to provide transverse beam shaping at the irradiation plane. Given the large irradiation field sizes specified, high-gradient quadrupoles are required to establish quasi-ballistic beam optics (strongly divergent cone towards the DUT), enabling controlled transverse beam broadening.\ A collimation section consisting of scrapers and masks positioned downstream of the triplet serves the dual purpose of localising beam losses and defining the transverse beam profile for delivery to the end-user stations. The optimal placement of the collimators, masks and scrapers will be studied in detail in future design iterations. The full beam optics layout is illustrated in Fig.~\ref{fig:OpticsTL-OptA}.

As an illustrative example, the final triplet is matched to produce a beam broad enough to cover the required $20 \times 20$~cm$^2$ field. Because the extracted beam has a small emittance, this requires $\beta$-functions at the DUT of tens to hundreds of kilometres ($\beta_x = \sigma_x^2/\epsilon_x$): the triplet focuses the beam to a small waist, from which it diverges freely over the remaining drift to the DUT (quasi-ballistic optics).

\begin{figure}
    \centering
    \includegraphics[width=\linewidth]{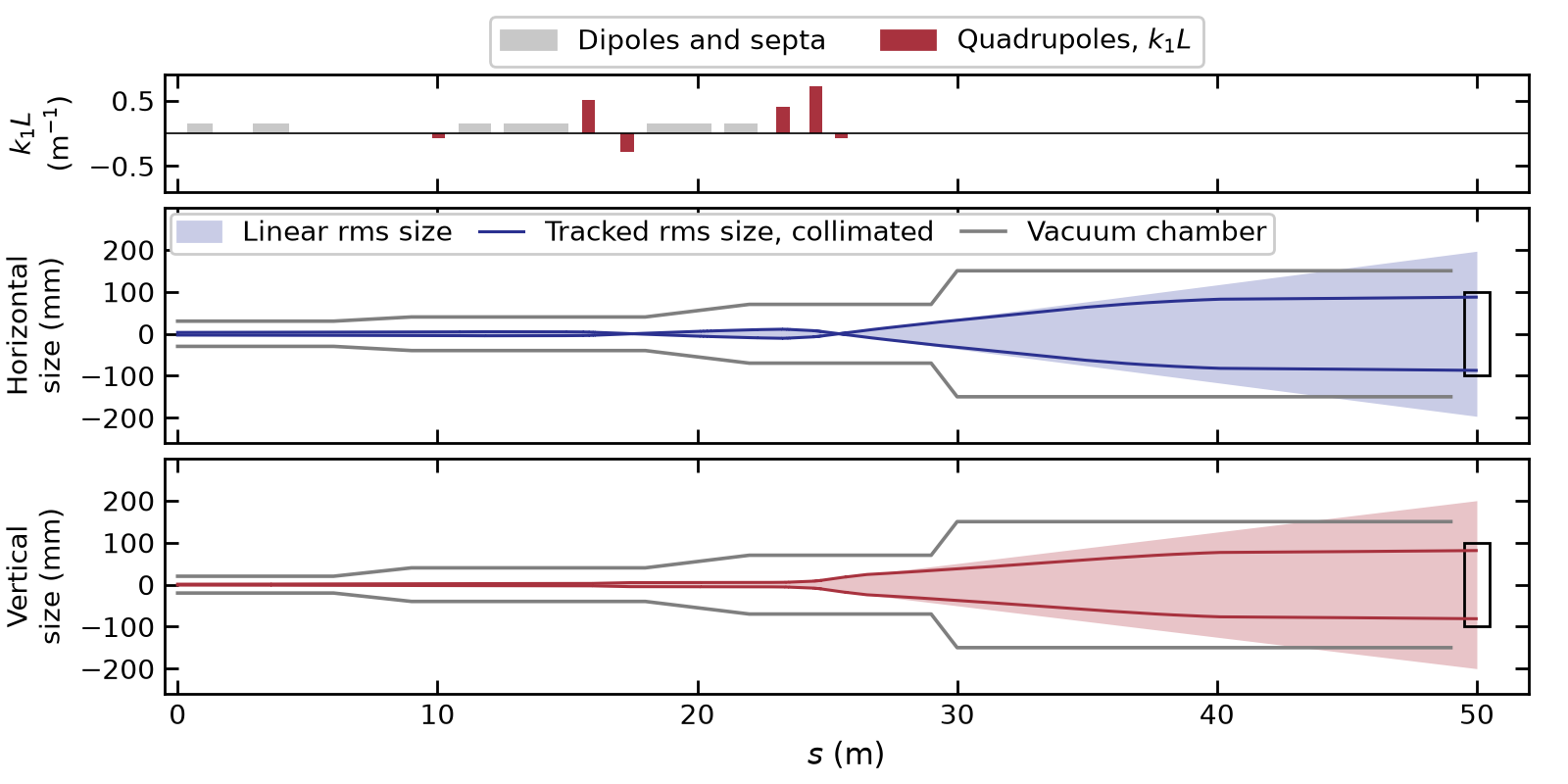}
    \caption{Beam transport of one RMS beam size along the baseline transfer line. The RMS beam spot is matched to 20cm in both planes. Colored shadows illustrate the linear transport without collimators/masks. The solid lines follow a collimated beam covering the irradiation area of 20x20cm$^2$. Magnetic elements follow the same color-coding as in Fig.\ref{fig:TL2HEARTS-Options}: grey — dipoles (bending magnets), red — quadrupoles (focusing/defocusing).}
    \label{fig:OpticsTL-Transport}
\end{figure}

\begin{figure}
    \centering
    \includegraphics[width=0.6\linewidth]{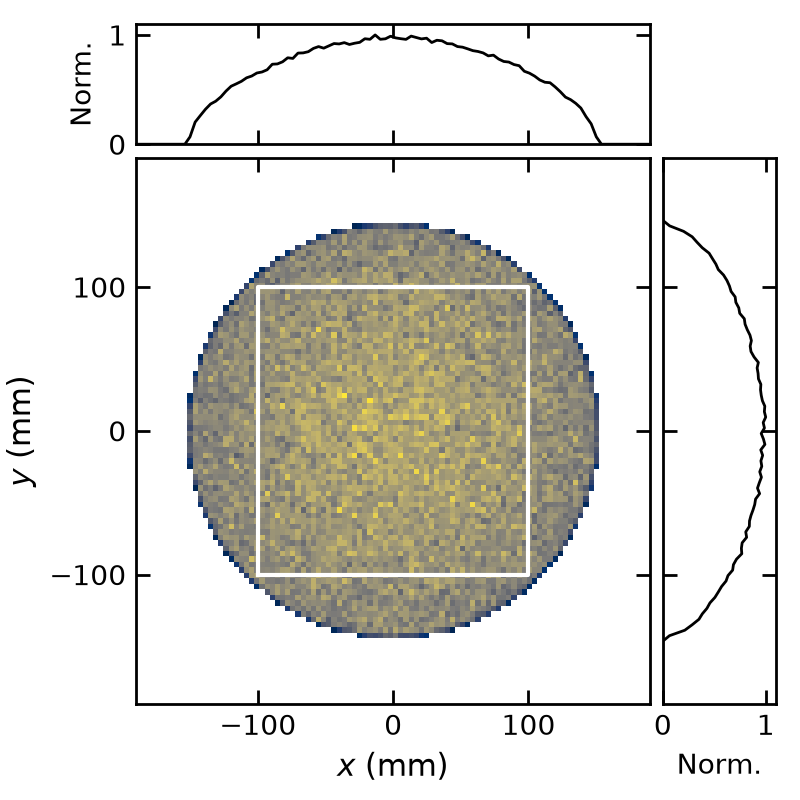}
    \caption{Exemplary beam spot at DUT to cover a 20cm x 20cm irradiation field (white square) for the minimalistic case. The top and right panels show the projections of the distribution.}
    \label{fig:BeamSpotAtDUT-Baseline}
\end{figure}

Beam transport efficiency and transverse profiles are computed via element-by-element tracking in Xsuite~\cite{Iadarola:2023fuk}. The transport of one RMS beam size is illustrated in Fig.~\ref{fig:OpticsTL-Transport}. Note that the vacuum chamber requires significant enlargement downstream of the final focus to allow beam passage. Figure~\ref{fig:BeamSpotAtDUT-Baseline} depicts the beam spot at the DUT covering the broadest irradiation area specified, noting that simulations were carried out with Gaussian beams.

\begin{figure}[htbp]
    \centering
    \includegraphics[width=\linewidth]{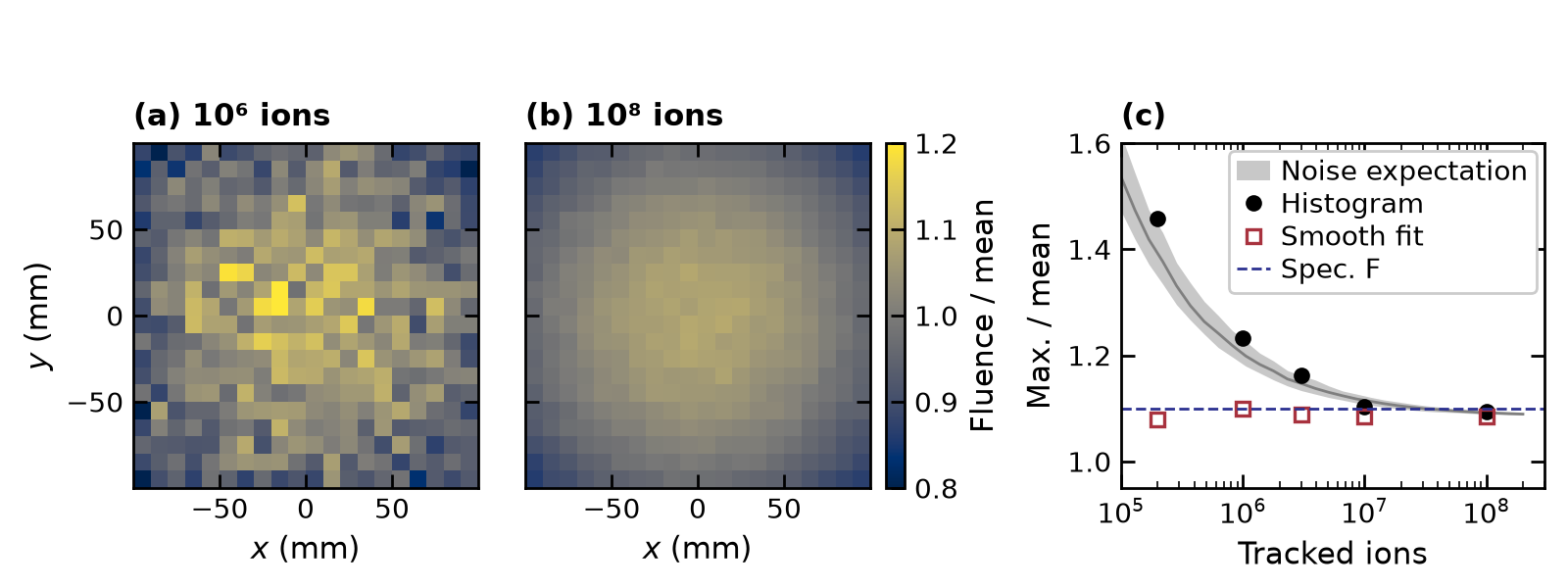}
    \caption{Homogeneity of the baseline scheme (blow-up and scraping) at the DUT for the $20 \times 20$~cm$^2$ field, in \SI{1}{\centi\meter\squared} bins normalised to the mean. (a) $10^{6}$ tracked ions, about 370 per bin in the field. (b) $10^{8}$ tracked ions, about 37\,000 per bin. (c) Ratio of maximum to mean fluence versus the number of tracked ions: histogram (circles), smooth fit with a fourth-order polynomial in $x$ and $y$ (squares), and the value expected from the distribution of (b) with the counting noise of each sample (line and band, median and 68\,\% range). Dashed: limit of Specification~F.}
    \label{fig:TLHomogeneity}
\end{figure}

Finally, the homogeneity specification has been taken into account to determine the required beam size blow-up and the associated beam losses, resulting in a beam-to-DUT transfer efficiency of 14.9\%.
The homogeneity is evaluated on the fluence of the tracked ions in \SI{1}{\centi\meter\squared} bins at the DUT (Fig.~\ref{fig:TLHomogeneity}). With $10^{6}$ tracked ions, about 370 fall in each bin of the field, so the histogram fluctuates by 5\,\% rms from counting statistics alone, and its maximum, 1.23 times the mean, is set by these fluctuations: a perfectly flat field with the same number of ions already gives 1.15. A smooth distribution fitted to the tracked ions, whose residuals are consistent with counting statistics, gives 1.10. A tracking of $10^{8}$ ions, about 37\,000 per bin, which takes \SI{10}{\minute} on 14 processor cores, reduces the fluctuations to 0.5\,\% and gives 1.09 directly from the histogram, in agreement with the fit. The fluence therefore reaches at most 1.09 times the mean, at the limit of Specification~F, and falls to 0.86 times the mean in the corners of the field. A measurement with the dedicated diagnostic counts about $10^{7}$~ions per \si{\centi\meter\squared}, for which the statistical fluctuations are negligible. The smooth fit to $10^{6}$ tracked ions gives 1.07 for the octupole option (Table~\ref{tab:TLoptionsComparison}).

Following these results, the expected particle fluxes at DUT with the available data shown in Table~\ref{tab:IonIntDetails} are summarized in Table~\ref{tab:fluxes_at_DUT}. The calculations are performed considering a total cycle length of 5BPs (6s), a conservative extraction efficiency of $\eta_{\text{SRE}} = 0.7$ (see Fig.~\ref{fig:extractionEfficiency}) and intensity values listed in Table~\ref{tab:IonIntDetails}, limited to the radiation-protection limit of Table~\ref{rp_charge_limits}. Note that all values shown stay at the desired level of 10$^5$~ions cm$^{-2}$ s$^{-1}$ (average over the whole cycle) for the irradiation areas of 12x12cm$^2$.

\begin{table}[h]
\centering
\caption{Fluxes at DUT of different ion species for intensities listed in Table~\ref{tab:IonIntDetails}, limited to the radiation-protection limit per cycle of Table~\ref{rp_charge_limits} for O$^{8+}$, Kr$^{22+}$ and Xe$^{40+}$; for Kr$^{22+}$ the limit of Kr$^{29+}$ at \SI{100}{MeV/nucleon} is applied, which is conservative for the lower energy reached by Kr$^{22+}$. The average flux is calculated considering an extraction efficiency of $\eta_{\text{SRE}} = 0.7$, beam transport efficiency of $\eta_{\text{TL}} = 0.15$ and a total cycle length of 6s.}
\begin{tabular}{cccccc}
    \toprule
    \multirow{2}{*}{Ion species} 
        & \multirow{2}{*}{\parbox{3.5cm}{\centering Max. intensity \\($10^9$ ions)}} 
        & \multicolumn{3}{c}{Average flux (10$^5$~ions cm$^{-2}$ s$^{-1}$)} \\
    \cmidrule(lr){3-5}
        & & 12x12cm$^2$ & 15x15cm$^2$ & 20x20cm$^2$\\%
    \midrule
    O$^{8+}$ & 3.88 & 4.7 & 3.0 & 1.7\\%
    Ar$^{11+}$ & 2.5 & 3.0 & 1.9 & 1.1\\%
    Ar$^{16+}$ & 1.84 & 2.2 & 1.4 & 0.8\\%
    Kr$^{22+}$ & 2.10 & 2.6 & 1.6 & 0.9\\%
    Kr$^{29+}$ & 0.93 & 1.1 & 0.7 & 0.5\\%
    Xe$^{39+}$ & 1.12 & 1.4 & 0.8 & 0.4\\%
    Xe$^{40+}$ & 2.30 & 2.8 & 1.8 & 1.0\\%
    Pb$^{54+}$ & 1.84 & 2.2 & 1.4 & 0.8\\%
    \bottomrule
\end{tabular}
\label{tab:fluxes_at_DUT}
\end{table}

\subsection{Homogenization with octupoles}

\begin{figure}
    \centering
    \includegraphics[width=\linewidth]{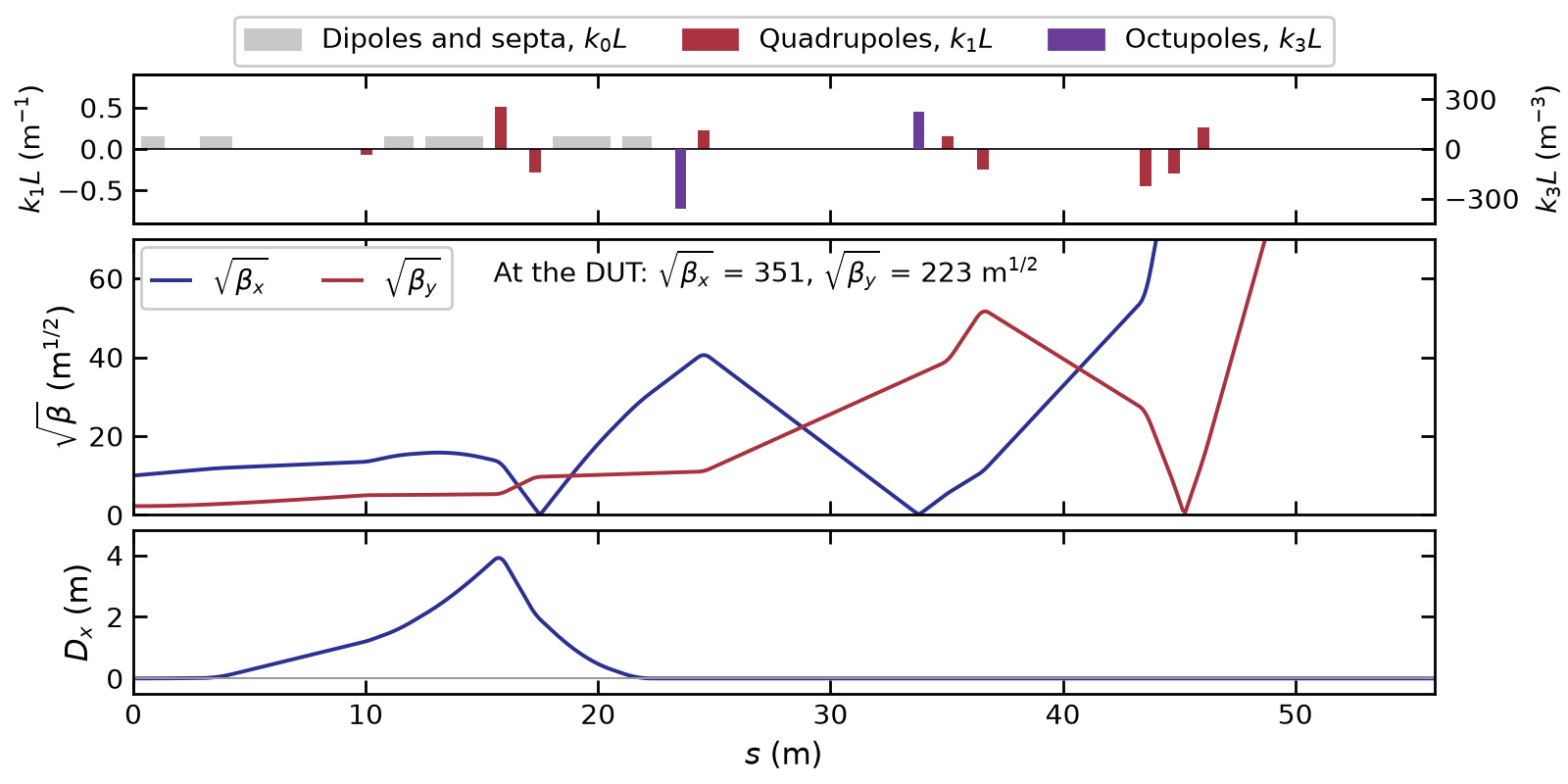}
    \caption{Beam optics for transverse homogenization using octupole pairs. Top: dipoles and septa (grey), integrated quadrupole strengths (red, left axis) and octupole strengths used in the tracking (purple, right axis), as in Fig.~\ref{fig:TL2HEARTS-Options}. Middle: square root of the $\beta$-functions, proportional to the beam size, shown up to \SI{70}{m^{1/2}}; downstream of the final focus it grows linearly in the drift to the DUT (values at the DUT in the panel). Bottom: horizontal dispersion.}
    \label{fig:TLwithOctupoles}
\end{figure}

In the second design variant, the beam transport up to the exit of the bending arc remains identical to the baseline, with achromatic transport preserved throughout.
Downstream of the arc, a dedicated insertion for transverse phase-space distribution tail folding is introduced, utilising a pair of octupole magnets. The first octupole acts primarily on the horizontal distribution, while the second addresses the vertical plane. This decoupled manipulation is achieved by exploiting asymmetric optics conditions, where the $\beta$-functions are tailored to maximise the independent influence of each octupole on the respective transverse plane. 
The final focus section is a five-quadrupole assembly, in which the three type-F final-focus quadrupoles of Table~\ref{tab:TLMagnetSpecs} work together with two of the type-T quadrupoles, providing the necessary degrees of freedom to match the phase advances and transverse beam sizes to the requirements of the irradiation field.
As a result of the increased optical complexity introduced by the octupole insertion and matching section, the total transfer line length extends to 55 m, with reduced flexibility for total length adjustment compared to the baseline design.
Further details on the octupole-based flux homogenisation technique are provided in Refs.~\cite{NSRL-Octupoles, PhysRevSTAB2007-Japan}. The full optical layout is illustrated in Fig.~\ref{fig:TLwithOctupoles}.

\begin{figure}
    \centering
    \includegraphics[width=\linewidth]{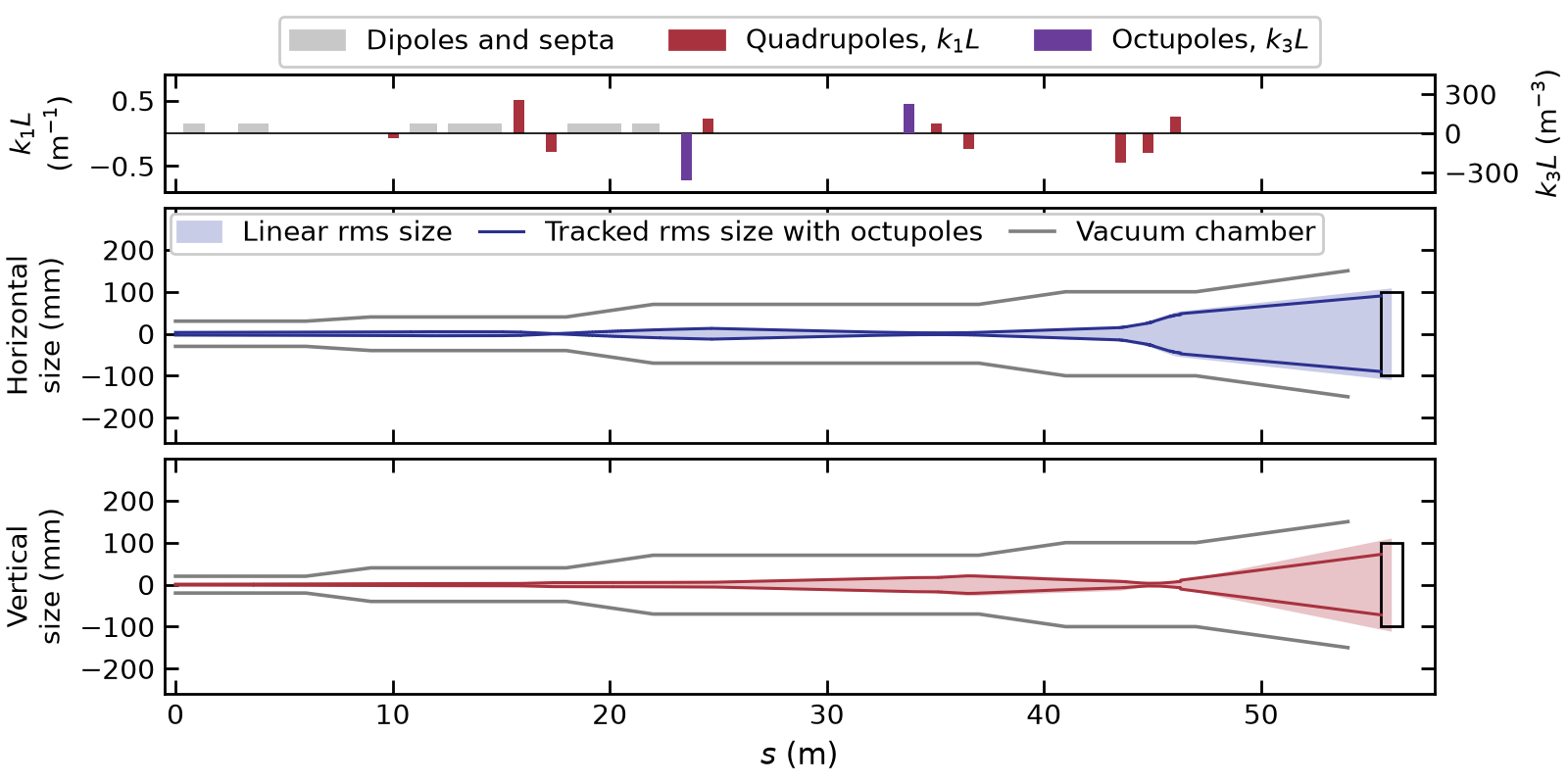}
    \caption{Beam transport of one RMS beam size with transverse homogenization using octupole pairs. The RMS beam spot is matched to 20cm in both planes. Colored shadows illustrate the linear transport without octupoles. The solid lines follow beam envelope with octupoles included. Magnetic elements follow the same color-coding as in Fig.\ref{fig:TL2HEARTS-Options}: grey — dipoles (bending magnets), red — quadrupoles (focusing/defocusing), purple — octupoles (transverse distribution homogenization).}
    \label{fig:OctBeamTransport}
\end{figure}

\begin{figure}
    \centering
    \includegraphics[width=0.6\linewidth]{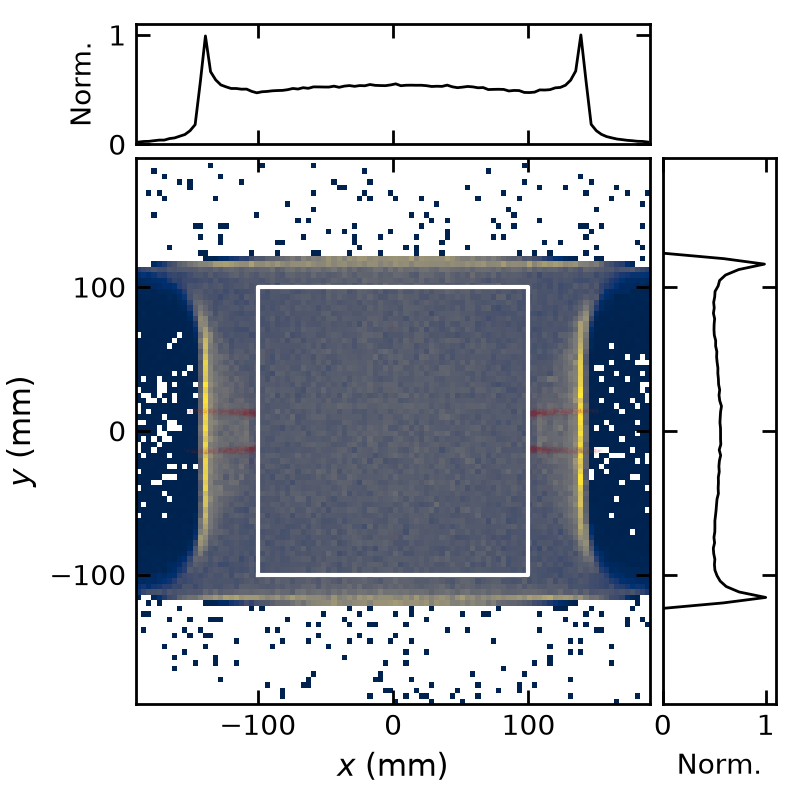}
    \caption{Exemplary beam spot shaped with octupoles at DUT to cover a 20cm x 20cm irradiation field (white square). The top and right panels illustrate the beam distribution projections. Red stripes represent particles lost along the transfer line.}
    \label{fig:DistAtIrradField20x20Oct}
\end{figure}

The beam transport of one RMS of a Gaussian particle distribution tracked along the transfer line is illustrated in Fig.~\ref{fig:OctBeamTransport}, with particle tracking performed using Xsuite~\cite{Iadarola:2023fuk}. The beam spot at the DUT is depicted in Fig.~\ref{fig:DistAtIrradField20x20Oct}. The expected beam transport efficiency for this configuration is 50\%, with beam losses predominantly localised at the DUT from particles outside the irradiation area. The average fluxes follow the values listed in Table~\ref{tab:fluxes_at_DUT}, increased by a factor of three due to the higher transport efficiency.

\subsection{Scanning with pencil beams}
\label{sec:TLscanning}

The two schemes described so far shape a beam that illuminates the whole irradiation field at every instant. A third approach delivers a small, focused spot --- a \emph{pencil beam} --- and sweeps it across the field, so that the required uniformity is obtained in the time-integrated fluence rather than in the instantaneous transverse profile. 
This is the painting or raster-scanning technique used at several therapy and irradiation facilities. It was developed at GSI for heavy-ion therapy~\cite{Haberer1993}, and a raster-scanning system of this kind is in use in Cave~A at the SIS18 synchrotron of GSI, where it serves, among other applications, for radiation-hardness testing of electronics with high-energy heavy ions~\cite{GSI-HEARTS-KT2023}. It has been evaluated here for the HEARTS@LEIR baseline lattice.
The scan patterns studied are taken from the scanning transmission electron microscopy literature~\cite{SpiralScans}, where the same problem arises --- delivering a uniform dose over an area with a scanned probe, while keeping the trajectory smooth enough for the deflection system to follow at speed --- and where the properties of the candidate spirals have already been characterised in detail.

The scheme is attractive because it does not discard the beam: instead of blowing the distribution up to the size of the field and scraping the excess, essentially all extracted particles are placed inside the irradiation area. It also requires no additional focusing elements. The final focus of the baseline line is used to produce the spot, which is then swept by a pair of wide-aperture sweeping correctors, one horizontal and one vertical, driven with the appropriate time dependence.

\subsubsection*{Scan pattern}

\begin{figure}
    \centering
    \includegraphics[width=0.52\linewidth]{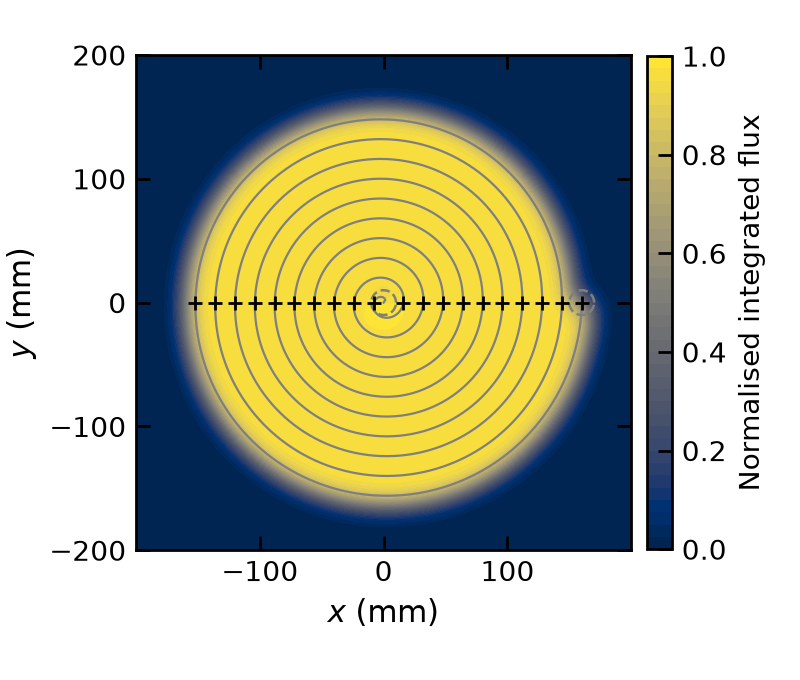}
    \caption{Constant-linear-velocity spiral trajectory of the pencil beam (grey line, with the sampled spot positions marked) superimposed on the resulting integrated flux. The spot is swept over a radius of approximately \SI{150}{\milli\meter} in order to cover the $20 \times 20$~cm$^2$ irradiation field.}
    \label{fig:SPBspiral}
\end{figure}

\begin{figure}
    \centering
    \includegraphics[width=0.52\linewidth]{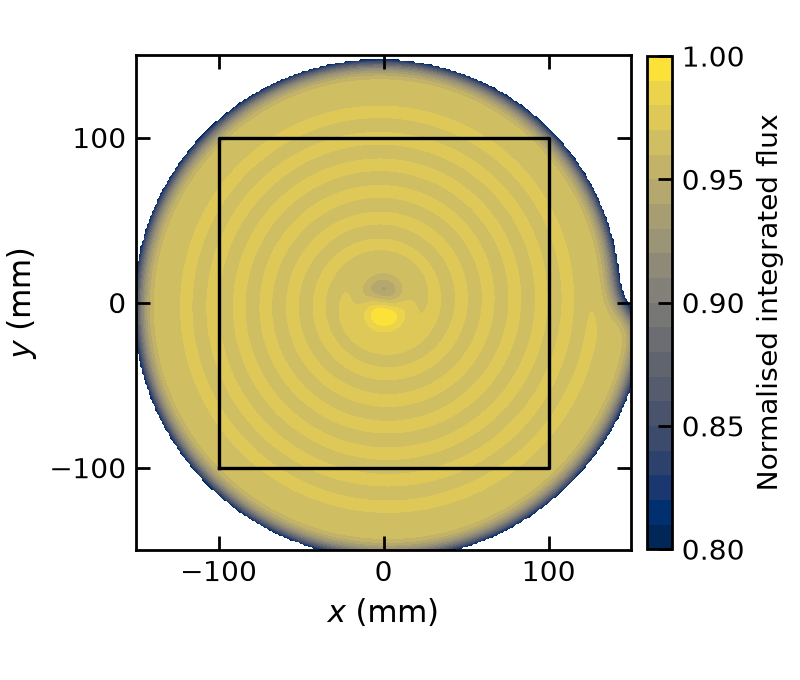}
    \caption{Normalised integrated flux delivered by the constant-linear-velocity spiral scan. The black square marks the $20 \times 20$~cm$^2$ irradiation field, within which the flux is uniform to better than 5\,\%.}
    \label{fig:SPBhomogeneity}
\end{figure}

The pattern chosen for the conceptual study is a spiral, which covers a circular field without the sharp turning points of a serpentine raster and therefore imposes a gentler $\mathrm{d}B/\mathrm{d}t$ on the sweeping correctors. The choice of spiral matters: an Archimedean spiral, swept at constant angular velocity, dwells longer near the centre and produces a markedly non-uniform integrated flux. A spiral traversed at constant linear velocity (CLV) distributes the dwell time uniformly over the field and yields an excellent result; the same conclusion is reached in Ref.~\cite{SpiralScans}, where Archimedean, Fermat and constant-linear-velocity spirals are compared for dose uniformity. Figure~\ref{fig:SPBspiral} shows the CLV trajectory together with the resulting integrated flux, and Fig.~\ref{fig:SPBhomogeneity} the corresponding uniformity over a $20 \times 20$~cm$^2$ field.

Tracking through the baseline lattice gives a transport efficiency of approximately 63\,\% for the largest specified field --- against 14.9\,\% for blow-up and scraping --- with a uniformity better than 5\,\% (max/mean in \SI{1}{\centi\meter\squared} bins), comfortably inside Specification~F. The exact efficiency depends on the number of spiral rotations per spill, which is a free parameter of the scheme.

\subsubsection*{Aperture requirement}

The scheme carries one unavoidable consequence for the hardware. Because the spot must reach every point of the field, the beam envelope downstream of the sweeping correctors is no longer the size of the beam but the size of the scan. Covering a $20 \times 20$~cm$^2$ field requires the trajectory to be swept over a radius of approximately \SI{150}{\milli\meter}, and the vacuum chamber must accommodate that envelope over the whole distance between the correctors and the DUT. In practice this means a large-diameter vacuum pipe over the last stretch of the line of order \SI{300}{\milli\meter} of free aperture, together with the correspondingly large vacuum window and pumping arrangement at the exit.

This requirement is common to any painting scheme and is not a defect of the particular pattern chosen; the octupole option needs a wide aperture for the same geometric reason, although a smaller one. It concerns the vacuum chamber only and leaves the magnet specifications unchanged, whereas the octupole option requires the large-aperture final-focus quadrupoles of Table~\ref{tab:TLMagnetSpecs}. The chamber diameter, its sectorisation, its supports and the associated cost are to be determined during the technical design phase, in coordination with TE-VSC and with the integration study of Chapter~\ref{sec:InfrastructureAndIntegration}. The corresponding cost is not included in the vacuum estimate of the present report, which assumes the baseline blow-up-and-scrape configuration.

\subsubsection*{Time structure of the delivered dose}

The second consequence is more fundamental. A scanned beam correlates the dose delivery with time: any given point of the DUT is irradiated only during the fraction of the scan in which the spot passes over it, at a local instantaneous flux far above the average. Depending on the number of rotations per spill, the ratio of instantaneous to average flux ranges from approximately unity to more than ten. For a large class of cumulative SEE measurements this is immaterial, since only the integrated fluence matters. For tests in which the time structure of the irradiation is itself a variable, such as rate-dependent effects, destructive latch-up and single-event burnout campaigns, or measurements synchronised with the device duty cycle, it may be unacceptable.

The specification governing instantaneous versus average flux (Requirement~D and Specification~H of Chapter~\ref{sec:BeamSpecs}) is therefore not yet sufficient to decide the question. The feasibility of the scheme and, more importantly, the user cases for which it is appropriate will be evaluated at a later stage of the project. It is retained here as an upgrade path rather than as part of the baseline: it needs no change to the lattice, so the decision can be deferred without prejudicing the design, provided the aperture question is settled in the TDR.

\subsection{Comparison of the options and magnet requirements}
\label{sec:TLcomparison}

Three schemes have been studied at conceptual level for delivering the specified irradiation field, all sharing the same achromatic arc: blow-up and scraping, octupole tail folding, and pencil-beam scanning. Their performance is compared in Table~\ref{tab:TLoptionsComparison}.

\begin{table}[htbp]
\centering
\caption{Comparison of the three flux-homogenisation schemes for a $20 \times 20$~cm$^2$ irradiation field. Efficiency is the fraction of extracted beam delivered inside the field; homogeneity is the ratio of maximum to mean fluence in \SI{1}{\centi\meter\squared} bins (Fig.~\ref{fig:TLHomogeneity}): from a tracking of $10^{8}$ ions for the baseline and from the smooth fluence distribution fitted to $10^{6}$ tracked ions for the octupole option; the histograms of $10^{6}$ tracked ions give 1.23 and 1.13, dominated by the statistical fluctuations of the simulation;  the last column is the ratio of instantaneous to average flux at a point of the device under test.}
\label{tab:TLoptionsComparison}
\small
\setlength{\tabcolsep}{4pt}
\begin{tabular}{@{}l>{\raggedright\arraybackslash}p{4.0cm}rrl@{}}
\toprule
\textbf{Scheme} & \textbf{Additional hardware} & \textbf{Effic.} & \textbf{Homog.} & \textbf{Inst./ave.} \\
 &  & \textbf{(\%)} & \textbf{(max/mean)} & \textbf{flux} \\
\midrule
Blow-up and scraping  & None (baseline)  & 14.9 & 1.09 & 1:1 \\
Octupole tail folding & 2 octupoles, additional final-focus quadrupoles, wider apertures & 50 & 1.07 & 1:1 \\
Pencil-beam scanning  & 1--2 pairs of wide sweeping correctors, large-aperture vacuum chamber & 63 & $<$1.05 & 1:1 to $>$10:1 \\
\bottomrule
\end{tabular}
\end{table}

The baseline design, blow-up and scraping, is adopted for its simplicity and expected reliability, at the cost of efficiency. The octupole version offers better transmission and allows beam losses to be conveniently localised at the DUT; however, it presents several technical challenges. First, the setup requires sophisticated optics matching, which limits operational flexibility, and a broader set of initial conditions will need to be considered to assess the impact on operability and commissioning. Second, broad apertures are required. Third, flexibility with respect to transfer line length is not guaranteed. At this stage of the design it cannot be confirmed that all technical and practical constraints have been fully accounted for, which represents a risk for the deployment of this option. The feasibility of the octupole version is not in question, but it represents a considerably higher technical challenge. The pencil-beam scheme is the most efficient of the three and needs no additional focusing elements, but carries the aperture and time-structure implications discussed in Section~\ref{sec:TLscanning}.

All three schemes assume Gaussian transverse distributions in the tracking presented here. The extracted beam is not Gaussian in the extraction plane, and the impact of the real distribution on efficiency and uniformity remains to be quantified for each option.

With these considerations in mind, the preliminary list of required magnets is given in Table~\ref{tab:TLMagnetSpecs}. A set of readily available magnet designs has been identified for consideration, as developing new designs could potentially introduce significant delays to the project timeline. Their design codes, as recorded in NORMA, the CERN database of normal-conducting magnets, are included in Table~\ref{tab:TLMagnetSpecs}.

\begin{table}[]
    \centering
    \caption{Preliminary list of transfer line magnets. Two transfer line options are considered: (a) the full list is required for the octupole folding homogenisation option, whereas for (b) only the dipoles, quadrupole types A, T and correctors are necessary for the baseline option.} 
    \begin{tabular}{l c c c}
    \hline
    Quadrupole type & A (Achromat) & T (Triplet) & F (Final focus)\\
    \hline
    NORMA design code & \multicolumn{2}{c}{\footnotesize Q130 (PXMQNCDNWP)} & \footnotesize PXMQNEQNWP$^{\dagger}$\\
    Quantity & 3 & 3 & 3\\
    Full aperture (mm) & 80 & 140 & 200\\
    Magnetic length (m) & 0.5 & 0.5 & 0.5 \\
    kL$_{\text{max}} ( $m$^{-1}$) & 0.6 & 0.7 & 0.6\\\\
    \hline
    \multicolumn{4}{c}{Octupole type MTE}\\
    \hline
    \multicolumn{2}{l}{NORMA design code} & \multicolumn{2}{l}{\footnotesize PXMONDAFWP$^{\dagger}$}\\
    \multicolumn{2}{l}{Quantity} & \multicolumn{2}{l}{2}\\
    \multicolumn{2}{l}{Full aperture (mm)} & \multicolumn{2}{l}{140}\\
    \multicolumn{2}{l}{Magnetic length (m)} & \multicolumn{2}{l}{0.5}\\
    \multicolumn{2}{l}{k$_3$L (10$^3$ m$^{-3}$)} & \multicolumn{2}{l}{1.235}\\
    \hline
    \multicolumn{4}{c}{Dipoles}\\
    \hline
    \multicolumn{2}{l}{Required BdL (Tm, max)} &\multicolumn{2}{l}{9.1}\\
    \multicolumn{2}{l}{Max./Min. rigidity (Tm)} & \multicolumn{2}{l}{5.8/2}\\
    \multicolumn{2}{l}{Full aperture (HxV, mm$^2$)} & \multicolumn{2}{l}{140x70}\\
    \multicolumn{2}{l}{Total deflection} & \multicolumn{2}{l}{90$^{\circ}$ ($4 \times 22.5^{\circ}$)$^{\S}$}\\
    \multicolumn{2}{l}{Bending radius (m)} & \multicolumn{2}{l}{4.24}\\
    \hline
    Tentative & \multicolumn{2}{c}{CNAO/MedAustron (MBH-C, MBH-D)}\\
    \hline
    NORMA design code & \multicolumn{2}{c}{\footnotesize PXMBHFCHWP, PXMBHFHCWP$^{\ddagger}$}\\
    Magnetic length (m) & \multicolumn{2}{c}{1.66 (iron 1.64, total 1.92)} \\
    Quantity & \multicolumn{2}{c}{4}\\
    Full aperture (HxV, mm$^2$) & \multicolumn{2}{c}{198x72}\\
    Max. B-field / BdL (T, Tm) & \multicolumn{2}{c}{1.5 / 2.495}\\
    B-field at 5.8/2 Tm (T) & \multicolumn{2}{c}{1.37/0.47}\\
    \hline
    Correctors & H & V\\
    \hline
    Quantity & 4 & 4\\
    Deflection (mrad) & \multicolumn{2}{c}{$\pm$10}\\
    \end{tabular}\\[3pt]
    {\footnotesize Design codes from NORMA, the CERN database of normal-conducting magnets. $^{\dagger}$\,One spare unit. $^{\ddagger}$\,MedAustron designs; no spare units. $^{\S}$\,Bending magnets only; the electrostatic and magnetic septa add about $11^{\circ}$. The optics of this chapter were computed with an arc of four dipoles of 1.25 and \SI{2.5}{\metre} with a bending radius of \SI{4.69}{\metre} and a total deflection of $91.6^{\circ}$; the geometry of the arc is to be adapted to the magnets above in the technical design phase.}
    \label{tab:TLMagnetSpecs}
\end{table}

\subsection{Additional hardware}

In addition to the magnetic elements for beam transport and optics, three BSGs are foreseen for beam commissioning and optimization. Power converters are required for all quadrupoles, while the bending dipoles can be connected in series. Safety elements will be installed in the transfer line to ensure reliable user access to the irradiation stage. To this end, one beam stopper, one beam dump, and one dual-purpose stopper-dump will be installed along the line. The vacuum chamber will extend until the end of the line.

\section{Preliminary cost estimate}
\label{sec:TLcost}

Only the baseline option is costed here. Table~\ref{tab:TransferLineCost} lists the beam-line equipment proper; the experimental area, its infrastructure, the access system, transport and handling, and the integration effort are covered in Chapters~\ref{sec:expArea} and~\ref{sec:InfrastructureAndIntegration}. Together these form work package~WP4, whose consolidated total of \SI{6702}{\kCHF} is reported in Table~\ref{tab:res_wp}. A rough estimate of the octupole version is expected to increase the material costs of the transfer line magnets and power converters by a factor two, but is subject to scrutiny given the limited information collected. 
\begin{table}[]
    \centering
    \caption{Preliminary cost estimate of the hardware required for the transfer line. Together with Table~\ref{tab:ExpAreaCost} it forms work package~WP4 of Table~\ref{tab:res_wp}; the experimental area, access system, handling and integration are costed in Chapter~\ref{sec:expArea}.}
    \small
    \setlength{\tabcolsep}{4pt}
    \begin{tabular}{lllr}
    \toprule
    Item & Composition & Group & Cost (kCHF) \\
    \midrule
    Transfer line magnets & 6 quadrupoles, 4 dipoles, 6 orbit correctors, & TE-MSC & 2\,112 \\
                          & 2 sweepers, $+$10\,\% ancillary systems & & \\
    Power converters      & 15 POLARIS~S, 2 BOREAL-IS-2P, spares & SY-EPC & 1\,970 \\
                          & and qualification & & \\
    Beam instrumentation  & 3 BSG & SY-BI & 225 \\
    Beam stoppers         & 1 beam stopper, 1 beam dump, & SY-STI & 210 \\
                          & 1 combined stopper-dump & & \\
    Vacuum systems        & Vacuum chamber, pumps and subsystems & TE-VSC & 500 \\
    \midrule
    \textbf{Total}        & & & \textbf{5\,017} \\
    \bottomrule
    \end{tabular}
    \label{tab:TransferLineCost}
\end{table}

\chapter{Experimental area and test station}
\label{sec:expArea}

The transfer line described in Chapter~\ref{sec:Beamline2HEARTS} terminates at the test station, which is the only part of the facility that users interact with directly. Its design therefore determines much of the perceived quality of the facility: how quickly a device can be mounted and aligned, how reliably the delivered fluence is known, and how much of a campaign is spent waiting for access. This chapter states the functional requirements of the test station, describes how the design departs from the existing HEARTS installation at the PS, and defines the layout and safety zoning of the area.

The experimental area houses the dedicated test station, which serves as the central facility through which users interact with the heavy-ion beam for electronics radiation-effects testing. The test station consolidates all functionality required for devices-under-test (DUTs) to be exposed to the beam in a remotely controlled fashion, minimising the need for personnel access to the irradiated zone during operations.
\section{Test station requirements}
The test station functionality is organised across three main categories: beam control, station operation, and monitoring.
Beam control encompasses the ability to switch the beam on and off — achieved through a combination of a dedicated beam stopper and bending magnet switching — as well as straightforward physical access to the irradiation area when required. Beam parameters including size, delivered flux and fluence, ion species, and energy shall be configurable as autonomously as possible, reducing operator burden and improving reproducibility across user runs.
Test station operation requires motorised DUT alignment and rotation, allowing precise and repeatable positioning of the device with respect to the beam axis without manual intervention in the irradiation zone.
Monitoring capabilities include live visualisation of beam parameters and the operational configuration via beam instrumentation (BI), as well as provision of timing signals (TTL) reflecting the time structure and spill sequence of the beam. These signals are essential for users synchronising their own data acquisition systems with the beam delivery.

A set of guidelines for the requirements can be found in~\cite{HEARTS_D51}.

\section{Design basis and evolution from HEARTS at the PS}
The test station design is directly informed by the operational experience gained at the existing HEARTS facility at the CERN PS, where a parasitic installation within the IRRAD controlled area has been used for heavy-ion irradiation campaigns. While functional, the current setup presents several limitations: space constraints restrict the flexibility of DUT mounting and ancillary equipment, no dedicated beam stopper is available, and beam instrumentation is partially located at some distance from the test station rather than in its immediate vicinity. Beam and instrument control are currently handled through the WRAP framework and a dedicated but separate graphical user interface (GUI).
The new facility addresses each of these shortcomings. The test station will be located in a supervised rather than a controlled area, reducing access complexity for users. Adequate space will be provisioned around the station, with proximity to the patch panel for signal routing. A dedicated beam stopper, sized on a worst-case scenario corresponding to the lightest ion species at the highest available energy, will provide instantaneous beam absorption and ensure a low level of error on the desired fluence. Beam instrumentation will be positioned to allow monitoring directly at the test station location. A unified GUI will integrate all beam control, station operation, and monitoring functionality into a single operator interface.
A particularly important capability targeted for the new facility is the ability to diagnose beam properties using the instrumentation without simultaneously irradiating the DUT. This is not achievable in the current HEARTS setup at the PS, where beam characterisation and device irradiation cannot be fully decoupled, and represents a meaningful improvement in operational flexibility and measurement quality.

The current baseline design includes a Multi-Wire Proportional Chamber (MWPC) for beam position and profile diagnostics, and two particle counter devices: a Secondary Emission Chamber (XSEC) for continuous flux and fluence monitoring, and a scintillator screen. 
Together, these instruments will enable live monitoring of particle flux and beam parameters, as well as beam position adjustment when required.

\section{Layout and safety zoning}
\label{sec:EALayout}
The experimental area layout follows a beamline configuration comprising, in sequence, a bending magnet, beam instrumentation, a beam stopper, and the DUT on the test station, with a beam dump located downstream. In the present design, the area is divided into two interlocked areas: the transfer-line interlocked area, covering the bending and steering elements, and the experimental interlocked area, covering the test station itself. Their protection follows the CERN concept of Important Safety Elements (EIS, \textit{\'El\'ements importants pour la s\'ecurit\'e})~\cite{Herty2025EIS}: EIS-access elements, such as doors and access devices, control entry into each interlocked area, while EIS-beam elements stop the beam upstream of it; for the experimental area, the dedicated beam stopper and the bending magnet, through its power converter, are foreseen as EIS-beam elements. Linked together in safety chains, these elements make beam delivery and access to the area mutually exclusive states enforced at the hardware level.
The beam dump follows a LEIR-based design with passive cooling, with a spare unit potentially available. Shielding will be installed on the wall behind the test station to contain secondary radiation. Where required by the beamline design, integration into the vacuum system may necessitate a large vacuum window in close proximity to the test station.
The detailed engineering of the test station, safety zoning, and associated infrastructure is being carried out in collaboration with the CERN groups concerned, and will be further refined during the technical design study phase.

\section{Preliminary cost estimate}
\label{sec:EAcost}

Table~\ref{tab:ExpAreaCost} collects the materials cost of the experimental area, its services and its access system. The technical justification of the infrastructure items is given in Chapter~\ref{sec:InfrastructureAndIntegration}; they are costed here so that the experimental area appears as a single line of the work breakdown. Together with the \SI{5017}{\kCHF} of beam-line equipment of Table~\ref{tab:TransferLineCost}, these \SI{1685}{\kCHF} make up the \SI{6702}{\kCHF} of work package~WP4 reported in Table~\ref{tab:res_wp}.

\begin{table}[htbp]
    \centering
    \caption{Preliminary cost estimate for the experimental area, test station and associated services. Beam-line equipment upstream of the area is costed separately in Table~\ref{tab:TransferLineCost}; together the two tables form work package~WP4 of Table~\ref{tab:res_wp}.}
    \label{tab:ExpAreaCost}
    \small
    \setlength{\tabcolsep}{4pt}
    \begin{tabular}{llr}
    \toprule
    Item & Group & Cost (kCHF) \\
    \midrule
    Secondary emission chamber (XSEC)        & SY-BI  &  70 \\
    Scintillator particle counter            & SY-BI  &  70 \\
    Multi-wire proportional chamber (MWPC)   & SY-BI  &  45 \\
    \midrule
    Experimental area infrastructure         & BE-EA  & 250 \\
    Control room                             & BE-EA  & 300 \\
    Gas system                               & BE-EA  & 270 \\
    Integration study (TDR and production)   & BE-EA  & 100 \\
    Configuration management                 & BE-EA  &  60 \\
    \midrule
    Access and interlock zone$^{*}$          & EN-AA  & 300 \\
    Material relocation, installation and shielding & EN-THE & 190 \\
    Additional radiation monitoring          & HSE-RP &  30 \\
    \midrule
    \textbf{Total}                           &        & \textbf{1\,685} \\
    \bottomrule
    \end{tabular}\\[2pt]
    \footnotesize{$^{*}$ Lower bound. A firm figure requires the risk analysis with HSE-RP and BE-DSO (Chapter~\ref{sec:InfrastructureAndIntegration}). No contingency is applied; BE-EA recommends 20\,\% on its own scope.}
\end{table}

\chapter{Infrastructure and integration}
\label{sec:InfrastructureAndIntegration}

The preceding chapters describe the beam-producing and beam-shaping systems. This chapter collects the services without which none of them can be installed or operated: vacuum, cooling and ventilation, electrical distribution, transport and handling, access and safety, and the 3D integration study that reconciles all of them within the available space. Each section states the technical requirement, the responsible group, and the associated cost, which is carried forward to Chapter~\ref{sec:ResourceEstimate}.
Unless stated otherwise, the materials costs quoted in this chapter fall in the construction phase; the effort required in the pre-TDR and TDR phases is given per group in Table~\ref{tab:res_personnel}.

\section{Vacuum systems}
The vacuum system for the HEARTS@LEIR facility requires a comprehensive redesign across both the Linac~3 and LEIR subsystems. 
Key deliverables include a new vacuum layout for the ion source and beam lines, new beam chambers for the bending magnets and spectrometer, updated gas injection systems for the ion sources, and new vacuum controls and supervisory control and data acquisition (SCADA) integration.
A particular challenge in the Linac~3 area concerns the system that injects the working gas into the ion source, which in its current form is obsolete — existing valves and controllers are no longer in production, and the system is limited to injecting a single gas species. A replacement system is under development and testing. The spectrometer chamber presents additional complexity, as it spans several magnets without independent alignment capability; its design must therefore be developed in conjunction with the magnet mechanical design, accounting for vacuum forces, sputtering, and magnetic permeability constraints.

LEIR operates at ultra-high vacuum conditions of approximately 10$^{-12}$~mbar, achieved through extensive use of Non-Evaporable Getter (NEG) coatings, strict material selection, mandatory vacuum firing of stainless steel components, and rigorous outgassing limits. Any intervention in LEIR requires a minimum bakeout period of three weeks in addition to the mechanical intervention time, and the number of vent/pump cycles per sector is limited due to NEG ageing. All equipment to be installed in LEIR must undergo rigorous material evaluation and quality control in coordination with TE-VSC prior to installation. The new extraction beam line to the experimental area will require a dedicated vacuum layout and sectorization, the design of which is pending the mechanical integration study.
These requirements are highlighted as a key risk area: any non-conforming material introduced into LEIR could compromise vacuum integrity and result in extended downtime. All groups contributing equipment to the LEIR insertion region are required to coordinate with TE-VSC from the earliest design stages.

The preliminary cost estimate for the vacuum systems totals \SI{718}{\kCHF} for the ion source and Linac~3 part of the project (WP2) and \SI{500}{\kCHF} for the LEIR and new transfer line section (WP4), giving \SI{1218}{\kCHF} in total, all of it in the construction phase. The associated TE-VSC personnel amounts to \SI{3.0}{\PY} of Graduate-like effort and \SI{1.1}{\PY} of staff effort; of this, \SI{0.15}{\PY} of staff effort falls in the pre-TDR and TDR phases and the remainder in construction.

\section{Cooling and ventilation}
The HEARTS@LEIR facility requires several upgrades to the existing cooling and ventilation infrastructure, coordinated by EN-CV.
The LEIR cooling distribution network must be extended to serve the new transfer line and the extraction septum, and a new cooling sub-station is needed for the ion sources. The transfer line and septum require approximately \SI{70}{\kilo\watt} of cooling capacity, supplied through an extension of the distribution network of approximately \SI{60}{\meter}, while each of the two ion sources, the existing one and the new one, requires \SI{150}{\kilo\watt} at a flow rate of approximately \SI{4.3}{\cubic\meter\per\hour}. 
A new dedicated cooling sub-station in the Linac~3 source area, separate from the LEIR cooling station discussed below, is required to serve both sources simultaneously, operating between \SI{15}{\celsius} and \SI{45}{\celsius} with a backup heat exchanger.
The LEIR cooling station itself requires an upgrade to accommodate the additional load introduced by HEARTS@LEIR operation. 
The total projected load with HEARTS@LEIR is approximately \SI{1150}{\kilo\watt} at a flow rate of \SI{205}{\cubic\meter\per\hour}. 

Two upgrade options are under consideration: 
\begin{itemize}
    \item Option 1 retains the existing cooling towers and replaces pumps and heat exchangers, achieving a demineralised-water supply temperature of approximately 28.5°C;
    \item Option 2 additionally replaces the cooling towers, achieving a lower supply temperature of approximately 27°C.
\end{itemize}  
A decision between these options will be made during the TDR phase. It is noted that the LEIR cooling towers are approaching end of life and will need consolidation, independent of the HEARTS@LEIR project.
No ventilation system currently exists in Building 150, and no local temperature or humidity control is in place. Ventilation user requirements for HEARTS@LEIR have not yet been defined; this represents an open item to be addressed during the TDR phase. The heating, ventilation and air-conditioning (HVAC) systems in Buildings 150 and 250 are reported to be in poor condition, and consolidation work may be required in parallel with HEARTS@LEIR infrastructure development.
Preliminary cost estimates for the cooling and ventilation infrastructure are approximately \SI{100}{\kCHF} for the LEIR cooling network extension (excluding power converters) and \SI{200}{\kCHF} for the source cooling sub-station. The LEIR cooling station upgrade is quoted by EN-CV at \SI{450}{\kCHF} under Option~1, rising to \SI{650}{\kCHF} if the cooling towers are also replaced (Option~2). The consolidated estimate of Chapter~\ref{sec:ResourceEstimate} carries the more conservative Option~2. Annual operational cost increases are estimated at approximately \SI{9}{\kCHF}/year for make-up water and water treatment and \SI{6}{\kCHF}/year for electricity, with no additional maintenance costs anticipated.

\section{Electrical infrastructure}

Because of the start of LS3 activities, EN-EL was not in a position to provide a cost estimate for the HEARTS@LEIR electrical infrastructure within the timescale of this conceptual design study. The electrical scope, and in particular the signal and power cabling for the new beam line together with the cooling of the associated power converters, is therefore omitted from the costing exercise presented in Chapter~\ref{sec:ResourceEstimate}. Establishing it is a priority for the early part of the technical design phase.

The electrical infrastructure requirements for the experimental area are included within the overall infrastructure estimate for the facility. Contributions to the experimental area budget include \SI{30}{\kCHF} for electrical installations and \SI{30}{\kCHF} for control room electrical works, with an additional \SI{5}{\kCHF} for IT infrastructure. Detailed electrical service definitions will be developed in collaboration with the EN-EL design office during the production phase. The quantification of the electrical loads associated with the power converters of the new beam line will be addressed at the beginning of the technical design phase, as noted in Section~\ref{sec:ResourceExclusions}.

\section{Transport and handling}

Transport and handling requirements, coordinated by EN-THE, are significant and span two main areas: the source region and the transfer line, including its beam dump.
In the source area, the low crane height in the relevant zone (approximately 3.2 m under the hook) combined with the height of magnets fitted with beam shutters (approximately 1.5–2 m) means that the new beam line, once installed, will obstruct the existing handling corridor used for equipment removal. It is therefore a requirement that a section of the beam line approximately 2 m wide be designed to be removable to maintain access. In addition, part of the new source installation falls outside the crane coverage area, necessitating a dedicated non-standard handling solution and bespoke tooling, estimated at approximately \SI{100}{\kCHF}. General handling activities in the source area, including installation of magnets and racks, are estimated at approximately \SI{12}{\kCHF} (3 persons over 2 weeks).\\

In the transfer line and beam dump area, a significant quantity of stored material must first be cleared and relocated elsewhere at CERN before installation can begin, estimated at approximately \SI{80}{\kCHF} (4 persons over 2 months). Installation of the new beam line, shielding walls, and roof is estimated at a further \SI{80}{\kCHF} under the same resource assumption. Removal and reinstallation of shielding around LEIR is estimated at approximately \SI{30}{\kCHF} (4 persons over 2 weeks). Additional handling associated with ancillary equipment such as ventilators, cables, and beam loss monitors has not yet been quantified, pending further definition of scope.
The total EN-THE resource requirement is estimated at approximately \SI{200}{\kCHF} for handling manpower and \SI{100}{\kCHF} for the dedicated handling job and associated tooling, giving a total of approximately \SI{300}{\kCHF}.

\section{Safety and access structure}
\label{sec:SafetyAccess}

In the present design, the facility downstream of the extraction is divided into two interlocked areas, the transfer-line and the experimental interlocked areas, protected by Important Safety Elements (EIS)~\cite{Herty2025EIS}, as described in Section~\ref{sec:EALayout}.
EIS-beam and EIS-access elements, linked in safety chains, enforce the mutual exclusivity of beam delivery and personnel access. The access and interlock system itself is costed with the experimental area (Table~\ref{tab:ExpAreaCost}) at \SI{300}{\kCHF}, a lower bound pending the risk analysis with HSE-RP and BE-DSO. The access control, fire detection and safety infrastructure of the control rooms are included in the experimental-area infrastructure budget, at \SI{5}{\kCHF} and \SI{2.5}{\kCHF} respectively.
Detailed safety zoning definitions will be developed in collaboration with HSE-RP-AS and EN-AA-AC.

\section{Integration}

\begin{figure}[htbp]
    \centering
    \includegraphics[width=0.85\linewidth]{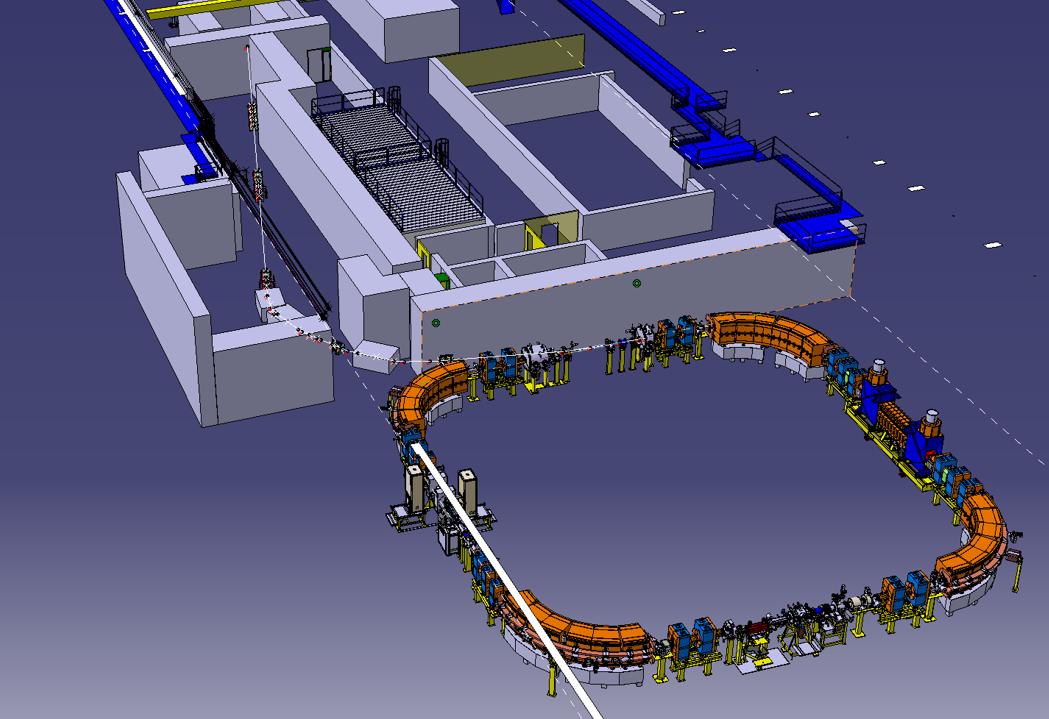}
    \caption{View of LEIR and of the HEARTS@LEIR extraction line in the CERN 3D integration model, with the shielding walls and the surrounding infrastructure; preliminary~\cite{SLAWGIntro2026}.}
    \label{fig:Integration3D}
\end{figure}

The integration of the HEARTS@LEIR facility is led by BE-EA and EN-ACE. The integration programme follows a staged approach aligned with the overall project timeline. A first 3D view of LEIR and of the new extraction line in its present surroundings, from the CERN 3D integration model, is shown in Fig.~\ref{fig:Integration3D}~\cite{SLAWGIntro2026}; it is the starting point of the integration study described below.

During the technical design phase, a conceptual 3D integration study at level of detail LOD200 (on the scale used for the CERN 3D integration models) will be produced, covering simplified models of all major components with maintenance and transport volumes defined. 
A 3D as-built scan of Building 150 (room R-011) and neighbouring areas will also be completed to provide an accurate baseline for subsequent design work. The integration study at TDR level is estimated at 800 hours of effort (approx.~\SI{50}{\kCHF}), covering the extraction line (including optics import from MAD-X to CATIA), shielding studies in coordination with the Radiation Protection group, irradiation zone mechanical support and services layout, and control room integration.

During the production and installation phases, the integration model will be developed to LOD350 and subsequently LOD500, with full 2D layout drawings, installation drawings, and a post-installation as-built scan. Infrastructure procurement specifications will be released for the control rooms, experimental bench, patch panels, shielding, and gas system. A post-TDR integration effort of equivalent scope (approx.~800 hours, approx.~\SI{50}{\kCHF}) is foreseen for updates to all sub-systems following design evolution during procurement.

Key dependencies for the integration programme include 3D equipment models from ABT (septa), MSC, VSC, and BI (progressing from LOD250 to LOD500), shielding simulations from the Radiation Protection group, and services definitions from the EL and CV design offices. The integration responsibilities of EN-ACE and BE-EA are split at the start of the experimental area: EN-ACE covers everything upstream of it, including the extraction line, and BE-EA takes over from that point.
It is noted that BE-EA's availability during LS3 will be constrained by concurrent commitments to other approved projects, and that formal resource commitments from the group are pending. This dependency is carried as a schedule risk in Chapter~\ref{sec:RiskAssessment}. The total infrastructure and integration budget from BE-EA, including the two integration studies above, is estimated at approximately \SI{980}{\kCHF} over the project lifetime (Table~\ref{tab:ExpAreaCost}), excluding the contingency of 20\,\% recommended by the group and not applied in Chapter~\ref{sec:ResourceEstimate}.

\section{Summary of costs}
\label{sec:InfraCost}

The infrastructure items are booked with the systems they serve, so that the cost table of each system chapter is complete: the vacuum system and the handling of the source area in Table~\ref{tab:IonSourceCost}, the vacuum system of the transfer line in Table~\ref{tab:TransferLineCost}, and the access system, the handling and shielding of the transfer-line area and the integration studies in Table~\ref{tab:ExpAreaCost}. Cooling and ventilation, which serve the whole facility, are booked in work package~WP1 (Table~\ref{tab:res_wp}). The electrical infrastructure is not yet costed (Section~\ref{sec:ResourceExclusions}).

\chapter{Radiation safety and radiation protection}
\label{sec:RadiationProtection}
\chapterauthors{M.~Widorski}

Radiation protection considerations place limits on how the LEIR facility can be operated. A maximum allowable beam intensity is defined for circulation in LEIR and for transport to the experimental area; this limit must be demonstrated to meet the intensity needs of the users. It also sets the requirements for shielding of the new transfer line and the experimental area, thereby influencing certain integration costs. This chapter covers both topics and specifies the beam-loss assumptions underlying the shielding evaluations.

\section{Shielding of LEIR and of the new areas}

The LEIR accelerator has lateral shielding provided by \SI{160}{cm}-thick walls, but it has no overhead shielding. As a result, the allowed beam power is restricted by limits on stray radiation in nearby accessible areas caused by sky-shine, as well as by the maximum radiation levels within LEIR due to limited physical measures preventing access.

The new transfer line leading to the experimental area, together with the experimental area itself, needs lateral and overhead shielding estimated at \SIrange{80}{160}{cm} of concrete. The exact shielding thickness will be defined during the technical design phase. The overhead shielding is necessary to limit intrusion risk and to reduce stray radiation levels in the adjacent building, which includes supervised and non-designated areas.

\section{Intensity limits}

LEIR typically runs with $^{208}$Pb$^{54+}$ ions at \SI{72}{MeV/nucleon}, delivering on average up to \SI{5.14E8}{} ions per second, assuming an upper limit of \SI{1E11}{} charges per 3.6-second cycle, which is approximately the space-charge limit for $^{208}$Pb$^{54+}$ in LEIR \cite{hans2026}. These operating conditions have been shown to be suitable from a radiation-protection perspective for LEIR and the adjacent areas.

Ion beams with energies from a few tens up to several hundred MeV per nucleon, potentially accelerated in LEIR, exhibit a supra-linear increase of ambient dose equivalent with energy. At identical intensity, 245 MeV/nucleon oxygen ions would generate nearly ten times higher radiation levels than 100 MeV/nucleon oxygen ions. In addition, for the same energies, the radiation hazard is expected to increase for lighter ions with respect to the Pb reference case.

\noindent The following operational parameters are assumed:
\begin{itemize}
    \item a cycle time of 5 basic periods of 1.2 seconds, i.e., 6 seconds
    \item a slow extraction over 500 milliseconds towards the experimental area
    \item a minimum number of \SI{1E9}{} ions per cycle at the LEIR flat top, as required by the DUT
\end{itemize}

Relative to the $^{208}$Pb$^{54+}$ reference case, the beam intensity limit is \SI{3.1E9}{} ions per 6-second cycle. This value is applicable to within about $\pm~30\%$ for all ions listed in Table~\ref{rp_charge_limits} at the specified energies, since their radiation production is similar to that of $^{208}$Pb$^{54+}$. Table~\ref{rp_charge_limits} specifies the radiation protection limit in charges per cycle, adjusted by an ion-energy-dependent correction factor relative to $^{208}$Pb$^{54+}$, and also includes the minimum number of charges needed for beam diagnostics as well as the charge required by the DUT (> \SI{1E9}{} ions at flat top in LEIR). It should be noted that the provided limit may be above the technical limitations of LEIR, mainly imposed by the space-charge limit.

The results indicate that the permissible beam intensities in LEIR are consistent with both the technical constraints for LEIR operation and the experimental requirements for the specified ion species and energies.

The technical approach for reliably restricting the energy and beam intensity injected into LEIR needs to be investigated and defined during the technical design phase.

\section{Beam-loss assumptions}

For loss scenarios in LEIR, it is generally assumed that beam losses are continuous at top energy with nominal intensity. Table~\ref{rp_loss_map} gives a more detailed, preliminary beam loss distribution from LEIR to the DUT; this can be refined further during the technical design phase and then used to adapt the shielding requirements for the transfer line and the experimental area.

\begin{table}[hbt!]
\centering
\setlength{\tabcolsep}{3pt}
\renewcommand{\arraystretch}{1.2}
\caption{Charge requirements and limits for different ion species in LEIR. The ion-energy correction factor accounts for the change in radiation production relative to $^{208}$Pb$^{54+}$.}
\begin{tabular}{
>{\raggedright\arraybackslash}p{1.6cm}
S[table-format=3.0, table-column-width=2.2cm]
S[table-format=1.1e2, table-column-width=2.3cm]
S[table-format=1.1e2, table-column-width=2.3cm]
S[table-format=1.1, table-column-width=1.3cm]
S[table-format=1.1e2, table-column-width=2.6cm]
}
\toprule
 &
{\parbox{2.2cm}{\centering Energy\\(MeV/nucleon)}} &
{\parbox{2.3cm}{\centering Required charges per cycle by LEIR}} &
{\parbox{2.3cm}{\centering Required charges per cycle by DUT}} &
{\parbox{1.3cm}{\centering Corr.\\factor}} &
{\parbox{2.6cm}{\centering Max. charges per cycle at flat top in LEIR (RP limit)}} \\
\midrule
$^{16}$O$^{8+}$    & 100 & 1.0e10 & 0.8e10 & 0.8 & 3.1e10 \\
$^{40}$Ar$^{16+}$  & 95  & 1.0e10 & 1.6e10 & 1.0 & 5.0e10 \\
$^{86}$Kr$^{29+}$  & 100  & 1.0e10 & 2.9e10 & 1.5 & 6.1E10 \\
$^{129}$Xe$^{40+}$ & 97  & 1.0e10 & 4.0e10 & 1.3 & 9.2e10 \\
$^{208}$Pb$^{54+}$ & 72  & 1.0e10 & 5.4e10 & 1.0 & 1.7e11 \\
\bottomrule
\end{tabular}\\[2pt]
\label{rp_charge_limits}
\end{table}

\begin{table}[hbt!]
\centering
\setlength{\tabcolsep}{3pt}
\renewcommand{\arraystretch}{1.2}
\caption{Beam intensity and corresponding losses at different locations. Reference is the beam intensity in LEIR at flat top. With the baseline blow-up-and-scrape scheme (Chapter~\ref{sec:Beamline2HEARTS}), most of the beam reaching the end of the line is stopped on the collimators and masks in front of the DUT; the total loss at the end of the line is unchanged.}
\begin{tabular}{
>{\raggedright\arraybackslash}p{4.4cm}
S[table-format=3.0, table-column-width=2.8cm]
S[table-format=2.0, table-column-width=2.8cm]
}
\toprule
{Location} &
{\parbox{2.2cm}{\centering Beam intensity (\%)}} &
{\parbox{2.8cm}{\centering Lost intensity at location (\%)}} \\
\midrule

Flat top in LEIR           & 100        & 0 \\
(After) extraction septum  & {70--95}   & {5--30} \\
(After) transfer line      & {50--67}   & {20--28} \\
On DUT                     & {50--67}   & {50--67} \\

\bottomrule
\end{tabular}
\label{rp_loss_map}\\[2pt]
\end{table}

The additional radiation monitoring required by the new areas, \SI{30}{\kCHF}, is costed with the experimental area (Table~\ref{tab:ExpAreaCost}); the shielding is included in the handling and installation items of the same table.

\FloatBarrier

\chapter{Resource estimate}
\label{sec:ResourceEstimate}

This chapter consolidates the resource and cost estimate of the HEARTS@LEIR upgrade. It gathers the item-level figures introduced in Chapters~\ref{sec:IonSources}--\ref{sec:RadiationProtection} into a single, internally consistent set of summary tables, and presents them by work package, by responsible CERN group, and as a function of time.

\section{Basis of estimate}
\label{sec:ResourceBasis}

The estimate is built bottom-up from an item list compiled with the contributing equipment and service groups between May and June 2026. Three resource categories are distinguished throughout:

\begin{itemize}
    \item \textbf{Materials} --- equipment, components, services and industrial support to be procured. Quoted in kCHF at 2026 price levels, excluding VAT and escalation.
    \item \textbf{Graduate-like personnel} --- fixed-term personnel to be recruited specifically for the project, costed at the group-dependent rates listed in Table~\ref{tab:res_rates}, which differ because each group requested a different experience level for its positions. Because these positions must be funded by the project, their cost is added to the materials cost to form the total funded envelope.
    \item \textbf{Staff} --- CERN staff effort, reported in person-years only. It is treated as an in-kind institutional contribution and is deliberately \emph{not} converted into a monetary figure.
\end{itemize}

Effort is quoted throughout in person-years (\si{\PY}), where one person-year is one full-time equivalent (FTE) working for one year. The two units, PY and FTE, are identical; \si{\PY} is used consistently in this report to avoid confusion with headcount, which is not tracked here.

\noindent The scope is organised in four work packages (WP), consistent with the breakdown used throughout this report:

\begin{itemize}[nosep]
    \item \textbf{WP1} --- Project coordination, knowledge transfer, and transversal technical infrastructure;
    \item \textbf{WP2} --- Second ion source, Linac~3 and beam preparation for injection into LEIR (Chapter~\ref{sec:IonSources});
    \item \textbf{WP3} --- Slow resonant extraction from LEIR (Chapter~\ref{sec:SRE});
    \item \textbf{WP4} --- New transfer line, experimental area and test station (Chapters~\ref{sec:Beamline2HEARTS}--\ref{sec:InfrastructureAndIntegration}).
\end{itemize}

For clarity's sake, this report keeps this original work-package structure of the CDR study, in which the estimate was built. In the HEARTS++ project, the technical design phase has since been restructured into work packages HEARTS++ WP5--WP7; Section~\ref{sec:ResourceRestructuring} describes the new structure and regroups the present estimate accordingly.
\par\medskip

Following common practice for infrastructure projects at this stage of maturity, the figures presented here carry a CDR-level uncertainty of the order of $\pm$30--50\,\%. They are intended to establish the order of magnitude of the investment and the shape of its time profile, not to serve as a procurement baseline. A firm cost book will be produced during the Technical Design Report (TDR) phase.

\begin{table}[htbp]
\centering
\caption{Unit cost assumptions used to convert Graduate-like person-years into funded cost.}
\label{tab:res_rates}
\begin{tabular}{lc}
\toprule
Contributing group & Cost per person-year (kCHF) \\
\midrule
ABT, BI, CEM, KT, OP, RF, RP & 110 \\
ABP, VSC & 92 \\
MSC & 90 \\
STI & 80 \\
\bottomrule
\end{tabular}
\end{table}

\section{Summary by work package}
\label{sec:ResourceByWP}

Table~\ref{tab:res_wp} summarises the estimate by work package. The total funded envelope, materials plus Graduate-like personnel, amounts to \SI{18.3}{\MCHF}, of which \SI{13.9}{\MCHF} is materials and \SI{4.4}{\MCHF} is fixed-term personnel. A further \SI{32.5}{\PY} of CERN staff effort is required as an in-kind contribution.

The distribution across work packages reflects the physical scope of the upgrade: WP4 dominates the materials cost, since the transfer line, its power converters and the experimental area are entirely new construction, whereas WP3 reuses the existing LEIR ring and adds only the extraction insertion. Conversely, WP2 and WP3 dominate the personnel demand, because the second source and the slow-extraction system require the most extensive design, beam-dynamics and commissioning effort.

\begin{table}[htbp]
\centering
\caption{Resource estimate by work package. Personnel assigned to more than one work package are distributed equally among them. Staff effort is foreseen as an in-kind contribution and is not costed.}
\label{tab:res_wp}
\small
\setlength{\tabcolsep}{4pt}
\begin{tabular}{llrrrrr}
\toprule
\multirow{2}{*}{\textbf{WP}} & \multirow{2}{*}{\textbf{Scope}}
 & \textbf{Materials} & \multicolumn{2}{c}{\textbf{Graduate-like}} & \textbf{Funded total} & \textbf{Staff} \\
\cmidrule(lr){4-5}
 & & \textbf{(kCHF)} & \textbf{(PY)} & \textbf{(kCHF)} & \textbf{(kCHF)} & \textbf{(PY)} \\
\midrule
WP1 & Coordination, transversal    &   650 &  7.0 &   770 &  1\,420 &  4.6 \\
WP2 & Second source, Linac~3       & 4\,045 & 16.1 & 1\,530 &  5\,575 & 14.4 \\
WP3 & LEIR slow extraction         & 2\,495 & 12.6 & 1\,352 &  3\,847 &  7.6 \\
WP4 & Transfer line, exp.\ area    & 6\,702 &  7.6 &   745 &  7\,447 &  6.0 \\
\midrule
\multicolumn{2}{l}{\textbf{Total}} & \textbf{13\,892} & \textbf{43.2} & \textbf{4\,398} & \textbf{18\,290} & \textbf{32.5} \\
\bottomrule
\end{tabular}\\[2pt]
\footnotesize{Columns may not add exactly to the totals because of rounding.}
\end{table}

\section{Materials by responsible group}
\label{sec:ResourceByGroup}

Table~\ref{tab:res_mat_group} breaks the materials cost down by the CERN group holding equipment ownership. Power converters (EPC) and magnets (MSC) together account for almost half of the materials cost, a direct consequence of the number of new magnetic elements in the transfer line and of the requirement for pulse-to-pulse modulated, spare-backed converters. The extraction hardware (ABT) and the vacuum systems (VSC) form the next tier.

The materials cost of the ECR source microwave generator is taken as \SI{910}{\kCHF}, the figure quoted by BE-ABP at the HEARTS@LEIR workshop of 11~May 2026~\cite{ABPinput2026} and subsequently presented to the IEFC in June~2026~\cite{IEFC2026}. An intermediate revision of the resource file carried \SI{1000}{\kCHF} for the generator together with a further \SI{500}{\kCHF} for tooling and maintenance equipment; this was a double count of a single line item and has been removed. Responsibility for the item is expected to transfer from BE-ABP to SY-RF, which is how it is booked here.

\begin{table}[htbp]
\centering
\caption{Materials cost by responsible group. Percentages refer to the total materials cost. Signal and power cabling and the cooling of the associated power converters are not included (Section~\ref{sec:ResourceExclusions}).}
\label{tab:res_mat_group}
\small
\setlength{\tabcolsep}{4pt}
\begin{tabular}{llrr}
\toprule
\textbf{Group} & \textbf{Main deliverable} & \textbf{Materials (kCHF)} & \textbf{Share (\%)} \\
\midrule
EPC & Power converters and spares            & 3\,513 & 25 \\
MSC & Magnets (source, Linac~3, LEIR, line)  & 3\,117 & 22 \\
ABT & Extraction septa and ancillaries       & 1\,400 & 10 \\
VSC & Vacuum systems and controls            & 1\,218 &  9 \\
EA  & Experimental area and integration      &    980 &  7 \\
BI  & Beam instrumentation                   &    943 &  7 \\
RF  & Source microwave generator             &    910 &  7 \\
CV  & Cooling and ventilation                &    650 &  5 \\
ABP & Ion source hardware                    &    319 &  2 \\
THE & Transport, handling and tooling        &    302 &  2 \\
AA  & Access and interlock zone              &    300 &  2 \\
STI & Beam stoppers and dump                 &    210 &  2 \\
RP  & Radiation monitoring                   &     30 &  0 \\
CEM & Facility controls                      & \multicolumn{2}{c}{not yet quantified} \\
\midrule
\multicolumn{2}{l}{\textbf{Total}} & \textbf{13\,892} & \textbf{100} \\
\bottomrule
\end{tabular}
\end{table}

\section{Personnel by group}
\label{sec:ResourcePersonnel}

Table~\ref{tab:res_personnel} reports the personnel demand by group, separating the full project from the technical design phase alone. The distinction matters for the decision at hand: the TDR phase requires \SI{19.2}{\PY} of Graduate-like effort and \SI{9.9}{\PY} of staff effort, i.e.\ 44\,\% and 30\,\% respectively of the full-project demand, concentrated in the design-intensive groups ABP, ABT, MSC and STI.

Staff effort is requested to be predominantly supervisory during the TDR phase, in compliance with existing LS3 commitments; the corresponding design and simulation work is carried by Graduate-like personnel recruited for the project.

\begin{table}[htbp]
\centering
\caption{Personnel estimate by group, in person-years. The TDR columns give the subset of the effort falling in the technical design phase.}
\label{tab:res_personnel}
\begin{tabular}{lrrrr}
\toprule
\multirow{2}{*}{\textbf{Group}} & \multicolumn{2}{c}{\textbf{Graduate-like (PY)}} & \multicolumn{2}{c}{\textbf{Staff (PY)}} \\
\cmidrule(lr){2-3}\cmidrule(lr){4-5}
 & \textbf{Total} & \textbf{TDR} & \textbf{Total} & \textbf{TDR} \\
\midrule
ABP & 10.0 & 7.0  & 10.6 & 4.65 \\
ABT &  9.0 & 3.0  &  4.9 & 0.90 \\
BI  &  6.0 & 1.0  &  0.8 & 0.20 \\
CEM &  5.0 & 2.0  &  1.0 & 0.40 \\
MSC &  3.0 & 2.0  &  8.0 & 2.00 \\
VSC &  3.0 & ---  &  1.1 & 0.15 \\
OP  &  2.0 & ---  &  --- & ---  \\
STI &  2.0 & 2.0  &  2.6 & 0.80 \\
KT  &  2.0 & 2.0  &  0.2 & 0.20 \\
RF  &  1.0 & ---  &  0.5 & 0.10 \\
RP  &  0.2 & 0.2  &  0.1 & 0.10 \\
EA  &  --- & ---  &  2.7 & 0.40 \\
\midrule
\textbf{Total} & \textbf{43.2} & \textbf{19.2} & \textbf{32.5} & \textbf{9.9} \\
\bottomrule
\end{tabular}
\end{table}

\section{Time profile}
\label{sec:ResourceProfile}

The estimate is phased, as established in the CDR study, in four stages: technical design, procurement, fabrication and testing (four years), installation (one year) and beam commissioning (one year). The timeline is tentative and subject to the funding of the HEARTS++ proposal, which foresees 54~months from its start to the delivery of a procurement-ready TDR: a pre-TDR phase, while the CERN groups remain committed to LS3, followed by the full TDR in the last two years of the project, tentatively 2030--2031. The procurement-ready TDR opens a construction period of approximately four years, driven by the procurement lead times of selected accelerator components: procurement, fabrication and testing in 2032--2035, installation in 2034--2035 and beam commissioning in 2035, with user operations starting in 2036. Some construction items, notably those of the second ion source (WP2), could be procured ahead of the construction period, subject to the approval of the Ion Complex Upgrade project.

Table~\ref{tab:res_profile} gives the corresponding annual profile. The resources estimated in the CDR study for the technical design (Section~\ref{sec:ResourceTDR}) are now distributed over the pre-TDR and TDR phases of HEARTS++; their annual distribution will be fixed at the start of the project, and they are therefore given as a single column. The construction stages are mapped onto the calendar: the four procurement years fall in 2032--2035, the installation year is shared equally between 2034 and 2035, and the commissioning year falls in 2035. The totals are unchanged; only their distribution over the calendar is affected. %

Two features of the profile are worth emphasising. 
First, the technical design is inexpensive in materials (\SI{176}{\kCHF}, essentially prototyping and radiation-monitoring items) but carries a large personnel effort, \SI{19.2}{\PY} of Graduate-like personnel and \SI{9.9}{\PY} of staff. 
Second, the materials spending peaks in the first year of procurement, well after the end of LS3 for the injectors, which is a deliberate consequence of the schedule constraints imposed by LS3 and the HL-LHC upgrade; the installation of the LEIR modifications, of the experimental line and of the facility follows in a single campaign in 2034--2035, before beam commissioning in 2035. Because procurement, installation and commissioning overlap in 2034--2035, the staff effort is concentrated in those two years, which will have to be confirmed with the groups during the TDR.

\begin{table}[htbp]
\centering
\caption{Annual resource profile, following the tentative schedule. Materials are given in MCHF, personnel in person-years, and the funded cost of Graduate-like personnel in kCHF. Materials figures are rounded to \SI{10}{\kCHF}.}
\label{tab:res_profile}
\footnotesize
\setlength{\tabcolsep}{4pt}
\begin{tabular}{lc|cccc|c}
\toprule
 & \textbf{Technical design} & \multicolumn{4}{c|}{\textbf{Construction}} & \\
 & pre-TDR and TDR$^{a}$ & 2032 & 2033 & 2034 & 2035 & \textbf{Total} \\
\midrule
Materials (MCHF)       & 0.18   & 4.13 & 2.92 & 3.64 & 3.02 & \textbf{13.89} \\
Graduate-like (PY)     & 19.2   & 7.25 & 4.75 & 5.88 & 6.12 & \textbf{43.2} \\
Graduate-like (kCHF)   & 1\,886 & 737  & 509  & 608  & 658  & \textbf{4\,398} \\
Staff (PY)             & 9.9    & 3.80 & 3.80 & 7.32 & 7.68 & \textbf{32.5} \\
\bottomrule
\end{tabular}\\[2pt]
\parbox{0.9\linewidth}{\footnotesize $^{a}$~Distributed over the pre-TDR and TDR phases of HEARTS++ (Section~\ref{sec:ResourceRestructuring}).}
\end{table}

\section{Technical design phase}
\label{sec:ResourceTDR}

Since the immediate decision concerns the advancement of the study from CDR to TDR, the resources required for that phase alone are isolated in Table~\ref{tab:res_tdr}. The TDR phase requires a funded envelope of approximately \SI{2.1}{\MCHF}, dominated by fixed-term personnel, together with \SI{9.9}{\PY} of CERN staff effort spread over eleven groups; the CDR study had foreseen two years for this phase. In the present schedule, this work would be carried out within the HEARTS++ project, spread over its pre-TDR and TDR phases (Section~\ref{sec:ResourceRestructuring}).

\begin{table}[htbp]
\centering
\caption{Resources required for the technical design phase.}
\label{tab:res_tdr}
\begin{tabular}{lrr}
\toprule
\textbf{Category} & \textbf{Effort (PY)} & \textbf{Funded cost (kCHF)} \\
\midrule
Materials and services  & ---  & 176 \\
Graduate-like personnel     & 19.2 & 1\,886 \\
\midrule
\textbf{Funded total}   &      & \textbf{2\,062} \\
\midrule
CERN staff (in kind)    & 9.9  & --- \\
\bottomrule
\end{tabular}
\end{table}

\section{Operational cost}
\label{sec:ResourceOperations}

The figures above cover construction only. The operational cost of the facility, derived from the operational model described in Chapter~\ref{sec:Operations}, is summarised in Table~\ref{tab:res_opex}. It is dominated by the permanent facility team and by electricity. A complete operational and business model, including CCC coverage, equipment-group on-call support and consumables, will be developed during the TDR phase in collaboration with OP, STI, CEM and KT.

\begin{table}[htbp]
\centering
\caption{Preliminary operational cost, for a nominal 15~weeks of operation per year.}
\label{tab:res_opex}
\footnotesize
\setlength{\tabcolsep}{4pt}
\begin{tabular}{lrl}
\toprule
\textbf{Item} & \textbf{kCHF/year} & \textbf{Basis} \\
\midrule
Permanent facility staff & 400 & Two level~6--7, one level~3--4 post at 75\,\% \\
Electricity              & 250 & $\approx$\,\SI{6}{\giga\watt\hour}/year \\
Make-up water and treatment &   9 & EN-CV estimate \\
Cooling-plant electricity &  6 & EN-CV estimate \\
\midrule
\textbf{Total (quantified)} & \textbf{665} & \\
\midrule
CCC coverage, on-call, consumables & \multicolumn{2}{l}{to be quantified during TDR} \\
\bottomrule
\end{tabular}
\end{table}

\section{Exclusions, open items and contingency}
\label{sec:ResourceExclusions}

No global contingency has been applied to the figures above; the CDR-level uncertainty of $\pm$30--50\,\% quoted in Section~\ref{sec:ResourceBasis} is intended to cover it. The following items are known to be absent from, or only partially captured by, the present estimate and shall be quantified during the TDR phase:

\begin{itemize}
    \item \textbf{Cabling and power-converter cooling.} Signal and power cabling for the new beam line, and the cooling of the associated power converters, are not costed: because of the start of LS3 activities, EN-EL was unable to provide an estimate within the timescale of this study. This is the largest single omission from the present costing exercise, and establishing it is a priority for the early technical design phase.
    \item \textbf{Group-level contingencies.} SY-BI has recommended a contingency of approximately 10\,\% on its \SI{943}{\kCHF} of materials, i.e.\ \SI{94}{\kCHF}, and BE-EA a contingency of 20\,\% on its \SI{980}{\kCHF}, i.e.\ \SI{196}{\kCHF}. Neither of these \SI{290}{\kCHF} is included in Table~\ref{tab:res_wp}.
    \item \textbf{Access and interlock zone.} The \SI{300}{\kCHF} carried for the access system is an explicit lower bound; a firm figure requires the risk analysis to be performed with HSE-RP and BE-DSO.
    \item \textbf{Linac~3 items not costed separately.} The upgrade of the stripper ITF.STRIP and the three slits IBE2.SLH01, ITL2.SLH01 and ITL2.SLV01 are carried at zero cost by SY-STI and are expected to be absorbed within existing activities.
    \item \textbf{Facility controls.} The materials budget for BE-CEM is not yet quantified; only the associated personnel is captured.
    \item \textbf{Harmonic sextupoles.} Full control of the resonance driving term and of the horizontal chromaticity in the slow-extraction optics requires a current rating of about \SI{20}{\ampere} for the four harmonic sextupoles, against \SI{10}{\ampere} at present (Chapter~\ref{sec:SRE}). Whether this requires new magnets or only new power converters and cooling depends on the magnetic and thermal performance of the present magnets at \SI{20}{\ampere}, which is under study with TE-MSC; the item is not yet costed.
    \item \textbf{Potential savings.} The \SI{500}{\kCHF} power converter for the extraction magnetic septum SMH31 may be recoverable from LHC spare stock, and the LEIR cooling-tower replacement may be partly covered by the CERN site consolidation programme, since the towers are approaching end of life independently of HEARTS@LEIR.
    \item \textbf{Ventilation of Building 150.} No user requirement has yet been defined; the HVAC systems of Buildings 150 and 250 may require consolidation in parallel with the project.
\end{itemize}

\section{Consistency with the June 2026 IEFC estimate}
\label{sec:ResourceReconciliation}

The resource figures presented to the IEFC in June~2026~\cite{IEFC2026} were derived from the same bottom-up item list. Table~\ref{tab:res_reconcile} reconciles the two sets of numbers so that the audit trail is preserved.

The materials cost agrees group by group and work package by work package to within rounding, once the double-counted RF line discussed in Section~\ref{sec:ResourceByGroup} is removed. The Graduate-like demand for the TDR phase now agrees exactly. Two differences remain, both in the personnel bookkeeping rather than in the scope:

\begin{itemize}
    \item \textbf{Knowledge-transfer effort.} The June~2026 presentation carried \SI{2}{\PY} for the business-model study during the TDR, as originally requested by the group, while an intermediate revision of the item list had reduced it to \SI{1}{\PY}. The requested \SI{2}{\PY} is adopted here. Because that effort falls entirely within the technical design phase, the TDR total becomes \SI{19.2}{\PY}, identical to the June figure, while the full-project total rises to \SI{43.2}{\PY} --- the June headline of \SI{42.2}{\PY} having been computed from the reduced item list even though the TDR chart on the same slide used \SI{2}{\PY}.
    \item \textbf{Staff total.} The annual staff profile underlying the June~2026 presentation is identical to that of the CDR study and sums to \SI{32.5}{\PY} (\SI{9.9}{\PY} for the technical design, Table~\ref{tab:res_tdr}, and \SI{22.6}{\PY} for construction, Table~\ref{tab:res_profile}); the quoted headline of \SI{28.8}{\PY} corresponds instead to the sum of the nominal quantities declared by the groups. That column was intended to record the number of people involved, not the effort, and was populated inconsistently. The value of \SI{32.5}{\PY} carried in this report is the one consistent with the year-by-year commitments requested from the groups, and is adopted as the reference.
\end{itemize}

\begin{table}[htbp]
\centering
\caption{Reconciliation of the present estimate with the figures presented to the IEFC in June 2026.}
\label{tab:res_reconcile}
\small
\setlength{\tabcolsep}{4pt}
\begin{tabular}{llrrl}
\toprule
\textbf{Quantity} & \textbf{Unit} & \textbf{IEFC} & \textbf{This} & \textbf{Comment} \\
 &  & \textbf{06/2026} & \textbf{CDR} & \\
\midrule
Materials, WP1      & MCHF & 0.7  & 0.65  & rounding \\
Materials, WP2      & MCHF & 4.0  & 4.05  & rounding \\
Materials, WP3      & MCHF & 2.5  & 2.50  & --- \\
Materials, WP4      & MCHF & 6.7  & 6.70  & --- \\
\textbf{Materials, total} & \textbf{MCHF} & \textbf{13.9} & \textbf{13.89} & \\
\midrule
Graduate-like, TDR      & PY & 19.2 & 19.2 & --- \\
Graduate-like, project  & PY & 42.2 & 43.2 & KT effort at \SI{2}{\PY} throughout \\
Staff, TDR          & PY & 10.0 &  9.9 & rounding \\
Staff, project      & PY & 28.8 & 32.5 & bookkeeping, see above \\
\bottomrule
\end{tabular}
\end{table}

The item list has since been restructured to remove the source of the staff discrepancy: hardware and personnel are held on separate sheets, the quantity column has been dropped from the personnel sheet, and the effort total of every personnel line is now computed from its annual profile, so that a person-year figure can enter the estimate in only one way.

\section{Restructuring of the technical design phase in HEARTS++}
\label{sec:ResourceRestructuring}

The HEARTS@LEIR timeline complies with Long Shutdown~3 (LS3) activities and the HL-LHC upgrade it carries, the top priority of the Organization: the groups needed for the design remain heavily committed to LS3 and to the preparation of the HL-LHC until its completion. The technical design phase is therefore part of the HEARTS++ proposal submitted to the European Commission under the HORIZON-CL4-2026-SPACE-03-85 call of Horizon Europe. Subject to the grant, the project foresees 54~months until the TDR completion. Its first part is a pre-TDR phase, which matures the technical baseline, settles the open design choices and issues the functional specifications. A TDR-readiness gate at month~30 opens the full TDR, and the integrated HEARTS@LEIR TDR is delivered at month~54, tentatively at the end of 2031. The construction stages of the present estimate follow the TDR: procurement, fabrication and testing in 2032--2035, installation in 2034--2035 and beam commissioning in 2035, with user operations starting in 2036 (Section~\ref{sec:ResourceProfile}). These later phases will be confirmed once the funding decision for construction has been made.

For clarity's sake, this report keeps the original work-package structure of the CDR study (WP1--WP4, Section~\ref{sec:ResourceBasis}), in which the estimate was built. In the HEARTS++ proposal, the technical design is organised in three work packages, HEARTS++ WP5--WP7, alongside four project-wide or facility-specific ones: HEARTS++ WP1 (management, scientific and technical coordination, safety and communication), WP2 (industrial requirements, competitiveness and validation, including the access framework and the business model), WP3 (HEARTS@PS) and WP4 (HEARTS@GSI). In this report, work-package numbers without the HEARTS++ prefix refer to the CDR structure. The tasks of HEARTS++ WP5 and WP6 follow a common three-phase structure: phase~0 (months~1--30) comprises precursor studies whose results already benefit HEARTS@PS; phase~1 (same period) prepares the TDR; phase~2 (months~31--54) produces it.

\begin{itemize}
    \item \textbf{HEARTS++ WP5 --- Ion source, Linac~3 and LEIR.} The precursor studies establish a beam-dynamics model of the Linac~3 low-energy transport, prototype diagnostics for very-low-energy, high-intensity beams, equip the present source with four independent gas-injection lines for cocktail-beam tests, validate the Linac~3 to LEIR transfer functions, and study LEIR beam dynamics and optics for the new species, including the aperture model. The pre-TDR phase confirms the choice of the second ion source, for which the current baseline is an in-house copy of the present source, with commercial sources as benchmark or fallback; it also defines a fast momentum-spread measurement for ion switching, assesses the consolidation needs of Linac~3 and LEIR and pre-integrates the new beam lines. The TDR phase covers the magnets, vacuum and instrumentation of Linac~3 and LEIR, the LEIR impedance, the Linac~3 hall modifications, the relocation of the RFQ amplifier and the source microwave generator, the controls for source selection and slow extraction, the integration from Linac~3 to LEIR and the commissioning plan.
    \item \textbf{HEARTS++ WP6 --- Slow extraction and beam transfer.} The precursor studies set the baseline machine and beam specifications for HEARTS@PS and HEARTS@LEIR and improve spill quality and extraction efficiency (amplitude modulation of the excitation, alternative flux regulation, benchmarking of the simulations of Chapter~\ref{sec:SRE}). The pre-TDR phase designs the slow-extraction working point and the transfer-line optics for all species, comparing RF-KO- and momentum-driven extraction and the two homogenisation schemes, and issues functional specifications for the septa, extraction bumpers and sextupoles, RF and transverse feedback, instrumentation, vacuum and the modifications of the fast-extraction kicker system. The TDR phase produces the engineering design of the septa and kicker modifications, the transfer-line magnets, instrumentation, beam-intercepting devices and vacuum, their integration and the commissioning plan.
    \item \textbf{HEARTS++ WP7 --- Experimental area, user interface and technical infrastructure.} It covers the experimental area and control room, user dosimetry and beam monitoring, beam control and the user interface, and the operational and user-service model from the start of the project; the power converters, electrical distribution and cabling, and cooling and ventilation from month~30; and the facility integration, transport and handling throughout. It thereby closes the electrical-infrastructure gap of the present estimate (Section~\ref{sec:ResourceExclusions}).
\end{itemize}

Table~\ref{tab:res_hpp_wp} relates these work packages to the chapters and work packages of this report, Table~\ref{tab:res_hpp_dm} lists their deliverables and milestones.

\begin{table}[htbp]
\centering
\caption{Correspondence between the HEARTS++ work packages for the technical design of HEARTS@LEIR and the present report.}
\label{tab:res_hpp_wp}
\small
\setlength{\tabcolsep}{4pt}
\begin{tabular}{@{}>{\raggedright\arraybackslash}p{3.3cm}>{\raggedright\arraybackslash}p{6.9cm}>{\raggedright\arraybackslash}p{3.0cm}@{}}
\toprule
\textbf{HEARTS++ work package} & \textbf{Scope in the proposal} & \textbf{This report} \\
\midrule
WP5 --- Ion source, Linac~3 and LEIR & Second ion source; Linac~3 and LEIR beam dynamics, optics and aperture model; LEIR impedance; magnets, vacuum, beam instrumentation and controls; integration from Linac~3 to LEIR & Chapters~\ref{sec:IonSources} and~\ref{sec:LEIRaccelerator}; WP2 \\
\addlinespace
WP6 --- Slow extraction and beam transfer & Slow-extraction working point; septa, extraction bump and kicker modifications; transfer-line optics and magnets; beam instrumentation, intercepting devices and vacuum; integration & Chapters~\ref{sec:SRE} and~\ref{sec:Beamline2HEARTS}; WP3 and the beam-line part of WP4 \\
\addlinespace
WP7 --- Experimental area, user interface and technical infrastructure & Experimental area and control room; user dosimetry, beam control and user interface; operational model; power converters and cabling; cooling and ventilation; transport and handling; facility integration & Chapters~\ref{sec:Operations}, \ref{sec:expArea} and~\ref{sec:InfrastructureAndIntegration}; the experimental-area part of WP4 and the transversal part of WP1 \\
\bottomrule
\end{tabular}
\end{table}

\begin{table}[htbp]
\centering
\caption{Deliverables and milestones of HEARTS++ WP5--WP7, with their due month counted from the project start. The integrated HEARTS@LEIR TDR is delivered by WP1 at month~54.}
\label{tab:res_hpp_dm}
\small
\setlength{\tabcolsep}{4pt}
\begin{tabular}{@{}l>{\raggedright\arraybackslash}p{6.6cm}>{\raggedright\arraybackslash}p{5.6cm}@{}}
\toprule
\textbf{HEARTS++ WP} & \textbf{Deliverables (month)} & \textbf{Milestones (month)} \\
\midrule
WP5 & D5.1 Injector improvements benefiting HEARTS@PS (36); D5.2 Linac~3 and LEIR readiness for the TDR (30); D5.3 TDR of the Linac~3 and LEIR upgrades (54) & MS5.1 Linac~3 source ready for cocktail-beam tests (24); MS5.3 pre-TDR activities completed (30); MS5.2 fast ion switch tested up to the PS (54) \\
\addlinespace
WP6 & D6.1 Slow-extraction improvements benefiting HEARTS@PS (30); D6.2 Slow extraction and transfer line readiness for the TDR (30); D6.3 TDR of the slow-extraction system and transfer line (54) & MS6.1 Slow-extraction scheme decided (24); MS6.2 pre-TDR activities completed (30) \\
\addlinespace
WP7 & D7.1 Experimental area and user interface design (42); D7.2 Technical infrastructure design (46); D7.3 Operational and user-service model (48); D7.4 Facility integration and installation design (52) & MS7.1 Operational model and cost basis transferred to the business model (28); MS7.2 Experimental-area concept validated for safety and radiation protection (46) \\
\bottomrule
\end{tabular}
\end{table}

\subsection*{The present estimate in the HEARTS++ structure}

Table~\ref{tab:res_hpp_regroup} regroups the items of the present estimate (Table~\ref{tab:res_wp}) by HEARTS++ work package; the totals are unchanged. Each item is assigned to the work package whose tasks cover the corresponding system in the proposal. Source, Linac~3 and LEIR-ring items go to WP5; the extraction septa, extraction bump, transfer-line magnets, instrumentation of the extraction and of the line, beam-intercepting devices and vacuum go to WP6; the experimental area, its instrumentation and access system, cooling and ventilation, transport and handling, and all power converters go to WP7, since the proposal designs the power converters in WP7. Project coordination, the business-model study and the radiation-protection effort go to the project-wide WP1 and WP2. Personnel follows the same rule group by group; for SY-BI and TE-MSC, which contribute to both WP5 and WP6, the work-package shares of the item list are carried over.

\begin{table}[htbp]
\centering
\caption{Resource estimate of Table~\ref{tab:res_wp} regrouped by HEARTS++ work package. Staff effort is foreseen as an in-kind contribution and is not costed.}
\label{tab:res_hpp_regroup}
\small
\setlength{\tabcolsep}{4pt}
\begin{tabular}{@{}>{\raggedright\arraybackslash}p{4.2cm}rrrrr@{}}
\toprule
\multirow{2}{*}{\textbf{HEARTS++ work package}} & \textbf{Materials} & \multicolumn{2}{c}{\textbf{Graduate-like}} & \textbf{Funded total} & \textbf{Staff} \\
\cmidrule(lr){3-4}
 & \textbf{(kCHF)} & \textbf{(PY)} & \textbf{(kCHF)} & \textbf{(kCHF)} & \textbf{(PY)} \\
\midrule
WP5 --- Ion source, Linac~3, LEIR & 3\,012 & 16.0 & 1\,523 & 4\,535 & 14.6 \\
WP6 --- Slow extraction, beam transfer & 4\,920 & 18.0 & 1\,923 & 6\,843 & 11.3 \\
WP7 --- Experimental area, infrastructure$^{*}$ & 5\,930 & 7.0 & 710 & 6\,640 & 3.7 \\
WP1--WP2 --- Coordination, business model, safety & 30 & 2.2 & 242 & 272 & 2.9 \\
\midrule
\textbf{Total} & \textbf{13\,892} & \textbf{43.2} & \textbf{4\,398} & \textbf{18\,290} & \textbf{32.5} \\
\bottomrule
\end{tabular}\\[2pt]
\footnotesize{$^{*}$ Includes \SI{3513}{\kCHF} of power converters. Booked with the systems they supply, as in Table~\ref{tab:res_wp}, the materials would be \SI{3933}{\kCHF} for WP5, \SI{7512}{\kCHF} for WP6 and \SI{2417}{\kCHF} for WP7.}
\end{table}

\FloatBarrier

\chapter{Risk assessment}
\label{sec:RiskAssessment}

This chapter collects the risks identified during the conceptual design study, together with their assessed likelihood and impact and the mitigation foreseen. It is a CDR-level register: it is deliberately broad rather than deep, and its main purpose is to identify which risks must be retired, and which merely monitored, during the technical design phase.

\section{Method}
\label{sec:RiskMethod}

Each risk is characterised by a likelihood $L$ and an impact $I$, both graded low (L), medium (M) or high (H). Impact is assessed against the three project objectives separately --- performance, schedule and cost --- and the highest of the three is quoted. The resulting severity follows the matrix of Table~\ref{tab:RiskMatrix}. Risks assessed as high severity are listed, and the principal ones discussed individually, in Section~\ref{sec:RiskTop}.

The register carries no monetised risk provision. Consistent with Chapter~\ref{sec:ResourceEstimate}, cost risk at this stage is covered by the CDR-level uncertainty of $\pm$30--50\,\% rather than by a line-by-line contingency.

\begin{table}[htbp]
\centering
\caption{Severity matrix. Rows are impact, columns likelihood.}
\label{tab:RiskMatrix}
\begin{tabular}{lccc}
\toprule
 & \textbf{L} & \textbf{M} & \textbf{H} \\
\midrule
\textbf{H} & Medium & High   & High   \\
\textbf{M} & Low    & Medium & High   \\
\textbf{L} & Low    & Low    & Medium \\
\bottomrule
\end{tabular}
\end{table}

\section{Technical risks}
\label{sec:RiskTechnical}

The technical risks are listed in Table~\ref{tab:RiskTechnical}. They fall into three groups: the maturity of the extraction hardware design, the performance of the second ion source, and the beam physics that has been simulated but not yet demonstrated at LEIR.

{\footnotesize
\setlength{\tabcolsep}{3pt}
\renewcommand{\arraystretch}{1.15}
\begin{longtable}{@{}p{0.7cm}p{4.2cm}ccp{5.6cm}@{}}
\caption{Technical risk register.}\label{tab:RiskTechnical}\\
\toprule
\textbf{ID} & \textbf{Risk} & \textbf{$L$} & \textbf{$I$} & \textbf{Mitigation} \\
\midrule
\endfirsthead
\multicolumn{5}{@{}l}{\textit{\small Table \thetable\ (continued)}}\\
\toprule
\textbf{ID} & \textbf{Risk} & \textbf{$L$} & \textbf{$I$} & \textbf{Mitigation} \\
\midrule
\endhead
\midrule
\multicolumn{5}{r@{}}{\textit{\small continued on next page}}\\
\endfoot
\bottomrule
\endlastfoot
T1 & Non-conforming material introduced into the LEIR ultra-high-vacuum sectors compromises vacuum integrity & M & H & Mandatory material evaluation and quality control with TE-VSC from the earliest design stage for every group contributing equipment to the insertion region; vacuum acceptance built into the procurement specifications \\
T2 & No design exists for the electrostatic and magnetic extraction septa; specifications may not be met within the available space & M & H & Specifications frozen early in the TDR; reuse of the SEH11 design to be confirmed; fallback of relocating the spare kicker KFH31 to relax the septum requirements; beam-based measurement of the SS30 aperture (Section~\ref{sec:MTProgramme}) \\
T3 & Modification of the KFH3234 tank proves incompatible with fast-ejection performance & L & H & Approach already validated in the BioLEIR study~\cite{BioLEIR}; fallback is a new, correctly dimensioned tank, costed as a variant \\
T4 & The \SI{14}{\kilo\ampere}/\SI{8}{\volt} converter for SMH31 cannot be recovered from LHC spare stock & M & M & Item carried at full cost (\SI{500}{\kCHF}) in the baseline estimate, so recovery is an upside rather than a dependency \\
T5 & Second source does not deliver the required intensity or stability for the noble-gas species & M & H & Splitting species between two sources removes the space-charge penalty of a cocktail; pulsed LEBT and automatic re-optimisation; measurement campaign early in the TDR \\
T6 & Stripping efficiencies and charge-state yields are calculated (Baron's formula), not measured, for O, Ar, Kr and Xe & H & M & Dedicated Linac~3 measurement runs during the TDR; the estimate already uses conservative intensities \\
T7 & Injection into LEIR at low rigidity is not achievable for all species, all of which are less rigid at injection than $^{208}$Pb$^{54+}$ & M & M & Beam-dynamics study of accumulation, cooling and RF capture prioritised in the TDR; species palette can be adjusted \\
T8 & Electron-cooling time for the lighter ions exceeds the time available in the cycle & L & M & Scaling from the measured Pb$^{54+}$ cooling (Section~\ref{sec:MTPb}) gives about \SI{1}{\second} for $^{16}$O$^{8+}$, within the accumulation window; Ne$^{5+}$ cools faster than this scaling predicts, which gives about \SI{0.4}{\second} for O$^{8+}$ (Section~\ref{sec:MTLightIons}); to be verified per species (Section~\ref{sec:MTProgramme}) \\
T9 & Spill quality does not meet Specification~H ($c_V < 1$ at 100~Hz sampling), for instance because of power-converter ripple or drifts of the resonance driving term & M & M & Ripple tolerances of the ring power converters to be specified in the TDR; RF-KO excitation through the transverse feedback (consolidation in the SY-RF scope) as actuator of a spill feedback loop; amplitude modulation of the excitation and alternative flux-regulation methods studied within HEARTS++; techniques established at therapy facilities; response of the beam to RF-KO excitation and time structure of the losses first measured in LEIR in 2026 (Sections~\ref{sec:MTRFKO} and~\ref{sec:MTSX}) \\
T10 & Non-Gaussian extracted distribution degrades the efficiency and uniformity obtained in tracking & M & M & Tracking to be repeated with the realistic distribution from extraction simulations for all three homogenisation schemes (Section~\ref{sec:TLcomparison}) \\
T11 & Octupole option cannot be matched within the space and aperture available & M & L & Not in the baseline; blow-up and scraping is the reference solution \\
T12 & Pencil-beam scanning requires a large-aperture vacuum chamber whose cost and integration are not yet established & M & M & Aperture and chamber design to be settled in the TDR; scheme runs on the baseline lattice, so the decision can be deferred (Section~\ref{sec:TLscanning}) \\
T13 & Radiation-protection intensity limit proves incompatible with user flux requirements & L & H & Analysis in Chapter~\ref{sec:RadiationProtection} shows compatibility for all baseline species; the delivery performance of Tables~\ref{tab:OpsDeliveryTimes} and~\ref{tab:fluxes_at_DUT} is computed with the intensities capped at this limit; technical implementation of the energy and intensity limitation to be designed in the TDR \\
T14 & The harmonic sextupoles cannot be operated at the \SI{20}{\ampere} needed to control both the resonance driving term and the horizontal chromaticity; present rating \SI{10}{\ampere}, thermal saturation observed on the bench at \SI{10}{\ampere} & M & M & Performance at \SI{20}{\ampere} under study with TE-MSC; fallback within the present rating is to power the defocusing sextupoles XDN as well and to work further from the resonance; replacement of the four magnets to be costed in the TDR \\
\end{longtable}
}

\section{Schedule and dependency risks}
\label{sec:RiskSchedule}

HEARTS@LEIR is not a stand-alone project. It depends on the Ion Complex Upgrade for the second source, on the ion complex consolidation for the systems it deliberately does not upgrade, and on the availability of groups that are simultaneously committed to LS3. These dependencies, listed in Table~\ref{tab:RiskSchedule}, are the dominant schedule risk.

{\footnotesize
\setlength{\tabcolsep}{3pt}
\renewcommand{\arraystretch}{1.15}
\begin{longtable}{@{}p{0.7cm}p{4.2cm}ccp{5.6cm}@{}}
\caption{Schedule and dependency risk register.}\label{tab:RiskSchedule}\\
\toprule
\textbf{ID} & \textbf{Risk} & \textbf{$L$} & \textbf{$I$} & \textbf{Mitigation} \\
\midrule
\endfirsthead
\multicolumn{5}{@{}l}{\textit{\small Table \thetable\ (continued)}}\\
\toprule
\textbf{ID} & \textbf{Risk} & \textbf{$L$} & \textbf{$I$} & \textbf{Mitigation} \\
\midrule
\endhead
\midrule
\multicolumn{5}{r@{}}{\textit{\small continued on next page}}\\
\endfoot
\bottomrule
\endlastfoot
S1 & The ion complex consolidation is not approved, or is delayed, removing the premise that existing systems need no upgrade & M & H & Explicitly stated as a prerequisite in Section~\ref{sec:OpsEnvelope}; HEARTS@LEIR investment decision sequenced after consolidation approval \\
S2 & The second ion source, shared with the ICU project, is descoped or delayed & M & H & Scope and responsibility split with ICU to be fixed during the TDR; without a second source, species switching within \SI{15}{\minute} cannot be met and Specification~B would have to be relaxed. The project will aim to take over the ICU items that are imprescindible for HEARTS@LEIR. \\
S3 & Equipment-group availability during LS3 is insufficient; BE-EA in particular is committed to other approved projects & H & M & Full TDR tentatively scheduled for 2030--2031, after a pre-TDR phase within HEARTS++; TDR executed primarily by Graduate-like personnel with CERN experts in supervisory roles only, in compliance with LS3 commitments; formal group commitments to be obtained before the TDR starts \\
S4 & Installation window in 2034--2035 is not available, or conflicts with injector-complex planning & M & M & Installation deliberately planned after LS3, in a single campaign in 2034--2035 (Section~\ref{sec:ResourceProfile}); the window to be agreed with the injector planning during the TDR \\
S5 & Long-lead procurement of magnets and power converters slips & M & M & Existing designs adopted wherever possible (CNAO/MedAustron dipoles, POLARIS and BOREAL converters) specifically to avoid new-design lead times \\
S6 & Space for the beam line, experimental area and power converters cannot be reserved & M & H & Treated in detail at TDR level; 3D as-built scan of Building~150 and LOD200 integration study in the TDR scope \\
S7 & LEIR cooling towers reach end of life before the project delivers & M & L & Replacement may be covered by the CERN site consolidation programme independently of HEARTS@LEIR; Option~2 of Section~\ref{sec:InfrastructureAndIntegration} carried in the baseline \\
\end{longtable}
}

\section{Cost risks}
\label{sec:RiskCost}

{\footnotesize
\setlength{\tabcolsep}{3pt}
\renewcommand{\arraystretch}{1.15}
\begin{longtable}{@{}p{0.7cm}p{4.2cm}ccp{5.6cm}@{}}
\caption{Cost risk register.}\label{tab:RiskCost}\\
\toprule
\textbf{ID} & \textbf{Risk} & \textbf{$L$} & \textbf{$I$} & \textbf{Mitigation} \\
\midrule
\endfirsthead
\multicolumn{5}{@{}l}{\textit{\small Table \thetable\ (continued)}}\\
\toprule
\textbf{ID} & \textbf{Risk} & \textbf{$L$} & \textbf{$I$} & \textbf{Mitigation} \\
\midrule
\endhead
\midrule
\multicolumn{5}{r@{}}{\textit{\small continued on next page}}\\
\endfoot
\bottomrule
\endlastfoot
C1 & The estimate carries a CDR-level uncertainty of $\pm$30--50\,\% & H & M & Inherent to the maturity of the design; a firm cost book is a deliverable of the TDR \\
C2 & Cabling and power-converter cooling are not costed --- the largest single omission --- because EN-EL could not provide an estimate before the start of LS3 activities & H & M & Electrical load inventory and cabling scope to be established with EN-EL and EN-CV at the beginning of the TDR \\
C3 & Group-level contingencies recommended by SY-BI (10\,\%) and BE-EA (20\,\%), totalling \SI{290}{\kCHF}, are not applied & H & L & Recorded explicitly in Section~\ref{sec:ResourceExclusions} and in the resource file, so they are visible rather than forgotten \\
C4 & The access and interlock zone is carried at a lower bound of \SI{300}{\kCHF} & M & M & Risk analysis with HSE-RP and BE-DSO to be performed in the TDR \\
C5 & Items carried at zero cost by SY-STI (stripper upgrade, three slits) prove not to be absorbable & L & L & Small absolute value; to be confirmed with the group \\
C6 & The BE-CEM materials budget is not yet quantified & M & L & To be established in the TDR together with the facility controls scope \\
C7 & Operational cost is only partly quantified (\SI{665}{\kCHF} per year); CCC coverage, on-call support and consumables are missing & M & M & Full operational and business model to be developed at TDR with OP, STI, CEM and KT \\
\end{longtable}
}

\section{Safety, radiation protection and compliance risks}
\label{sec:RiskSafety}

{\footnotesize
\setlength{\tabcolsep}{3pt}
\renewcommand{\arraystretch}{1.15}
\begin{longtable}{@{}p{0.7cm}p{4.2cm}ccp{5.6cm}@{}}
\caption{Safety, radiation protection and compliance risk register.}\label{tab:RiskSafety}\\
\toprule
\textbf{ID} & \textbf{Risk} & \textbf{$L$} & \textbf{$I$} & \textbf{Mitigation} \\
\midrule
\endfirsthead
\multicolumn{5}{@{}l}{\textit{\small Table \thetable\ (continued)}}\\
\toprule
\textbf{ID} & \textbf{Risk} & \textbf{$L$} & \textbf{$I$} & \textbf{Mitigation} \\
\midrule
\endhead
\midrule
\multicolumn{5}{r@{}}{\textit{\small continued on next page}}\\
\endfoot
\bottomrule
\endlastfoot
R1 & The shielding required for the new line and experimental area exceeds the estimated \SIrange{80}{160}{\centi\meter} of concrete, with consequences for space and cost & M & M & Shielding simulations by HSE-RP during the TDR, using the tentative loss map of Table~\ref{rp_loss_map}; shielding thickness treated as an integration constraint from the outset \\
R2 & Absence of top shielding at LEIR limits the admissible beam power because of sky-shine effects & M & M & Operating point kept within the intensity envelope already demonstrated for $^{208}$Pb$^{54+}$ operation (Chapter~\ref{sec:RadiationProtection}) \\
R3 & The technical implementation of a reliable energy and intensity limitation for injection into LEIR is not yet designed & H & M & Design is an explicit TDR deliverable; without it the RP limits cannot be enforced by hardware \\
R4 & Activation of the experimental area restricts user access and turnaround & M & M & Test station located in a supervised rather than a controlled area; beam stopper and dump sized on worst case; loss localisation at the DUT and at dedicated masks \\
R5 & The access and interlock architecture (EIS-beam, EIS-access) is not yet specified to the level needed for a user facility with 24/7 operation & M & M & Developed with EN-AA-AC and HSE-RP-AS during the TDR; two interlocked zones already defined at the conceptual level \\
\end{longtable}
}

\section{Programmatic and funding risks}
\label{sec:RiskProgrammatic}

{\footnotesize
\setlength{\tabcolsep}{3pt}
\renewcommand{\arraystretch}{1.15}
\begin{longtable}{@{}p{0.7cm}p{4.2cm}ccp{5.6cm}@{}}
\caption{Programmatic and funding risk register.}\label{tab:RiskProgrammatic}\\
\toprule
\textbf{ID} & \textbf{Risk} & \textbf{$L$} & \textbf{$I$} & \textbf{Mitigation} \\
\midrule
\endfirsthead
\multicolumn{5}{@{}l}{\textit{\small Table \thetable\ (continued)}}\\
\toprule
\textbf{ID} & \textbf{Risk} & \textbf{$L$} & \textbf{$I$} & \textbf{Mitigation} \\
\midrule
\endhead
\midrule
\multicolumn{5}{r@{}}{\textit{\small continued on next page}}\\
\endfoot
\bottomrule
\endlastfoot
P1 & The present EU call does not accept a partial response consisting of a TDR only, since it targets objectives achievable by the end of the activity & H & H & Addressed in the HEARTS++ proposal, which combines the TDR (HEARTS++ WP5--WP7) with the industrialisation of the operating HEARTS facilities at the PS and at GSI; alternative sources of external co-funding to be investigated should the proposal not be retained; Technology Infrastructures in FP10 identified as the main option for the construction phase \\
P2 & External co-funding for the TDR is not secured & M & H & In the HEARTS++ proposal the TDR is carried by HEARTS++ WP5--WP7, with EU funding complemented by matching resources; a staged or partially CERN-funded TDR remains possible \\
P3 & The construction decision, deferred to after the TDR, is not taken & M & H & Inherent to the staged approach; the TDR is scoped to produce a decision-quality cost book and a demonstrated design \\
P4 & The business model for paid access is not viable, or the CERN policy on paid access is not in place in time & M & M & Business case analysis by KT during the TDR; access restricted to civil applications per KT and Legal Service provisions \\
P5 & Operational resources from OP are not committed & M & M & OP involved during the TDR to render the facility as operable as possible; operational model a TDR deliverable \\
P6 & The demand assumed for the facility does not materialise & L & H & Demonstrated demand at HEARTS at the CERN PS, already at the limit of what T8 can provide, with the EU call targeting a factor ten more \\
\end{longtable}
}

\section{Principal risks}
\label{sec:RiskTop}

Fourteen risks are assessed as high severity: T1, T2, T5 and T6; S1, S2, S3 and S6; C1 and C2; R3; and P1, P2 and P3. Nine of them are managed through the mitigations listed in the registers: the design and measurement work of the technical design phase (T2, T6, S6, C2 and R3; Section~\ref{sec:RiskTDR}), the scheduling of the TDR after LS3 (S3), and the staged approach to the cost estimate and to the funding (C1, P2 and P3). The remaining five deserve explicit management attention and are discussed below.

\textbf{P1 --- compatibility of the TDR with the present EU call.} This is the only risk that can stop the project immediately rather than degrade it. The call targets the completion of its objectives, including delivered beam hours, by the end of the activity, whereas what is proposed here is a technical design study with the construction decision taken afterwards. The HEARTS++ proposal addresses this by combining the technical design of HEARTS@LEIR with the industrialisation of the operating HEARTS facilities at the PS and at GSI, whose objectives, including delivered beam hours, are achievable within the action. Should the proposal not be retained, external co-funding must be found elsewhere before the TDR can start.

\textbf{S1 and S2 --- dependency on the ion complex consolidation and on ICU.} The entire cost argument of this report rests on fitting HEARTS@LEIR operation inside the existing capacity of the accelerator systems, as quantified in Section~\ref{sec:OpsEnvelope}. That premise holds only if the consolidation already foreseen for the ion physics programme goes ahead. Similarly, the second ion source is shared scope with ICU; without it the \SI{15}{\minute} species-switching specification is not attainable, and the facility loses much of its distinctive value. Neither risk can be mitigated from within the project; both must be managed at programme level.

\textbf{T1 --- vacuum non-conformity in LEIR.} A single non-conforming component introduced into a NEG-coated sector can cost weeks of bakeout and vent-pump cycles, and the number of such cycles per sector is itself limited by NEG ageing. The mitigation is procedural rather than technical, and it must be in place from the first design iteration: every group contributing equipment to the insertion region coordinates with TE-VSC before, not after, the design is fixed.

\textbf{T5 --- second-source performance for noble gases.} The intensities quoted throughout this report for Ar, Kr and Xe rest on a small number of test runs, in some cases a decade old, and on calculated stripping efficiencies. Should the second source fall significantly short, the flux specification would be met only for the lighter species or only over reduced irradiation fields. A dedicated measurement campaign early in the TDR is the natural mitigation, and it is also the cheapest way to retire risk T6.

\section{Risks to be retired during the technical design phase}
\label{sec:RiskTDR}

The technical design phase should be scoped so that, at its end, the following are no longer open: the septa specifications and their integration in SS30 (T2, T3); the measured performance of the second source and of the stripping process (T5, T6); the beam-dynamics chain of accumulation, cooling and RF capture for every baseline species (T7, T8); the choice of homogenisation scheme, including the aperture consequences of the scanning option (T10, T11, T12); the shielding thickness and the hardware implementation of the intensity limitation (R1, R3); the electrical and cabling scope (C2); and the space reservation (S6). The programmatic risks P1 to P3 are, by construction, decided outside the project.

\section{Machine tests as risk mitigation}
\label{sec:MTProgramme}

Several of the technical risks of Table~\ref{tab:RiskTechnical} concern the beam physics in LEIR and can be reduced by measurements in the present machine, in continuation of the tests reported in Chapter~\ref{sec:MachineTests}. The 2026 run ended in August~2026, and no further beam time is available before the injector complex restarts after Long Shutdown~3, expected in the second half of 2028. Until then, the analysis of the existing data will be completed and the extraction simulations will be benchmarked against the slow-extraction test of Section~\ref{sec:MTSX}. The 2028 and 2029 runs fall within the pre-TDR phase of HEARTS++ (months 1 to 30, Section~\ref{sec:ResourceRestructuring}). The measurements proposed below provide the experimental input for deliverables D5.2 and D6.2 and for the choice of the slow-extraction scheme (milestone MS6.1). Since MD time in the first year after the restart will be limited by the recommissioning of the complex, the measurements that do not depend on each other should be prepared so that they can be carried out in parallel or parasitically.

\begin{enumerate}
    \item \textbf{Thermal envelope.} HEARTS@LEIR operation will respect the RMS envelope of present operation, so as not to add stress on ageing equipment (Section~\ref{sec:MTFlattop}); no further extension of the flat top at the maximum rigidity is therefore needed. The longer flat tops foreseen at lower rigidities (Table~\ref{tab:DutyCycles}) are to be verified within this envelope, with logging of the coil and cooling-plant temperatures until thermal equilibrium, and the RMS limits of the other LEIR circuits, in particular the sextupoles and the orbit correctors, are to be evaluated. These measurements consolidate the operating envelope of Section~\ref{sec:OpsEnvelope}.
    \item \textbf{Beam preparation for all species.} Emittance, momentum spread and cooling time on the flat top for every species that Linac~3 can deliver, and the transmission from RF capture to the flat top on the 4BP cycle (Sections~\ref{sec:MTEmittance} and~\ref{sec:MTSX}). For Pb$^{54+}$, the vertical emittance on the flat top with set-up~(b) of the 4BP cycle, for which no profiles were recorded, and the coupling of the ring at the crossing of the difference resonance, a candidate cause of the vertical emittance growth observed with set-up~(a) (Section~\ref{sec:MTPb}). These measurements address risks T7 and T8 and provide the initial conditions of the simulations of Chapter~\ref{sec:SRE}.
    \item \textbf{Aperture and optics.} The vertical aperture scan; the aperture of SS30 with the bump based on the measured orbit response; a beam-based calibration of the BPMs, bumps with their maximum at the suspected elements and a review of the aperture database against the mechanical drawings, to locate the restrictions missing in the database (Section~\ref{sec:MTAperture}); the optics functions and the dispersion with a model matched to the measured tunes; the chromaticity and the tune spread on the flat top; a systematic correction of the flat-top momentum drift (Sections~\ref{sec:MTMomentum} to~\ref{sec:MTOptics}). The SS30 aperture enters directly into the specification of the septa (risk T2).
    \item \textbf{RF-KO excitation.} A calibration of the kick voltage as a function of the gain; scans at low gain ($A \lesssim 0.1$) on cycles with longer flat tops at reduced rigidity, within the present RMS envelope, to extend the spill to \SI{1}{\second} and then to several seconds; amplitude modulation of the excitation and a first closed-loop spill control (Section~\ref{sec:MTRFKO}). These tests provide the experimental basis for the comparison between RF-KO- and momentum-driven extraction and address risk T9.
    \item \textbf{Loss and spill instrumentation.} A permanent loss monitor in SS30 and a calibrated detector with sufficient dynamic range and bandwidth to evaluate Specification~H on the spill of the slow-extraction test, complementing the Timepix detector (risk T9).
    \item \textbf{Benchmarking.} A repetition of the slow-extraction test under conditions reproduced in the simulations (measured emittance, aperture, optics and loss locations), in order to benchmark the extraction simulations before the septum design is frozen (HEARTS++ WP6, phase~0; risk T2).
\end{enumerate}

\bibliographystyle{abbrv}
\bibliography{references}

\begin{thebibliography}{10}

\bibitem{cern_home}
{HEARTS innovates to foster European access to space}.
\newblock Available at:
  \url{https://home.cern/news/news/knowledge-sharing/hearts-innovates-foster-european-access-space}.

\bibitem{hearts_project}
{HEARTS Project Website}.
\newblock Available at: \url{https://hearts-project.eu/}.

\bibitem{Alemany-Fernandez:IPAC2018-TUPAF020}
R.~Alemany-Fernandez et~al.
\newblock {P}erformance of the {CERN} {L}ow {E}nergy {I}on {R}ing ({LEIR}) with
  {X}enon beams.
\newblock In {\em Proc. 9th International Particle Accelerator Conference
  (IPAC'18), Vancouver, BC, Canada, April 29-May 4, 2018}, number~9 in
  International Particle Accelerator Conference, pages 705--708, Geneva,
  Switzerland, June 2018. JACoW Publishing.
\newblock https://doi.org/10.18429/JACoW-IPAC2018-TUPAF020.

\bibitem{ABPinput2026}
R.~Alemany~Fern\'andez, D.~K\"uchler, R.~Scrivens, and M.~S{\l}upecki.
\newblock Input from {BE-ABP}: preliminary material cost estimate for the
  second ion source.
\newblock HEARTS@LEIR workshop, CERN, 11 May, 2026.
\newblock \url{https://indico.cern.ch/event/1679820/}.

\bibitem{ArrutiaDiffusive}
P.~A. Arrutia~Sota, E.~C. Cort\'es~Garc\'ia, and V.~A. Sansipersico.
\newblock A diffusive model for radio-frequency knock-out slow extraction,
  2025.
\newblock arXiv:2511.08068 [physics.acc-ph]; submitted to Phys. Rev. Accel.
  Beams.

\bibitem{BARON1993177}
E.~Baron, M.~Bajard, and C.~Ricaud.
\newblock Charge exchange of very heavy ions in carbon foils and in the
  residual gas of {GANIL} cyclotrons.
\newblock {\em Nuclear Instruments and Methods in Physics Research Section A:
  Accelerators, Spectrometers, Detectors and Associated Equipment},
  328(1):177--182, 1993.

\bibitem{LHCDesignReport}
M.~Benedikt, P.~Collier, V.~Mertens, J.~Poole, and K.~Schindl.
\newblock {\em {LHC Design Report}}.
\newblock CERN Yellow Reports: Monographs. CERN, Geneva, 2004.

\bibitem{BiancacciZampetakis2024}
N.~Biancacci and M.~Zampetakis.
\newblock {Simulation and measurement of the interplay of electron cooling,
  impedance, intrabeam scattering, and space charge for coasting beams}.
\newblock {\em Phys. Rev. Accel. Beams}, 27:014201, 2024.

\bibitem{LEIRTFB2008}
A.~Blas, M.~Chanel, and A.~Findlay.
\newblock Description of the {LEIR} transverse damping system.
\newblock Internal note, CERN AB-RF, 2008.

\bibitem{Bodart2025}
D.~Bodart.
\newblock Impact of longer flat tops to magnets operation ({LEIR}, {HEARTS}).
\newblock Internal presentation, TE-MSC, CERN, 25 August, 2025.

\bibitem{SLAWGMagnets2026}
D.~Bodart.
\newblock {LEIR} magnet temperature measurements and {RMS} constraints.
\newblock Slow Extraction Working Group meeting ``HEARTS \#1: slow extraction
  test at {LEIR}'', CERN, 19 November 2026,
  \url{https://indico.cern.ch/event/1744007/}; calculation spreadsheet EDMS
  3557931, 2026.

\bibitem{AccumulationLeadIons}
J.~Bosser, C.~Carli, M.~Chanel, C.~Hill, A.~M. Lombardi, R.~MacCaferri,
  S.~Maury, D.~Möhl, G.~Molinari, S.~Rossi, E.~Tanke, G.~Tranquille, and
  M.~Vretenar.
\newblock {Experimental investigation of electron cooling and stacking of lead
  ions in a low energy accumulation ring}.
\newblock {\em Part. Accel.}, 63:171--210, 1999.

\bibitem{MD20250708}
E.~C. Cort\'es~Garc\'ia.
\newblock Summary of {MD} 2025-07-08 at {LEIR}.
\newblock Internal presentation, CERN, 11 July, 2025.

\bibitem{SXMockupPrep2026}
E.~C. Cort\'es~Garc\'ia.
\newblock Slow extraction test at {LEIR}, {MD} preparation: aperture
  measurements and tracking simulations.
\newblock Internal presentation, CERN, 24 March, 2026.

\bibitem{FourBP2026}
E.~C. Cort\'es~Garc\'ia, T.~Argyropoulos, M.~S{\l}upecki, O.~Hans, and
  M.~Jebramcik.
\newblock Summary of the {4BP} set-up for the slow-extraction test at {LEIR}.
\newblock Internal presentation, CERN, 4 September, 2026.

\bibitem{FlattopExtension2025}
E.~C. Cort\'es~Garc\'ia and M.~A. Fraser.
\newblock {LEIR} flat-top extension for a specified {RMS} current for the total
  cycle.
\newblock Internal presentation, CERN, 25 November, 2025.

\bibitem{FlattopMinutes2025}
E.~C. Cort\'es~Garc\'ia and M.~A. Fraser.
\newblock {LEIR} power limits considerations: minutes of the meetings of 25
  {August} and 11 {November} 2025.
\newblock Internal note, CERN, 2025.
\newblock \url{https://indico.cern.ch/event/1594797/}.

\bibitem{CycleComposition2026}
E.~C. Cort\'es~Garc\'ia, M.~A. Fraser, T.~Argyropoulos, and M.~S{\l}upecki.
\newblock {LEIR} cycle composition.
\newblock Internal presentation, CERN, 28 April, 2026.

\bibitem{SLAWGBLM2026}
J.~C. Esteban~Felipe, S.~Morales~Vigo, B.~Salvachua~Ferrando, and E.~Effinger.
\newblock {BLM} installation at {LEIR}: layout, positioning and displacements.
\newblock Slow Extraction Working Group meeting ``HEARTS \#1: slow extraction
  test at {LEIR}'', CERN, 19 November 2026,
  \url{https://indico.cern.ch/event/1744007/}, 2026.

\bibitem{Franchi2007}
A.~Franchi, E.~M\'etral, and R.~Tom\'as.
\newblock {E}mittance sharing and exchange driven by linear betatron coupling
  in circular accelerators.
\newblock {\em Phys. Rev. ST Accel. Beams}, 10:064003, 2007.

\bibitem{HEARTS_D51}
S.~Francola, S.~Gerardin, and R.~Mangeret.
\newblock Finalised list of beam parameter requirements concurring to establish
  a {TRL} 6-7 for the {HEARTS} facilities, deliverable {D5.1}, June 2023.
\newblock Available at: \url{https://hearts-project.eu/}.

\bibitem{ICUproject}
{Future Ions Working Group}.
\newblock Upgrade and consolidation of the {CERN} ion injector complex: Phase
  {I}, Feb. 2025.
\newblock To be published.

\bibitem{SLAWGIntro2026}
R.~Garc\'ia~Al\'ia.
\newblock Intro -- {HEARTS} and {HEARTS++}.
\newblock Slow Extraction Working Group meeting ``HEARTS \#1: slow extraction
  test at {LEIR}'', CERN, 19 November 2026,
  \url{https://indico.cern.ch/event/1744007/}, 2026.

\bibitem{IEFC2026}
R.~Garc\'ia~Al\'ia, E.~C. Cort\'es~Garc\'ia, M.~A. Fraser, M.~S{\l}upecki, and
  A.~Waets.
\newblock {HEARTS@LEIR}: resources and person-power update, {IEFC} proposal for
  the {TDR}.
\newblock Injectors and Experimental Facilities Committee, CERN, 5 June, 2026.
\newblock \url{https://indico.cern.ch/event/1686894/}.

\bibitem{BioLEIR}
S.~Ghithan, G.~Roy, and S.~Schuh.
\newblock {\em {Feasibility Study for BioLEIR}}.
\newblock CERN Yellow Reports: Monographs. CERN, Geneva, 2017.
\newblock 183 pages.

\bibitem{Grote:2003}
H.~Grote and F.~Schmidt.
\newblock {MAD-X : An Upgrade from MAD8}.
\newblock Technical Report CERN-AB-2003-024-ABP, CERN, Geneva, 2003.

\bibitem{Haberer1993}
T.~Haberer, W.~Becher, D.~Schardt, and G.~Kraft.
\newblock Magnetic scanning system for heavy ion therapy.
\newblock {\em Nucl. Instrum. Methods Phys. Res. A}, 330:296--305, 1993.

\bibitem{hans2026}
O.~Hans, R.~Alemany-Fernández, T.~Argyropoulos, P.~Arrutia~Sota, H.~Bartosik,
  G.~Bellodi, M.~Bozzolan, S.~Cettour~Cave, E.~C. Cortés~García, D.~Cotte,
  H.~Damerau, M.~Delrieux, A.~Huschauer, M.~Jebramcik, D.~Küchler, A.~Lasheen,
  K.~Li, E.~Mahner, B.~Mikulec, G.~Papotti, B.~Rodríguez~Mateos,
  E.~Sarafoglou, E.~Marshall, R.~Scrivens, M.~Slupecki, A.~Spierer, R.~Wegner,
  C.~Wetton, M.~Zafeiropoulou, and P.~Zisopoulos.
\newblock Performance and operational overview of the 2025 lead ion run in the
  {{LHC}} injectors.
\newblock In {\em Proc.\ 17th International Particle Accelerator Conference
  (IPAC'26)}, 2026.

\bibitem{Herty2025EIS}
A.~Herty.
\newblock Important safety element documentation: unchanged concept, improved
  data handling.
\newblock Internal presentation, CERN, 22 September, 2025.

\bibitem{ExpwGTSLHC}
C.~E. Hill, D.~K{\"u}chler, C.~Mastrostefano, M.~O'Neil, R.~Scrivens, et~al.
\newblock Experience with the {GTS-LHC} ion source, 2006.
\newblock Proc. 3rd LHC Project Workshop, Chamonix XV, pp. 239--241.

\bibitem{Iadarola:2023fuk}
G.~Iadarola et~al.
\newblock {Xsuite: An Integrated Beam Physics Simulation Framework}.
\newblock {\em JACoW}, HB2023:TUA2I1, 2024.

\bibitem{SLAWGOptics2026}
M.~A. Jebramcik.
\newblock {LEIR} optics measurements for {HEARTS}.
\newblock Slow Extraction Working Group meeting ``HEARTS \#1: slow extraction
  test at {LEIR}'', CERN, 19 November 2026,
  \url{https://indico.cern.ch/event/1744007/}, 2026.

\bibitem{LEIRNLO2026}
M.~A. Jebramcik.
\newblock The next steps: {FCC-ee}, {LEP3}, {LEIR}.
\newblock BE-ABP NLO section meeting, 1 July 2026, 2026.

\bibitem{LEIRDipoleModel2026}
M.~A. Jebramcik and H.~Deveci.
\newblock {ORM} measurements and update on the {LEIR} model.
\newblock LEIR Machine Performance Committee, 4 March 2026,
  \url{https://indico.cern.ch/event/1654080/}, 2026.

\bibitem{KALVAS2017205}
T.~Kalvas, A.~Javanainen, H.~Kettunen, H.~Koivisto, O.~Tarvainen, and
  A.~Virtanen.
\newblock Application and development of ion-source technology for
  radiation-effects testing of electronics.
\newblock {\em Nuclear Instruments and Methods in Physics Research Section B:
  Beam Interactions with Materials and Atoms}, 406:205--209, 2017.
\newblock Proceedings of the 12th European Conference on Accelerators in
  Applied Research and Technology (ECAART12).

\bibitem{Kuchler:2916870}
D.~Kuchler, B.~Bhaskar, G.~Bellodi, M.~Slupecki, and R.~Scrivens.
\newblock Light ions from the {GTS-LHC} ion source for future physics at
  {CERN}.
\newblock {\em JACoW}, ECRIS2024:MOP10, 2024.

\bibitem{Kuechler_2026}
D.~Küchler, R.~A. Fernández, G.~Bellodi, B.~Bhaskar, E.~Mahner, R.~Scrivens,
  and M.~Slupecki.
\newblock Overview of the heavy ion physics programme at {CERN} -- current and
  future ion source requirements and challenges (invited).
\newblock {\em Journal of Physics: Conference Series}, 3237(1):012042, May
  2026.

\bibitem{CERNLayoutDB}
{CERN} layout database, {Linac~3} machines {ITL}, {IAQ}, {ITM}, {IA}, {ITF} and
  {ITH}, version {YETS} 2025-2026.
\newblock \url{https://layout.cern.ch}, 2026.

\bibitem{EPC2025Flattop}
G.~Le~Godec and C.~Mutin.
\newblock {LEIR\_HEART}: beamless flattop extension test.
\newblock Technical note, SY-EPC, EDMS 3352430 v.0.1 (draft), 2025.
\newblock \url{https://edms.cern.ch/document/3352430/1}.

\bibitem{accmodelsLEIR}
{LEIR} optics repository (acc-models-leir).
\newblock CERN GitLab, 2026.
\newblock \url{https://acc-models.web.cern.ch/acc-models/leir/}, repository
  \url{https://gitlab.cern.ch/acc-models/acc-models-leir}.

\bibitem{MontoyaSoto2024}
G.~R. Montoya~Soto.
\newblock {\em $^{208}${Pb}$^{29+}$ beam physics studies at the {CERN} {Linac3}
  in the 2021 run}.
\newblock PhD thesis, Universidad de Guanajuato, 2024.
\newblock CERN-THESIS-2023-344.

\bibitem{SLAWGTimepix2026}
D.~Prelipcean.
\newblock {Timepix3} at {HEARTS@LEIR}.
\newblock Slow Extraction Working Group meeting ``HEARTS \#1: slow extraction
  test at {LEIR}'', CERN, 19 November 2026,
  \url{https://indico.cern.ch/event/1744007/}, 2026.

\bibitem{Timepix2}
D.~Prelipcean, G.~Lerner, and R.~Garcia~Alia.
\newblock {Characterization of the Radiation Field in LHC IR4 Using a Timepix
  Radiation Monitor Measurements and FLUKA Simulations}.
\newblock {\em IEEE Trans. Nucl. Sci.}, 73(8):3004--3015, 2026.

\bibitem{Timepix1}
D.~Prelipcean, G.~Lerner, I.~Slipukhin, D.~Lucsanyi, H.~Sandberg, J.~Storey,
  P.~Martin-Holgado, A.~Romero-Maestre, Y.~Morilla~García, and
  R.~García~Alía.
\newblock Towards a {Timepix3} radiation monitor for the accelerator mixed
  radiation field: characterisation with protons and alphas from 0.6 {MeV} to
  5.6 {MeV}.
\newblock {\em Applied Sciences}, 14(2), 2024.

\bibitem{SpiralScans}
X.~Sang, A.~R. Lupini, R.~R. Unocic, M.~Chi, A.~Y. Borisevich, S.~V. Kalinin,
  E.~Endeve, R.~K. Archibald, and S.~Jesse.
\newblock Dynamic scan control in {STEM}: spiral scans.
\newblock {\em Advanced Structural and Chemical Imaging}, 2(6), 2016.
\newblock \url{https://doi.org/10.1186/s40679-016-0020-3}.

\bibitem{Slupecki_2025}
M.~Slupecki, R.~Alemany-Fernandez, T.~Argyropoulos, H.~Bartosik, G.~Bellodi,
  R.~Bruce, F.~Carlier, H.~Damerau, D.~Gamba, O.~Hans, A.~Huschauer, P.~Kruyt,
  D.~Kuchler, A.~Lasheen, E.~Mahner, R.~Scrivens, R.~Wegner, M.~Widorski,
  B.~Woolley, and E.~Waagaard.
\newblock Development of new ion beams at the {CERN} ion injector complex for
  future physics programmes.
\newblock {\em Journal of Physics: Conference Series}, 3094(1):012006, Sept.
  2025.

\bibitem{ESA-TRLHandbook}
{TEC-SHS}.
\newblock Technology readiness levels handbook for space applications, Sept.
  2008.
\newblock Available at:
  \url{https://connectivity.esa.int/sites/default/files/TRL_Handbook.pdf}.

\bibitem{accmodelsTLS}
{CERN} transfer-line optics repository (acc-models-tls), {LEIR} injection line
  {ITH-ITE-ETL-EI}.
\newblock CERN GitLab, 2026.
\newblock \url{https://gitlab.cern.ch/acc-models/acc-models-tls}.

\bibitem{TOMIZAWA1993}
M.~Tomizawa, M.~Yoshizawa, K.~Chida, J.~Yoshizawa, Y.~Arakaki, R.~Nagai,
  A.~Mizobuchi, A.~Noda, K.~Noda, M.~Kanazawa, A.~Ando, H.~Muto, and
  T.~Hattori.
\newblock {Slow beam extraction at TARN II}.
\newblock {\em Nuclear Instruments and Methods in Physics Research Section A:
  Accelerators, Spectrometers, Detectors and Associated Equipment},
  326(3):399--406, 1993.

\bibitem{NSRL-Octupoles}
N.~Tsoupas, L.~Ahrens, S.~Bellavia, R.~Bonati, K.~A. Brown, I.-H. Chiang, C.~J.
  Gardner, D.~Gassner, S.~Jao, W.~W. Mackay, I.~Marneris, W.~Meng, D.~Phillips,
  P.~Pile, R.~Prigl, A.~Rusek, L.~Snydstrup, and K.~Zeno.
\newblock Uniform beam distributions at the target of the {NASA Space Radiation
  Laboratory}'s beam line.
\newblock {\em Phys. Rev. ST Accel. Beams}, 10:024701, Feb 2007.

\bibitem{GSI-HEARTS-KT2023}
U.~Weber and C.~Schuy.
\newblock Definition of terms and discussion for goals and precision.
\newblock HEARTS WP4 Knowledge Transfer Meeting, GSI, Darmstadt, 20 April 2023,
  2023.

\bibitem{YangIPAC2022}
J.~Yang, P.~Forck, T.~Giacomini, P.~Niedermayer, R.~Singh, and S.~Sorge.
\newblock {I}mprovement of spill quality for slowly extracted ions at
  {GSI}-{SIS18} via transverse emittance exchange.
\newblock In {\em Proc. 13th International Particle Accelerator Conference
  (IPAC'22), Bangkok, Thailand}, pages 2093--2095. JACoW Publishing, 2022.

\bibitem{PhysRevSTAB2007-Japan}
Y.~Yuri, N.~Miyawaki, T.~Kamiya, W.~Yokota, K.~Arakawa, and M.~Fukuda.
\newblock Uniformization of the transverse beam profile by means of nonlinear
  focusing method.
\newblock {\em Phys. Rev. ST Accel. Beams}, 10:104001, Oct 2007.

\end{thebibliography}

\end{document}